%% file: oi_survey.tex
\documentclass[twocolumn,tighten]{aastex62}

\newcommand{\lya}{Ly$\alpha$}

\newcommand{\hi}{\ion{H}{1}}

\newcommand{\oi}{\ion{O}{1}}
\newcommand{\cii}{\ion{C}{2}}
\newcommand{\ciii}{\ion{C}{3}}
\newcommand{\civ}{\ion{C}{4}}
\newcommand{\siii}{\ion{Si}{2}}
\newcommand{\siiii}{\ion{Si}{3}}
\newcommand{\siiv}{\ion{Si}{4}}

\newcommand{\mgii}{\ion{Mg}{2}}

\newcommand{\nv}{\ion{N}{5}}

\newcommand{\kms}{km~s$^{-1}$}

\newcommand{\lam}{$\lambda$}

\shorttitle{\oi\ survey}
\shortauthors{Becker et al.}

\begin{document}

\title{The Evolution of \oi\ over $3.2 < z < 6.5$: Reionization of the Circumgalactic Medium}

\correspondingauthor{George Becker}
\email{george.becker@ucr.edu}

\author[0000-0003-2344-263X]{George D. Becker}
\affiliation{Department of Physics \& Astronomy, University of California, Riverside, CA, 92521, USA}
\author[0000-0002-5139-4359]{Max Pettini}
\affiliation{Institute of Astronomy, University of Cambridge, Madingley Road, Cambridge CB3 0HA, UK}
\author[0000-0002-9946-4731]{Marc Rafelski}
\affiliation{Space Telescope Science Institute, Baltimore, MD 21218, USA}
\affiliation{Department of Physics \& Astronomy, Johns Hopkins University, Baltimore, MD 21218, USA}
\author[0000-0003-3693-3091]{Valentina D'Odorico}
\affiliation{INAF-Osservatorio Astronomico di Trieste, Via Tiepolo 11, I-34143 Trieste, Italy}
\affiliation{Scuola Normale Superiore Piazza dei Cavalieri, 7 I-56126 Pisa, Italy}
\author[0000-0002-8340-6537]{Elisa Boera}
\affiliation{Department of Physics \& Astronomy, University of California, Riverside, CA, 92521, USA}
\author[0000-0001-8415-7547]{Lise Christensen}
\affiliation{DARK, Niels Bohr Institute, University of Copenhagen, Lyngbyvej 2, 2100 Copenhagen, Denmark}
\author[0000-0002-6830-9093]{Guido Cupani}
\affiliation{INAF-Osservatorio Astronomico di Trieste, Via Tiepolo 11, I-34143 Trieste, Italy}
\author[0000-0002-1768-1899]{Sara L. Ellison}
\affiliation{Department of Physics and Astronomy, University of Victoria, Victoria, British Columbia, V8P 1A1, Canada}
\author[0000-0002-6822-2254]{Emanuele P. Farina}
\affiliation{Max-Planck-Institut f\"ur Astronomie, K\"onigstuhl 17, D-69117 Heidelberg, Germany}
\author[0000-0001-6676-3842]{Michele Fumagalli}
\affiliation{Centre for Extra-galactic Astronomy (CEA), Durham University, South Road, Durham DH1 3LE, UK}
\affiliation{Institute for Computational Cosmology (ICC), Durham University, South Road, Durham DH1 3LE, UK}
\affiliation{Dipartimento di Fisica G. Occhialini, Universit\`a degli Studi di Milano Bicocca, Piazza della Scienza 3, 20126 Milano, Italy}
\author[0000-0003-0389-0902]{Sebastian L\'opez}
\affiliation{Departamento de Astronom\'ia, Universidad de Chile, Casilla 36-D, Santiago, Chile}
\author[0000-0002-9838-8191]{Marcel Neeleman}
\affiliation{Max-Planck-Institut f\"ur Astronomie, K\"onigstuhl 17, D-69117 Heidelberg, Germany}
\author[0000-0002-5360-8103]{Emma V. Ryan-Weber}
\affiliation{Centre for Astrophysics and Supercomputing, Swinburne University of Technology, PO Box 218, Hawthorn, VIC 3122, Australia}
\affiliation{ARC Centre of Excellence for All Sky Astrophysics in 3 Dimensions (ASTRO 3D), Australia}
\author[0000-0003-0960-3580]{G\'abor Worseck}
\affiliation{Max-Planck-Institut f\"ur Astronomie, K\"onigstuhl 17, D-69117 Heidelberg, Germany}
\affiliation{Institut f\"ur Physik und Astronomie, Universit\"at Potsdam, Karl-Liebknecht-Str. 24/25, D-14476 Potsdam, Germany}

\begin{abstract}

We present a survey for metal absorption systems traced by neutral oxygen over $3.2 < z < 6.5$.  Our survey uses Keck/ESI and VLT/X-Shooter spectra of 199  QSOs with redshifts  up to 6.6.  In total we detect 74 \oi\ absorbers, of which 57 are separated from the background QSO by more than 5000~\kms.  We use a maximum likelihood approach to fit the distribution of \oi~\lam1302 equivalent widths in bins of redshift, and from this determine the evolution in number density of absorbers with $W_{1302} > 0.05$~\AA.  We find that the number density does not monotonically increase with decreasing redshift, as would naively be expected from the buildup of metal-enriched circumgalactic gas with time.  The number density over $4.9 < z < 5.7$ is a factor of 1.7--4.1 lower (68\% confidence) than over $5.7 < z < 6.5$, with a lower value at $z < 5.7$ favored with 99\% confidence.  This decrease suggests that the fraction of metals in a low-ionization phase is larger at $z \sim 6$ than at lower redshifts.  Absorption from highly ionized metals traced by \civ\ is also weaker in higher-redshift \oi\ systems, supporting this picture.  The evolution of \oi\ absorbers implies that metal-enriched circumgalactic gas at $z \sim 6$ is undergoing an ionization transition driven by a strengthening ultraviolet background.  This in turn suggests that the reionization of the diffuse intergalactic medium may still be ongoing at or only recently ended by this epoch.
\end{abstract}


\section{Introduction} \label{sec:intro}

Metal absorption lines in the spectra of background QSOs are a versatile probe of the gas around galaxies.  Their kinematics trace the gas inflows and outflows that help regulate star formation.  Their chemical abundances reflect the stellar populations from which the metals were produced.  They offer a means to study faint galaxies that can be well below the detection thresholds of galaxy emission surveys.  Moreover, the wide range of ionization potentials of the absorbing species means that metals can be used to constrain the ionization state of the absorbing gas, and hence, for photoionized gas, the nature of the ionizing radiation field \citep[for a review see][]{tumlinson2017}.

The sensitivity of metal lines to the ionization of circumgalactic gas is particularly useful near the reionization epoch.  As the surrounding diffuse intergalactic medium (IGM) is ionized, the gas around galaxies becomes exposed to ionizing ultraviolet background (UVB) radiation from distant sources.  If the photo-ionization of the circumgalactic medium (CGM) is driven mainly by the UVB, rather than by photons produced locally by the host galaxy, then  we should see an increase in the ionization of the CGM during and/or shortly after reionization, as the intensity of the UVB increases.  If the CGM gas is metal-enriched, then the species producing metal absorption lines will transition from being predominantly neutral or singly ionized to being more highly ionized.  Neutral or low-ionization metal absorbers can therefore potentially be used to trace regions of the IGM that have not yet reionized or where the UVB is still weak \citep[e.g.,][]{oh2002,furlanetto2003,oppenheimer2009,finlator2013,finlator2015,finlator2018,keating2014}.

Multiple surveys have now traced metal absorbers with a range of ionization potentials out to $z \sim 6$--7.  The comoving number and mass density of highly ionized metals traced by \civ\ increases significantly from $z \sim 6$ to 3 \citep{becker2009,ryan-weber2009,simcoe2011a,dodorico2013,codoreanu2018,meyer2019}.  There is some direct evidence that the ionization balance of these absorbers is changing; for example, \civ\ absorbers tend to show more \siiv\ at higher redshifts \citep{dodorico2013}, which potentially constraints the shape of the high-redshift UVB \citep[e.g.,][]{finlator2016,doughty2018}.  The general trend in \civ, however, reflects an overall increase in CGM metallicity towards lower redshifts from enriched galaxy outflows \citep{oppenheimer2006,oppenheimer2009,finlator2015,garcia2017}.  The number density of strong \mgii\ systems (with \mgii~\lam2796 rest equivalent width $W_{2796} > 1.0$~\AA) also increases with decreasing redshift over $2 < z < 7$ \citep{matejek2012,chen2017}, roughly tracing the global star formation rate density \citep[see also][]{menard2011}.  On the other hand, the total number density of weaker \mgii\ systems remains relatively constant, albeit with significant uncertainties at $z \gtrsim 6$ \citep{chen2017,bosman2017,codoreanu2017}.  \mgii\ absorption can arise from either neutral or ionized gas; nevertheless, the weaker \mgii\ systems must either already be largely in place by $z \sim 7$, or their evolution must include some change in the ionization of the absorbers.  Recently, \citet{cooper2019} showed that metal absorbers at $z > 5.7$ generally tend to exhibit less absorption from high-ionization species than those at lower redshifts, with a larger fraction of systems showing absorption from low-ionization species alone \citep[see also][]{codoreanu2018}.  This suggests that the ionization states of metal absorbers may indeed be evolving, although changes in total enrichment could also be playing a role.

One of the most direct probes of low-ionization, metal-enriched gas is absorption from neutral oxygen.  Oxygen has a first ionization potential nearly identical to that of hydrogen, and is locked in charge exchange equilibrium with hydrogen for temperatures above $\sim$10$^3$~K \citep[e.g.,][]{chambaud1980,stancil1999}.  The presence of \oi\ absorption therefore typically indicates significantly neutral gas.  Previous studies have hinted that the number density of \oi\ systems may be larger at $z \sim 6$ than at lower redshifts \citep{becker2006,becker2011a}, but the surveys have been too small to be conclusive.  In addition, self-consistent searches for \oi\ have not been conducted over a wide enough redshift range to determine how the number density at $z \sim 6$ compares to the number density at lower redshifts.  At $z < 5$, absorption systems with \oi\ are typically found via their strong \hi\ absorption; most \oi\ absorbers are either damped ($N_{\rm H\,I} > 10^{20.3}~{\rm cm^{-2}}$) or sub-damped ($10^{19}~{\rm cm^{-2}} < N_{\rm H\,I} < 10^{20.3}~{\rm cm^{-2}}$) \lya\ systems \citep[DLAs or sub-DLAs; e.g.,][]{dessauges-zavadsky2003,wolfe2005,rafelski2012,rafelski2014}.  At $z > 5$, however, general \lya\ forest absorption increases to the point where it becomes difficult to identify individual strong \hi\ absorbers, and so metal absorbers must be identified using the metal lines alone.  Variations in the metal enrichment and ionization of \hi-selected systems will complicate a comparison between the number density of DLAs and sub-DLAs at $z < 5$ and \oi\ systems at higher redshifts.  A more robust approach is to search for \oi\ systems independent of \hi\ at all redshifts.

In this work we present the first self-consistent survey for \oi\ absorbers with a large enough sample (199 QSO lines of sight) and long enough redshift baseline ($3.2 < z < 6.5$) to robustly determine how the number density at $z \sim 6$ compares to that at lower redshifts.  We identify \oi\ absorbers solely via their metal lines, and can therefore apply the same selection technique at all redshifts.  Our focus here is mainly on the number density evolution of \oi.  We also examine how highly ionized metals associated with these systems evolve with redshift, but leave a detailed analysis of their kinematics and chemical abundances for future work.  

We note that many of the absorbers presented here have been identified in previous surveys that either selected on different ions or were significantly smaller.  Nearly all of the systems at $z \le 4.5 $ were identified in the XQ-100 surveys for DLAs and sub-DLAs, which use \hi\ selection \citep[][and in prep]{sanchez-ramirez2016,berg2016}.  The systems over $4.7 \le z \le 5.3$ will largely be included in an upcoming survey for DLAs near $z \sim 5$ (Rafelski et al., in prep).  Several of the absorbers at $z \ge 5.6$ have also been published previously in smaller \oi\ surveys \citep{becker2006,becker2011a} or studies of other metal lines \citep{ryan-weber2009,simcoe2011,dodorico2013,dodorico2018,chen2017,meyer2019,cooper2019}.  We refer the reader to these papers for further details on individual systems.

The rest of the paper is organized as follows.  In Section~\ref{sec:data} we describe the spectra used for the survey.  The selection of \oi\ systems is described in Section~\ref{sec:survey}, where we also derive their number density evolution based on the \oi~\lam1302 rest-frame equivalent width distributions.  High-ionization metal lines associated with the \oi\ absorbers are examined in Section~\ref{sec:highions}.  We briefly look at clustering of \oi\ absorbers in Section~\ref{sec:clustering}.  We then discuss the implications of the redshift evolution in the \oi\ number density for the reionization\footnote{We recognize that ``re''-ionization may be somewhat of a misnomer when applied to the CGM.  Nevertheless, we will refer to a global transition of circumgalactic gas from a neutral to an ionized state as a reionization event in order to highlight the potential connection to the reionization of the IGM.}  of circumgalactic gas in Section~\ref{sec:discussion} before summarizing our results in Section~\ref{sec:summary}.  Throughout this paper we assume a $\Lambda$CDM cosmology with $\Omega_{m} = 0.3$, $\Omega_{\Lambda} = 0.7$, and $h = 0.7$.  All wavelengths are given in angstroms.  All equivalent widths are rest-frame and are given in angstroms except where noted.

\section{The Data}\label{sec:data}

\subsection{Sample}\label{sec:sample}

Our survey includes 199 QSOs with redshifts $3.52 \le z_{\rm em} \le 6.65$ observed with either the X-Shooter spectrograph on the Very Large Telescope (VLT) \citep{vernet2011} or the Echellette Spectrograph and Imager (ESI) on Keck \citep{sheinis2002}.  The full sample is made up of three sub-samples.  The low-redshift end ($3.52 < z_{\rm em} < 4.81$) is made up of 100 QSOs from the XQ-100 survey \citep{lopez2016}.  To this we add 42 QSOs over $5.0 < z_{\rm em} < 5.7$ from X-Shooter program 098.A-0111 (PI: M. Rafelski), a survey originally intended to perform a blind search for DLAs at $z \gtrsim 5$.  Finally, we include 67 QSOs over $5.62 < z_{\rm em} < 6.65$ drawn from our own programs and the VLT and Keck archives.  All of our spectra have a signal-to-ratio of at least 10 (and often much higher) per 30 \kms\ interval at a rest-frame wavelength of 1285 \AA.  The data are summarized in Table~\ref{tab:qsos}.

In all cases the targets were selected independently of intervening absorbers.  Objects in the XQ-100 and 098.A-0111 datasets were selected based on their redshift and continuum luminosity, independent of any known absorption systems.  In some cases we substituted ESI spectra for the 098.A-0111 objects if the signal-to-noise ratio for ESI was higher.  The $z_{\rm em} > 5.6$ sample is more heterogeneous; however, objects at this redshift are generally targeted with these instruments based on their redshift and luminosity in an effort to build up samples that can be used for unbiased studies of the \lya\ forest and/or metal lines, or to study the QSOs themselves.  Prior information about any metal absorbers is generally very limited and, to our knowledge, none of the objects in this group were targeted because of known metal lines.  

We note that in some cases where \oi\ systems were detected in ESI spectra, we used  additional data to cover metal lines that fell in the near infrared.  This was true for J1332+2208, SDSS J2054-0005, and SDSS J2315-0023, for which we used X-Shooter NIR spectra, and SDSS J1148+5251, for which we used a Keck NIRSPEC echelle spectrum from \citet{becker2009}.

\startlongtable
\begin{deluxetable}{clcccc}
\tabletypesize{\footnotesize}
\tablecaption{QSO spectra used in this work\label{tab:qsos}}
\tablehead{
   \colhead{\#}  &  \colhead{QSO}  &  \colhead{$z_{\rm forest}$}  &  \colhead{Instrument}  &  \colhead{$S/N$}  &  \colhead{$W_{1302}^{50\%}$} }
\colnumbers
\startdata
  1  &  J1442+0920        &  3.524  &  X-Shooter  &   47.0  &  0.032  \\
  2  &  J1024+1819        &  3.525  &  X-Shooter  &   32.1  &  0.042  \\
  3  &  J1332+0052        &  3.525  &  X-Shooter  &   59.3  &  0.035  \\
  4  &  J1445+0958        &  3.527  &  X-Shooter  &   47.5  &  0.029  \\
  5  &  J0100-2708        &  3.528  &  X-Shooter  &   32.0  &  0.042  \\
  6  &  J1018+0548        &  3.530  &  X-Shooter  &   39.5  &  0.061  \\
  7  &  J1201+1206        &  3.530  &  X-Shooter  &   76.0  &  0.024  \\
  8  &  J1517+0511        &  3.570  &  X-Shooter  &   31.4  &  0.041  \\
  9  &  J1202-0054        &  3.599  &  X-Shooter  &   27.6  &  0.053  \\
 10  &  J1524+2123        &  3.599  &  X-Shooter  &   43.8  &  0.027  \\
 11  &  J1416+1811        &  3.602  &  X-Shooter  &   19.6  &  0.067  \\
 12  &  J1103+1004        &  3.607  &  X-Shooter  &   41.7  &  0.041  \\
 13  &  J1117+1311        &  3.618  &  X-Shooter  &   45.0  &  0.027  \\
 14  &  J0920+0725        &  3.623  &  X-Shooter  &   59.9  &  0.027  \\
 15  &  J0056-2808        &  3.624  &  X-Shooter  &   39.5  &  0.033  \\
 16  &  J1126-0124        &  3.628  &  X-Shooter  &   21.2  &  0.059  \\
 17  &  J1037+0704        &  3.628  &  X-Shooter  &   61.4  &  0.020  \\
 18  &  J1042+1957        &  3.630  &  X-Shooter  &   41.5  &  0.037  \\
 19  &  J1304+0239        &  3.655  &  X-Shooter  &   49.2  &  0.030  \\
 20  &  J0057-2643        &  3.655  &  X-Shooter  &   58.8  &  0.023  \\
 21  &  J1053+0103        &  3.658  &  X-Shooter  &   41.9  &  0.028  \\
 22  &  J1020+0922        &  3.660  &  X-Shooter  &   28.1  &  0.048  \\
 23  &  J1249-0159        &  3.666  &  X-Shooter  &   44.7  &  0.028  \\
 24  &  J0755+1345        &  3.669  &  X-Shooter  &   40.9  &  0.029  \\
 25  &  J1108+1209        &  3.670  &  X-Shooter  &   48.9  &  0.029  \\
 26  &  J1503+0419        &  3.670  &  X-Shooter  &   53.4  &  0.028  \\
 27  &  J0818+0958        &  3.692  &  X-Shooter  &   46.8  &  0.032  \\
 28  &  J1421-0643        &  3.695  &  X-Shooter  &   48.6  &  0.027  \\
 29  &  J1352+1303        &  3.698  &  X-Shooter  &   15.2  &  0.094  \\
 30  &  J0937+0828        &  3.699  &  X-Shooter  &   32.8  &  0.047  \\
 31  &  J1621-0042        &  3.700  &  X-Shooter  &   61.8  &  0.034  \\
 32  &  J1248+1304        &  3.714  &  X-Shooter  &   57.9  &  0.024  \\
 33  &  J1320-0523        &  3.715  &  X-Shooter  &   62.9  &  0.025  \\
 34  &  J0833+0959        &  3.718  &  X-Shooter  &   42.7  &  0.036  \\
 35  &  J1552+1005        &  3.735  &  X-Shooter  &   54.4  &  0.028  \\
 36  &  J1126-0126        &  3.744  &  X-Shooter  &   31.1  &  0.037  \\
 37  &  J1312+0841        &  3.746  &  X-Shooter  &   43.8  &  0.034  \\
 38  &  J1658-0739        &  3.759  &  X-Shooter  &   31.6  &  0.038  \\
 39  &  J0935+0022        &  3.760  &  X-Shooter  &   32.9  &  0.042  \\
 40  &  J1013+0650        &  3.796  &  X-Shooter  &   45.0  &  0.034  \\
 41  &  J1336+0243        &  3.801  &  X-Shooter  &   42.3  &  0.029  \\
 42  &  J0124+0044        &  3.817  &  X-Shooter  &   52.2  &  0.033  \\
 43  &  J1331+1015        &  3.852  &  X-Shooter  &   44.8  &  0.038  \\
 44  &  J1135+0842        &  3.856  &  X-Shooter  &   65.1  &  0.019  \\
 45  &  J0042-1020        &  3.859  &  X-Shooter  &   72.1  &  0.022  \\
 46  &  J1111-0804        &  3.927  &  X-Shooter  &   54.4  &  0.022  \\
 47  &  J1330-2522        &  3.953  &  X-Shooter  &   60.9  &  0.025  \\
 48  &  J0211+1107        &  3.968  &  X-Shooter  &   36.1  &  0.036  \\
 49  &  J0800+1920        &  3.970  &  X-Shooter  &   57.9  &  0.026  \\
 50  &  J1542+0955        &  3.970  &  X-Shooter  &   45.7  &  0.026  \\
 51  &  J0137-4224        &  3.972  &  X-Shooter  &   33.3  &  0.040  \\
 52  &  J0214-0517        &  3.977  &  X-Shooter  &   51.4  &  0.029  \\
 53  &  J1054+0215        &  3.982  &  X-Shooter  &   25.9  &  0.051  \\
 54  &  J0255+0048        &  3.992  &  X-Shooter  &   41.9  &  0.038  \\
 55  &  J2215-1611        &  3.995  &  X-Shooter  &   58.8  &  0.040  \\
 56  &  J0835+0650        &  3.997  &  X-Shooter  &   47.2  &  0.031  \\
 57  &  J1032+0927        &  4.008  &  X-Shooter  &   39.8  &  0.034  \\
 58  &  J0244-0134        &  4.047  &  X-Shooter  &   55.3  &  0.027  \\
 59  &  J0311-1722        &  4.049  &  X-Shooter  &   57.1  &  0.025  \\
 60  &  J1323+1405        &  4.058  &  X-Shooter  &   36.2  &  0.035  \\
 61  &  J0415-4357        &  4.066  &  X-Shooter  &   27.9  &  0.062  \\
 62  &  J0959+1312        &  4.071  &  X-Shooter  &   86.9  &  0.023  \\
 63  &  J0048-2442        &  4.106  &  X-Shooter  &   29.9  &  0.046  \\
 64  &  J1037+2135        &  4.119  &  X-Shooter  &   67.5  &  0.023  \\
 65  &  J0121+0347        &  4.131  &  X-Shooter  &   44.6  &  0.026  \\
 66  &  J1057+1910        &  4.133  &  X-Shooter  &   32.7  &  0.041  \\
 67  &  J0003-2603        &  4.136  &  X-Shooter  &  110.7  &  0.020  \\
 68  &  J1110+0244        &  4.144  &  X-Shooter  &   52.5  &  0.026  \\
 69  &  J0747+2739        &  4.151  &  X-Shooter  &   30.9  &  0.060  \\
 70  &  J0132+1341        &  4.152  &  X-Shooter  &   46.1  &  0.028  \\
 71  &  J2251-1227        &  4.157  &  X-Shooter  &   44.9  &  0.033  \\
 72  &  J0133+0400        &  4.170  &  X-Shooter  &   78.1  &  0.031  \\
 73  &  J0529-3552        &  4.181  &  X-Shooter  &   26.9  &  0.070  \\
 74  &  J0030-5129        &  4.183  &  X-Shooter  &   34.6  &  0.039  \\
 75  &  J0153-0011        &  4.195  &  X-Shooter  &   27.7  &  0.051  \\
 76  &  J2349-3712        &  4.221  &  X-Shooter  &   42.0  &  0.033  \\
 77  &  J0839+0318        &  4.226  &  X-Shooter  &   31.6  &  0.041  \\
 78  &  J0403-1703        &  4.233  &  X-Shooter  &   44.9  &  0.030  \\
 79  &  J0117+1552        &  4.242  &  X-Shooter  &   72.5  &  0.023  \\
 80  &  J0247-0556        &  4.255  &  X-Shooter  &   43.8  &  0.033  \\
 81  &  J1034+1102        &  4.288  &  X-Shooter  &   55.0  &  0.027  \\
 82  &  J0234-1806        &  4.305  &  X-Shooter  &   44.0  &  0.033  \\
 83  &  J0034+1639        &  4.324  &  X-Shooter  &   49.5  &  0.032  \\
 84  &  J0426-2202        &  4.325  &  X-Shooter  &   43.5  &  0.030  \\
 85  &  J0113-2803        &  4.339  &  X-Shooter  &   46.4  &  0.040  \\
 86  &  J1058+1245        &  4.349  &  X-Shooter  &   44.0  &  0.031  \\
 87  &  J2344+0342        &  4.351  &  X-Shooter  &   58.9  &  0.028  \\
 88  &  J1633+1411        &  4.372  &  X-Shooter  &   54.0  &  0.027  \\
 89  &  J0529-3526        &  4.416  &  X-Shooter  &   45.0  &  0.030  \\
 90  &  J1401+0244        &  4.432  &  X-Shooter  &   28.7  &  0.045  \\
 91  &  J0248+1802        &  4.433  &  X-Shooter  &   50.6  &  0.026  \\
 92  &  J0955-0130        &  4.437  &  X-Shooter  &   51.1  &  0.043  \\
 93  &  J0525-3343        &  4.437  &  X-Shooter  &   65.4  &  0.026  \\
 94  &  J0714-6455        &  4.484  &  X-Shooter  &   62.0  &  0.030  \\
 95  &  J2216-6714        &  4.496  &  X-Shooter  &   55.9  &  0.030  \\
 96  &  J1036-0343        &  4.507  &  X-Shooter  &   62.7  &  0.021  \\
 97  &  J0006-6208        &  4.522  &  X-Shooter  &   35.3  &  0.040  \\
 98  &  J1723+2243        &  4.549  &  X-Shooter  &   77.8  &  0.024  \\
 99  &  J2239-0552        &  4.566  &  X-Shooter  &  103.0  &  0.020  \\
100  &  J0307-4945        &  4.813  &  X-Shooter  &   99.4  &  0.026  \\
101  &  J0251+0333        &  4.987  &  X-Shooter  &   29.3  &  0.061  \\
102  &  J2344+1653        &  4.988  &  X-Shooter  &   10.4  &  0.170  \\
103  &  J1200+1817        &  5.004  &  X-Shooter  &   28.3  &  0.051  \\
104  &  SDSS J0221-0342   &  5.019  &  X-Shooter  &   18.9  &  0.082  \\
105  &  J0846+0800        &  5.022  &  X-Shooter  &   15.0  &  0.113  \\
106  &  SDSS J0017-1000   &  5.024  &  ESI        &   57.9  &  0.051  \\
107  &  SDSS J0338+0021   &  5.028  &  ESI        &   81.4  &  0.044  \\
108  &  J0025-0145        &  5.048  &  X-Shooter  &   28.2  &  0.050  \\
109  &  J1423+1303        &  5.051  &  X-Shooter  &   29.5  &  0.059  \\
110  &  J2202+1509        &  5.060  &  X-Shooter  &   27.7  &  0.087  \\
111  &  J1601-1828        &  5.064  &  X-Shooter  &   20.3  &  0.076  \\
112  &  J0835+0537        &  5.066  &  X-Shooter  &   18.6  &  0.108  \\
113  &  J1004+2025        &  5.075  &  X-Shooter  &   17.0  &  0.106  \\
114  &  J0115-0253        &  5.076  &  X-Shooter  &   16.7  &  0.120  \\
115  &  J2226-0618        &  5.077  &  X-Shooter  &   16.0  &  0.095  \\
116  &  J2201+0302        &  5.099  &  X-Shooter  &   32.5  &  0.045  \\
117  &  J1332+2208        &  5.117  &  ESI        &   36.2  &  0.055  \\
118  &  J0957+1016        &  5.137  &  X-Shooter  &   11.7  &  0.139  \\
119  &  J2228-0757        &  5.148  &  ESI        &   30.2  &  0.065  \\
120  &  J0957+0610        &  5.167  &  ESI        &   40.6  &  0.055  \\
121  &  SDSS J0854+2056   &  5.177  &  X-Shooter  &   17.2  &  0.099  \\
122  &  J0131-0321        &  5.183  &  X-Shooter  &   23.1  &  0.066  \\
123  &  J0241+0435        &  5.186  &  X-Shooter  &   19.1  &  0.087  \\
124  &  J0902+0851        &  5.224  &  X-Shooter  &   14.2  &  0.104  \\
125  &  J2325-0553        &  5.232  &  X-Shooter  &   15.9  &  0.096  \\
126  &  J0216+2304        &  5.238  &  X-Shooter  &   17.0  &  0.099  \\
127  &  J0747+1153        &  5.248  &  X-Shooter  &   48.1  &  0.040  \\
128  &  J2351-0459        &  5.248  &  X-Shooter  &   18.7  &  0.099  \\
129  &  J1436+2132        &  5.249  &  X-Shooter  &   24.8  &  0.063  \\
130  &  J2225+0330        &  5.255  &  X-Shooter  &   25.4  &  0.070  \\
131  &  J1147-0109        &  5.264  &  X-Shooter  &   16.3  &  0.089  \\
132  &  J2358+0634        &  5.299  &  X-Shooter  &   26.3  &  0.055  \\
133  &  J2330+0957        &  5.305  &  X-Shooter  &   12.9  &  0.133  \\
134  &  J0812+0440        &  5.306  &  X-Shooter  &   25.7  &  0.063  \\
135  &  J0116+0538        &  5.356  &  X-Shooter  &   23.2  &  0.055  \\
136  &  J0155+0415        &  5.379  &  X-Shooter  &   26.0  &  0.056  \\
137  &  J0306+1853        &  5.395  &  X-Shooter  &   56.4  &  0.024  \\
138  &  J1022+2252        &  5.471  &  X-Shooter  &   26.5  &  0.061  \\
139  &  J2207-0416        &  5.529  &  X-Shooter  &   41.6  &  0.042  \\
140  &  J0108+0711        &  5.577  &  X-Shooter  &   28.7  &  0.067  \\
141  &  J1335-0328        &  5.693  &  X-Shooter  &   29.9  &  0.047  \\
142  &  SDSS J0927+2001   &  5.768  &  X-Shooter  &   66.4  &  0.026  \\
143  &  PSO J215-16       &  5.782  &  X-Shooter  &   39.2  &  0.040  \\
144  &  SDSS J0836+0054   &  5.801  &  X-Shooter  &   72.3  &  0.028  \\
145  &  SDSS J2147+0107   &  5.812  &  X-Shooter  &   14.0  &  0.119  \\
146  &  SDSS J0002+2550   &  5.820  &  ESI        &   92.6  &  0.053  \\
147  &  SDSS J1044-0125   &  5.829  &  X-Shooter  &   36.8  &  0.040  \\
148  &  SDSS J0005-0006   &  5.847  &  ESI        &   24.1  &  0.073  \\
149  &  SDSS J0840+5624   &  5.849  &  ESI        &   41.0  &  0.060  \\
150  &  ULAS J0203+0012   &  5.856  &  ESI        &   12.1  &  0.173  \\
151  &  SDSS J1335+3533   &  5.902  &  ESI        &   11.2  &  0.154  \\
152  &  SDSS J1411+1217   &  5.916  &  ESI        &   46.0  &  0.067  \\
153  &  SDSS J2053+0047   &  5.926  &  X-Shooter  &   14.9  &  0.111  \\
154  &  SDSS J0841+2905   &  5.950  &  ESI        &   10.7  &  0.180  \\
155  &  PSO J056-16       &  5.960  &  X-Shooter  &   34.8  &  0.042  \\
156  &  PSO J007+04       &  5.981  &  X-Shooter  &   16.5  &  0.116  \\
157  &  SDSS J2310+1855   &  5.992  &  X-Shooter  &   30.2  &  0.047  \\
158  &  PSO J009-10       &  5.995  &  X-Shooter  &   14.3  &  0.111  \\
159  &  SDSS J0818+1722   &  5.997  &  X-Shooter  &   91.6  &  0.024  \\
160  &  ULAS J0148+0600   &  5.998  &  X-Shooter  &  111.4  &  0.021  \\
161  &  PSO J340-18       &  5.999  &  X-Shooter  &   32.1  &  0.069  \\
162  &  ATLAS J029-36     &  6.021  &  X-Shooter  &   11.7  &  0.129  \\
163  &  VIK J0046-2837    &  6.021  &  X-Shooter  &   15.4  &  0.110  \\
164  &  SDSS J1306+0356   &  6.024  &  X-Shooter  &   62.7  &  0.035  \\
165  &  SDSS J1137+3549   &  6.026  &  ESI        &   27.6  &  0.076  \\
166  &  ULAS J1207+0630   &  6.031  &  X-Shooter  &   25.0  &  0.083  \\
167  &  SDSS J2054-0005   &  6.039  &  ESI        &   29.1  &  0.082  \\
168  &  SDSS J1630+4012   &  6.055  &  ESI        &   17.4  &  0.106  \\
169  &  ATLAS J158-14     &  6.055  &  X-Shooter  &   20.8  &  0.072  \\
170  &  SDSS J0842+1218   &  6.069  &  X-Shooter  &   35.3  &  0.035  \\
171  &  SDSS J1602+4228   &  6.080  &  ESI        &   33.9  &  0.065  \\
172  &  SDSS J0303-0019   &  6.081  &  X-Shooter  &   14.4  &  0.102  \\
173  &  CFHQS J2100-1715  &  6.084  &  X-Shooter  &   14.7  &  0.113  \\
174  &  CFHQS J1509-1749  &  6.114  &  X-Shooter  &   51.2  &  0.031  \\
175  &  SDSS J2315-0023   &  6.124  &  ESI        &   25.0  &  0.099  \\
176  &  ULAS J1319+0950   &  6.125  &  X-Shooter  &   71.1  &  0.031  \\
177  &  VIK J2318-3029    &  6.139  &  X-Shooter  &   20.6  &  0.085  \\
178  &  SDSS J0353+0104   &  6.152  &  ESI        &   21.7  &  0.097  \\
179  &  SDSS J1250+3130   &  6.154  &  ESI        &   53.0  &  0.050  \\
180  &  PSO J359-06       &  6.171  &  X-Shooter  &   26.8  &  0.068  \\
181  &  PSO J065-26       &  6.186  &  X-Shooter  &   26.7  &  0.083  \\
182  &  PSO J308-21       &  6.245  &  X-Shooter  &   25.9  &  0.123  \\
183  &  CFHQS J0050+3445  &  6.254  &  ESI        &   27.3  &  0.153  \\
184  &  SDSS J1623+3112   &  6.255  &  ESI        &   15.8  &  0.187  \\
185  &  SDSS J1030+0524   &  6.300  &  X-Shooter  &   59.9  &  0.048  \\
186  &  ATLAS J025-33     &  6.318  &  X-Shooter  &   86.7  &  0.027  \\
187  &  SDSS J0100+2802   &  6.326  &  X-Shooter  &  237.2  &  0.020  \\
188  &  ATLAS J332-32     &  6.329  &  X-Shooter  &   17.8  &  0.109  \\
189  &  ULAS J1148+0702   &  6.347  &  X-Shooter  &   13.3  &  0.152  \\
190  &  VIK J1152+0055    &  6.363  &  X-Shooter  &   11.1  &  0.194  \\
191  &  PSO J159-02       &  6.381  &  X-Shooter  &   13.3  &  0.164  \\
192  &  SDSS J1148+5251   &  6.411  &  ESI        &   63.8  &  0.056  \\
193  &  VIK J2318-3113    &  6.446  &  X-Shooter  &   13.6  &  0.116  \\
194  &  PSO J247+24       &  6.479  &  X-Shooter  &   13.8  &  0.155  \\
195  &  VDES J0224-4711   &  6.504  &  X-Shooter  &   19.0  &  0.118  \\
196  &  PSO J036+03       &  6.539  &  X-Shooter  &   33.5  &  0.049  \\
197  &  PSO J323+12       &  6.585  &  X-Shooter  &   14.9  &  0.114  \\
198  &  VIK J0305-3150    &  6.597  &  X-Shooter  &   12.4  &  0.214  \\
199  &  PSO J338+29       &  6.647  &  X-Shooter  &   10.5  &  0.202  \\
\enddata
\end{deluxetable}
\vspace{-0.9cm}
\tablecomments{Columns: (1) QSO index number, (2) QSO name, (3) QSO redshift based on the apparent start of the \lya\ forest, (4) instrument used for the \oi\ search, (5) signal-to-noise ratio per 30~\kms\ near rest wavelength 1285 \AA, (6) \oi~\lam1302 rest-frame equivalent width, in angstroms, at which the \oi\ search is 50\% complete.}
% \tablecomments is usually placed within the \deluxetable environment, but here it causes the text to overrun the column width.

\subsection{Data reduction}

The data were uniformly reduced using a custom pipeline similar to the one described in \citet{becker2012} and \citet{lopez2016}.  For each exposure, optimal sky subtraction was performed on the un-rectified frame following \citet{kelson2003}.  X-Shooter NIR frames were processed without nod subtraction.  Instead, a high-$S/N$ composite dark frame was subtracted from each exposure to remove dark current and other detector features prior to sky modeling.  For all exposures, a preliminary one-dimensional spectrum was then extracted using optimal weighting \citep{horne1986} and flux calibration derived from a standard star.  Correction for telluric absorption was computed for the individual preliminary spectra using models based on the Cerro Paranal Advanced Sky Model \citep{noll2012,jones2013}.  The telluric corrections were then propagated back to the corresponding two-dimensional arrays.  Finally, a single one-dimensional spectrum was extracted simultaneously from all exposures of an object with a given instrument (the X-Shooter arms extracted separately) to optimize the rejection of bad pixels.  Continuum fitting over wavelengths redward of the start of the \lya\ forest was done by hand using a slowly varying cubic spline.  While our sample excludes strong broad absorption line (BAL) QSOs, it does include some weak or moderate BALs.  In these cases the continuum was drawn through the smooth BAL trough.  Spectral resolution was estimated from the fits to the telluric absorption.  There is some variation within each instrument, but generally we found that the resolution was somewhat better than the nominal slit-limited values, which suggests that the seeing FWHM was often smaller than the projected slit width.  We adopted resolution ${\rm FWHM} \simeq 45$~\kms for ESI and 25~\kms\ for the VIS arm of X-Shooter.

Our sample includes the ultra-luminous $z=6.30$ QSO SDSS~J0100+2802 \citep{wu2015}, whose line of sight contains four \oi\ systems, as noted below \citep[see also][]{cooper2019}.  In addition to deep X-Shooter observations, we obtained a high-resolution Keck High Resolution Echelle Spectrometer \citep[HIRES;][]{vogt1994} spectrum of this QSO.  We use the HIRES data in this work only to identify an \oi\ system outside of our statistical sample that could not be confirmed with the X-Shooter spectrum alone (see Section~\ref{sec:clustering}).  We nevertheless briefly describe the HIRES data here.  The object was observed for 5.0 hours split between two grating settings.  We used the C2 decker, which has a 0\farcs86 width slit and delivers a resolution ${\rm FWHM} \simeq 6$~\kms.  The reduction process generally followed the steps outlined above, with the exception of the flux calibration.  Because HIRES is notoriously difficult to flux calibrate, a custom response function was generated separately for each exposure by matching the raw extracted flux from each order to our X-Shooter spectrum of the object.  This allowed us to extract a single, flux-calibrated spectrum prior to continuum fitting.  For more details, see \citet{boera2019}.  We note that we also use the HIRES spectrum of SDSS~J1148+5251 from \citet{becker2011a} to measure equivalent widths for some of the absorbers along that line of sight (see Appendix~\ref{app:details} for details).

\vspace{0.1cm} % Needed to create white space before section heading.

\section{\oi\ Survey}\label{sec:survey}

\subsection{Identification}\label{sec:identification}

\begin{figure*}
   \centering
   \begin{minipage}{\textwidth}
   \begin{center}
   \includegraphics[width=0.9\textwidth]{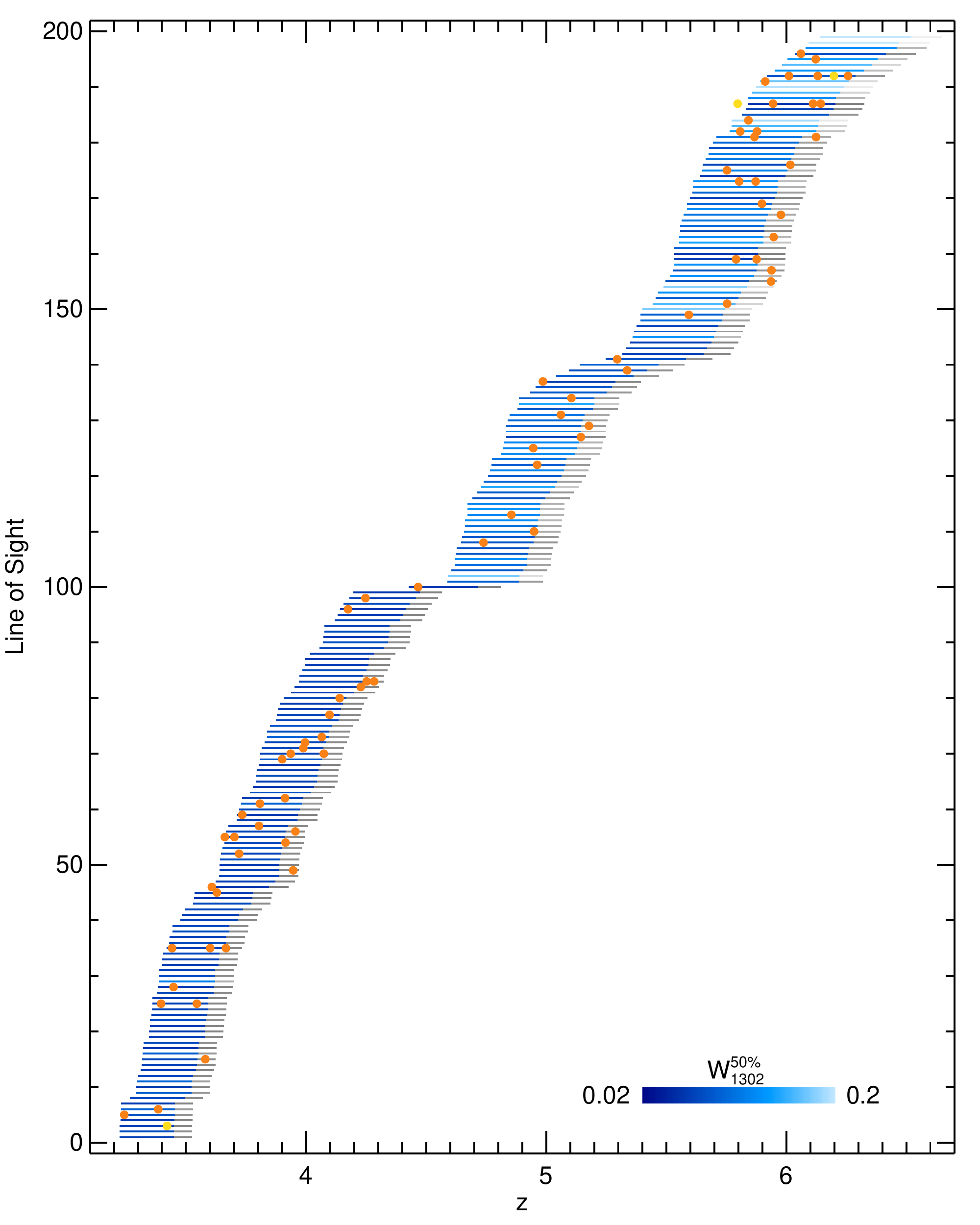}
   \caption{Summary of the survey results.  The horizontal lines span the redshift interval over which each lines of sight was surveyed for \oi.  The grey shaded region on the right hand side of each line marks the proximate region within 5000~\kms\ of the QSO redshift.  The lines are shaded according to the 50\% detection completeness limit in \oi\ equivalent width.  \oi\ systems identified in our survey of ESI and X-Shooter spectra are marked with orange filled circles.  The yellow circles in lines 187 (SDSS J0100+0524) and 191 (SDSS J1148+5251) are additional \oi\ systems that were identified in Keck HIRES data only and are not part of our statistical sample.  The yellow circle in line 3 (J1132+0052) is a probable \oi\ system whose \oi\ absorption is heavily blended, and is also not part of our statistical sample.\label{fig:los}}
   \end{center}
   \end{minipage}
   \vspace{0.1in}
\end{figure*}

Each line of sight was surveyed for metal absorption systems traced by \oi\ both visually and using the automated algorithm described in Section~\ref{sec:completeness}.  The systems were identified via the coincidence of \oi~\lam1302 in redshift with lines from other low-ionization metal ions, primarily \siii\ and \cii.  A detection required there to be significant absorption in \oi~\lam1302 and either \siii~\lam1260 or \cii~\lam1334.  The velocity profiles were also required to be self consistent, taking into account occasional blends with unrelated absorption lines or contamination from strong skyline residuals.  Both silicon and carbon should be mostly singly ionized in absorbers where hydrogen and oxygen are mostly neutral.  Due to their higher ionization potentials, moreover, \siii\ (16.3 eV) and \cii\ (24.4 eV) will tend to be ionized  less easily than \oi\ (13.6 eV).  Our requirement that either \siii\ or \cii\ be detected along with \oi\ should therefore be robust to relative variations in line strength due to ionization.  We could in principle miss systems due to large variations in relative abundances, but such variations are not generally seen \citep{cooke2011a,becker2012}.  In practice, both \siii~\lam1260 (in cases where it falls redward of the \lya\ forest) and \cii~\lam1334 were always detected along with \oi~\lam1302.     \oi~\lam1302, \siii~\lam1260, and \cii~\lam1334 tend to have comparable equivalent widths (though see Section~\ref{sec:highions}).  We also searched for weaker lines such as \siii~\lam1304 and \siii~\lam1526.  These tend to be present for stronger systems but were not required for detections.  Similarly, we included the \mgii~\lam\lam2796,2803 doublet in systems for which we had the necessary wavelength coverage, but did not use these lines for the initial identification.  We also measured the high-ionization doublets \civ~\lam\lam1548,1550 and \siiv~\lam\lam1393,1402 for  detected \oi\ systems as our wavelength coverage permitted.

Each line of sight was surveyed for \oi\ systems between the QSO redshift and the redshift where \oi~\lam1302 enters the \lya\ forest.  In all cases our spectra cover the primary low-ionization lines, including \cii~\lam1334, up to the QSO redshift.  The minimum redshift is defined as $1 + z_{\rm min} = (1+z_{\rm forest})\lambda_{\rm Ly\alpha}/\lambda_{\rm O\,I}$, where $\lambda_{\rm Ly\alpha}$ and $\lambda_{\rm O\,I}$ are the rest-frame wavelengths of \hi\ \lya\ and \oi~\lam1302, respectively, and $z_{\rm forest}$ is the redshift of the apparent start of the \lya\ forest, determined visually for each line of sight (see Table~\ref{tab:qsos}).  We note that $z_{\rm forest}$ is generally very similar to published values for the QSO emission redshifts.  In our formal analysis we excluded ``proximate'' systems within 5,000~\kms\ of $z_{\rm forest}$.  The total redshift interval surveyed over $3.2 < z < 6.5$ is $\Delta z = 57.7$ (76.5 including the proximity zones).  

In order to more easily evaluate the evolution of \oi\ systems with redshift, we generally quantify the survey pathlength in terms of an absorption pathlength interval $\Delta X$ \citep{bahcall1969}, defined as
\begin{equation}\label{eq:dzdX}
\Delta X = \int_{z_{1}}^{z_{2}} (1+z)^2 \frac{H_0}{H(z)} dz \, .
\end{equation}
Here, $z_1$ and $z_2$ are the redshift bounds of the survey interval, and $H(z)$ is the Hubble parameter at redshift $z$.  A non-evolving population of absorbers with constant comoving number density and proper cross-sectional area will have a constant line-of-sight number density $dn/dX$.  Our total absorption pathlength interval is $\Delta X = 250.1$ (332.7 including the proximity zones).

In total we detected 74 \oi\ systems, 57 of which are separated from the QSO by more than 5000~\kms.  A graphic depiction of the survey is shown in Figure~\ref{fig:los}.  An additional system at $z = 3.421$ towards J1332+0052 displays prominent low- and high-ionization lines; however, the \oi~\lam1302 line itself is fully obscured by unrelated absorption lines.  We therefore do not  include it as part of our sample, even though it is probably an \oi\ absorber.\footnote{The loss of survey pathlength due to blends is taken into account in our completeness estimates; see Section~\ref{sec:completeness}.}  We note that the raw (not corrected for completeness) number densities of proximate and non-proximate absorbers are similar, $dn/dX \sim 0.2$ averaged over the entire redshift range.  This contrasts with the enhanced number of high-ionization proximate absorbers typically seen along QSO lines of sight \citep[e.g.,][]{weymann1979,nestor2008,wild2008,perrotta2016}.  A summary of the properties of each system is given in Table~\ref{tab:ew}.  Line profiles are plotted in Figures~\ref{fig:z3p2429550_J0100-2708}--\ref{fig:z6p2575350_SDSSJ1148+5251}.

\subsection{Equivalent Width Measurements}\label{sec:measurements}

We measured rest-frame equivalent widths ($W$) for up to eleven ionic transitions.  For each system, a single velocity interval over which to integrate the absorption from low-ionization species (\oi, \cii, \siii, \mgii) was chosen via inspection of the line profiles.  These intervals  typically spanned less than 250~\kms\ but extended up to 670~\kms\ in some cases.  A separate interval was chosen for the high-ionization lines (\siiv\ and \civ).  When detected, the high-ionization lines often spanned a larger velocity interval than the low-ionization lines.  In cases where no high-ionization lines are visually apparent, the equivalent widths for these lines were integrated over $\pm$100~\kms\ of the nominal redshift of the low-ionization lines.  We note that ESI and X-Shooter will generally not resolve the narrow ($b \lesssim 10$~\kms) components that are common for low-ionization absorbers, making it difficult to obtain column densities in many cases.  We could in principle determine column densities for optically thin lines, or for species with multiple lines falling on different parts of the curve of growth.  We  chose to focus on equivalent widths, however, which are independent of resolution.  

Blended lines were identified visually based on the line strength, velocity profile, and proximity of other absorption lines.  We generally report the equivalent width for these systems as upper limits.  For some mild blends, however, we measured an equivalent width after removing the blended line.  For example, the \oi~\lam1302 line at $z=3.804$ towards J1032+0927 is blended with \civ~\lam1550 at $z=3.034$ (Figure~\ref{fig:z3p8039370_J1032+0927}).  In this case we inferred the \civ~\lam1550 profile by rescaling the \civ~\lam1548 line according to the ratio of the optical depths for these transitions.  This pixel-by-pixel approach becomes problematic when the intervening lines are unresolved or optically thick.  In general, however, we expect the de-blended equivalent widths to be accurate enough for the analysis considered below.  Blending must also be considered in the case of doublets when the velocity extent of the absorption profile exceeds the doublet separation.  We observed this in four cases for \civ\ ($z=3.7013$ towards J2215-1611, Figure~\ref{fig:z3p7013160_J2215-1611};  $z=3.8039$ towards J1032+0927, Figure~\ref{fig:z3p8039370_J1032+0927}; $z=3.9557$ towards J0835+0650, Figure~\ref{fig:z3p9556800_J0835+0650}; and $z=4.0742$ towards J0132+1341, Figure~\ref{fig:z4p0741980_J0132+1341}), but never in other doublets, for which the intrinsic velocity separation is larger.  To correct this self-blending in \civ\ we used pixel-by-pixel optical depth rescaling to infer the portion of the blended profile from one transition from the corresponding un-blended portion of the other transition.   That is, we calculated the red (blended) portion of the \lam1548 profile from the red (unblended) portion of the \lam1550 profile, and the blue (blended) portion of the \lam1550 profile from the blue (unblended) portion of the \lam1548 profile.  Here again, although this procedure is not perfect, we do not expect the errors to significantly impact our analysis.

Notes on individual absorbers are given in Appendix~\ref{app:details}.

\subsection{Completeness}\label{sec:completeness}

We estimated our completeness by randomly inserting artificial absorption systems into the data and assessing whether they would be detected.  The artificial absorbers were modeled as Voigt profiles convolved with the instrumental resolution, and were described by three parameters: a redshift, an \oi\ column density, and a Doppler parameter.  For each line of sight, 10$^4$ absorbers were inserted in separate trials over the redshift interval where \oi\ falls redward of the \lya\ forest, matching our survey range.  The \oi\ column density was drawn randomly over the logarithmic interval $13 < \log{(N_{\rm O\,I}/{\rm cm^{-2}})} < 16$.  The Doppler parameter was drawn randomly over the interval $b_{\rm min} < b < b_{\rm max}$, where $b_{\rm min} = 10$~\kms and $b_{\rm max}$ increased linearly with $\log{N_{\rm O\,I}}$ from 10~\kms\ at $\log{(N_{\rm O\,I}/{\rm cm^{-2}})} = 13$ to 100~\kms\ at $\log{(N_{\rm O\,I}/{\rm cm^{-2}})} = 16$.  These ranges in $\log{N_{\rm O\,I}}$ and $b$ were guided by Voigt profile fits to a selection of our observed systems, and were meant to roughly span the range in equivalent width and velocity width of the full observed sample.  We note that a single Voigt profile does not capture the full kinematic complexity of many of the observed systems; however, the detectability of a system often depends on the strength of a dominant component.  This is particularly true for weaker systems, for which completeness corrections are more important.

For each system we generated absorption lines in \oi, \cii, and \siii.  The \cii\ and \siii\ column densities were scaled from the \oi\ values as $\log{N_{\rm C\, II}} = \log{N_{\rm O\, I}} - 0.54$ and $\log{N_{\rm Si\, II}} = \log{N_{\rm O\, I}} - 1.26$.  These scalings were adopted from the relative abundances of metal-poor DLAs and sub-DLAs over $2 \lesssim z \lesssim 4$ \citep{cooke2011a,dessauges-zavadsky2003,peroux2007} and \oi\ systems at $z > 5$ where column density measurements from high-resolution spectra are available \citep{becker2012}.  

We note that adopting fixed column density ratios ignores variations due to differences in relative abundance or, perhaps more significantly,  ionization effects.  As noted above, at a given \oi\ equivalent width a partially ionized absorber will tend to have stronger \siii\ and \cii\ than one that is fully neutral (in hydrogen), making it easier to detect.  We argue below that the number density of \oi\ absorbers is higher over $5.7 < z < 6.5$ than over $4.9 < z < 5.7$, a conclusion that depends partly on our completeness estimates.  Assuming fixed column density ratios when determining completeness is conservative with respect to this conclusion in that if there are undetected weak \oi\ absorbers with stronger \siii\ and/or \cii, then the total \oi\ number density should increase the most at $z > 5.7$, where our sensitivity to weak systems is lowest.  In practice, however, there is limited evidence for a large population of weak, partially ionized \oi\ systems.  The observed ratios of \oi\ and \cii\ equivalent widths tend to be near unity, particularly at $z > 4.9$ (see Figure~\ref{fig:ions}), with some exceptions noted below (Section~\ref{sec:highions}).

In order to increase efficiency we used an automated detection algorithm for our completeness trials.  The algorithm was developed to roughly mimic the process of identifying systems by eye, with detection criteria established using the real \oi\ absorbers as a training set.  The algorithm was also tested against by-eye identifications for artificial systems over the relevant range of absorber properties, spectral resolutions, and signal-to-noise ratios.  Briefly, a detection required there to be significant absorption in \oi~\lam1302 and at least one other line, and the kinematic profiles of the lines needed to be consistent with one another.  The other lines examined were \siii~\lam1260, when it fell redward of the \lya\ forest, and  \cii~\lam1334, which were the primary lines used when identifying \oi\ systems visually.  

For each of the 10$^4$ artificial absorbers, the automated algorithm stepped across the nominal redshift in increments of 5~\kms\ and examined the regions around the expected positions of each available line using the following steps.  It first determined whether there was significant ($>$3$\sigma$) absorption after smoothing the spectrum by the instrumental resolution.  If significant detection existed for all lines that fall redward of the \lya\ forest then it fit a Voigt profile over $\pm$200~\kms\ of that redshift independently to each available line.  The continuum was allowed to vary by up to 5\%, providing a mechanism to deal with small continuum errors as well as with nearby weak absorption lines.  The flux cross-correlation between each pair of lines was also computed over the same velocity interval.  In all cases, a detection required that, for at least two lines, the FWHM of the fitted profiles agreed to within a factor of 1.5, the equivalent widths of the fits agreed to within a factor of 4.0, and the ratio of the maximum absorption depths to the FWHM (in \kms) of the fits were greater than 0.002.  The last condition was meant to reject spurious wide, weak lines.  A detection was recorded if the centroids of the fits aligned to within 5~\kms, the FWHM of both Voigt profile fits were less than 200~\kms, and the reduced $\chi^2$ of the fits over the central FWHM were less than 5.0.  Alternatively, a detection was recorded if the centroids of the Voigt profile fits aligned to within 30~\kms and the cross-correlation was greater than 0.75.  

\begin{figure}
   \centering
   \includegraphics[width=0.42\textwidth]{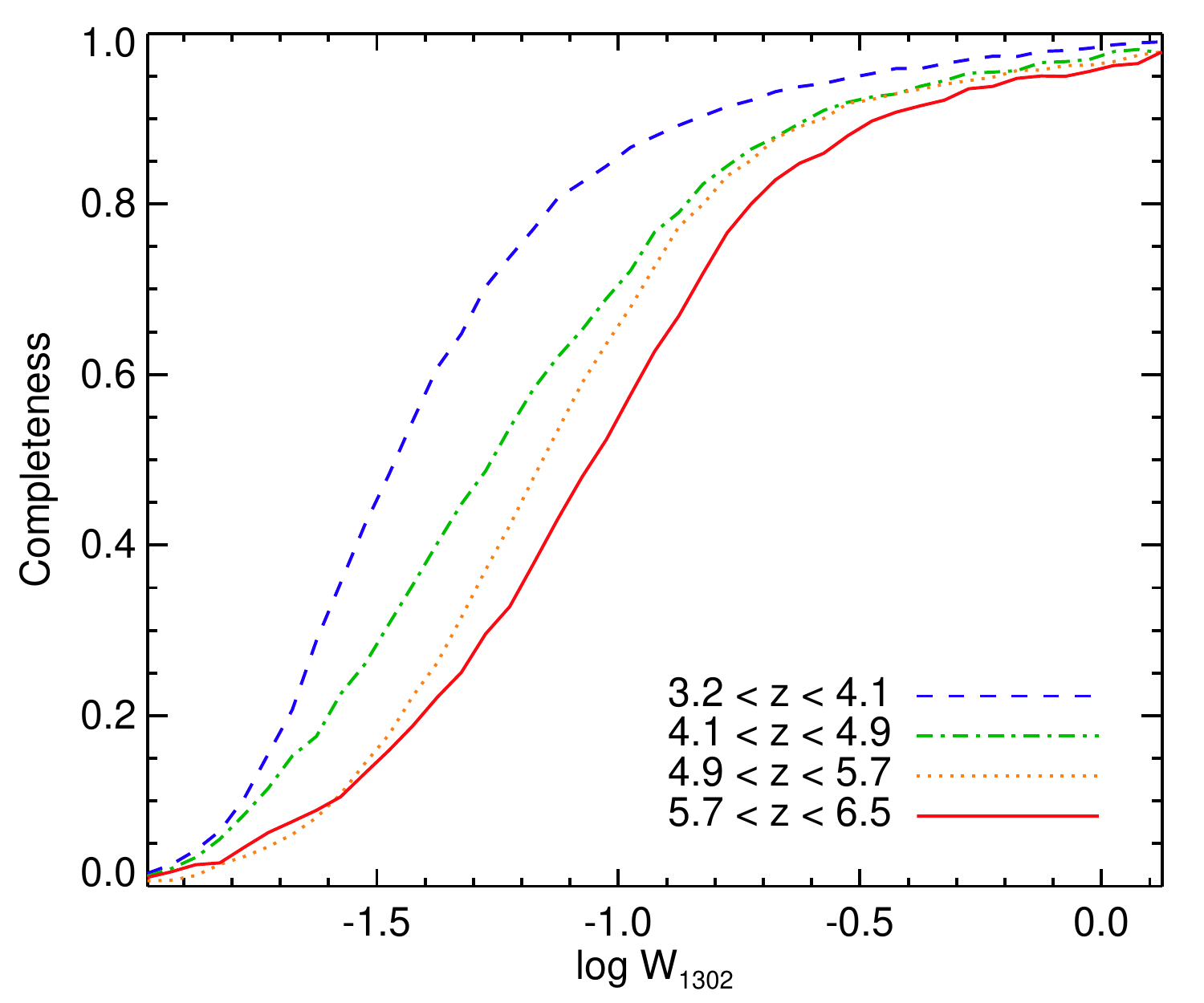}
   \caption{Survey completeness as a function of \oi~\lam1302 equivalent width.  Our four redshift intervals are plotted with the line styles indicated.\label{fig:completeness}}
\end{figure}

Our completeness estimates as a function of \oi\ equivalent width are plotted in Figure~\ref{fig:completeness}.  For each redshift bin we computed the pathlength-weighted mean completeness of all contributing lines of sight, excluding the redshift intervals immediately around known \oi\ systems.  As expected, the completeness tends to decrease with increasing redshift.  This is largely driven by a decrease with redshift in the typical $S/N$ of our spectra (see Table~\ref{tab:qsos}).  The increased number of strong residuals from sky emission line subtraction and telluric absorption correction towards redder wavelengths also contributes to this trend.

We expect some error in our completeness estimates due to the challenge of designing an automated detection algorithm that is robust to the inherent variations in real \oi\ absorbers.  The abundance ratios of different ions in these systems are not necessarily constant, and their kinematic profiles do not always perfectly match (for example, \cii\ can have components not present in \oi\ due to ionization effects; see discussion above).  We nevertheless believe that errors in our completeness estimates are not strongly affecting our results.  Out of the 74 real systems identified visually, the automated algorithm identified 68 (92\%).  Of the six systems missed, two have \oi~\lam1302 equivalent widths below the cutoff of $W_{1302} > 0.05$~\AA\ we impose for the analysis in Section~\ref{sec:distributions}.  The remaining four were readily identified by eye but were missed by the automated algorithm due to complexities in the line profiles or the presence of nearby strong lines, which caused the algorithm to fail in some cases.  Of these four, three had $W_{1302} > 0.2$~\AA, where our completeness estimate is $>$80\% at all redshifts (Figure~\ref{fig:completeness}).  The remaining system was one out of 32 real \oi\ systems in the range $0.05~{\rm \AA} < W_{1302} < 0.20~{\rm \AA}$, where the completeness corrections are more significant.  We therefore expect the overall errors in our completeness estimates to be small compared to the statistical errors described below.  Completeness corrections are discussed further in Section~\ref{sec:density}.

False positive detections should only be a minor concern for this work.  In principle there can be false detections from the chance alignment of unrelated lines.  In practice, however, although we used only \oi~\lam1302, \siii~\lam1260, and \cii~\lam1334 to visually identify \oi\ absorbers, nearly all of our systems were detected in at least three lines.  If \siii~\lam1260 fell in the \lya\ forest, then \siii~1304 and \siii~\lam1526 were generally available, or the system was detected in \mgii\ or higher ionization lines (\siiv\ and/or \civ).  In the single case where the system was only detected in \oi~\lam1302 and \cii~\lam1334 (the $z=5.7533$ system towards SDSS J2315-0023, Figure~\ref{fig:z5p7532610_SDSSJ2315-0023}), the asymmetric kinematic profiles are distinct enough that a false positive from unrelated lines is unlikely.  We therefore expect that essentially all of our \oi\ detections are genuine.

\subsection{Equivalent Width Distributions}\label{sec:distributions}

Our primary goal is to determine how the number density of \oi\ systems evolves with redshift.  In order to extend this analysis to equivalent widths where we are significantly incomplete, we need to adopt a functional form for the distribution of equivalent widths.  Exponential and power law distributions are commonly adopted for metal lines (see also the Schechter function used by \citet{mathes2017} and \citet{bosman2017}).  Here we adopt an exponential distribution, which we find provides a reasonable fit to the observed distribution.  We note, however, that our final conclusions do not depend sensitively on this choice.  We repeated the analysis below using a power law fit to the equivalent width distribution and obtained very similar results for the integrated number density. 

We fit an exponential distribution of the form
\begin{equation}\label{eq:fW}
   f(W) = \frac{\partial^2 n}{\partial W \partial X} = \frac{A}{W_0} e^{-W/W_0} \, ,
\end{equation}
where $W_0$ is the exponential cutoff scale and $A$ is the number density per unit pathlength $X$ integrated over $0 < W < \infty$.  $A$ and $W_{0}$ were fit simultaneously in four redshift bins using a maximum likelihood approach similar to the one described in \citet{bosman2017}.  We used a forward modeling approach in which likelihood values were derived from the model intrinsic $f(W)$ multiplied by our completeness (Figure~\ref{fig:completeness}).  We divided our survey into roughly equal bins in redshift: $3.2 < z < 4.1$, $4.1 < z < 4.9$, $4.9 < z < 5.7$, and $5.7 < z < 6.5$ (but see Appendix~\ref{app:binning}).  These bins contain 18, 5, 8, and 18 intervening systems with $W_{1302} > 0.05$~\AA, respectively.  In order to minimize our statistical uncertainties while limiting our sensitivity to large completeness corrections we only fit over equivalent widths where we are $>$25\% complete at all redshifts, $W_{1302} > 0.05$~\AA.  We comment further on this choice below.

\begin{figure}
   \centering
   \includegraphics[width=0.42\textwidth]{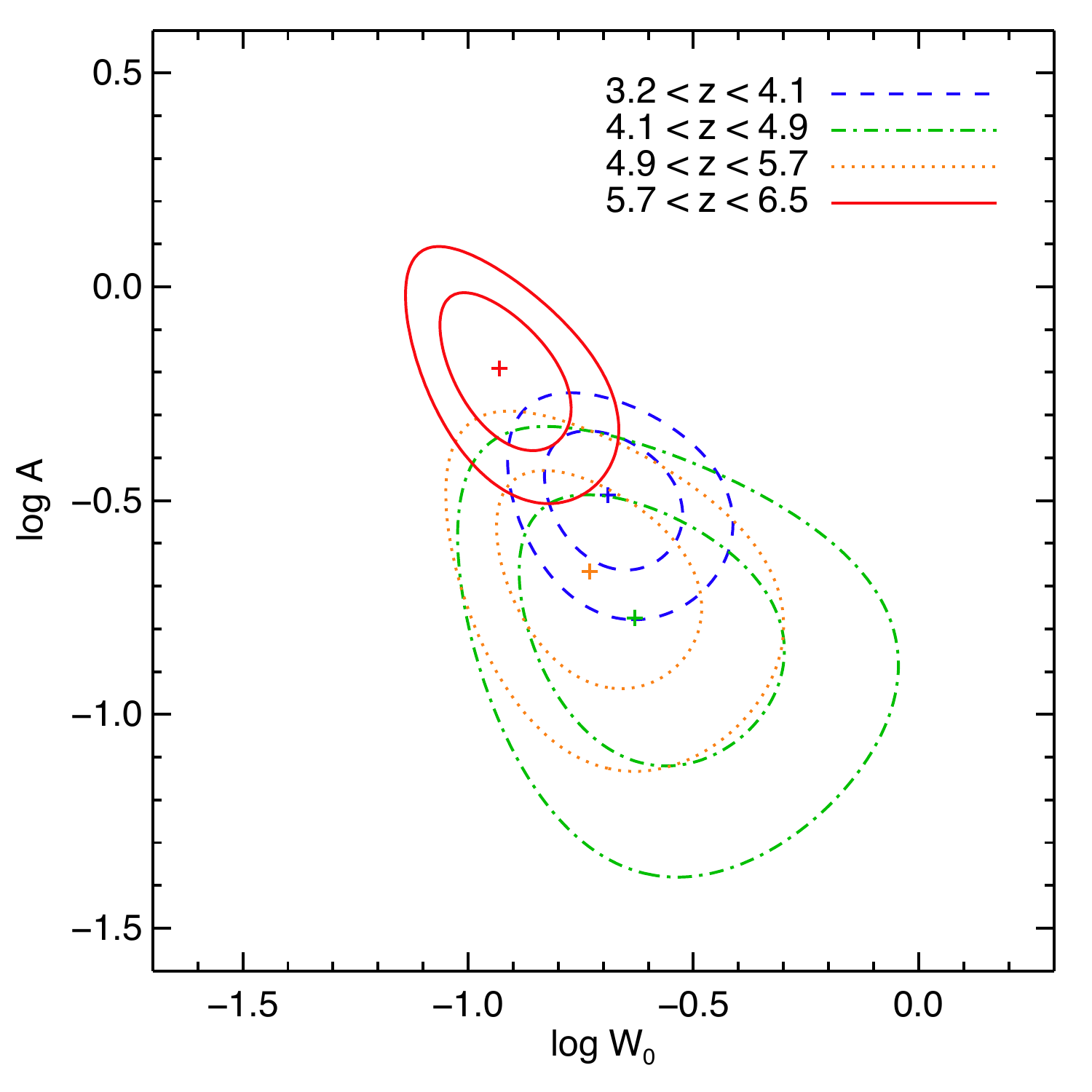}
   \caption{Results of the exponential fit to the distribution of \oi~\lam1302 equivalent widths for $W_{1302} > 0.05$~\AA.  The contours show the 68\% and 95\% likelihood bounds over each redshift interval for the parameters in Equation \ref{eq:fW}.  Crosses mark the highest probability values.\label{fig:exp_L}}
\end{figure}

\begin{deluxetable}{lccccc}
\tablecaption{Summary of results\label{tab:results}}
\tablehead{
   \colhead{$z$}  &  \colhead{$\Delta X$}  &  \colhead{$n$}  &  \colhead{$\log{A}$}  &  \colhead{$\log{W_0}$}  &   \colhead{$dn/dX$} }
\colnumbers
\startdata
3.2--4.1  &  79.0  &  28,22,18  &  $-0.49^{+0.09}_{-0.13}$  &  $-0.69^{+0.11}_{-0.09}$  &  $ 0.255^{+0.057}_{-0.061}$  \\
4.1--4.9  &  44.1  &  9,6,5  &  $-0.77^{+0.16}_{-0.26}$  &  $-0.63^{+0.24}_{-0.15}$  &  $ 0.136^{+0.059}_{-0.059}$  \\
4.9--5.7  &  62.5  &  11,9,8  &  $-0.67^{+0.13}_{-0.20}$  &  $-0.73^{+0.17}_{-0.13}$  &  $ 0.165^{+0.055}_{-0.058}$  \\
5.7--6.5  &  66.3  &  26,20,18  &  $-0.19^{+0.10}_{-0.14}$  &  $-0.93^{+0.11}_{-0.08}$  &  $ 0.421^{+0.098}_{-0.101}$  \\
\enddata
\tablecomments{Columns: (1) redshift range, (2) absorption pathlength interval, (3) number of \oi\ systems: total, non-proximate, and non-proximate with $W_{1302} > 0.05$~\AA, (3) constraints on $\log{A}$, (4) constraints on $\log{W_0}$, where $W_0$ is in \AA, (5) constraints on $dn/dX$.  All constraints are for \oi\ systems with $W > 0.05$~\AA.  Errors are marginalized 68\% confidence intervals.  The errors in $\log{A}$ and $\log{W_0}$ are correlated (Figure~\ref{fig:exp_L}).}
\end{deluxetable}

Our likelihood contours are shown in Figure~\ref{fig:exp_L}, and the results are summarized in Table~\ref{tab:results}.  For all redshifts there is a degeneracy between $A$ and $W_0$.  The fits are least constrained over $4.1 < z < 4.9$ and $4.9 < z < 5.7$, where there are the fewest detections.  These bins are consistent with a similar cutoff but lower normalization than $3.2 < z < 4.1$, although the uncertainties are significant.  Over $5.7 < z < 6.5$ our fits prefer a somewhat lower cutoff and a higher normalization.  The 95\% confidence intervals overlap between the highest and two middle redshift bins.  Overall, however, there is evidence that equivalent width distribution evolves with redshift, which we explore further below.  

\begin{figure}
   \centering
   \includegraphics[width=0.42\textwidth]{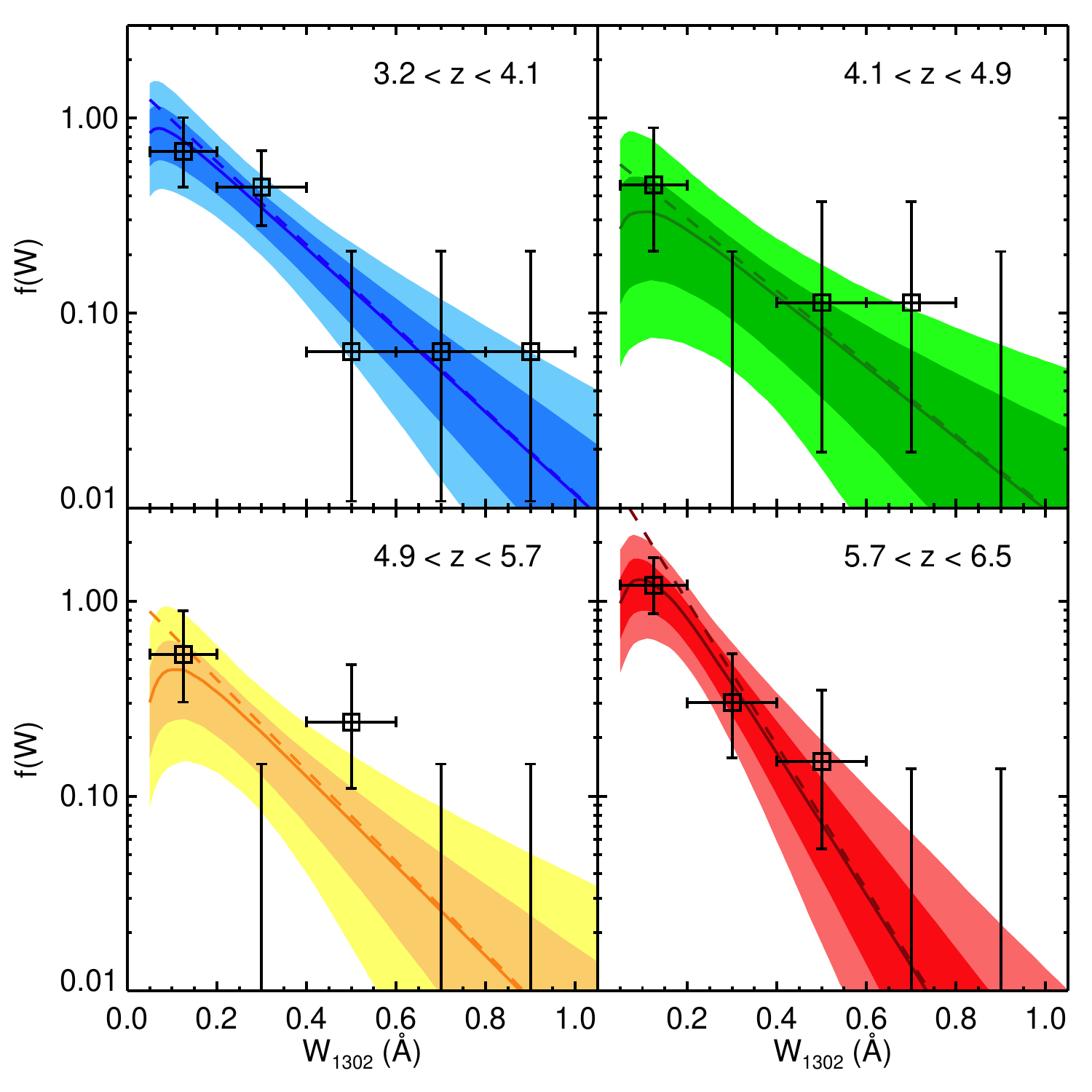}
   \caption{Binned \oi~\lam1302 equivalent width distributions.  Bins containing detected systems are plotted with squares and two-sided 68\% Poisson confidence intervals.  Non-detections are shown with one-sided 84\% upper limits.  The data are not corrected for completeness.  The dark and light shaded regions in each panel show the 68\% and 95\% confidence intervals to the exponential fit, derived from the likelihood distributions in Figure~\ref{fig:exp_L} and multiplied by the completeness functions in Figure~\ref{fig:completeness}.  Dashed and solid lines show the best fitting exponential distributions before and after multiplying by the completeness functions, respectively.\label{fig:binned_distributions}}
\end{figure}

In Figure~\ref{fig:binned_distributions} we compare our fits to the observed equivalent width distributions in bins of $W_{1302}$.  Confidence intervals for $f(W)$ were determined by sampling the full posterior distribution for $A$ and $W_0$.  We do not correct the binned data for completeness, which requires knowing the underlying shape of the distribution.  Instead, we multiply the model fits by our completeness estimates.  We emphasize that Figure~\ref{fig:binned_distributions} is for visualization only; the parameters for $f(W)$ were fit to the unbinned data.  Nevertheless, it shows that our choice of an exponential distribution gives a reasonable fit to the data.  

\subsection{Number Density}\label{sec:density}

\begin{figure}
   \centering
   \includegraphics[width=0.42\textwidth]{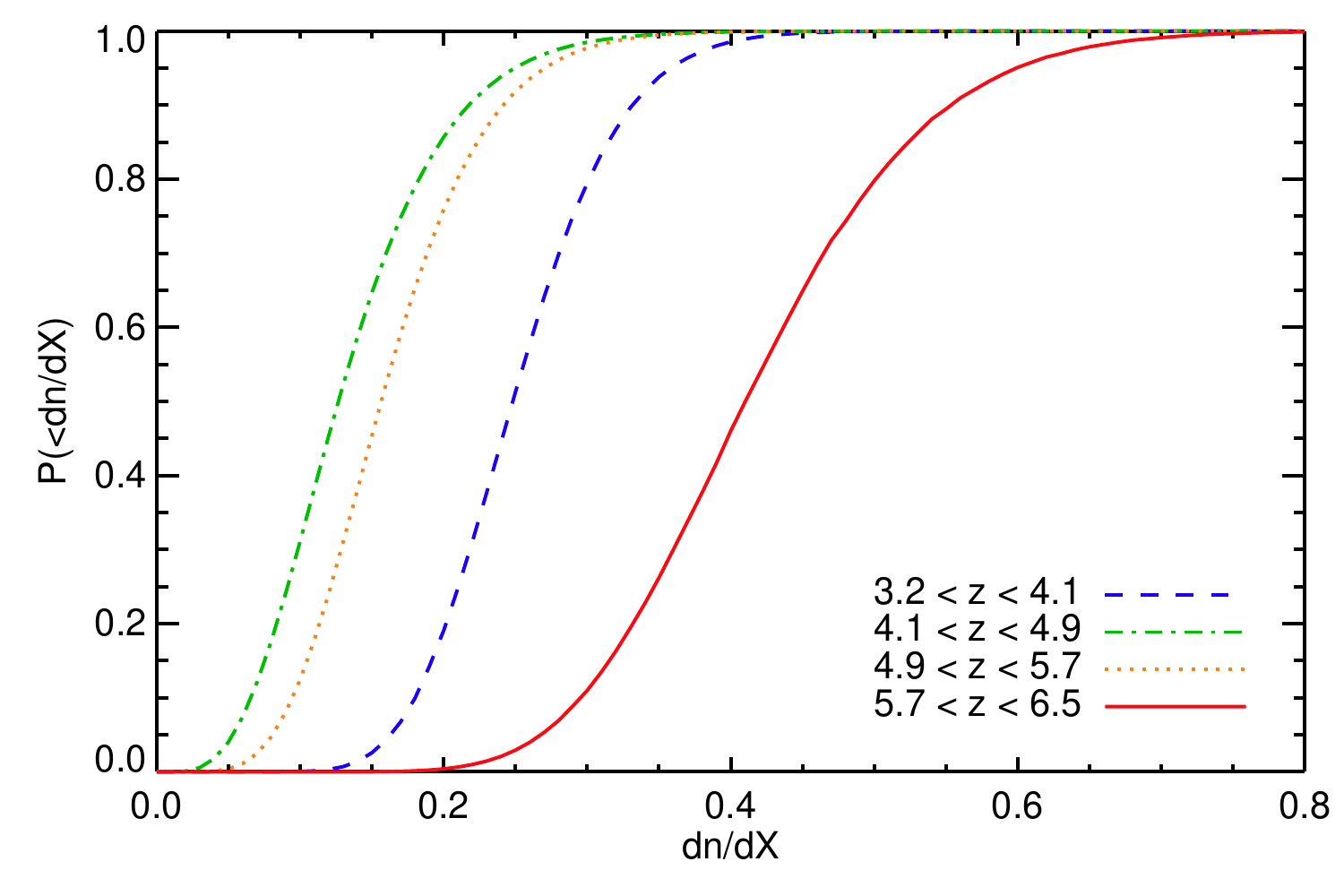}
   \caption{Cumulative probability distributions for the comoving number density of \oi\ systems with $W_{1302} > 0.05$~\AA.  The distributions were computed by integrating Equation~\ref{eq:fW}, marginalizing over the parameter distributions in Figure~\ref{fig:exp_L}.\label{fig:CPDF}}
\end{figure}

\begin{figure*}
   \centering
   \begin{minipage}{0.7\textwidth}
   \begin{center}
   \includegraphics[width=0.9\textwidth]{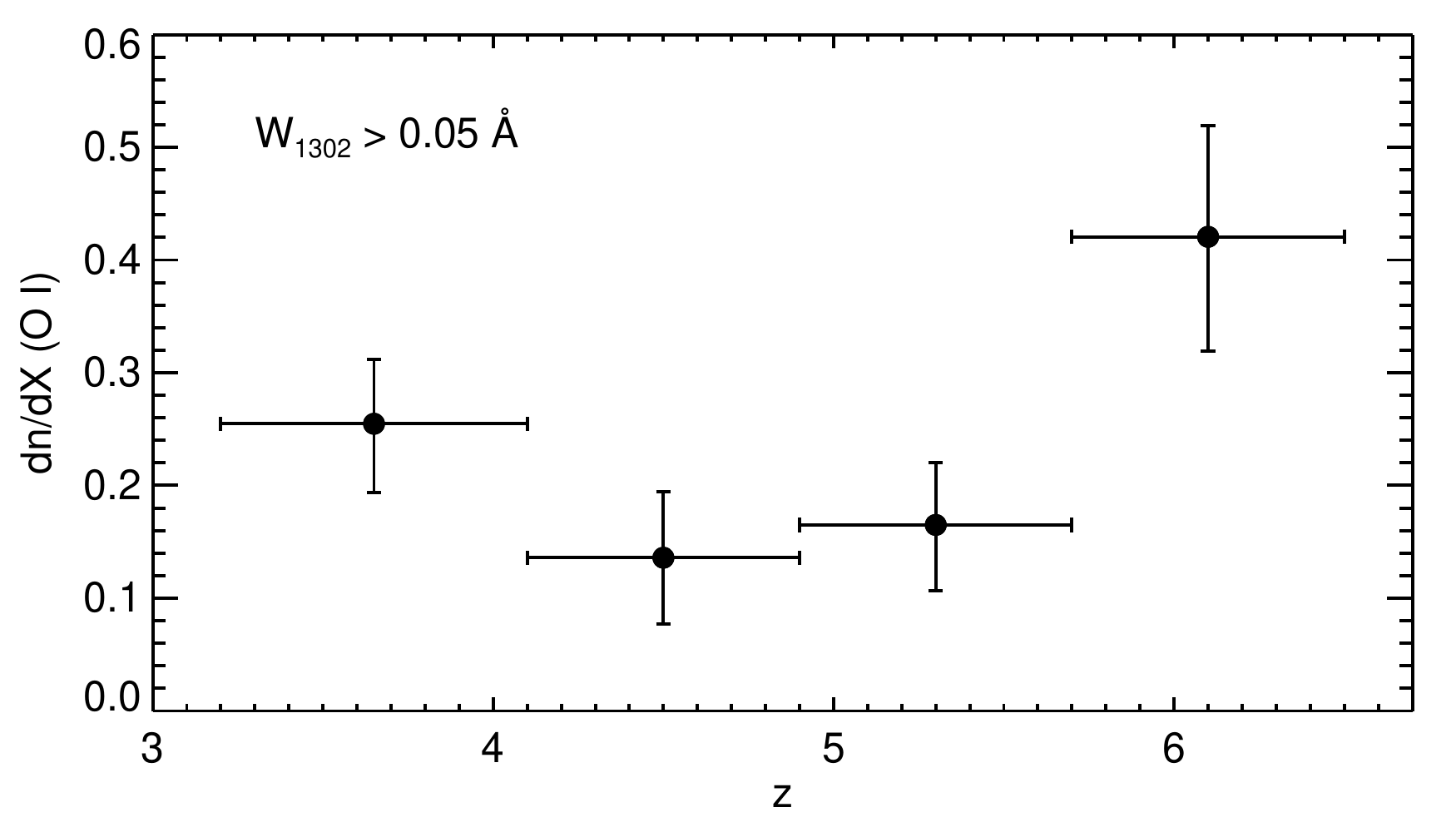}
   \caption{Comoving number density of \oi\ systems with $W_{1302} > 0.05$~\AA\ as a function of redshift.  The results were obtained by integrating Equation~\ref{eq:fW} using the parameter distribution shown in Figure~\ref{fig:exp_L}.  Filled symbols show the most likely values at each redshift.  Vertical errors bars show the 68\% confidence intervals taken from Figure~\ref{fig:CPDF}.  Results for smaller redshift bins are shown in Figure~\ref{fig:OI_dndX_dz0p4}.\label{fig:dndX}}
   \end{center}
   \end{minipage}
   \vspace{0.1in}
\end{figure*}

\begin{figure}
   \centering
   \includegraphics[width=0.47\textwidth]{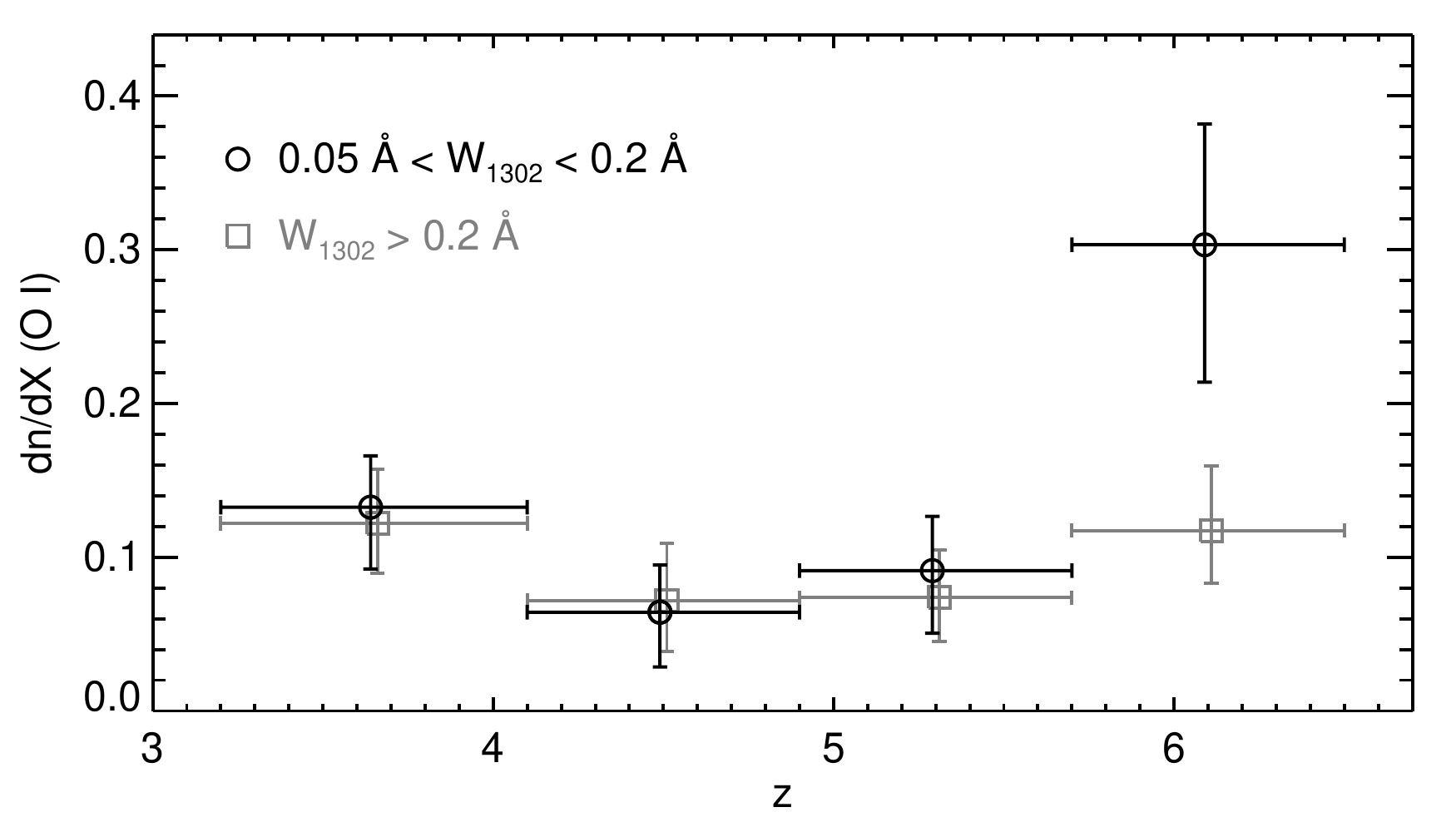}
   \caption{Comoving number density of \oi\ systems as a function of redshift, divided into equivalent width ranges.  The results were obtained by integrating Equation~\ref{eq:fW} using the parameter distribution shown in Figure~\ref{fig:exp_L}.  Circles show the most likely values for systems with $0.05~{\rm \AA} < W_{1302} < 0.2~{\rm \AA}$, while squares are for $W_{1302} > 0.2~{\rm \AA}$.  Vertical errors bars show the 68\% confidence intervals.  Points are slightly offset in redshift for clarity\label{fig:OI_dndX_EW_ranges}}
\end{figure}

We now turn to computing the integrated line-of-sight number density of \oi\ systems as a function of redshift.  In each redshift bin we computed $dn/dX$ by integrating Equation~\ref{eq:fW} over $W_{1302} \ge 0.05$~\AA, the range over which $f(W)$ was fit.  We constructed a probability distribution for $dn/dX$ from the full posterior distributions for $A$ and $W_0$ shown in Figure~\ref{fig:exp_L}.  The cumulative distributions for our four redshift bins are shown in Figure~\ref{fig:CPDF}.  

In Figure~\ref{fig:dndX} we plot the most probable values and 68\% confidence intervals for $dn/dX$ as a function of redshift.  The results are summarized in Table~\ref{tab:results}.  The evolution over $3.2 < z < 4.9$ is consistent with a moderate increase with decreasing redshift.  This is the expected behavior if the number density of \oi\ systems is driven mainly by the ongoing metal enrichment of the CGM with time.  The number densities over $4.1 < z < 4.9$ and $4.9 < z < 5.7$ are similar, albeit with significant uncertainties.  The number density over $5.7 < z < 6.5$, however, is notably higher than over $4.9 < z < 5.7$.  Using the cumulative probability distributions plotted in Figure~\ref{fig:CPDF}, we find that $dn/dX$ over $5.7 < z < 6.5$ is a factor of $2.5^{+1.6}_{-0.8}$ greater than over $4.9 < z < 5.7$ at 68\% confidence, with a probability that the $dn/dX$ is larger at $z > 5.7$ of 98.9\%.  This increases to 99.7\% if we compare $dn/dX$ over $5.7 < z < 6.5$ to that inferred from a single fit to $f(W)$ over $4.1 < z < 5.7$.  This decline with decreasing redshift runs contrary to the naive expectation that the number density of \oi\ systems should monotonically trace the buildup of CGM metals with time.  

In Figure~\ref{fig:OI_dndX_EW_ranges} we divide the $dn/dX$ results into two equivalent width ranges, $0.05~{\rm \AA} < W_{1302} < 0.2~{\rm \AA}$ and $W_{1302} > 0.2~{\rm \AA}$.  We caution that the values in Figure~\ref{fig:OI_dndX_EW_ranges} were calculated using the fits to Equation~\ref{eq:fW} computed over the full equivalent width range ($W_{1302} > 0.05$~\AA; Figure~\ref{fig:exp_L}), rather than from separate fits over these smaller ranges.   Nevertheless, the results demonstrate how the shape of $f(W)$ evolves with redshift.  In the three redshift bins over $3.2 < z < 5.7$ the number density of systems in the two $W_{1302}$ ranges is roughly equal.  Over $5.7 < z <   6.5$, however, the number density of systems with $0.05~{\rm \AA} < W_{1302} < 0.2~{\rm \AA}$ is a factor of $\sim$3 higher than those with $W_{1302} > 0.2~{\rm \AA}$.  The number density of stronger systems is nominally somewhat lower over $4.1 < z < 5.7$ that at $z > 5.7$, but is consistent within the 68\% confidence intervals with no evolution over the entire redshift range studied.  Most of the evolution occurs among the weaker systems, where $dn/dX$ declines by a factor of $3.3^{+2.8}_{-1.2}$ (68\% confidence) from $5.7 < z < 6.5$ to $4.9 < z < 5.7$.

As noted above, our results depend partly on completeness corrections, which increase towards smaller values of $W_{1302}$.  For the best fits to $f(W)$ in Table~\ref{tab:results}, our total corrections on $dn/dX$ for $W_{1302} > 0.05$~\AA\ are factors of 1.2 over $3.2 < z < 4.1$, 1.4 over $4.1 < z < 4.9$, 1.5 over $4.9 < z < 5.7$, and 1.9 over $5.7 < z < 6.5$.  We can test whether our results may be driven by errors in the completeness estimates by varying the range in $W_{1302}$ over which we fit $f(W)$.  Increasing or decreasing the minimum $W_{1302}$ by a factor of two produces nominal values for $A$ and $W_0$ that are well within the 68\% uncertainties in Figure~\ref{fig:exp_L}.  The more conservative limit of $W_{1302} > 0.1$~\AA\ gives a minimum completeness of $>$55\% at all redshifts (Figure~\ref{fig:completeness}), and smaller total completeness corrections for the nominal values of $dn/dX$, factors of 1.1, 1.2, 1.2, and 1.4 in order of increasing redshift.  We nevertheless recover the same trends in $dn/dX$ for $W_{1302} > 0.1$~\AA, albeit at somewhat lower statistical significance; a decrease in $dn/dX$ from $5.7 < z < 6.5$ to $4.9 < z < 5.7$ is still favored at 96\% confidence.  Our results therefore do appear to be driven by errors in the completeness estimates for small values of $W_{1302}$.

We discuss the implications of the evolution in $dn/dX$ below, but first examine the high-ionization metal species associated with our \oi\ absorbers.

\section{High-Ionization Components}\label{sec:highions}

\begin{figure*}
   \centering
   \begin{minipage}{\textwidth}
   \begin{center}
   \includegraphics[width=0.9\textwidth]{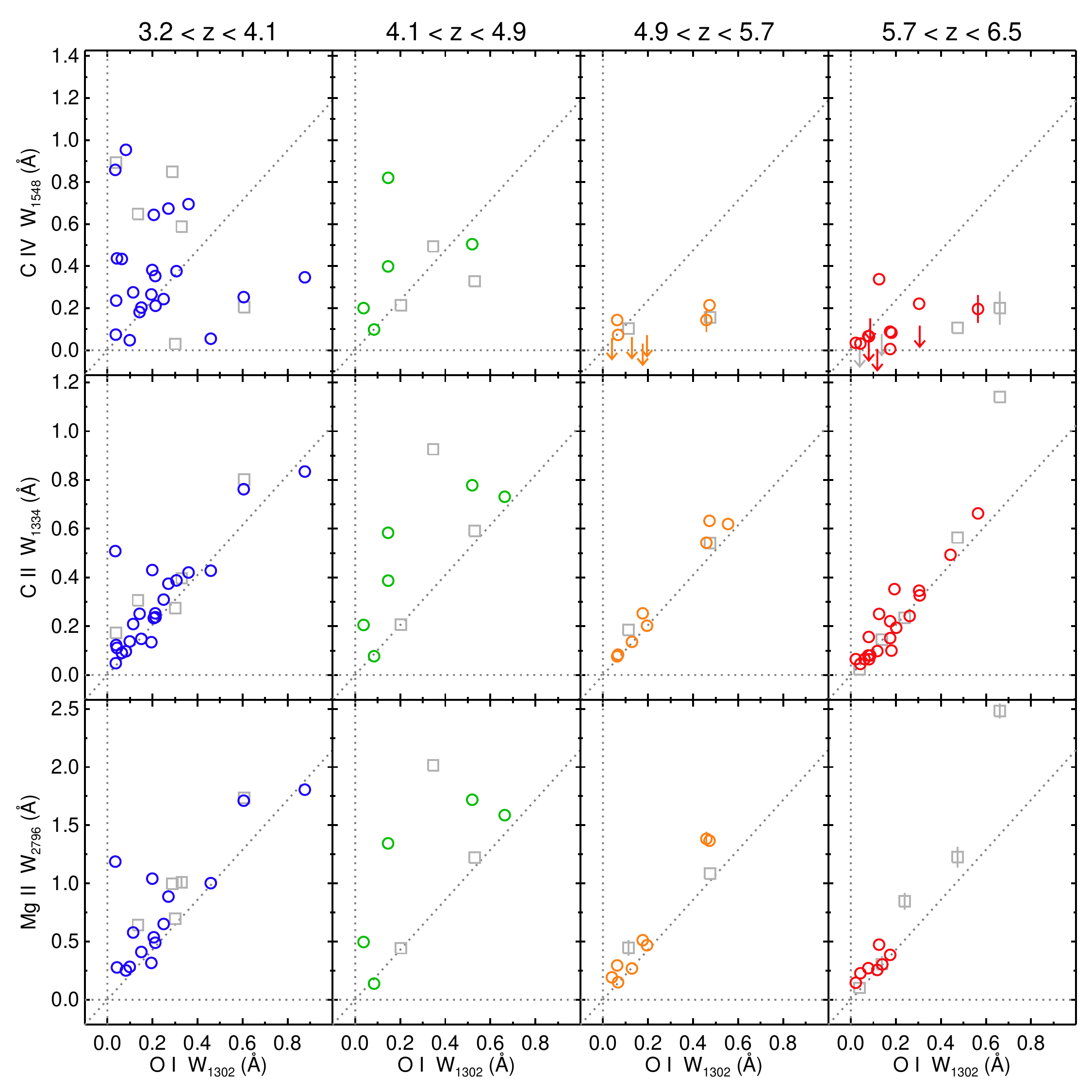}
   \caption{\civ~\lam1548 (top row), \cii~\lam1334 (middle row), and \mgii~\lam2796 (bottom row) equivalent widths as a function of \oi~\lam1302 equivalent width for the systems in our survey.  The sample is divided into redshift bins as indicated along the top.  In each panel we plot only cases where measurements were obtained for both ions.  Downward arrows are 2$\sigma$ upper limits on non-detections.  68\% errors bars are shown when they are larger than 0.05~\AA.  In some cases, points with error bars were measured from noisy data where there is visually no obvious detection, and should therefore be treated with caution.  Light grey squares denote proximate absorbers.  The diagonal line in each panel shows where the two ions have equal equivalent widths in dimensionless units.\label{fig:ions}}
   \end{center}
   \end{minipage}
   \vspace{0.1in}
\end{figure*}

In the top panels of Figure~\ref{fig:ions} we compare the rest-frame equivalent widths of the high-ionization line \civ~\lam1548 to \oi~\lam1302 for our detected \oi\ systems.  The diagonal dotted line in each panel represents equal \civ\ and \oi\ equivalent widths in dimensionless units (i.e., with the rest-frame wavelengths factored out).    A  clear trend of increasing \civ\ strength towards lower redshifts is seen.  All of the \oi\ systems at $z < 4.9$ have detected \civ, and for a large majority of these the \civ\ absorption is similar to or stronger than \oi.  At $z > 4.9$, in contrast, \civ\ is typically weak or not detected.  In some cases with high-$S/N$ data (e.g., the $z=6.1115$ and 6.1436 systems towards SDSS~J010+2802, Figures~\ref{fig:z6p1115230_SDSSJ0100+2802} and \ref{fig:z6p1435510_SDSSJ0100+2802}) the 3$\sigma$ upper limits on \civ\ are extremely low ($W_{1548} < 0.005$~\AA).  Similar results were found by \citet{cooper2019}.

We note that where high-ionization components are detected they tend not to align kinematically with \oi.  In cases where they can be clearly identified, the \civ\ and \siiv\ components are often broader, more numerous, and/or offset in velocity from \oi.  As others have noted \citep[e.g.,][]{fox2007}, this implies that, in many cases, \oi\ and \civ\ are likely to arise from separate phases of the CGM.  This potentially complicates the role that high-ionization lines can play in interpreting the redshift evolution of \oi, a point we return to below.

For comparison, we also compare the low-ionization transitions \cii~\lam1334 and \mgii~\lam2796 to \oi\ in Figure~\ref{fig:ions}.  Here there is noticeably less redshift evolution.  In the large majority of systems, the rest-frame equivalent widths (in dimensionless units) of \cii\ and \mgii\ are comparable to \oi.  Cases where $W_{1334}$ and $W_{2796}$ are substantially higher than $W_{1302}$, which mainly occur at $z < 4.9$, are likely to be partially ionized absorbers.  This is reflected in the fact that absorption components with strong \cii\ compared to \oi\ also tend to appear in \siiv\ and \civ\ (e.g., the components at $\Delta v \simeq 60$ and 170~\kms\ in the $z=3.3844$ system towards J1018+0548, Figure \ref{fig:z3p3844720_J1018+0548}; and the $z=3.6287$ system towards J0042-1020, Figure~\ref{fig:z3p6287100_J0042-1020}).

Finally, we note that proximate absorbers (light grey squares in Figure~\ref{fig:ions}) show similar trends in the relative strength of high- and low-ionization lines as non-proximate absorbers.  This suggests that many of the proximate absorbers may be far enough away from the background QSO that ionizing radiation from the QSO does not strongly affecting the ionization balance, in contrast with the trends seen for absorbers selected via \civ\ \citep{perrotta2016}.  Alternatively, the similarity may result from a combination of trends in metallicity and ionization that are a function of proximity to the background QSO \citep[e.g.,][]{ellison2010,ellison2011}.

\section{Potential Clustering of \oi\ Absorbers}\label{sec:clustering}

A notable feature of Figure~\ref{fig:los} is that detections of multiple \oi\ systems along a single line of sight seem to be more common near $z \sim 6$ than at lower redshifts.  This was previously seen by \citet{becker2006} in the case of SDSS J1148+5251, which contains four \oi\ systems within a span of $\Delta z = 0.25$ (100 comoving Mpc).  The weakest of these, marked by a yellow circle in Figure~\ref{fig:los}, is detected only in high-resolution Keck HIRES data \citep[Figure 6 of][]{becker2011a}.  Here we find that SDSS J0100+2802 also contains four \oi\ systems over a similar interval ($\Delta z = 0.31$, 130 comoving Mpc).  The redshift of one of these, $z=5.7975$, falls just below our nominal survey pathlength for this object and is not included in the statistical sample.  The \oi\ line falls in the proximity zone region of the \lya\ forest but is clearly identified by its narrow width in HIRES data (Figure~\ref{fig:z5p79750_SDSSJ0100+2802}).  Three other $z \sim 6$ lines of sight (SDSS J0818+1722, CFHQS J2100-1715, and PSO J3008-21) contain two \oi\ systems outside the proximity zone.  In contrast, multiple detections outside the proximity zone are seen towards only two lower-redshift QSOs (J1108+1209 and J2215-1611).

We caution that this apparent increase in \oi\ multiplicity with redshift could be misleading for two reasons.  First, the redshift interval $\Delta z$ between the edge of the proximity zone and the redshift where \oi~\lam1302 enters the \lya\ forest increases with redshift.  This can be seen as a lengthening of the survey paths towards higher redshift in Figure~\ref{fig:los}.  In addition, the absorption pathlength per unit redshift, $dX/dz$, also increases with redshift (Equation~\ref{eq:dzdX}).  The combination of these factors means that the absorption pathlength interval $\Delta X$ over which we searched for \oi\ is a factor of 1.7 larger for a QSO at $z = 6$ than for one at $z = 4$.  This may partially explain the greater incidence rate of multiple detections towards higher redshifts. 

It is nevertheless worth examining whether the \oi\ systems near $z \sim 6$ are clustered.  Some amount of clustering at any redshift is naturally expected due to galaxy clustering.  If, as we propose below, the incidence of \oi\ at $z > 5.7$ is higher than at lower redshift because of a lower ionizing UVB, then additional clustering at these redshifts may be expected if there are also fluctuations in the UVB amplitude \citep[e.g.,][]{finlator2015}.  UVB fluctuations may indeed be present, as they are broadly expected near the tail end of reionization \citep[e.g.,][]{mesinger2009,crociani2011,mcquinn2011a,davies2016,finlator2018,daloisio2018,kulkarni2019,keating2019}, and may be driving the large observed scatter in IGM \lya\ opacity near $z \sim 6$ \citep{fan2006a,becker2015,bosman2018,eilers2018}.  

Here we focus on whether there is significant evidence for clustering, leaving a more sophisticated analysis of the underlying correlation function for future work.  We tested the null hypothesis of no clustering, where \oi\ systems are distributed randomly along the QSO lines of sight, using a Monte Carlo approach to generate mock data sets.  In each of $10^5$ trials we assigned a random number of systems along each line of sight drawn from a Poisson distribution with a mean value equal to $\Delta X$ for that line of sight multiplied by the number density expected from integrating over the best-fitting equivalent width distribution $f(W)$ for $5.7 < z < 6.5$ ($\log{W_0} = -0.93$ and $\log{A} = -0.19$; Table~\ref{tab:results}).  We integrated down to $W_{1302} = 0.02$~\AA, or somewhat lower than the weakest \oi\ detection in our statistical sample.  The systems were assigned equivalent widths by randomly drawing from the $f(W)$ distribution, and random redshifts within the survey interval for each QSO.  We then randomly determined whether the systems were detected using the completeness function for that QSO.  We found that at least five lines of sight yielded at least two \oi\ detections at $z > 5.7$, similar to the observed data, in 18\% of trials.  Two or more lines of sight yielded three or more \oi\ detections at $z > 5.7$, similar to SDSS J1148+5251 and SDSS J0100+2802, in 7\% of the trials.  There is therefore some hint that \oi\ systems at $z \sim 6$ may be clustered, although this test does not strongly rule out the null hypothesis of no clustering.  Stronger constraints may come from a larger and/or more sensitive survey.

\section{Discussion}\label{sec:discussion}

In this section we consider the implications of our observations for the evolution of metal-enriched circumgalactic gas.  Two facts about \oi\ systems are apparent from the data:
\begin{enumerate}
\item{The number density of \oi\ systems is substantially (a factor of $\sim$2--4) lower over $4.1 < z < 5.7$ than over $5.7 < z < 6.5$.}
\item{Over the redshift range of this study ($3.2 < z < 6.5$), \oi\ systems show increasing amounts of \civ\ absorption towards lower redshifts.}
\end{enumerate}
The first point contrasts with the overall trend of metal enrichment expected for the CGM, namely that the metal content of circumgalactic gas should increase with time as metals are driven into the CGM by galactic outflows.  If the ionization balance of these metals remained constant with time, therefore, we would expect the number density of all species, including \oi, to increase with decreasing redshift.  The fact that \oi\ {\it decreases} at $z < 5.7$ presumably then means that a substantial fraction of the circumgalactic metals are transitioning from a relatively neutral state to higher ionization states where  \oi\ is less favored.  In other words, the CGM of galaxies at $z\sim6$ appears to be undergoing reionization. 

The fact that the ionization transition at $z \sim 6$ is relatively rapid (see also Appendix~\ref{app:binning}) suggests that circumgalactic metals are generally ionized by an external radiation field.  There are no obvious changes in the global properties of galaxies at that epoch that would rapidly drive {\it inside-out} ionization of the CGM.  The evolving meta-galactic radiation field during or shortly after reionization is a more likely culprit.  Indeed, the intensity of hydrogen ionizing background is inferred to increase by roughly an order of magnitude from $z \sim 6$ to 5 based on measurements of the opacity of the \lya\ forest \citep{wyithe2011,davies2018b} and the extent of QSO proximity zones \citep{calverley2011}.  As shown in Figure~\ref{fig:OI_dndX_EW_ranges}, we find that much of the evolution in \oi\ absorbers near $z \sim 6$ occurs among weaker systems ($W_{1302} < 0.2$~\AA).  The weaker \oi\ systems may be more sensitive to changes in the UVB if they correspond to lower-density gas.

The reionization of the metal-enriched CGM may occur contemporaneously with reionization of the surrounding IGM as an ionization front sweeps through.  Alternatively, the circumgalactic gas, being more dense, may remain self-shielded for some time after the local IGM becomes ionized.  In the latter case the CGM would become increasingly ionized as the surrounding UV background strengthens.  This should occur near the tail end of reionization as the local mean free path of ionizing photons increases, exposing a given region to ionizing photons from more distant sources \citep[e.g.,][]{mesinger2009,crociani2011,mcquinn2011a}.  In either case, the reionization of the CGM should be coupled to the reionization of the IGM.  Careful modeling is needed to precisely constrain the timing of IGM reionization using metal absorption lines \citep[e.g.,][Doughty, in prep.]{keating2014,finlator2015,finlator2018}.  Broadly speaking, however, the observed evolution in \oi\ suggests that a significant phase of intergalactic---as well as circumgalactic---reionization may have occurred at or not long before $z \sim 6$.  

The evolution of \civ\ in our \oi\ systems supports a picture in which highly ionized metals make up a larger proportion of circumgalactic metals towards lower redshifts.  While some \oi\ absorbers at $z \sim 6$ must transition to more highly ionized states at lower redshifts, however, it is not obvious that a large fraction of the gas producing \oi\ absorption at $z \sim 6$ produces \civ\ absorption at $z < 5$.  For a C/O number density ratio of 0.3, typical of low-metallicity DLAs and high-redshift \oi\ systems \citep{cooke2011a,becker2012}, a fully neutral absorber with \oi~\lam1302 optical depth $\tau_{1302} = 1.0$ that becomes ionized will have a \civ~\lam1548 optical depth $\tau_{1548} = 1.4$ (i.e., easily detectable) if half of the carbon is in \civ.  The \civ\ fraction will depend on the gas density and the spectrum of the ionizing radiation, however, and may be much lower \citep[e.g.,][]{simcoe2011}.  It is possible that some $z \sim 6$ \oi\ systems become mildly ionized absorbers that appear in \cii, \siii, and/or \mgii\ but not \oi.  The \oi\ systems may also initially become absorbers dominated by \ciii\ and \siiii, whose transitions fall in the heavily absorbed \lya\ forest.  In any case, the evolutionary link between \oi\ at $z \sim 6$ and \civ\ at lower redshifts is unclear.  The buildup of \civ\ in \oi\ absorbers with time may simply reflect the ongoing enrichment of the CGM, with \civ\ at $z < 5$ largely tracing metals that were not yet in place at $z\sim6$.

\begin{figure}
   \centering
   \includegraphics[width=0.47\textwidth]{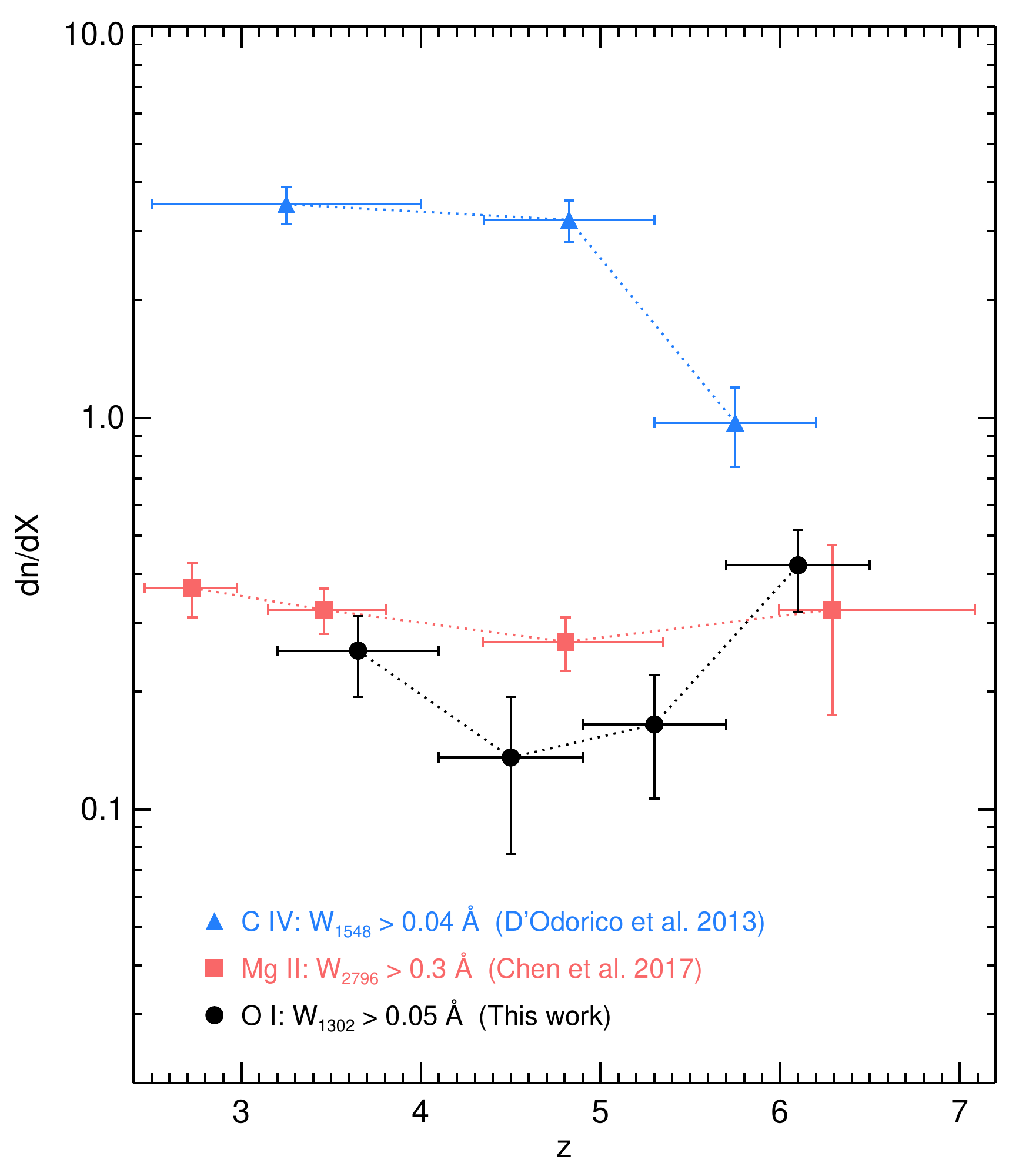}
   \caption{Number density of  \oi\ (circles), \mgii\ (squares), and \civ\ (triangles) systems over $3 \lesssim z \lesssim 7$ down to current equivalent width limits.  Vertical error bars are 68\% confidence intervals.  The results for \oi\ systems with $W_{1302} > 0.05$~\AA\ are from this work.  Results for \mgii\ systems with $W_{2796} > 0.3$~\AA\ are from \citet{chen2017}.  The \civ\ results for systems with $W_{1548} > 0.04$~\AA\ were derived from \citet{dodorico2013} \citep[see also][]{codoreanu2018}.  See text for details.\label{fig:ions_dndX}}
\end{figure}

In Figure~\ref{fig:ions_dndX} we summarize the number density evolution of multiple ions from different surveys out to $z \sim 7$.  The results for \mgii\ systems with \mgii~\lam2796 equivalent width $W_{2796} > 0.3$~\AA\ are from \citet{chen2017}.  For \civ\ we integrated the column density distributions from \citet{dodorico2013} over $\log{(N_{\rm C\, IV}/{\rm cm^{-2}})} > 13.0$, and converted the lower bound in column density into a \civ~\lam1548 equivalent width limit of $W_{1548} > 0.04$~\AA\ assuming Doppler parameters $ b > 10$~\kms.  The rise in \civ\ from $z \sim 6$ to 5 \citep[similar to that found by][]{codoreanu2018} contrasts with the drop in \oi\ over the same redshifts, while \mgii\ remains relatively flat.

We caution that the equivalent width limits in Figure~\ref{fig:ions_dndX}, which are generally set by the quality of the data at the highest redshifts, may complicate the comparison of different ions.  For example, the number density of \oi\ systems we measure at $z \sim 6$ is nominally somewhat higher than the number density of \mgii\ systems found by \citet{chen2017}, but this may be due to differences in sensitivity.  For an absorber with a column density ratio $N_{\rm O\,I} / N_{\rm Mg\, II}$ equal to the solar O/Mg ratio \citep{asplund2009}, an optically thin system would have an \mgii~\lam2796 equivalent width, in angstroms, roughly twice that of \oi~\lam1302.  In that sense the \citet{chen2017} limit of $W_{2796} > 0.3$ is only a factor of $\sim$3 above our \oi\ limit of $W_{1302} > 0.05$~\AA\ for weak low-ionization systems.  According to our best fit to Equation~\ref{eq:fW}, increasing our \oi\ limit by a factor of three, from $W > 0.05$~\AA\ to $W > 0.15$~\AA, would yield a factor of $\sim$2 fewer \oi\ systems at $z \sim 6$, i.e., somewhat lower than the number density of \mgii\ systems with $W_{2796} > 0.3$~\AA\ found by \citet{chen2017} though still within the error bars.  There may be significant numbers of weak \mgii\ absorbers at $z \sim 6$ without \oi, but further work will be needed to determine whether this is the case.  \citet{bosman2017} found that  an abundant population of \mgii\ systems at $z \sim 6$ may exist below the detection limit of \citet{chen2017}, although the evidence comes from only one line of sight.    

We also emphasize that the number densities plotted in Figure~\ref{fig:ions_dndX} are dominated by the weakest systems, and do not necessarily reflect the evolution in the total mass density.  A more comprehensive picture would come from examining how the full equivalent width (or column density) distributions of these ions evolve with redshift.  Nevertheless, these trends should already place strong constraints on models of CGM enrichment and ionization.

Finally, we briefly comment on two possible implications of our \oi\ results for the radiation emitted from high-redshift galaxies.  The higher \oi\ number density at $z > 5.7$ implies that, globally, the mean projected cross-section of largely neutral, optically thick gas around galaxies is higher at $z \gtrsim 6$ than at $z \sim 4$--5.  This could imply a smaller escape fraction of ionizing photons at $z \sim 6$ compared with lower redshifts, at odds with reionization models that require higher average escape fractions at higher redshifts \citep[e.g.,][]{kuhlen2012,haardt2012}.  On the other hand, ionizing photons may largely escape from a galaxy's interstellar and circumgalactic media through channels of low \hi\ column density \citep[e.g.,][]{ma2016,steidel2018}.  The number of such channels will depend on the three-dimensional geometry of the optically thick gas \citep[e.g.,][]{fernandez2011}, which can vary for a given two-dimensional cross-section.  The hosts of \oi\ absorbers and the galaxies that dominate the ionizing emissivity may also be separate populations.  A higher global cross-section of neutral gas could also have implications for galaxy \lya\ emission.  If a significant fraction of the \lya\ emission from a galaxy is scattered within optically thick regions of the CGM \citep[e.g.,][]{rauch2008,steidel2011,wisotzki2018}, then a larger neutral cross-section could potentially correspond to a more extended, lower surface brightness \lya\ halo.  This would make galaxy \lya\ emission more difficult to detect, and could be partially responsible for the lower fraction of galaxies that appear as \lya\ emitters at $z > 6$ \citep[e.g.,][and references therein]{hoag2019,mason2019}.  The significance of these effects is difficult to determine without further study.

\section{Summary}\label{sec:summary}

We conducted a survey for metal absorbers traced by \oi\ over $3.2 < z < 6.5$.  Using moderate-resolution spectra of a 199 QSOs, we find that the number density of systems with \oi\ equivalent width $W_{1302} > 0.05$~\AA\ decreases by a factor of $2.5^{+1.6}_{-0.8}$ (68\% confidence) from $5.7 < z < 6.5$ to $4.9 < z < 5.7$, with a decrease at some level favored with 99\% confidence.  Much of the decline occurs among weak ($W_{1302} < 0.2$~\AA) absorbers.  The number density then inflects towards an increasing trend with decreasing redshift over $3.2 < z < 5.7$.  

The decrease in \oi\ at $z < 5.7$ runs contrary to the general expectation that the overall metal content of circumgalactic gas should increase with time, and implies that metal-enriched gas at $z \sim 6$ tends to be in a more neutral state compared to lower redshifts.  Supporting this picture, we find that the amount of absorption by highly ionized metals traced by \civ\ associated with \oi\ systems increases with decreasing redshift \citep[see also][]{cooper2019}.

Our \oi\ results suggests that the metal-enriched gas around galaxies undergoes an ionization transition near $z \sim 6$ driven by a strengthening metagalactic ionizing background.  Such an increase in the UVB is expected near the end of hydrogen reionization.  The reionization of the CGM seen in \oi\ therefore adds to the growing observational evidence that the reionization of the IGM may have been ongoing or had only recently ended at $z \sim 6$.  The evolution in the CGM neutral fraction may also carry implications for the Lyman continuum and/or \lya\ emission from galaxies at $z \gtrsim 6$.

Further observations of \oi\ and other ions will help to clarify how metal enrichment and ionization proceed near reionization.  Larger surveys would help to determine the rate at which circumgalactic metals undergo the ionization transition detected here near $z \sim 6$.  More sensitive data at $z > 5$ would give further insight into the weak \oi\ systems that seem to evolve the most strongly.  Finally, pushing the search for \oi\ and other metals to even higher redshifts would help to better understand the connection between the evolution of the CGM and the reionization of the IGM.  The growing number of QSOs being discovered at $ z > 7$ \citep[e.g.,][]{banados2017} should make this possible.

\acknowledgments

We thank Kristian Finlator, Anson D'Aloisio, and Simeon Bird for helpful discussions and suggestions.  GB and EB were supported by the National Science Foundation through grants AST-1615814 and AST-1751404.  LC is supported by the Independent Research Fund Denmark (DFF Ð4090-00079).  MF acknowledges support by the Science and Technology Facilities Council [grant number  ST/P000541/1]. This project has received funding from the European Research Council (ERC) under the European Union's Horizon 2020 research and innovation program (grant agreement No 757535).  SL was funded by projects UCh/VID-ENL18/18 and FONDECYT  1191232.  MN acknowledges support from ERC Advanced grant 740246 (Cosmic\_Gas).  ERW acknowledges the Australian Research Council Centre of Excellence for All Sky Astrophysics in 3 Dimensions (ASTRO 3D), through project number CE170100013

This work is based in part on observations made with ESO Telescopes at the La Silla Paranal Observatory under program IDs 060.A-9024, 084.A-0360, 084.A-0390, 084.A-0550, 085.A-0299, 086.A-0162, 086.A-0574, 087.A-0607, 087.A-0890, 088.A-0897, 091.C-0934, 096.A-0095, 096.A-0418, 097.B-1070, 098.A-0111, 098.B-0537, 0100.A-0243, 0100.A-0625, 0102.A-0154, 189.A-0424, and 294.A-5031.  Further observations were made at the W.M. Keck Observatory, which is operated as a scientific partnership between the California Institute of Technology and the University of California; it was made possible by the generous support of the W.M. Keck Foundation.  The authors wish to recognize and acknowledge the very significant cultural role and reverence that the summit of Maunakea has always had within the indigenous Hawaiian community.  We are most fortunate to have the opportunity to conduct observations from this mountain.  Finally, this research has made use of the Keck Observatory Archive (KOA), which is operated by the W. M. Keck Observatory and the NASA Exoplanet Science Institute (NExScI), under contract with the National Aeronautics and Space Administration.

\appendix

\twocolumngrid

\section{Alternate Redshift Binning}\label{app:binning}

\begin{figure}
   \centering
   \includegraphics[width=0.47\textwidth]{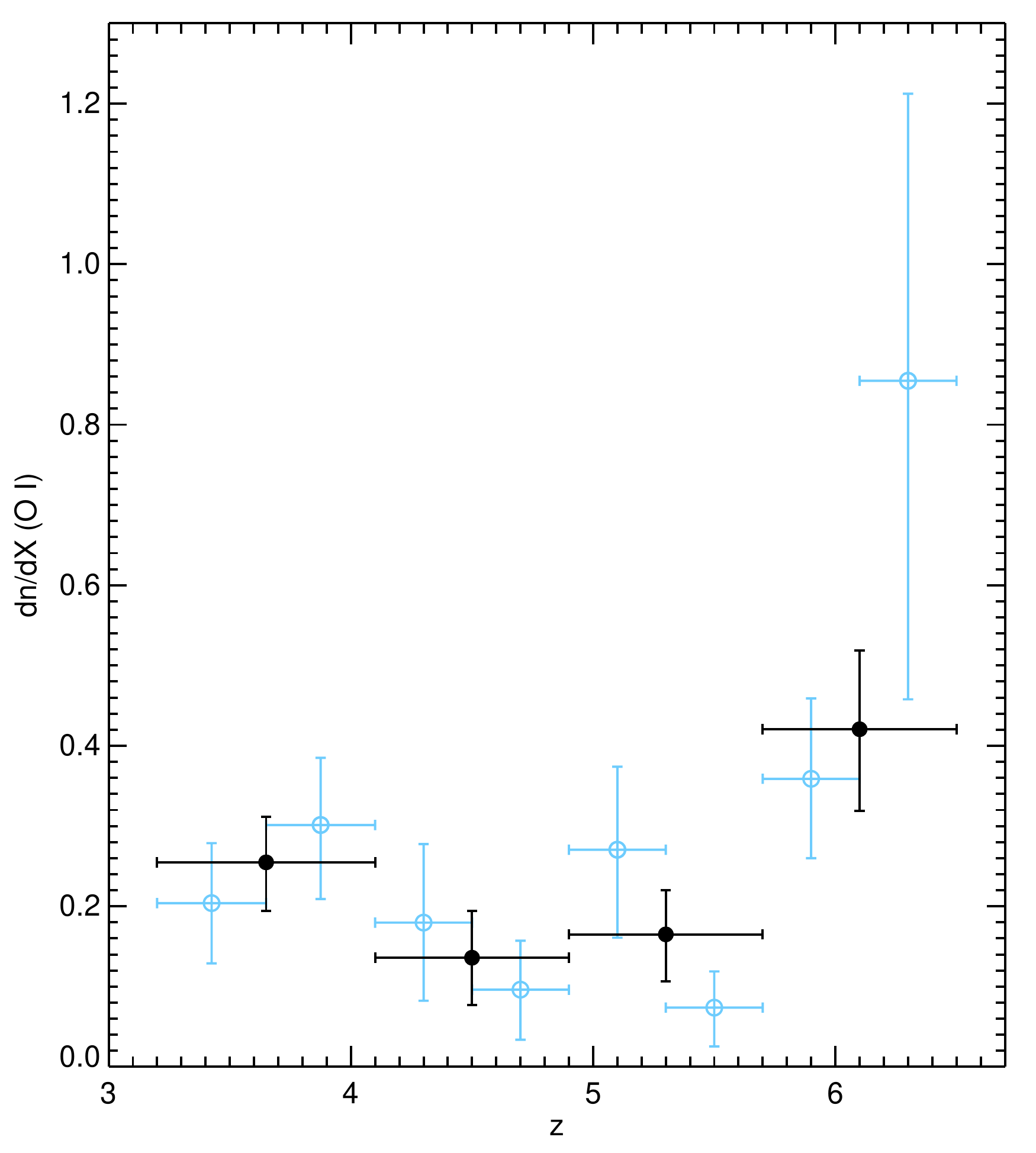}
   \caption{Comoving number density of \oi\ systems with $W_{1302} > 0.05$~\AA\ as a function of redshift.  Filled circles are the same as in Figure~\ref{fig:dndX} and use bin sizes $\Delta z = 0.8$--0.9.  Open circles use redshift bins $\Delta z = 0.4$--0.45.\label{fig:OI_dndX_dz0p4}}
\end{figure}

In this appendix we explore the use of smaller redshift bins for tracing the number density of \oi\ systems.  We repeated the procedure described in Sections~\ref{sec:distributions} and \ref{sec:density}, but divided each of our nominal redshift bins into two such that the bins sizes are $\Delta z = 0.45$ over $3.2 < z < 4.1$ and $\Delta z = 0.4$ over $4.1 < z < 6.5$.  The $dn/dX$ results for absorbers with $W_{1302} > 0.05$~\AA\ are shown in Figure~\ref{fig:OI_dndX_dz0p4}.  Over $3.2 < z < 5.7$ we see the same general trend of a flat or increasing number density with decreasing redshift, albeit with larger errors.  There is some evidence that the decline with decreasing redshift near $z \sim 6$ may be steeper than suggested by the $\Delta z \simeq 0.8$ bins.  This comes primarily from the high nominal value of $dn/dX$ at $z = 6.3$, though the uncertainty on this point is large.  There are five \oi\ systems over $6.1 < z < 6.5$ in our statistical sample, all of which have $W_{1302} < 0.17$ (although the proximate system at $z = 6.1242$ towards PSO~J065-26 has $W_{1302} \simeq 0.7$).  Completeness corrections are necessarily large in this bin, a factor of 3.2 overall in $dn/dX$ for the nominal fit to $f(W)$.  With such a small sample it is also unclear whether a single exponential is a reasonable model for $f(W)$.  While the finer redshift binning gives some hint that the evolution near $z \sim 6$ may be even more substantial than indicated by the nominal $\Delta z \simeq 0.8$ bins, therefore, the results push the limits of what can be learned from the current data.  More detailed constraints on \oi\ evolution at these redshifts will require a larger and/or more sensitive survey.

\section{Details on Individual Systems}\label{app:details}

Here we provide more detailed information on individual absorption systems.  The systems are plotted in Figures~\ref{fig:z3p2429550_J0100-2708}--\ref{fig:z6p2575350_SDSSJ1148+5251}.  Solid shading marks the region over which an equivalent width was integrated.  Hatched shading denotes lines that are either blended with an unrelated absorber or contaminated by strong telluric residuals.  
 
Equivalent width measurements for all ions are given in Table~\ref{tab:ew}.  Notes on individual systems are given below.  Blended lines are mentioned in the notes only in cases where a correction for blending has been made (see Section~\ref{sec:measurements}) or the blend is questionable.  Other blends are reported in Table~\ref{tab:ew} as upper limits on the equivalent width.

\begin{itemize}

\item{$z=3.6287$ towards J0042-1020 (Figure~\ref{fig:z3p6287100_J0042-1020}): \oi~\lam1302 is extremely weak compared to \cii~\lam1334\ and \siii~\lam1526, an indication that the gas traced by the low-ionization lines is significantly ionized in this system.  Among the XQ-100 lines of sight, this is the only intervening (not proximate) absorber not identified as a DLAs or sub-DLA \citep[][and in prep]{sanchez-ramirez2016,berg2016}.}

\item{$z=3.7013$ towards J2215-1611 (Figure~\ref{fig:z3p7013160_J2215-1611}): This system exhibits self-blending in \civ.  Deblended equivalent widths measured are reported in Table~\ref{tab:ew}.  We note that \mgii\ and the high-ionization lines exhibit two extended ($\Delta v \sim 100$~\kms) components separated by 500~\kms, which is very similar to the intrinsic \civ\ doublet separation.    A similar separation is seen in the \civ\ components of the $z=3.9557$ towards J0835+0650 (Figure~\ref{fig:z3p9556800_J0835+0650}).  Although it may occur by chance, this separation is reminiscent of a line-locking signature sometimes seen in radiately driven outflows \citep[e.g.,][]{milne1926,scargle1973,bowler2014}.  It is unclear whether \oi\ or \cii\ are present in the $+$500~\kms\ component.  \oi\ is blended with the \siii~\lam1304 component near $+$0~\kms.  There is a potential \cii\ line, but it does not match the velocity profile of \mgii.  This component is therefore not included in the equivalent widths measurements for \oi\ and \cii, although it would be a relatively small addition in both cases.}

\item{$z=3.8039$ towards J1032+0927 (Figure~\ref{fig:z3p8039370_J1032+0927}): \oi\ is blended with weak \civ~\lam1550 at $z=3.0337$.  The \civ\ lines also exhibit self-blending.  In both cases deblended equivalent widths are reported in Table~\ref{tab:ew}.}

\item{$z=3.8079$ towards J0415-4357 (Figure~\ref{fig:z3p8079200_J0415-4357}): \oi\ for this system falls near a complex of \nv\ absorption near the redshift of the QSO.  The equivalent width reported in Table~\ref{tab:ew} has been corrected for blending with mild \nv~\lam1242 absorption at $z=4.0383$ and moderate \nv~\lam1238 absorption at $z=4.0526$.}

\item{$z=3.9557$ towards J0835+0650 (Figure~\ref{fig:z3p9556800_J0835+0650}): This system exhibits self-blending in \civ.  Deblended equivalent widths are reported in Table~\ref{tab:ew}.  We note that, as with the $z=3.701$ system towards J2215-1611 (Figure~\ref{fig:z3p7013160_J2215-1611}), the high-ionization lines exhibit two extended components separated by 500~\kms, which is very similar to the intrinsic \civ\ doublet separation.}

\item{$z=4.0742$ towards J0132+1341 (Figure~\ref{fig:z4p0741980_J0132+1341}): This system exhibits self-blending in \civ.  Deblended equivalent widths are reported in Table~\ref{tab:ew}.}

\item{$z=4.1401$ towards J0247-0556 (Figure~\ref{fig:z4p1401050_J0247-0556}): This system contains multiple weak low-ionization components spanning $\sim$600~\kms.  The reddest component is somewhat tentative, but appears to be detected in \cii, \oi, and \siii~\lam1526.   There may also be weak high-ionization lines present in the reddest component, but they do not add significantly to the overall equivalent width.}

\item{$z=4.8555$ towards J1044+2025 (Figure~\ref{fig:z4p8554890_J1004+2025}): \siii~\lam1260 is detected but falls right at the start of the QSO proximity zone.  Contamination from \lya\ and significant continuum uncertainties are therefore possible for this line.}

\item{$z=5.0615$ towards J1147-01095 (Figure~\ref{fig:z5p0615260_J1147-0109}): \siiv~\lam1394 is a possible blend based on the lack of obvious absorption in \civ.}

\item{$z=5.2961$ towards J1335-0328 (Figure~\ref{fig:z5p2960670_J1335-0328}): This system is somewhat tentative as only \mgii~\lam2796 is present and apparently un-blended.  \mgii~\lam2803 and \cii~\lam1334\ are blended with skyline residuals.  \oi\ is blended with \civ~\lam1548 at $z=4.2962$.  The deblended \oi\ equivalent width reported in Table~\ref{tab:ew}, $W_{1302} =  0.040 \pm 0.007$~\AA, falls below our cutoff of 0.05~\AA\ for constraining $f(W)$.  Although it is tentative, therefore, this system does not impact our results.}

\item{$z=5.8786$ towards PSOJ308-21 (Figure~\ref{fig:z5p8785790_PSOJ308-21}): \oi\ is blended with \nv~\lam1238 at $z=6.2304$.  A deblended \oi\ equivalent width is reported in Table~\ref{tab:ew}.  The \siii~\lam1304 line appears to be a blend based on the lack of a stronger \siii~\lam1526 line; the blend may be with \civ~\lam1550 at $z=4.7952$.}

\item{$z=5.8987$ towards ATLASJ158-14 (Figure~\ref{fig:z5p8986670_ATLASJ158-14}): \oi\ falls in a patch of \civ~\lam1548 lines near $z=4.8$, with \siii~\lam1334 in the corresponding patch of \civ~\lam1550.  The relative weakness of the \siii~\lam1526 line suggests that the \siii~\lam1334 line is dominated by contamination.  We therefore de-blend the \oi\ line by assuming all of the absorption near \siii~\lam1334 is \civ~\lam1550.  The deblended equivalent width is reported in Table~\ref{tab:ew}.}

\item{$z=5.9450$ towards SDSS J0100+2802 (Figure~\ref{fig:z5p9449860_SDSSJ0100+2802}): \cii\ is blended with weak \civ~\lam1550 at $z=4.9766$.  The deblended equivalent width measured from the X-Shooter spectrum is reported in Table~\ref{tab:ew}.}

\item{$z=6.0114$ towards SDSS J1148+5251 (Figure~\ref{fig:z6p0113980_SDSSJ1148+5251}): \siii~\lam1304 is blended in the ESI spectrum but unblended in HIRES.  We therefore use the HIRES data to measure its equivalent width \citep[see][]{becker2011a}.}

\item{$z=6.1242$ towards PSO J065-26 (Figure~\ref{fig:z6p1242130_PSOJ065-26}): The equivalent width for \siii~\lam1304 does not include the component at $\Delta v \simeq 160$~\kms, which appears to be an unrelated intervening line.}

\item{$z=6.1314$ towards SDSS J1148+5251 (Figure~\ref{fig:z6p1313540_SDSSJ1148+5251}): The equivalent width for \siii~\lam1304 is measured from the HIRES spectrum, where it is better resolved from an adjacent line \citep[see][]{becker2011a}.}

\item{$z=6.2575$ towards SDSS J1148+5251 (Figure~\ref{fig:z6p2575350_SDSSJ1148+5251}): The equivalent width for \siii~\lam1260 is measured from the HIRES spectrum, where it is resolved from an adjacent line \citep[see][]{becker2011a}.}

\end{itemize}

\begin{longrotatetable}
\begin{deluxetable*}{llcccccccccccc}
\setlength{\tabcolsep}{4pt}
\tabletypesize{\tiny}
\tablecaption{Rest-frame equivalent width measurements\label{tab:ew}}
\tablehead{
   \colhead{$z$}  &  \colhead{QSO}  &  \colhead{\oi~\lam1302}  &   \colhead{\cii~\lam1334}  &   \colhead{\siii~\lam1260}  &   \colhead{\siii~\lam1304}  &   \colhead{\siii~\lam1526}  &   \colhead{\mgii~\lam2796}  &  \colhead{\mgii~\lam2803}  &  \colhead{\siiv~\lam1393}  &   \colhead{\siiv~\lam1402}  &   \colhead{\civ~\lam1548}  &   \colhead{\civ~\lam1550}  &  \colhead{Figure}}
\colnumbers
\startdata
\input{EW_table.tex}
\enddata
\tablecomments{Columns: (1) absorber redshift, (2) QSO name, (3)--(13) rest-frame equivalent widths, in \AA, (14) figure number}
\tablenotetext{a}{Proximate absorber with 5000~\kms\ of the QSO redshift}
\tablenotetext{b}{See notes on this system in Appendix~\ref{app:details}.}
\tablenotetext{c}{Blended line}
\tablenotetext{d}{Measurement from a de-blended line}
\tablenotetext{e}{Missing data due to contamination from telluric residuals}
\tablenotetext{f}{\mgii~\lam2796 equivalent width derived from \mgii~\lam2803 line}
\tablenotetext{g}{\civ~\lam1548 equivalent width derived from \civ~\lam1550 line}
\tablenotetext{h}{Formal detection with large error.  Should be treated with caution.}
\end{deluxetable*}
\end{longrotatetable}

\input{system_figures.tex}

\bibliographystyle{aasjournal} \bibliography{oi_survey_refs}

\end{document}

%% file: EW_table.tex
    3.2430  &  J0100-2708          &  0.099$\pm$0.010  &  0.138$\pm$0.004  &  \nodata  &  0.057$\pm$0.010  &  0.064$\pm$0.007  &  0.283$\pm$0.021  &  $<$0.309\tablenotemark{c}  &  0.017$\pm$0.006  &  0.013$\pm$0.006  &  0.032$\pm$0.005  &  0.017$\pm$0.006  &  \ref{fig:z3p2429550_J0100-2708}  \\
    3.3845  &  J1018+0548          &  0.114$\pm$0.007  &  0.209$\pm$0.008  &  0.207$\pm$0.005  &  $<$0.016  &  0.068$\pm$0.008  &  0.578$\pm$0.029  &  0.290$\pm$0.061  &  0.207$\pm$0.009  &  0.115$\pm$0.009  &  0.275$\pm$0.008  &  0.141$\pm$0.008  &  \ref{fig:z3p3844720_J1018+0548}  \\
    3.3963  &  J1108+1209          &  0.195$\pm$0.002  &  0.135$\pm$0.004  &  \nodata  &  $<$0.104\tablenotemark{c}  &  $<$0.116\tablenotemark{c}  &  0.317$\pm$0.012  &  0.221$\pm$0.043  &  0.186$\pm$0.007  &  0.138$\pm$0.008  &  0.266$\pm$0.007  &  0.156$\pm$0.008  &  \ref{fig:z3p3962790_J1108+1209}  \\
    3.4423  &  J1552+1005          &  0.043$\pm$0.004  &  0.111$\pm$0.004  &  \nodata  &  $<$0.290\tablenotemark{c}  &  0.039$\pm$0.004  &  0.278$\pm$0.023  &  0.177$\pm$0.009  &  0.240$\pm$0.006  &  $<$0.260\tablenotemark{c}  &  0.437$\pm$0.007  &  0.276$\pm$0.007  &  \ref{fig:z3p4422810_J1552+1005}  \\
    3.4484  &  J1421-0643          &  0.271$\pm$0.005  &  0.374$\pm$0.006  &  \nodata  &  0.156$\pm$0.005  &  0.242$\pm$0.006  &  0.887$\pm$0.014  &  0.705$\pm$0.017  &  0.381$\pm$0.010  &  0.241$\pm$0.011  &  0.674$\pm$0.018\tablenotemark{c}\tablenotemark{g}  &  0.399$\pm$0.010  &  \ref{fig:z3p4483820_J1421-0643}  \\
    3.5454  &  J1108+1209          &  0.876$\pm$0.005  &  0.835$\pm$0.005  &  0.832$\pm$0.003  &  0.590$\pm$0.007  &  0.820$\pm$0.006  &  1.806$\pm$0.020  &  1.808$\pm$0.023  &  0.308$\pm$0.008  &  0.166$\pm$0.008  &  0.347$\pm$0.006  &  0.195$\pm$0.007  &  \ref{fig:z3p5453680_J1108+1209}  \\
    3.5804\tablenotemark{a}  &  J0056-2808          &  0.136$\pm$0.008  &  0.306$\pm$0.006  &  $<$0.410\tablenotemark{c}  &  0.161$\pm$0.007  &  0.223$\pm$0.005  &  0.643$\pm$0.030  &  0.604$\pm$0.011  &  0.371$\pm$0.012  &  0.263$\pm$0.011  &  0.649$\pm$0.007  &  0.438$\pm$0.007  &  \ref{fig:z3p5803980_J0056-2808}  \\
    3.6009  &  J1552+1005          &  0.459$\pm$0.004  &  0.428$\pm$0.004  &  0.425$\pm$0.003  &  0.293$\pm$0.004  &  0.353$\pm$0.004  &  1.002$\pm$0.011  &  0.894$\pm$0.013  &  0.062$\pm$0.004  &  0.038$\pm$0.004  &  0.055$\pm$0.008\tablenotemark{c}\tablenotemark{g}  &  0.028$\pm$0.004  &  \ref{fig:z3p6008980_J1552+1005}  \\
    3.6078  &  J1111-0804          &  0.206$\pm$0.001  &  0.234$\pm$0.002  &  \nodata  &  0.124$\pm$0.002  &  0.165$\pm$0.003  &  0.537$\pm$0.011  &  0.472$\pm$0.012  &  0.395$\pm$0.007  &  0.197$\pm$0.007  &  0.644$\pm$0.007  &  0.366$\pm$0.007  &  \ref{fig:z3p6077840_J1111-0804}  \\
    3.6287\tablenotemark{b}  &  J0042-1020          &  0.035$\pm$0.006  &  0.508$\pm$0.006  &  \nodata  &  $<$0.294\tablenotemark{c}  &  0.203$\pm$0.007  &  1.187$\pm$0.031  &  0.939$\pm$0.031  &  0.664$\pm$0.006  &  0.461$\pm$0.007  &  0.858$\pm$0.006  &  0.567$\pm$0.007  &  \ref{fig:z3p6287100_J0042-1020}  \\
    3.6619  &  J2215-1611          &  0.250$\pm$0.001  &  0.309$\pm$0.003  &  \nodata  &  $<$0.185\tablenotemark{c}  &  0.150$\pm$0.005  &  0.651$\pm$0.006  &  0.549$\pm$0.007  &  0.354$\pm$0.007  &  0.143$\pm$0.007  &  0.243$\pm$0.008  &  0.157$\pm$0.009  &  \ref{fig:z3p6618780_J2215-1611}  \\
    3.6666\tablenotemark{a}  &  J1552+1005          &  0.607$\pm$0.006  &  0.802$\pm$0.006  &  0.816$\pm$0.005  &  0.278$\pm$0.007  &  0.523$\pm$0.006  &  1.736$\pm$0.024  &  1.293$\pm$0.035  &  0.086$\pm$0.006  &  0.030$\pm$0.006  &  0.205$\pm$0.005  &  0.102$\pm$0.006  &  \ref{fig:z3p6665670_J1552+1005}  \\
    3.7013\tablenotemark{b}  &  J2215-1611          &  0.199$\pm$0.003  &  0.430$\pm$0.004  &  \nodata  &  0.125$\pm$0.003  &  0.135$\pm$0.006  &  1.041$\pm$0.020  &  0.756$\pm$0.014  &  0.437$\pm$0.009  &  0.220$\pm$0.010  &  0.382$\pm$0.014  &  0.188$\pm$0.010  &  \ref{fig:z3p7013160_J2215-1611}  \\
    3.7212  &  J0214-0517          &  0.213$\pm$0.003  &  0.253$\pm$0.008  &  \nodata  &  0.136$\pm$0.003  &  0.226$\pm$0.004  &  0.490$\pm$0.015  &  \nodata\tablenotemark{e}  &  0.352$\pm$0.006  &  0.170$\pm$0.006  &  0.353$\pm$0.007  &  0.183$\pm$0.007  &  \ref{fig:z3p7212440_J0214-0517}  \\
    3.7343  &  J0311-1722          &  0.151$\pm$0.003  &  0.148$\pm$0.003  &  \nodata  &  0.036$\pm$0.003  &  0.046$\pm$0.004  &  0.411$\pm$0.028  &  0.265$\pm$0.009  &  0.073$\pm$0.007  &  0.047$\pm$0.008  &  0.203$\pm$0.009  &  0.103$\pm$0.010  &  \ref{fig:z3p7342600_J0311-1722}  \\
    3.8039\tablenotemark{b}  &  J1032+0927          &  0.082$\pm$0.005\tablenotemark{d}  &  0.097$\pm$0.005  &  \nodata  &  0.014$\pm$0.005  &  0.038$\pm$0.005  &  0.252$\pm$0.016  &  0.164$\pm$0.018  &  0.203$\pm$0.014  &  0.138$\pm$0.015  &  0.954$\pm$0.016  &  0.651$\pm$0.011  &  \ref{fig:z3p8039370_J1032+0927}  \\
    3.8079\tablenotemark{b}  &  J0415-4357          &  0.605$\pm$0.019\tablenotemark{d}  &  0.762$\pm$0.011  &  \nodata  &  $<$0.719\tablenotemark{c}  &  0.405$\pm$0.023  &  1.711$\pm$0.031  &  1.467$\pm$0.061  &  $<$0.311\tablenotemark{c}  &  0.138$\pm$0.016  &  0.253$\pm$0.015  &  0.118$\pm$0.014  &  \ref{fig:z3p8079200_J0415-4357}  \\
    3.9007  &  J0747+2739          &  0.360$\pm$0.007  &  0.420$\pm$0.010  &  \nodata  &  0.140$\pm$0.008  &  0.233$\pm$0.010  &  \nodata\tablenotemark{e}  &  \nodata\tablenotemark{e}  &  0.236$\pm$0.014  &  $<$0.146\tablenotemark{c}  &  0.695$\pm$0.017  &  \nodata\tablenotemark{e}  &  \ref{fig:z3p9006820_J0747+2739}  \\
    3.9124  &  J0959+1312          &  0.143$\pm$0.002  &  0.251$\pm$0.002  &  0.214$\pm$0.001  &  0.046$\pm$0.002  &  0.079$\pm$0.004  &  \nodata\tablenotemark{e}  &  \nodata\tablenotemark{e}  &  $<$0.423\tablenotemark{c}  &  0.067$\pm$0.003  &  0.182$\pm$0.008\tablenotemark{c}\tablenotemark{g}  &  0.099$\pm$0.004  &  \ref{fig:z3p9124130_J0959+1312}  \\
    3.9146\tablenotemark{a}  &  J0255+0048          &  0.301$\pm$0.006  &  0.274$\pm$0.005  &  0.297$\pm$0.003  &  0.232$\pm$0.006  &  0.281$\pm$0.005  &  0.697$\pm$0.037  &  0.560$\pm$0.018  &  0.080$\pm$0.005  &  0.069$\pm$0.005  &  0.030$\pm$0.014  &  0.059$\pm$0.005  &  \ref{fig:z3p9146190_J0255+0048}  \\
    3.9362  &  J0132+1341          &  0.214$\pm$0.004  &  0.237$\pm$0.004  &  \nodata  &  0.094$\pm$0.004  &  0.141$\pm$0.005  &  \nodata\tablenotemark{e}  &  \nodata\tablenotemark{e}  &  0.173$\pm$0.007  &  $<$0.503\tablenotemark{c}  &  0.212$\pm$0.014  &  0.057$\pm$0.012  &  \ref{fig:z3p9362010_J0132+1341}  \\
    3.9465\tablenotemark{a}  &  J0800+1920          &  0.330$\pm$0.004  &  0.397$\pm$0.004  &  0.349$\pm$0.004  &  0.123$\pm$0.005  &  0.201$\pm$0.004  &  1.010$\pm$0.052\tablenotemark{e}\tablenotemark{f}  &  0.781$\pm$0.038  &  0.207$\pm$0.005  &  0.135$\pm$0.005  &  0.588$\pm$0.004  &  0.351$\pm$0.003  &  \ref{fig:z3p9464860_J0800+1920}  \\
    3.9557\tablenotemark{a}\tablenotemark{b}  &  J0835+0650          &  0.289$\pm$0.005  &  $<$0.442\tablenotemark{c}  &  0.355$\pm$0.004  &  0.130$\pm$0.005  &  0.237$\pm$0.005  &  0.998$\pm$0.039\tablenotemark{e}\tablenotemark{f}  &  0.784$\pm$0.031  &  0.591$\pm$0.011  &  0.287$\pm$0.011  &  0.849$\pm$0.014  &  0.500$\pm$0.010  &  \ref{fig:z3p9556800_J0835+0650}  \\
    3.9887  &  J2251-1227          &  0.037$\pm$0.004  &  0.048$\pm$0.004  &  0.062$\pm$0.001  &  0.013$\pm$0.004  &  0.034$\pm$0.007  &  \nodata\tablenotemark{e}  &  \nodata\tablenotemark{e}  &  0.029$\pm$0.007  &  $<$0.074\tablenotemark{c}  &  0.074$\pm$0.005  &  0.041$\pm$0.005  &  \ref{fig:z3p9886920_J2251-1227}  \\
    3.9961  &  J0133+0400          &  0.307$\pm$0.003  &  0.388$\pm$0.003  &  0.413$\pm$0.002  &  0.101$\pm$0.004  &  0.166$\pm$0.007  &  \nodata\tablenotemark{e}  &  \nodata\tablenotemark{e}  &  $<$0.500\tablenotemark{c}  &  0.208$\pm$0.006  &  0.376$\pm$0.005  &  0.205$\pm$0.006  &  \ref{fig:z3p9960810_J0133+0400}  \\
    4.0656  &  J0529-3552          &  0.039$\pm$0.008  &  0.123$\pm$0.007  &  0.122$\pm$0.005  &  0.047$\pm$0.008  &  0.144$\pm$0.008  &  \nodata\tablenotemark{e}  &  \nodata\tablenotemark{e}  &  0.241$\pm$0.011  &  0.159$\pm$0.011  &  0.236$\pm$0.016  &  0.129$\pm$0.019  &  \ref{fig:z4p0655950_J0529-3552}  \\
    4.0742\tablenotemark{a}\tablenotemark{b}  &  J0132+1341          &  0.039$\pm$0.005  &  0.173$\pm$0.004  &  0.143$\pm$0.004  &  0.065$\pm$0.005  &  0.084$\pm$0.005  &  \nodata\tablenotemark{e}  &  \nodata\tablenotemark{e}  &  0.471$\pm$0.011  &  0.291$\pm$0.010  &  0.893$\pm$0.016  &  0.551$\pm$0.011  &  \ref{fig:z4p0741980_J0132+1341}  \\
    4.0979  &  J0839+0318          &  0.064$\pm$0.006  &  0.091$\pm$0.006  &  0.076$\pm$0.003  &  $<$0.012  &  $<$0.018  &  \nodata\tablenotemark{e}  &  \nodata\tablenotemark{e}  &  0.398$\pm$0.011  &  0.230$\pm$0.012  &  0.434$\pm$0.015  &  $<$0.425\tablenotemark{c}  &  \ref{fig:z4p0979260_J0839+0318}  \\
    4.1401\tablenotemark{b}  &  J0247-0556          &  0.146$\pm$0.015  &  0.387$\pm$0.014  &  $<$0.507\tablenotemark{c}  &  0.115$\pm$0.015  &  0.068$\pm$0.019  &  \nodata\tablenotemark{e}  &  \nodata\tablenotemark{e}  &  0.515$\pm$0.011  &  0.645$\pm$0.010  &  0.820$\pm$0.011  &  0.595$\pm$0.013  &  \ref{fig:z4p1401050_J0247-0556}  \\
    4.1748  &  J1036-0343          &  0.084$\pm$0.003  &  0.077$\pm$0.003  &  \nodata  &  0.011$\pm$0.003  &  $<$0.008  &  0.139$\pm$0.029  &  \nodata\tablenotemark{e}  &  0.061$\pm$0.004  &  0.033$\pm$0.005  &  0.099$\pm$0.006  &  0.032$\pm$0.006  &  \ref{fig:z4p1747810_J1036-0343}  \\
    4.2281\tablenotemark{a}  &  J0234-1806          &  0.346$\pm$0.008  &  0.926$\pm$0.008  &  0.915$\pm$0.005  &  0.296$\pm$0.008  &  0.516$\pm$0.010  &  2.013$\pm$0.038  &  1.929$\pm$0.088\tablenotemark{h}  &  0.561$\pm$0.010  &  0.348$\pm$0.010  &  0.494$\pm$0.009  &  0.332$\pm$0.009  &  \ref{fig:z4p2281300_J0234-1806}  \\
    4.2475  &  J1723+2243          &  0.037$\pm$0.002  &  0.205$\pm$0.002  &  \nodata  &  0.039$\pm$0.002  &  0.053$\pm$0.006  &  0.496$\pm$0.009  &  0.348$\pm$0.015  &  0.168$\pm$0.004  &  0.097$\pm$0.005  &  0.200$\pm$0.004  &  0.087$\pm$0.005  &  \ref{fig:z4p2474650_J1723+2243}  \\
    4.2524\tablenotemark{a}  &  J0034+1639          &  0.203$\pm$0.003  &  0.207$\pm$0.004  &  0.195$\pm$0.003  &  0.075$\pm$0.004  &  0.098$\pm$0.006  &  0.444$\pm$0.009  &  0.452$\pm$0.012  &  0.114$\pm$0.007  &  0.084$\pm$0.007  &  0.214$\pm$0.005  &  0.127$\pm$0.005  &  \ref{fig:z4p2524280_J0034+1639}  \\
    4.2837\tablenotemark{a}  &  J0034+1639          &  0.531$\pm$0.004  &  0.591$\pm$0.004  &  0.565$\pm$0.004  &  0.382$\pm$0.005  &  0.489$\pm$0.004  &  1.222$\pm$0.025  &  1.005$\pm$0.009  &  0.215$\pm$0.005  &  0.139$\pm$0.005  &  0.328$\pm$0.003  &  0.231$\pm$0.004  &  \ref{fig:z4p2836630_J0034+1639}  \\
    4.4669  &  J0307-4945          &  0.520$\pm$0.002  &  0.778$\pm$0.002  &  \nodata  &  0.401$\pm$0.002  &  0.503$\pm$0.005  &  1.719$\pm$0.023  &  1.560$\pm$0.035  &  0.613$\pm$0.005  &  $<$0.615\tablenotemark{c}  &  0.505$\pm$0.005  &  0.304$\pm$0.005  &  \ref{fig:z4p4668560_J0307-4945}  \\
    4.7392  &  J0025-0145          &  0.664$\pm$0.007  &  0.731$\pm$0.015  &  \nodata  &  $<$0.558\tablenotemark{c}  &  0.446$\pm$0.023  &  1.587$\pm$0.032  &  1.417$\pm$0.029  &  0.045$\pm$0.021  &  $<$0.034  &  \nodata\tablenotemark{e}  &  0.064$\pm$0.021  &  \ref{fig:z4p7391580_J0025-0145}  \\
    4.8555\tablenotemark{b}  &  J1004+2025          &  0.146$\pm$0.036  &  0.583$\pm$0.019  &  $<$0.694\tablenotemark{c}  &  0.075$\pm$0.037  &  0.209$\pm$0.023  &  1.344$\pm$0.044  &  \nodata\tablenotemark{e}  &  0.145$\pm$0.036  &  $<$0.065  &  0.399$\pm$0.021  &  0.236$\pm$0.022  &  \ref{fig:z4p8554890_J1004+2025}  \\
    4.9464  &  J2325-0553          &  0.473$\pm$0.014  &  0.632$\pm$0.025  &  \nodata  &  $<$0.301\tablenotemark{c}  &  0.333$\pm$0.027  &  1.367$\pm$0.049  &  1.198$\pm$0.054  &  0.144$\pm$0.041  &  \nodata\tablenotemark{e}  &  0.214$\pm$0.035  &  $<$0.081  &  \ref{fig:z4p9463820_J2325-0553}  \\
    4.9499  &  J2202+1509          &  0.176$\pm$0.009  &  0.253$\pm$0.010  &  0.259$\pm$0.006  &  0.064$\pm$0.015  &  0.101$\pm$0.012  &  0.512$\pm$0.024  &  0.414$\pm$0.024  &  0.076$\pm$0.010  &  $<$0.337\tablenotemark{c}  &  0.028$\pm$0.012  &  $<$0.029  &  \ref{fig:z4p9499140_J2202+1509}  \\
    4.9626  &  J0131-0321          &  0.063$\pm$0.007  &  0.076$\pm$0.010  &  0.097$\pm$0.004  &  0.020$\pm$0.008  &  $<$0.032  &  0.295$\pm$0.015  &  0.181$\pm$0.021  &  0.061$\pm$0.013  &  0.029$\pm$0.014  &  0.144$\pm$0.016  &  0.081$\pm$0.016  &  \ref{fig:z4p9625770_J0131-0321}  \\
    4.9866  &  J0306+1853          &  0.067$\pm$0.004  &  0.083$\pm$0.003  &  \nodata  &  0.041$\pm$0.003  &  0.046$\pm$0.006  &  0.150$\pm$0.007  &  0.132$\pm$0.007  &  0.023$\pm$0.006  &  $<$0.036\tablenotemark{c}  &  0.073$\pm$0.008  &  0.051$\pm$0.009  &  \ref{fig:z4p9865800_J0306+1853}  \\
    5.0615\tablenotemark{b}  &  J1147-0109          &  0.459$\pm$0.023  &  0.543$\pm$0.020  &  0.551$\pm$0.020  &  0.203$\pm$0.019  &  0.300$\pm$0.029  &  1.383$\pm$0.058\tablenotemark{e}\tablenotemark{f}  &  1.112$\pm$0.047  &  $<$0.242\tablenotemark{c}  &  0.108$\pm$0.026  &  0.153$\pm$0.056  &  0.127$\pm$0.041  &  \ref{fig:z5p0615260_J1147-0109}  \\
    5.1052  &  J0812+0440          &  0.195$\pm$0.010  &  0.202$\pm$0.010  &  0.183$\pm$0.003  &  0.035$\pm$0.010  &  0.099$\pm$0.038  &  0.468$\pm$0.048  &  0.344$\pm$0.038  &  \nodata\tablenotemark{e}  &  $<$0.026  &  $<$0.061  &  $<$0.047  &  \ref{fig:z5p1052330_J0812+0440}  \\
    5.1448\tablenotemark{a}  &  J0747+1153          &  0.476$\pm$0.007  &  0.542$\pm$0.005  &  0.503$\pm$0.003  &  0.356$\pm$0.009  &  0.474$\pm$0.035  &  1.086$\pm$0.012  &  1.023$\pm$0.015  &  0.162$\pm$0.008  &  0.099$\pm$0.009  &  0.156$\pm$0.011  &  0.111$\pm$0.015  &  \ref{fig:z5p1447510_J0747+1153}  \\
    5.1783\tablenotemark{a}  &  J1436+2132          &  0.113$\pm$0.016  &  0.185$\pm$0.016  &  0.272$\pm$0.010  &  $<$0.031  &  \nodata\tablenotemark{e}  &  0.447$\pm$0.065  &  0.466$\pm$0.064  &  0.080$\pm$0.013  &  0.058$\pm$0.021  &  0.099$\pm$0.024  &  0.079$\pm$0.025  &  \ref{fig:z5p1782590_J1436+2132}  \\
    5.2961\tablenotemark{b}  &  J1335-0328          &  0.040$\pm$0.007\tablenotemark{d}  &  \nodata\tablenotemark{e}  &  \nodata  &  $<$0.042\tablenotemark{c}  &  $<$0.032  &  0.192$\pm$0.015  &  \nodata\tablenotemark{e}  &  0.016$\pm$0.007  &  $<$0.018  &  $<$0.035  &  $<$0.029  &  \ref{fig:z5p2960670_J1335-0328}  \\
    5.3374  &  J2207-0416          &  0.128$\pm$0.006  &  0.136$\pm$0.008  &  0.159$\pm$0.003  &  0.056$\pm$0.006  &  0.066$\pm$0.017  &  0.269$\pm$0.013  &  0.290$\pm$0.012  &  $<$0.021  &  \nodata\tablenotemark{e}  &  $<$0.039  &  $<$0.039  &  \ref{fig:z5p3374230_J2207-0416}  \\
    5.5944  &  SDSS J0840+5624     &  0.555$\pm$0.009  &  0.619$\pm$0.010  &  \nodata  &  0.410$\pm$0.009  &  \nodata\tablenotemark{e}  &  \nodata  &  \nodata  &  0.110$\pm$0.014  &  $<$0.035  &  \nodata\tablenotemark{e}  &  \nodata\tablenotemark{e}  &  \ref{fig:z5p5944320_SDSSJ0840+5624}  \\
    5.7533  &  SDSS J2315-0023     &  0.261$\pm$0.010  &  0.242$\pm$0.012  &  \nodata  &  $<$0.246\tablenotemark{c}  &  \nodata  &  \nodata  &  \nodata  &  0.148$\pm$0.073  &  $<$0.144  &  \nodata  &  \nodata  &  \ref{fig:z5p7532610_SDSSJ2315-0023}  \\
    5.7538  &  SDSS J1335+3533     &  0.202$\pm$0.040  &  0.194$\pm$0.032  &  0.172$\pm$0.023  &  $<$0.055  &  \nodata  &  \nodata  &  \nodata  &  $<$0.162  &  $<$0.148  &  \nodata  &  \nodata  &  \ref{fig:z5p7538330_SDSSJ1335+3533}  \\
    5.7911  &  SDSS J0818+1722     &  0.174$\pm$0.005  &  0.220$\pm$0.004  &  0.261$\pm$0.002  &  0.064$\pm$0.004  &  0.047$\pm$0.014  &  \nodata\tablenotemark{e}  &  \nodata\tablenotemark{e}  &  0.055$\pm$0.013  &  0.027$\pm$0.010  &  0.088$\pm$0.011  &  0.052$\pm$0.014  &  \ref{fig:z5p7911030_SDSSJ0818+1722}  \\
    5.8039  &  CFHQS J2100-1715    &  0.565$\pm$0.024  &  0.663$\pm$0.023  &  \nodata  &  0.217$\pm$0.031  &  \nodata\tablenotemark{e}  &  \nodata\tablenotemark{e}  &  \nodata\tablenotemark{e}  &  \nodata\tablenotemark{e}  &  0.152$\pm$0.039  &  0.262$\pm$0.077  &  0.155$\pm$0.075  &  \ref{fig:z5p8039450_CFHQSJ2100-1715}  \\
    5.8085  &  PSO J308-21         &  0.303$\pm$0.005  &  0.346$\pm$0.006  &  \nodata  &  $<$0.268\tablenotemark{c}  &  0.166$\pm$0.031  &  \nodata\tablenotemark{e}  &  \nodata\tablenotemark{e}  &  0.122$\pm$0.015  &  0.065$\pm$0.016  &  0.221$\pm$0.024  &  0.103$\pm$0.025  &  \ref{fig:z5p8085010_PSOJ308-21}  \\
    5.8425  &  SDSS J1623+3112     &  0.442$\pm$0.015  &  0.493$\pm$0.023  &  \nodata  &  0.158$\pm$0.020  &  \nodata  &  \nodata  &  \nodata  &  \nodata\tablenotemark{e}  &  $<$0.202  &  \nodata  &  \nodata  &  \ref{fig:z5p8424870_SDSSJ1623+3112}  \\
    5.8677  &  PSO J065-26         &  0.306$\pm$0.010  &  0.326$\pm$0.014  &  \nodata  &  0.122$\pm$0.014  &  0.127$\pm$0.038  &  \nodata\tablenotemark{e}  &  \nodata\tablenotemark{e}  &  \nodata\tablenotemark{e}  &  $<$0.034  &  $<$0.117  &  $<$0.060  &  \ref{fig:z5p8677130_PSOJ065-26}  \\
    5.8726  &  CFHQS J2100-1715    &  0.087$\pm$0.010  &  0.080$\pm$0.012  &  0.135$\pm$0.012  &  $<$0.023  &  $<$0.077  &  \nodata\tablenotemark{e}  &  \nodata\tablenotemark{e}  &  $<$0.037  &  $<$0.043  &  $<$0.079  &  $<$0.089  &  \ref{fig:z5p8726170_CFHQSJ2100-1715}  \\
    5.8767  &  SDSS J0818+1722     &  0.078$\pm$0.002  &  0.080$\pm$0.003  &  0.102$\pm$0.003  &  0.014$\pm$0.003  &  $<$0.013  &  0.272$\pm$0.022  &  \nodata\tablenotemark{e}  &  0.027$\pm$0.005  &  0.034$\pm$0.005  &  0.067$\pm$0.008  &  0.024$\pm$0.008  &  \ref{fig:z5p8767010_SDSSJ0818+1722}  \\
    5.8786\tablenotemark{b}  &  PSO J308-21         &  0.125$\pm$0.014\tablenotemark{d}  &  0.250$\pm$0.009  &  \nodata  &  $<$0.125\tablenotemark{c}  &  0.109$\pm$0.026  &  0.474$\pm$0.048  &  \nodata\tablenotemark{e}  &  0.142$\pm$0.016  &  0.060$\pm$0.019  &  0.338$\pm$0.028  &  0.193$\pm$0.028  &  \ref{fig:z5p8785790_PSOJ308-21}  \\
    5.8987\tablenotemark{b}  &  ATLAS J158-14       &  0.080$\pm$0.017\tablenotemark{d}  &  0.156$\pm$0.012  &  0.244$\pm$0.006  &  $<$0.122\tablenotemark{c}  &  0.083$\pm$0.020  &  \nodata\tablenotemark{e}  &  \nodata\tablenotemark{e}  &  0.054$\pm$0.020  &  $<$0.040  &  0.042$\pm$0.020  &  0.072$\pm$0.021  &  \ref{fig:z5p8986670_ATLASJ158-14}  \\
    5.9127  &  PSO J159-02         &  0.194$\pm$0.021  &  0.352$\pm$0.018  &  \nodata  &  0.074$\pm$0.009  &  0.224$\pm$0.064  &  \nodata\tablenotemark{e}  &  \nodata\tablenotemark{e}  &  0.080$\pm$0.033  &  $<$0.099  &  \nodata\tablenotemark{e}  &  \nodata\tablenotemark{e}  &  \ref{fig:z5p9127050_PSOJ159-02}  \\
    5.9366\tablenotemark{a}  &  PSO J056-16         &  0.474$\pm$0.010  &  0.564$\pm$0.013  &  0.470$\pm$0.007  &  0.229$\pm$0.011  &  0.289$\pm$0.038  &  1.226$\pm$0.089\tablenotemark{h}\tablenotemark{e}\tablenotemark{f}  &  0.945$\pm$0.085\tablenotemark{h}  &  $<$0.059  &  $<$0.045  &  0.106$\pm$0.020  &  0.136$\pm$0.021  &  \ref{fig:z5p9365800_PSOJ056-16}  \\
    5.9387\tablenotemark{a}  &  SDSS J2310+1855     &  0.137$\pm$0.007  &  0.145$\pm$0.008  &  0.126$\pm$0.007  &  $<$0.110\tablenotemark{c}  &  0.071$\pm$0.021  &  0.307$\pm$0.067  &  \nodata\tablenotemark{e}  &  0.056$\pm$0.022  &  $<$0.147\tablenotemark{c}  &  $<$0.044  &  0.052$\pm$0.024  &  \ref{fig:z5p9387060_SDSSJ2310+1855}  \\
    5.9450\tablenotemark{b}  &  SDSS J0100+2802     &  0.022$\pm$0.001  &  0.065$\pm$0.001\tablenotemark{d}  &  \nodata  &  0.010$\pm$0.001  &  0.012$\pm$0.001  &  0.146$\pm$0.005  &  0.149$\pm$0.003  &  0.040$\pm$0.002  &  $<$0.034\tablenotemark{c}  &  0.035$\pm$0.002  &  0.026$\pm$0.002  &  \ref{fig:z5p9449860_SDSSJ0100+2802}  \\
    5.9480\tablenotemark{a}  &  VIK J0046-2837      &  0.240$\pm$0.019  &  0.235$\pm$0.027  &  0.359$\pm$0.013  &  0.072$\pm$0.020  &  $<$0.146  &  0.846$\pm$0.072  &  0.593$\pm$0.145\tablenotemark{h}  &  \nodata\tablenotemark{e}  &  \nodata\tablenotemark{e}  &  \nodata\tablenotemark{e}  &  \nodata\tablenotemark{e}  &  \ref{fig:z5p9479820_VIKJ0046-2837}  \\
    5.9777\tablenotemark{a}\tablenotemark{b}  &  SDSS J2054-0005     &  0.120$\pm$0.012  &  \nodata\tablenotemark{e}  &  0.203$\pm$0.008  &  0.046$\pm$0.011  &  \nodata\tablenotemark{e}  &  \nodata\tablenotemark{e}  &  \nodata\tablenotemark{e}  &  \nodata\tablenotemark{e}  &  \nodata\tablenotemark{e}  &  \nodata\tablenotemark{e}  &  \nodata\tablenotemark{e}  &  \ref{fig:z5p9776540_SDSSJ2054-0005}  \\
    6.0114\tablenotemark{b}  &  SDSS J1148+5251     &  0.180$\pm$0.002  &  0.100$\pm$0.004  &  \nodata  &  0.035$\pm$0.001  &  $<$0.041  &  \nodata  &  \nodata  &  0.048$\pm$0.019  &  0.037$\pm$0.019  &  0.083$\pm$0.038\tablenotemark{c}\tablenotemark{g}  &  0.055$\pm$0.020  &  \ref{fig:z6p0113980_SDSSJ1148+5251}  \\
    6.0172\tablenotemark{a}  &  ULAS J1319+0950     &  0.040$\pm$0.003  &  0.024$\pm$0.005  &  0.046$\pm$0.003  &  0.007$\pm$0.003  &  $<$0.026  &  0.102$\pm$0.030  &  0.071$\pm$0.022  &  $<$0.009  &  $<$0.010  &  $<$0.014  &  $<$0.012  &  \ref{fig:z6p0172180_ULASJ1319+0950}  \\
    6.0611  &  PSO J036+03         &  0.042$\pm$0.002  &  0.045$\pm$0.008  &  \nodata  &  $<$0.102\tablenotemark{c}  &  0.042$\pm$0.013  &  0.228$\pm$0.024  &  \nodata\tablenotemark{e}  &  0.034$\pm$0.007  &  $<$0.023  &  0.032$\pm$0.010  &  \nodata\tablenotemark{e}  &  \ref{fig:z6p0610940_PSOJ036+03}  \\
    6.1115  &  SDSS J0100+2802     &  0.118$\pm$0.001  &  0.098$\pm$0.001  &  0.099$\pm$0.001  &  0.032$\pm$0.001  &  $<$0.060\tablenotemark{c}  &  0.256$\pm$0.002  &  0.182$\pm$0.002  &  0.009$\pm$0.003  &  $<$0.007  &  $<$0.005  &  $<$0.005  &  \ref{fig:z6p1115230_SDSSJ0100+2802}  \\
    6.1228  &  VDES J0224-4711     &  0.140$\pm$0.004  &  \nodata\tablenotemark{e}  &  \nodata  &  0.048$\pm$0.005  &  $<$0.049  &  0.303$\pm$0.026  &  0.237$\pm$0.035  &  $<$0.056  &  $<$0.057  &  \nodata\tablenotemark{e}  &  $<$0.048  &  \ref{fig:z6p1228350_VDESJ0224-4711}  \\
    6.1242\tablenotemark{a}\tablenotemark{b}  &  PSO J065-26         &  0.661$\pm$0.017  &  1.141$\pm$0.039  &  1.025$\pm$0.014  &  0.394$\pm$0.019  &  \nodata\tablenotemark{e}  &  2.482$\pm$0.064  &  1.953$\pm$0.068  &  0.218$\pm$0.052  &  $<$0.111  &  0.201$\pm$0.078\tablenotemark{e}\tablenotemark{g}  &  0.127$\pm$0.040  &  \ref{fig:z6p1242130_PSOJ065-26}  \\
    6.1314\tablenotemark{b}  &  SDSS J1148+5251     &  0.081$\pm$0.002  &  0.065$\pm$0.005  &  \nodata  &  0.020$\pm$0.002  &  0.065$\pm$0.017  &  \nodata  &  \nodata  &  0.025$\pm$0.012  &  $<$0.032  &  0.077$\pm$0.018  &  $<$0.029  &  \ref{fig:z6p1313540_SDSSJ1148+5251}  \\
    6.1436  &  SDSS J0100+2802     &  0.175$\pm$0.001  &  0.151$\pm$0.001  &  0.159$\pm$0.001  &  0.024$\pm$0.002  &  0.049$\pm$0.001  &  0.386$\pm$0.003  &  0.205$\pm$0.013  &  0.009$\pm$0.003  &  $<$0.006  &  0.004$\pm$0.002  &  $<$0.003  &  \ref{fig:z6p1435510_SDSSJ0100+2802}  \\
    6.2575\tablenotemark{b}  &  SDSS J1148+5251     &  0.061$\pm$0.006  &  0.065$\pm$0.007  &  0.045$\pm$0.001  &  0.020$\pm$0.005  &  $<$0.032  &  \nodata  &  \nodata  &  $<$0.038  &  0.024$\pm$0.009  &  \nodata\tablenotemark{e}  &  \nodata\tablenotemark{e}  &  \ref{fig:z6p2575350_SDSSJ1148+5251}  \\

%% file: system_figures.tex
\begin{figure}
   \centering
   \includegraphics[height=0.40\textheight]{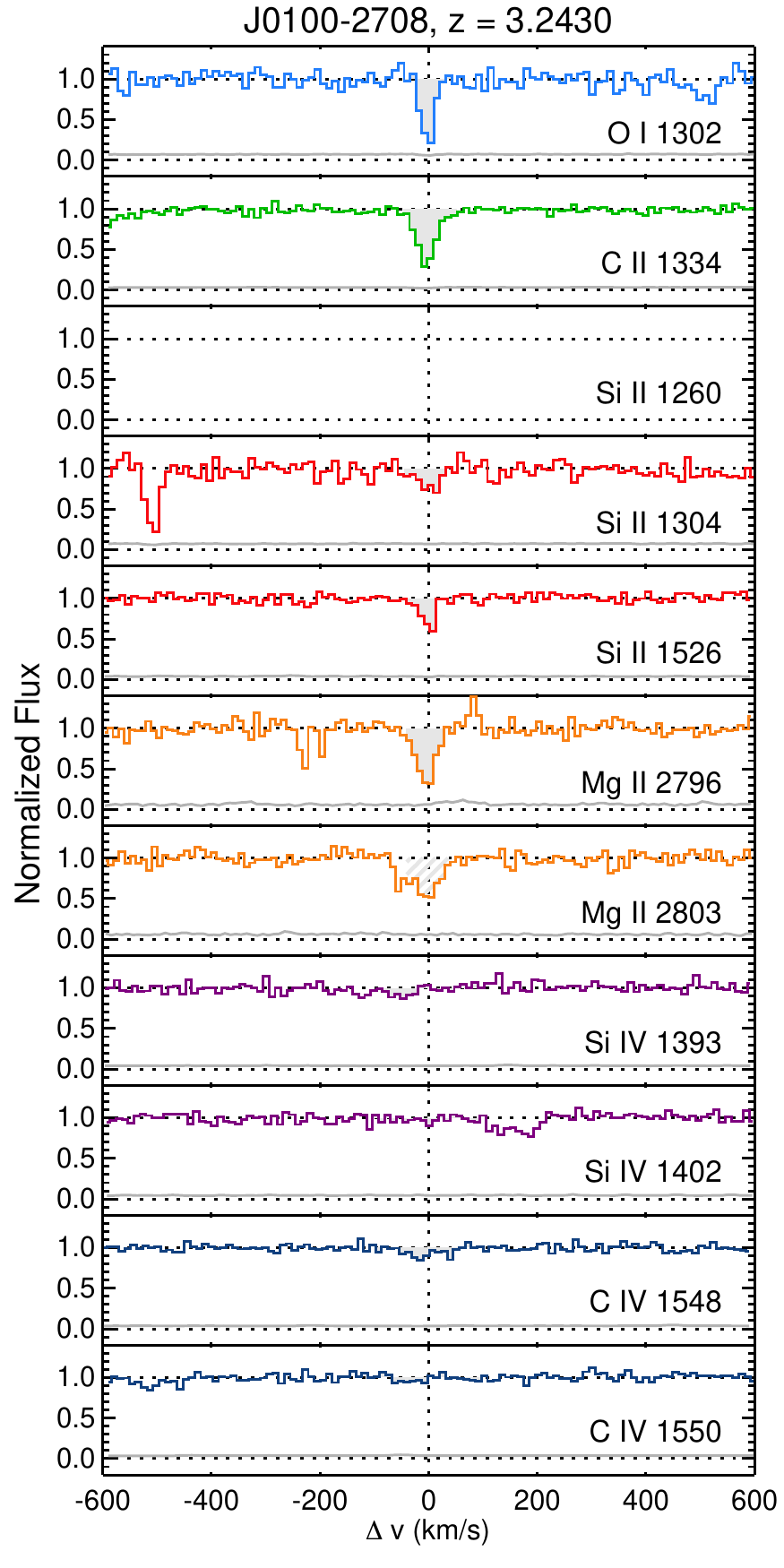}
   \caption{Stacked velocity plot for the $z=3.2430$ system towards J0100-2708.  Solid histograms show the normalized flux as a function of velocity offset from the nominal redshift.  The thin solid line at the bottom of each panel marks the 1$\sigma$ flux uncertainty.  Solid shading marks the intervals over which equivalent widths were measured.  Hatched shading (in this and other figures) denotes lines that are either blended with an unrelated absorber or contaminated by strong telluric residuals.\label{fig:z3p2429550_J0100-2708}}
\end{figure}
 
\begin{figure}
   \centering
   \includegraphics[height=0.40\textheight]{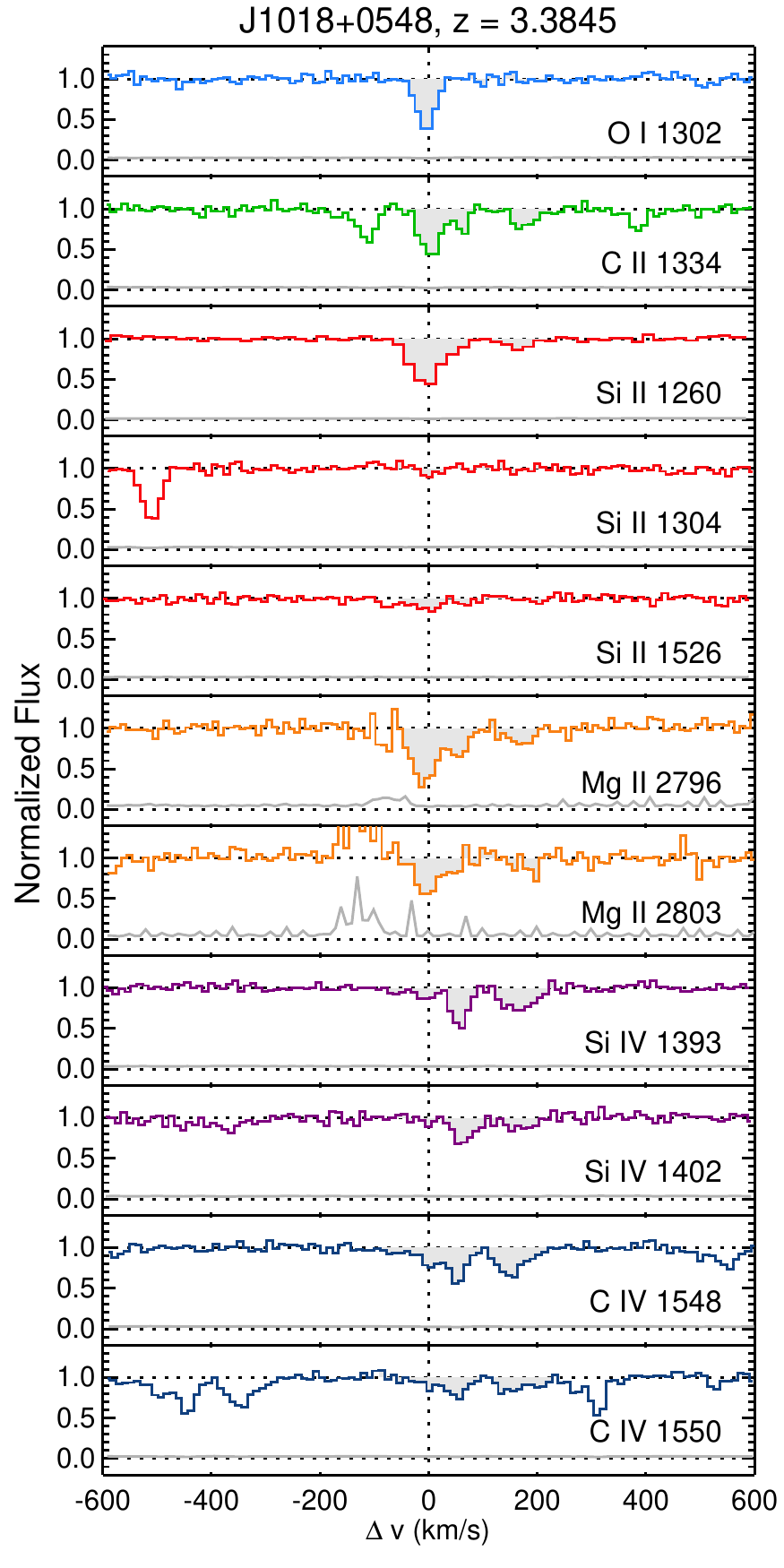}
   \caption{Stacked velocity plot for the $z=3.3845$ system towards J1018+0548.  Lines and shading are as described in Figure~\ref{fig:z3p3844720_J1018+0548}.\label{fig:z3p3844720_J1018+0548}}
\end{figure}
 
\begin{figure}
   \centering
   \includegraphics[height=0.40\textheight]{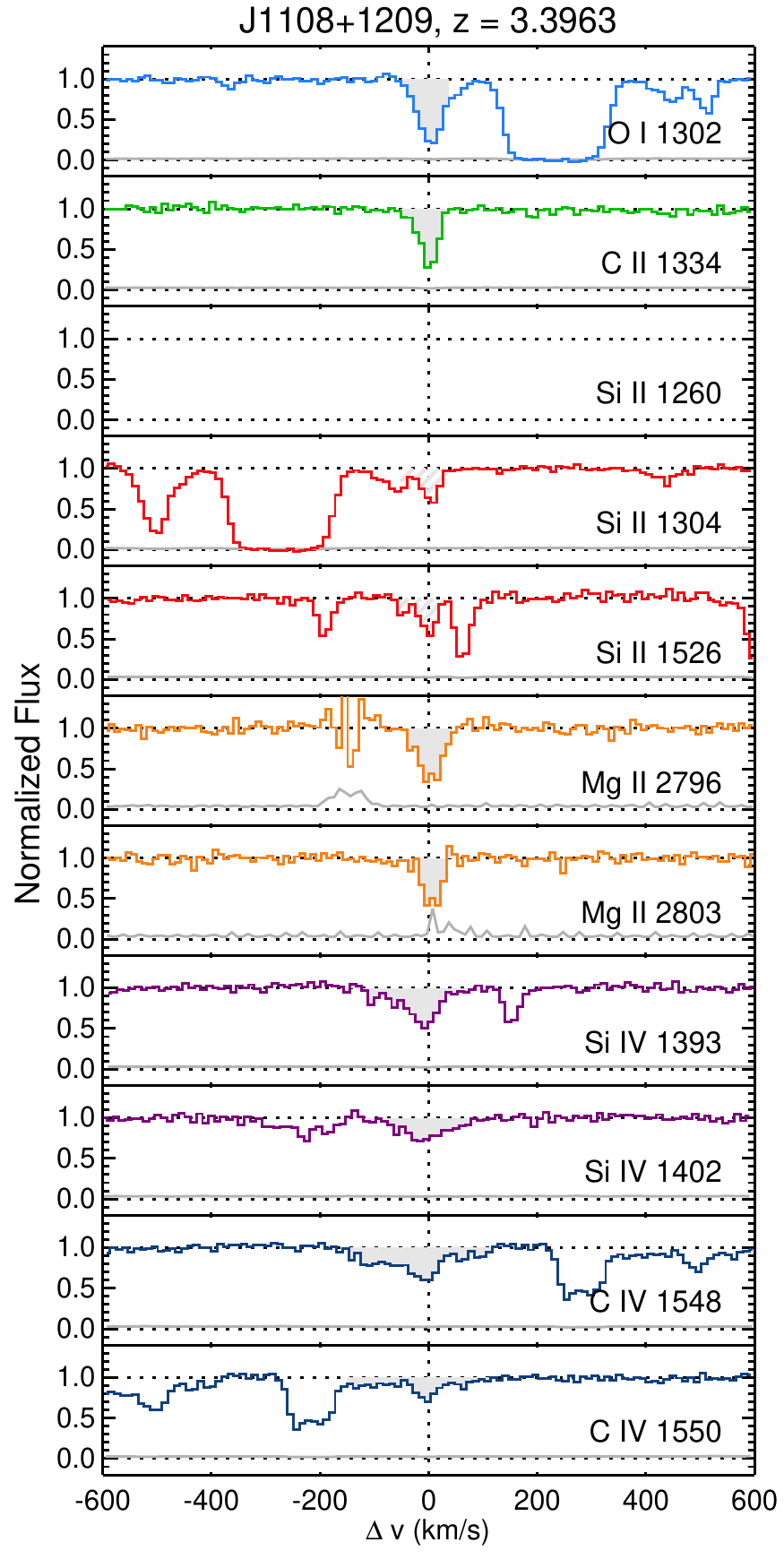}
   \caption{Stacked velocity plot for the $z=3.3963$ system towards J1108+1209.  Lines and shading are as described in Figure~\ref{fig:z3p3844720_J1018+0548}.\label{fig:z3p3962790_J1108+1209}}
\end{figure}
 
\clearpage
 
\begin{figure}[!t]
   \centering
   \includegraphics[height=0.40\textheight]{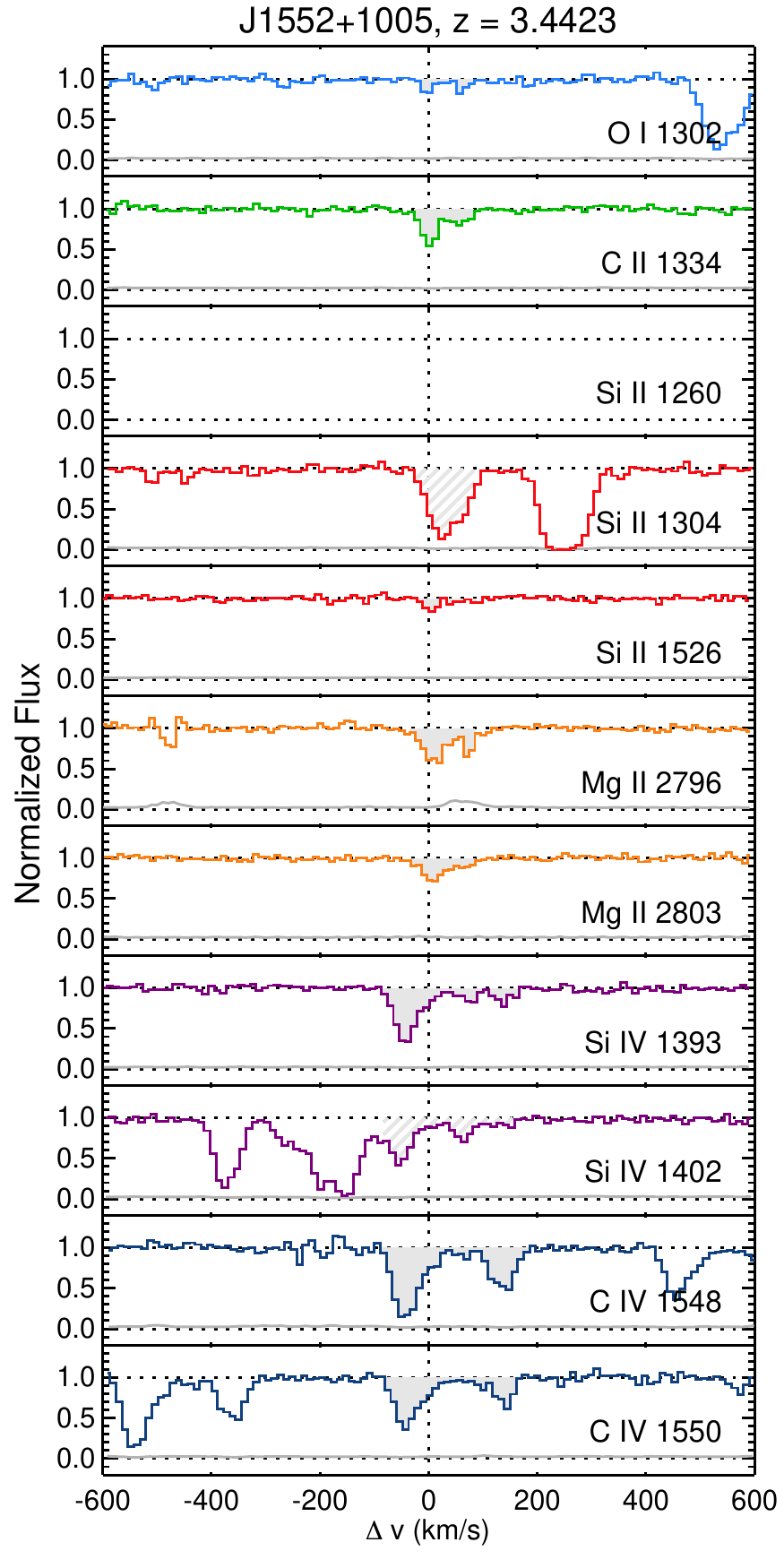}
   \caption{Stacked velocity plot for the $z=3.4423$ system towards J1552+1005.  Lines and shading are as described in Figure~\ref{fig:z3p3844720_J1018+0548}.\label{fig:z3p4422810_J1552+1005}}
\end{figure}
 
\begin{figure}[!b]
   \centering
   \includegraphics[height=0.40\textheight]{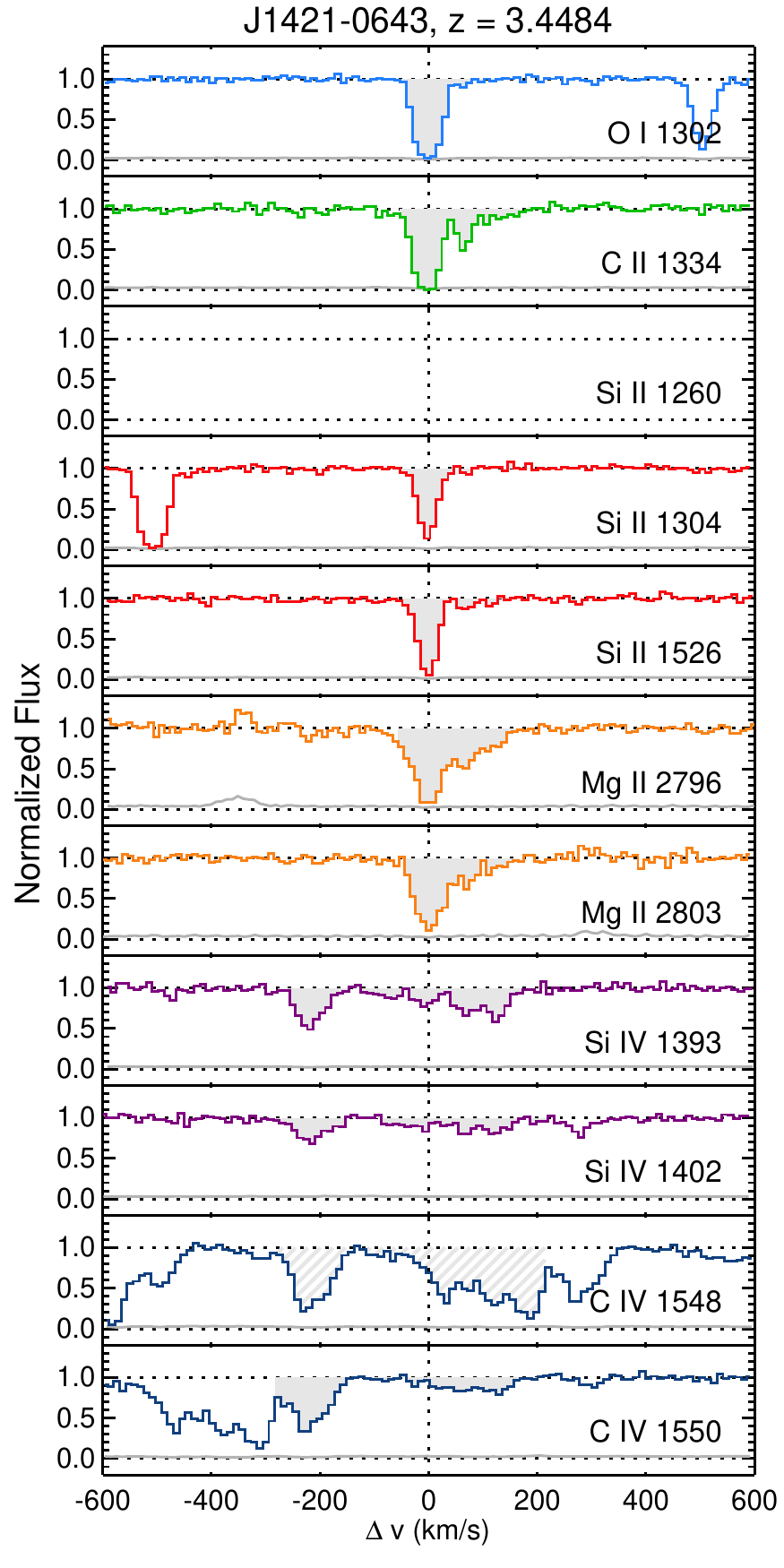}
   \caption{Stacked velocity plot for the $z=3.4484$ system towards J1421-0643.  Lines and shading are as described in Figure~\ref{fig:z3p3844720_J1018+0548}.\label{fig:z3p4483820_J1421-0643}}
\end{figure}
 
\begin{figure}[!t]
   \centering
   \includegraphics[height=0.40\textheight]{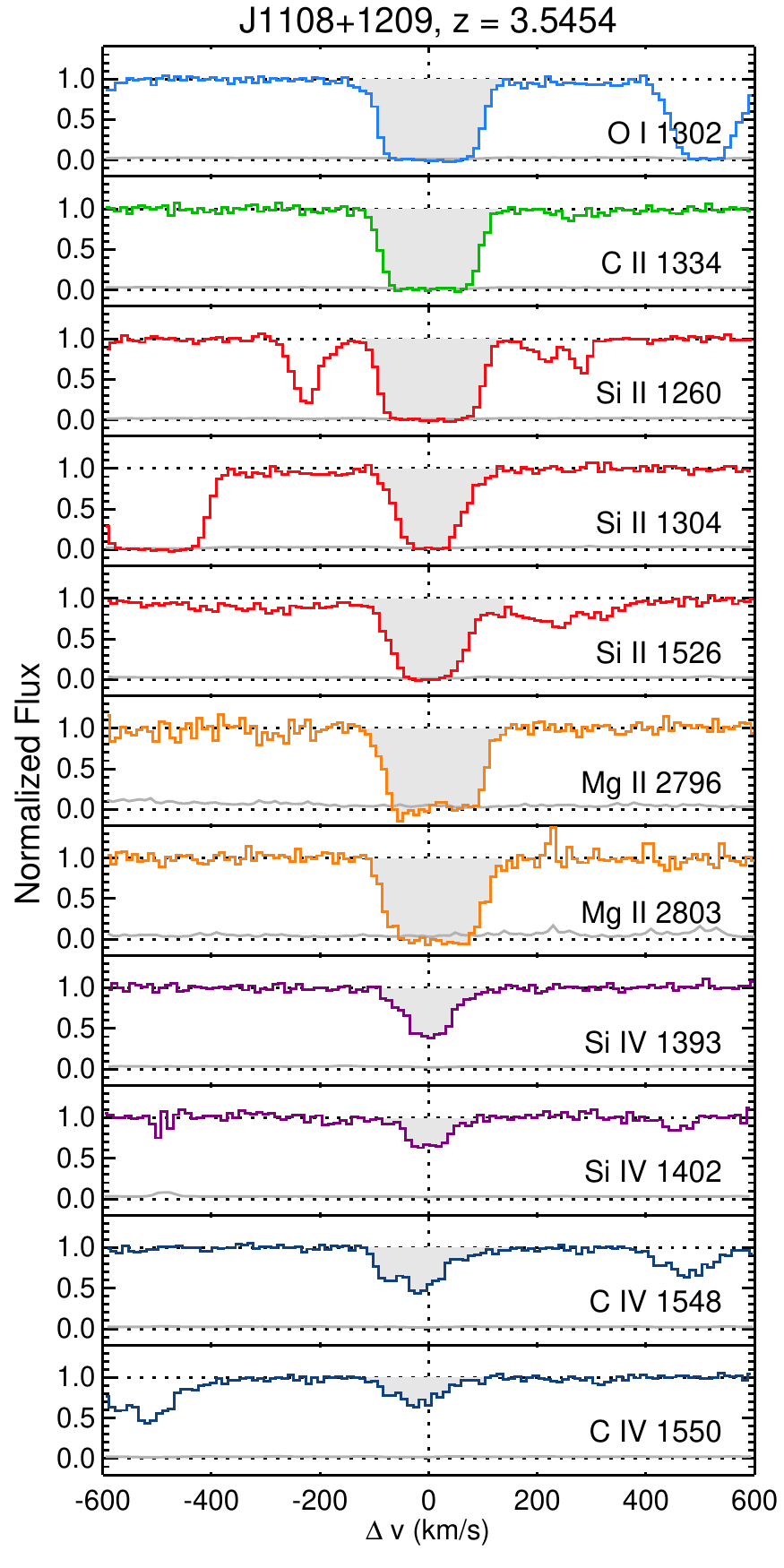}
   \caption{Stacked velocity plot for the $z=3.5454$ system towards J1108+1209.  Lines and shading are as described in Figure~\ref{fig:z3p3844720_J1018+0548}.\label{fig:z3p5453680_J1108+1209}}
\end{figure}
 
\begin{figure}[!b]
   \centering
   \includegraphics[height=0.40\textheight]{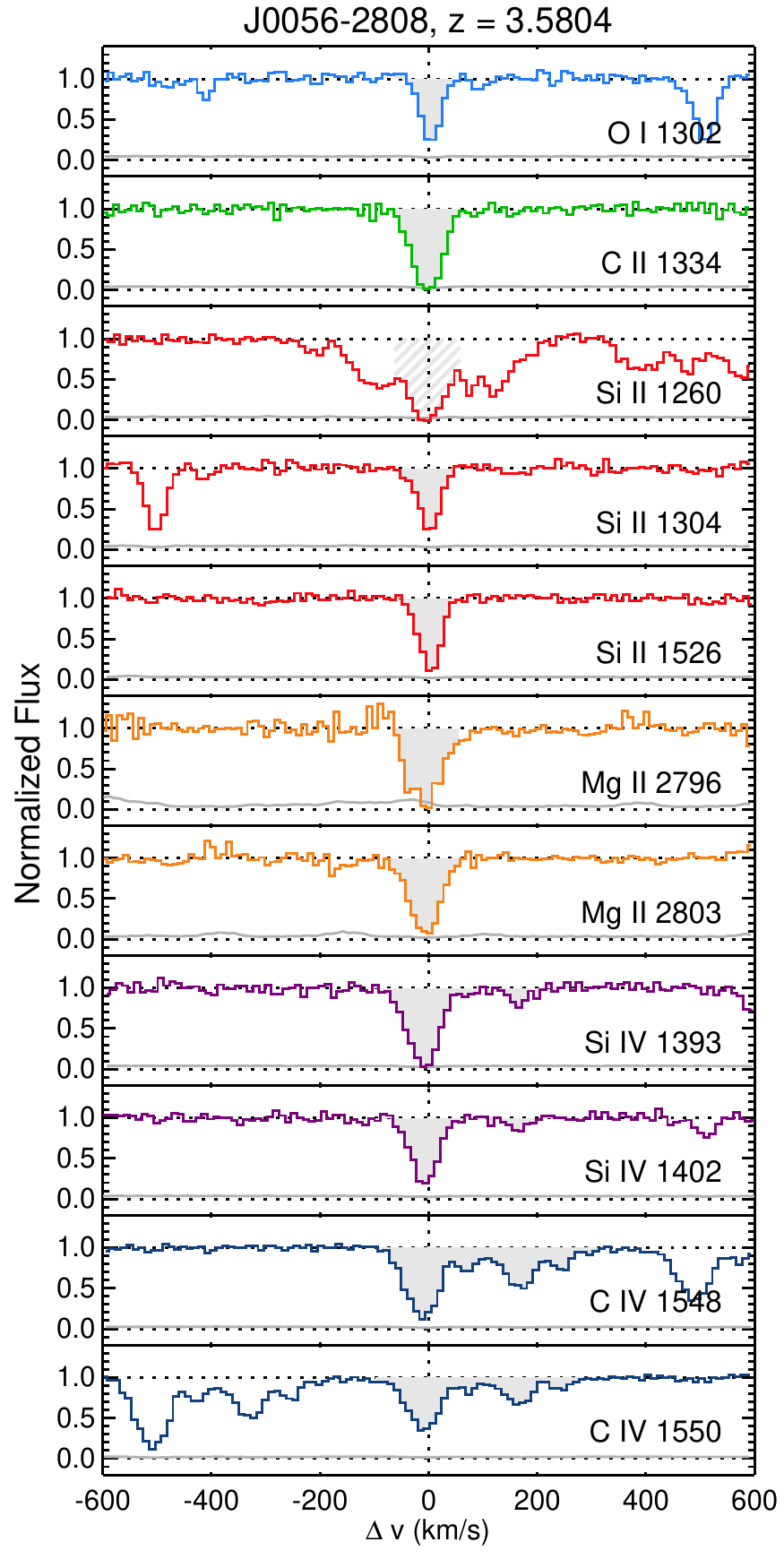}
   \caption{Stacked velocity plot for the $z=3.5804$ system towards J0056-2808.  Lines and shading are as described in Figure~\ref{fig:z3p3844720_J1018+0548}.\label{fig:z3p5803980_J0056-2808}}
\end{figure}
 
\clearpage
 
\begin{figure}[!t]
   \centering
   \includegraphics[height=0.40\textheight]{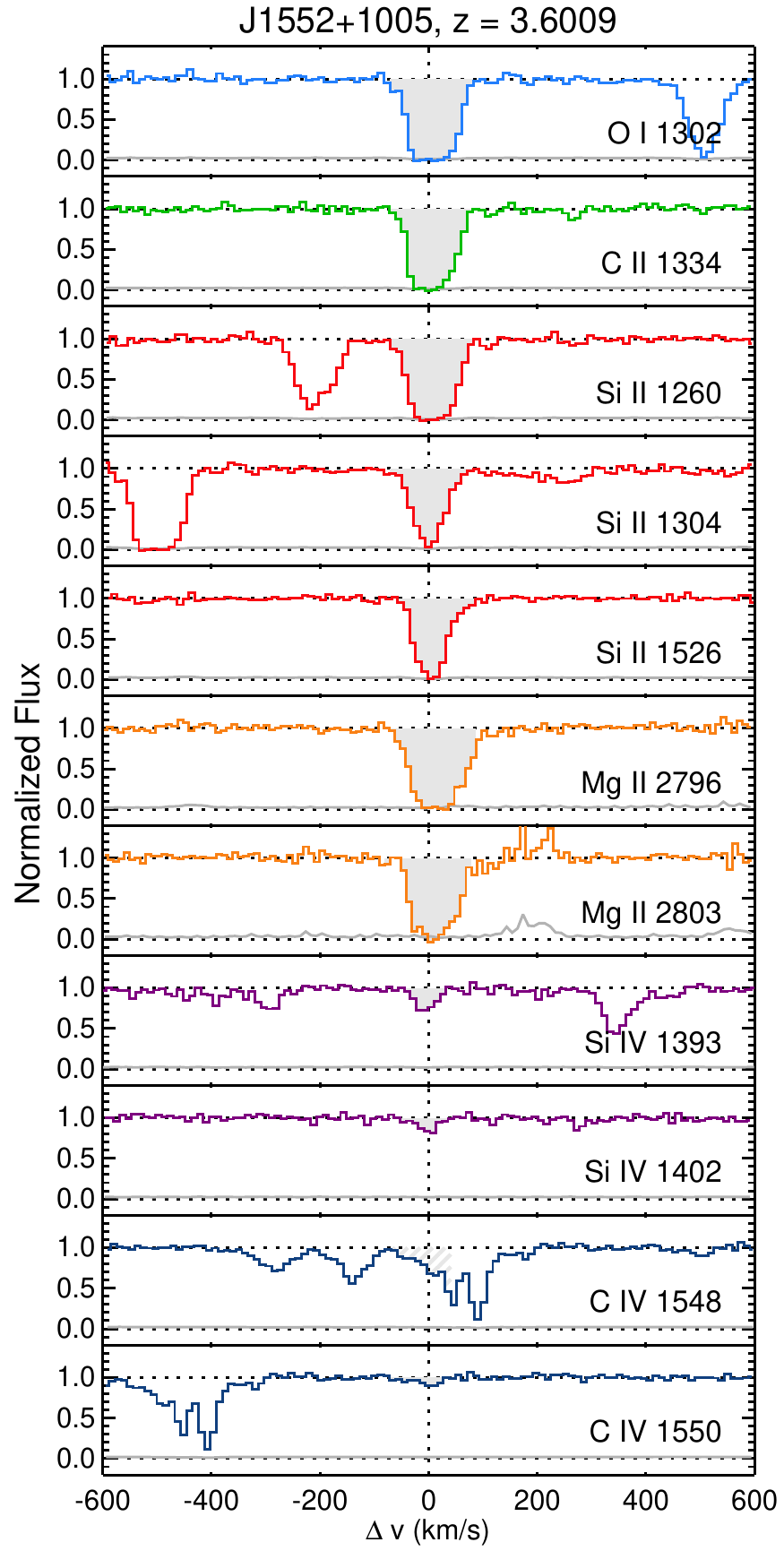}
   \caption{Stacked velocity plot for the $z=3.6009$ system towards J1552+1005.  Lines and shading are as described in Figure~\ref{fig:z3p3844720_J1018+0548}.\label{fig:z3p6008980_J1552+1005}}
\end{figure}
 
\begin{figure}[!b]
   \centering
   \includegraphics[height=0.40\textheight]{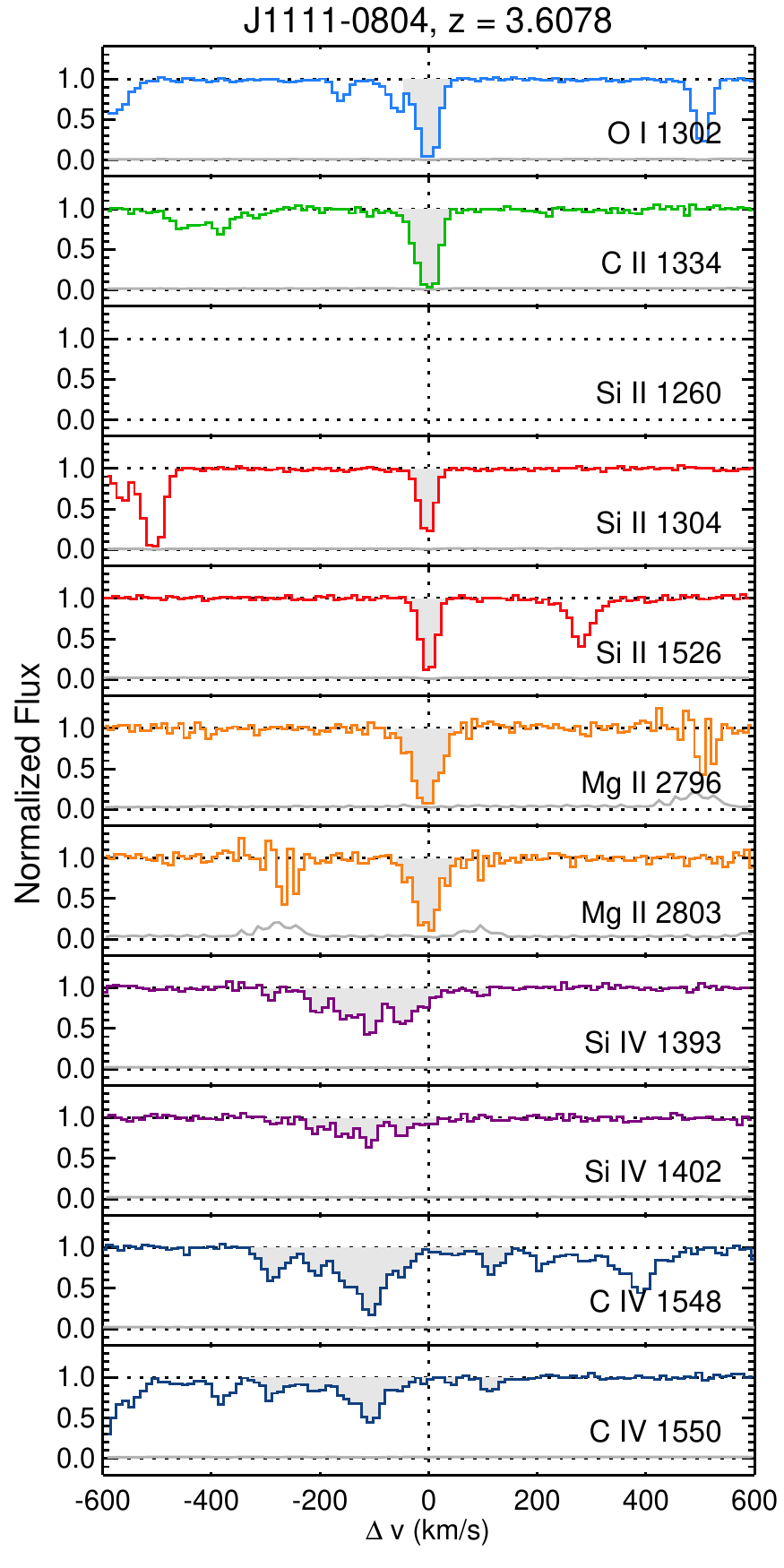}
   \caption{Stacked velocity plot for the $z=3.6078$ system towards J1111-0804.  Lines and shading are as described in Figure~\ref{fig:z3p3844720_J1018+0548}.\label{fig:z3p6077840_J1111-0804}}
\end{figure}
 
\begin{figure}[!t]
   \centering
   \includegraphics[height=0.40\textheight]{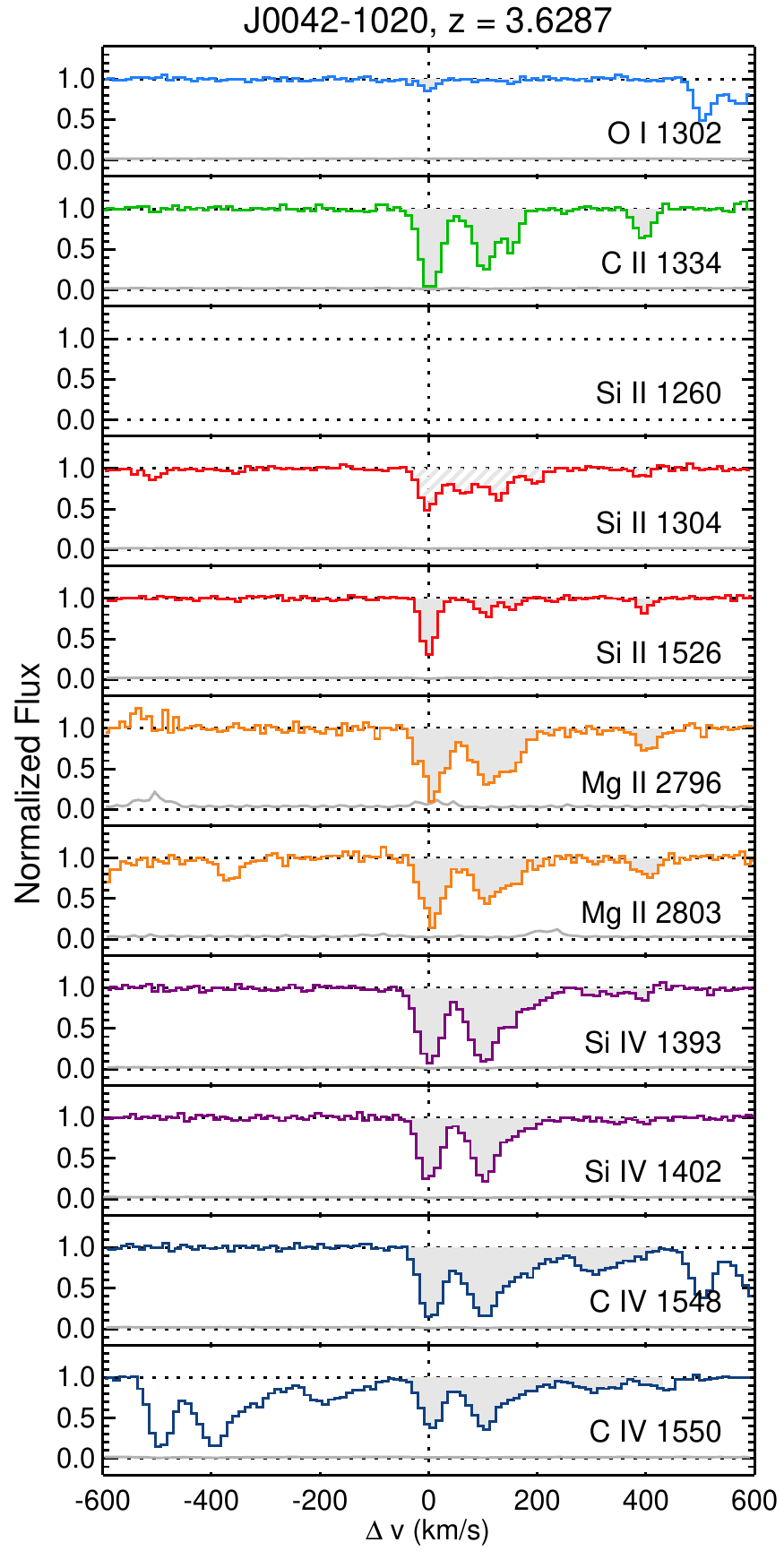}
   \caption{Stacked velocity plot for the $z=3.6287$ system towards J0042-1020.  Lines and shading are as described in Figure~\ref{fig:z3p3844720_J1018+0548}. See notes on this system in Appendix~\ref{app:details}.\label{fig:z3p6287100_J0042-1020}}
\end{figure}
 
\begin{figure}[!b]
   \centering
   \includegraphics[height=0.40\textheight]{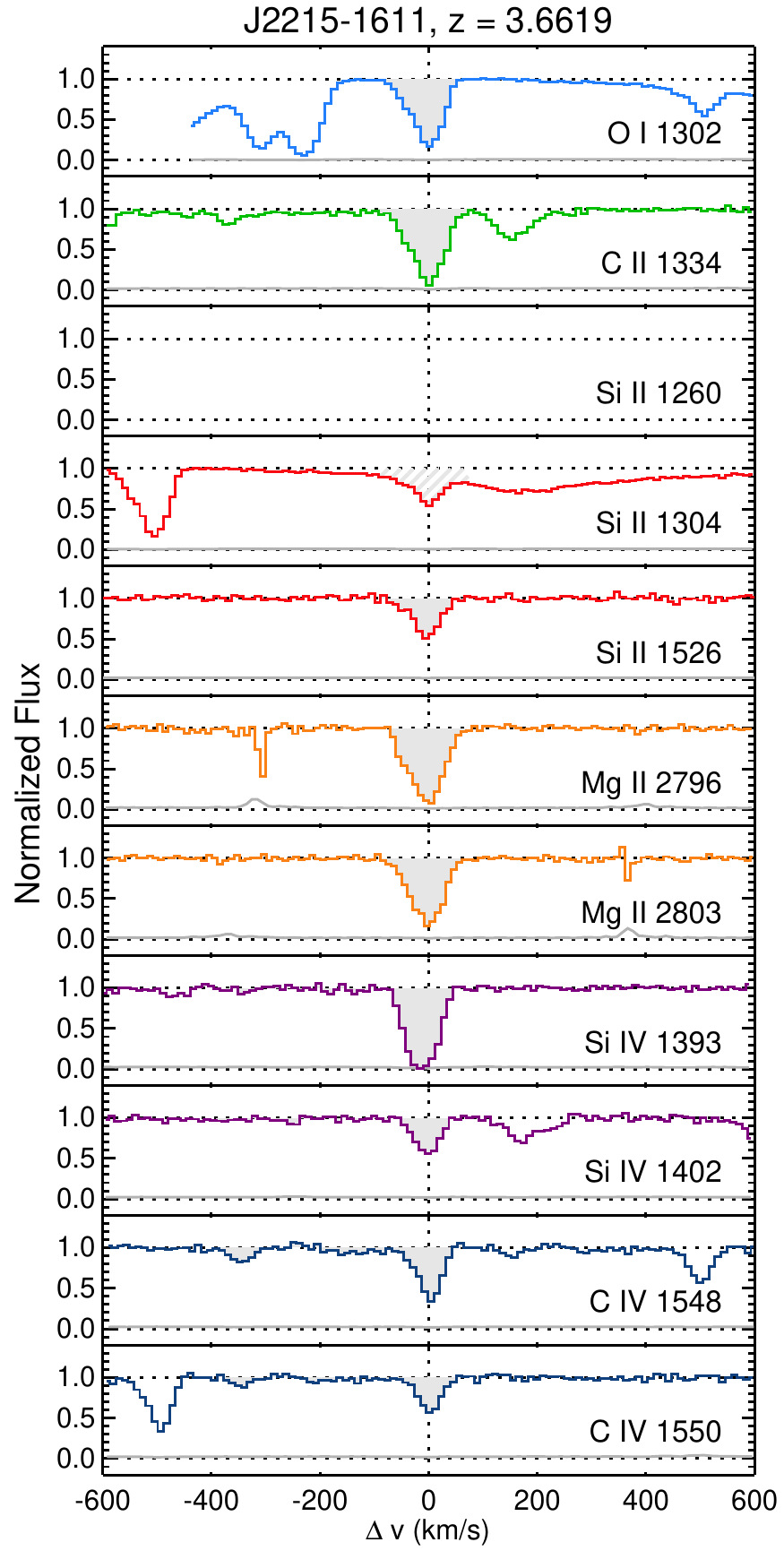}
   \caption{Stacked velocity plot for the $z=3.6619$ system towards J2215-1611.  Lines and shading are as described in Figure~\ref{fig:z3p3844720_J1018+0548}.\label{fig:z3p6618780_J2215-1611}}
\end{figure}
 
\clearpage
 
\begin{figure}[!t]
   \centering
   \includegraphics[height=0.40\textheight]{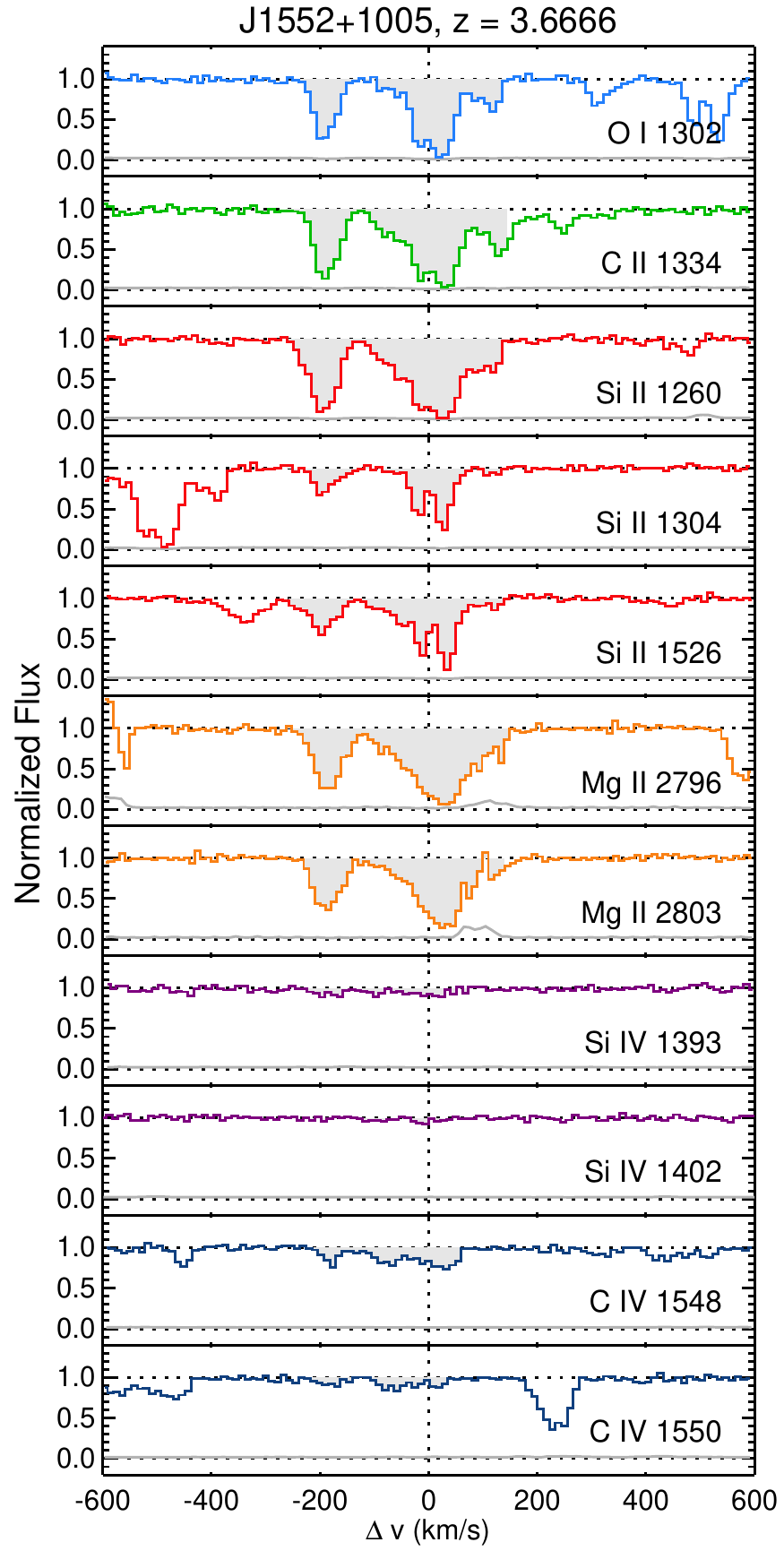}
   \caption{Stacked velocity plot for the $z=3.6666$ system towards J1552+1005.  Lines and shading are as described in Figure~\ref{fig:z3p3844720_J1018+0548}.\label{fig:z3p6665670_J1552+1005}}
\end{figure}
 
\begin{figure}[!b]
   \centering
   \includegraphics[height=0.40\textheight]{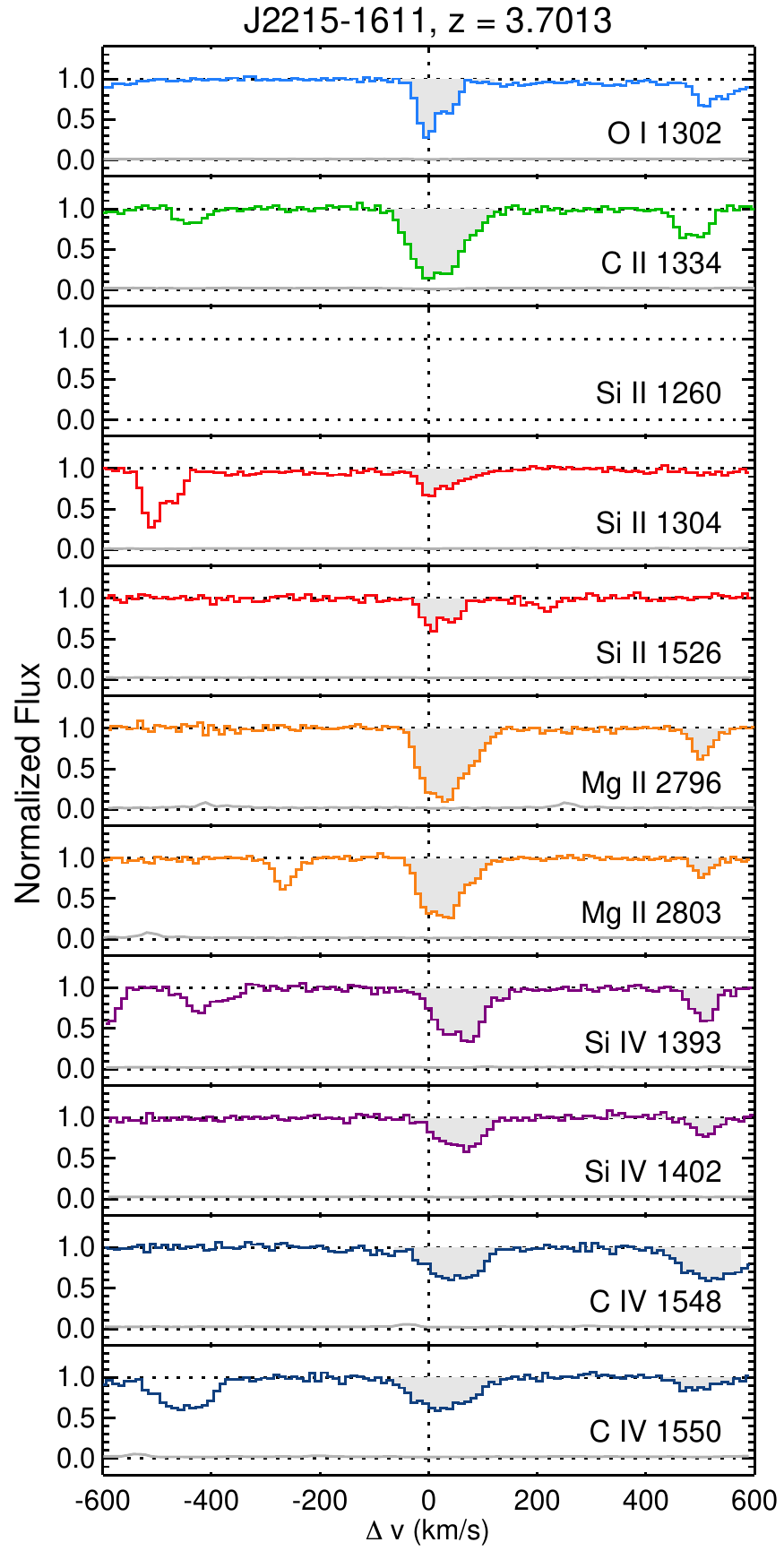}
   \caption{Stacked velocity plot for the $z=3.7013$ system towards J2215-1611.  Lines and shading are as described in Figure~\ref{fig:z3p3844720_J1018+0548}. See notes on this system in Appendix~\ref{app:details}.\label{fig:z3p7013160_J2215-1611}}
\end{figure}
 
\begin{figure}[!t]
   \centering
   \includegraphics[height=0.40\textheight]{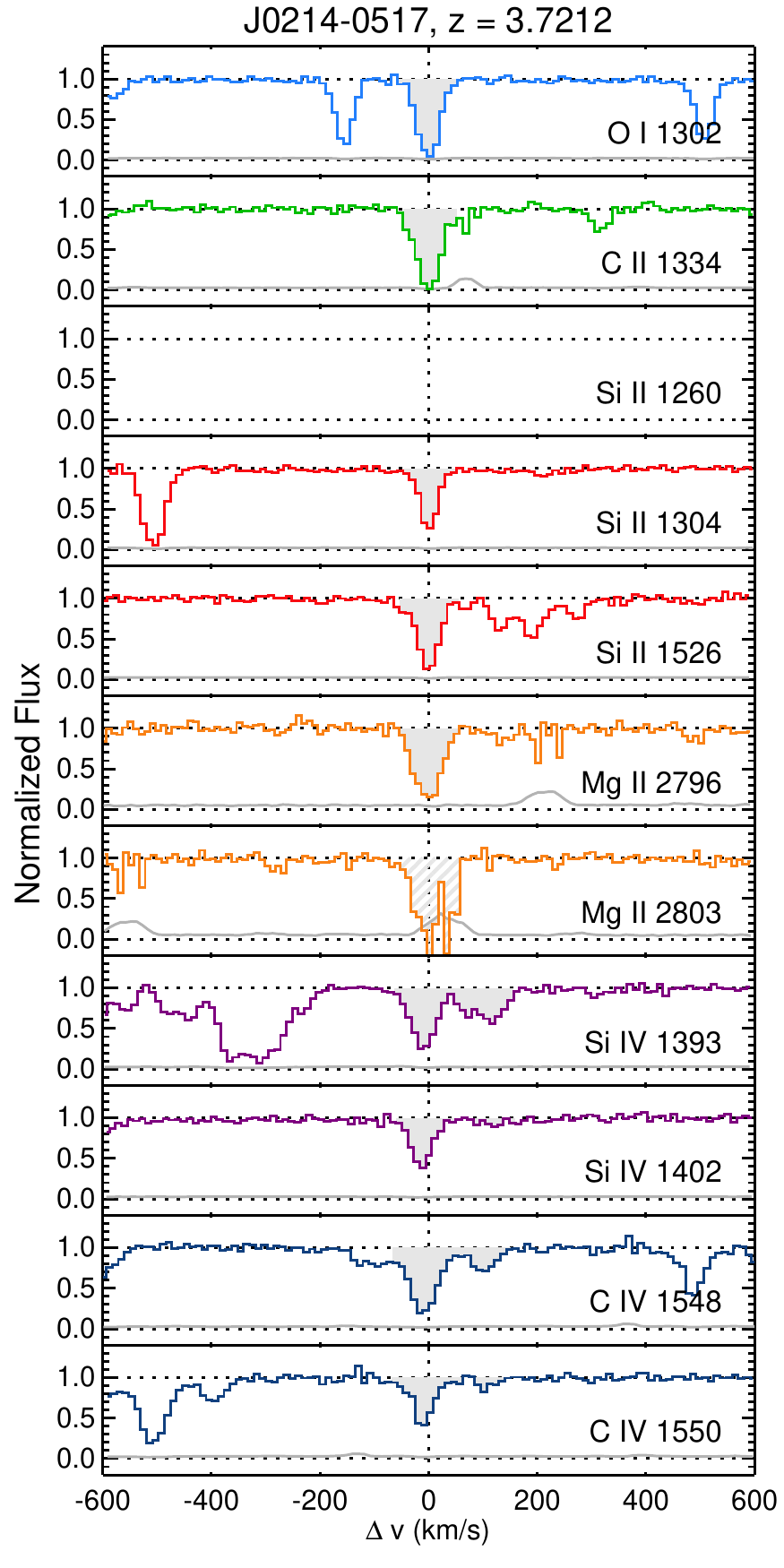}
   \caption{Stacked velocity plot for the $z=3.7212$ system towards J0214-0517.  Lines and shading are as described in Figure~\ref{fig:z3p3844720_J1018+0548}.\label{fig:z3p7212440_J0214-0517}}
\end{figure}
 
\begin{figure}[!b]
   \centering
   \includegraphics[height=0.40\textheight]{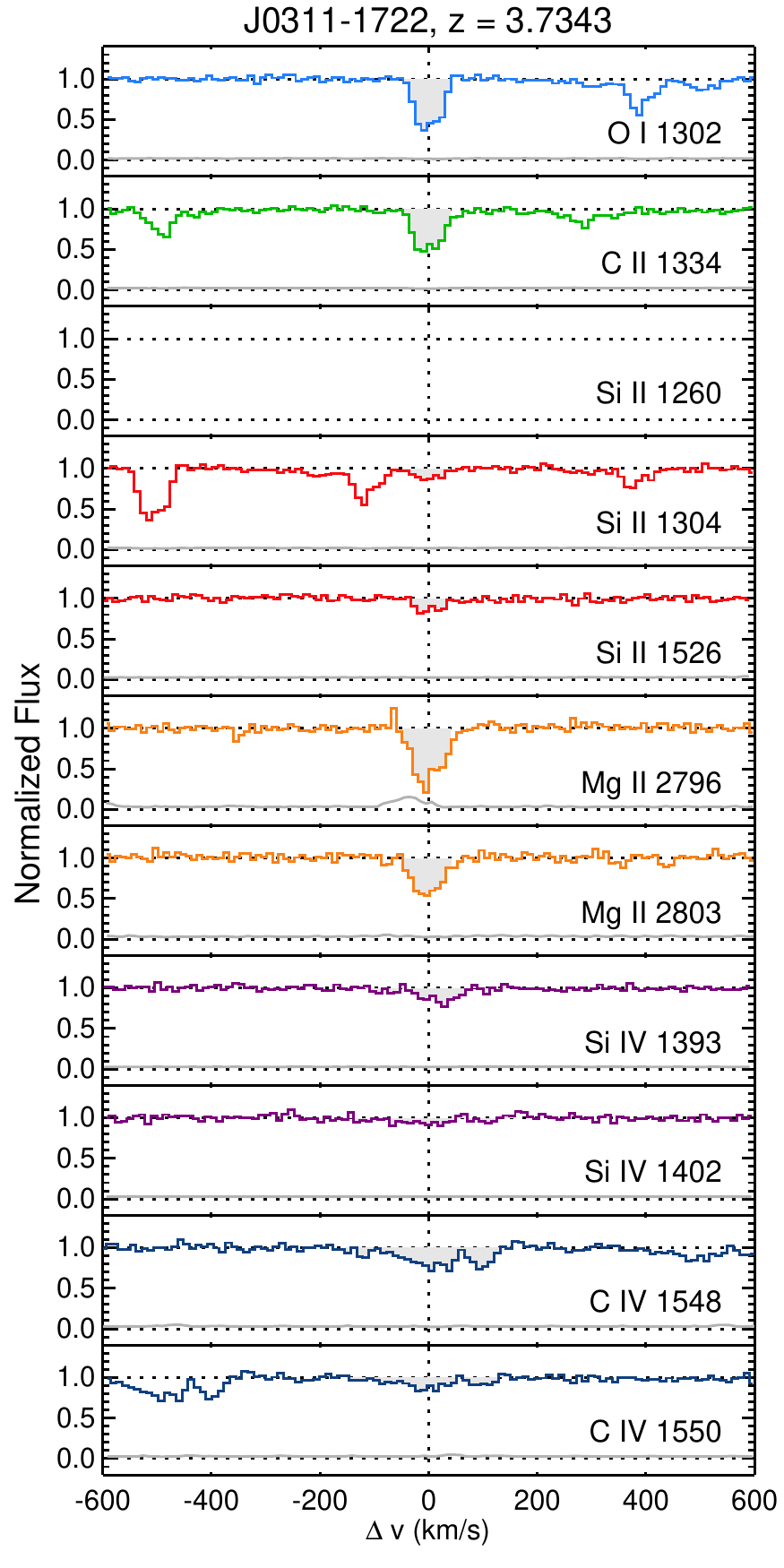}
   \caption{Stacked velocity plot for the $z=3.7343$ system towards J0311-1722.  Lines and shading are as described in Figure~\ref{fig:z3p3844720_J1018+0548}.\label{fig:z3p7342600_J0311-1722}}
\end{figure}
 
\clearpage
 
\begin{figure}[!t]
   \centering
   \includegraphics[height=0.40\textheight]{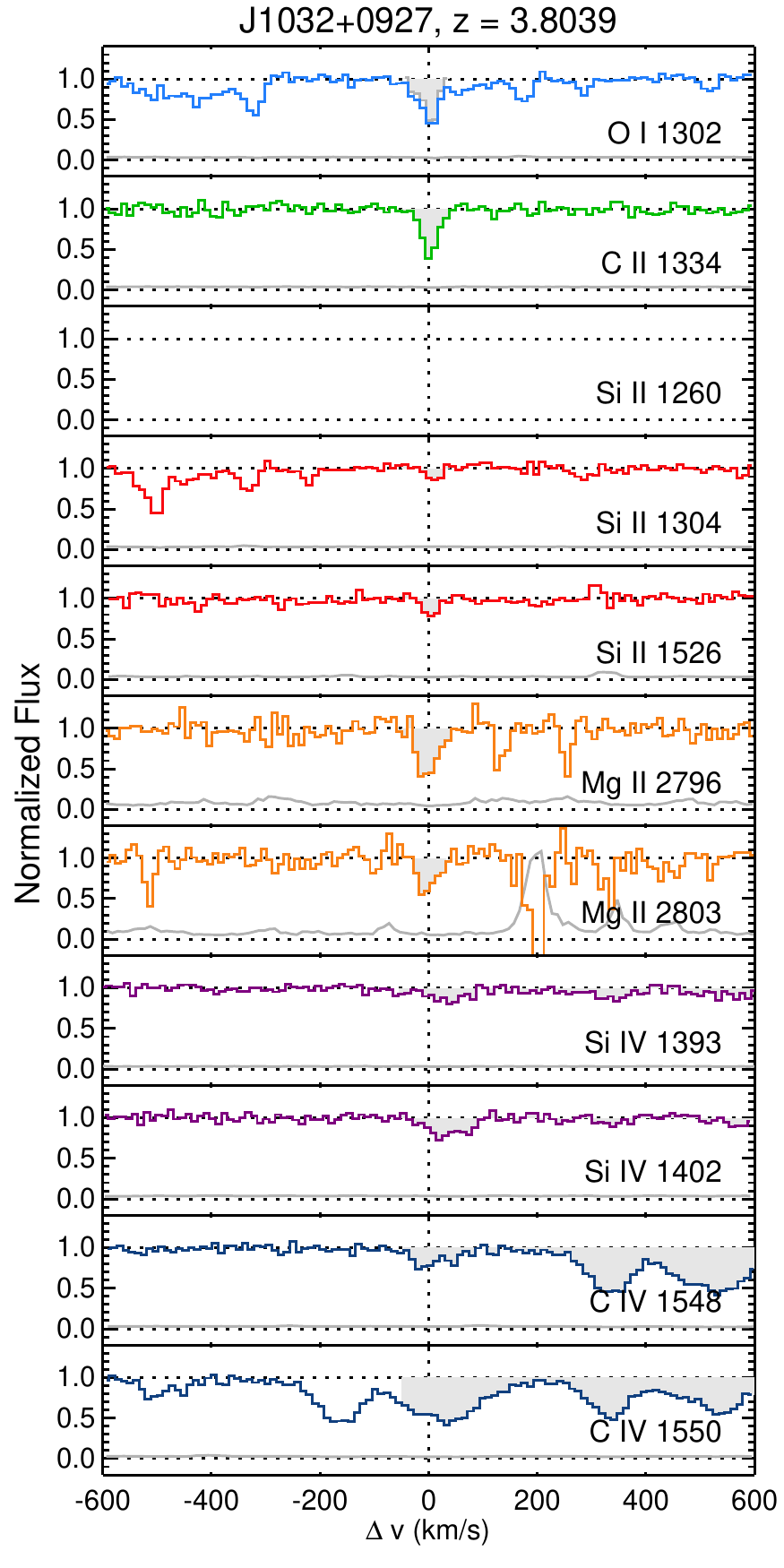}
   \caption{Stacked velocity plot for the $z=3.8039$ system towards J1032+0927.  Lines and shading are as described in Figure~\ref{fig:z3p3844720_J1018+0548}.  The grey histogram in the \oi~\lam1302 panel is the deblended flux. See notes on this system in Appendix~\ref{app:details}.\label{fig:z3p8039370_J1032+0927}}
\end{figure}
 
\begin{figure}[!b]
   \centering
   \includegraphics[height=0.40\textheight]{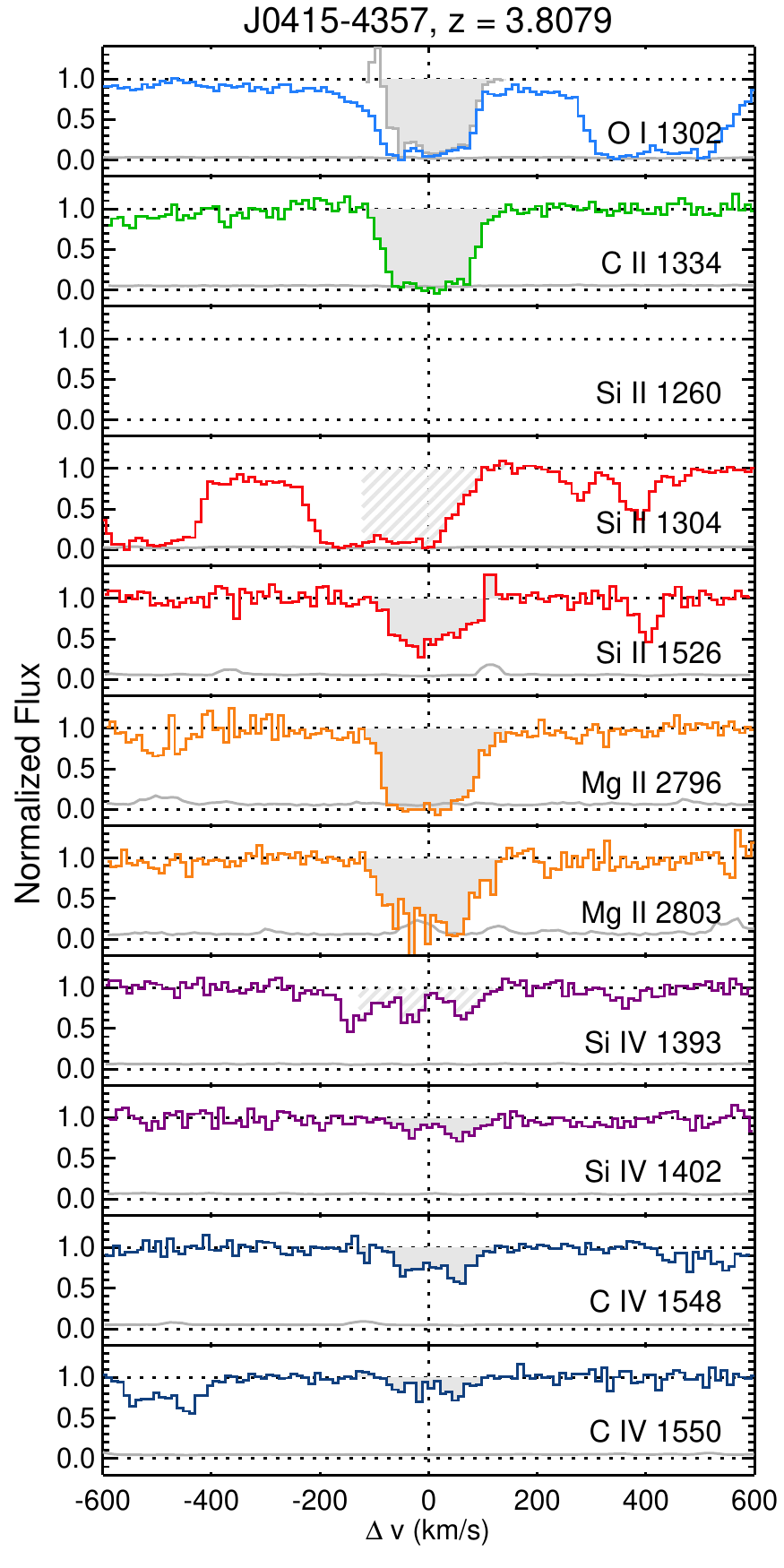}
   \caption{Stacked velocity plot for the $z=3.8079$ system towards J0415-4357.  Lines and shading are as described in Figure~\ref{fig:z3p3844720_J1018+0548}.  The grey histogram in the \oi~\lam1302 panel is the deblended flux. See notes on this system in Appendix~\ref{app:details}.\label{fig:z3p8079200_J0415-4357}}
\end{figure}
 
\begin{figure}[!t]
   \centering
   \includegraphics[height=0.40\textheight]{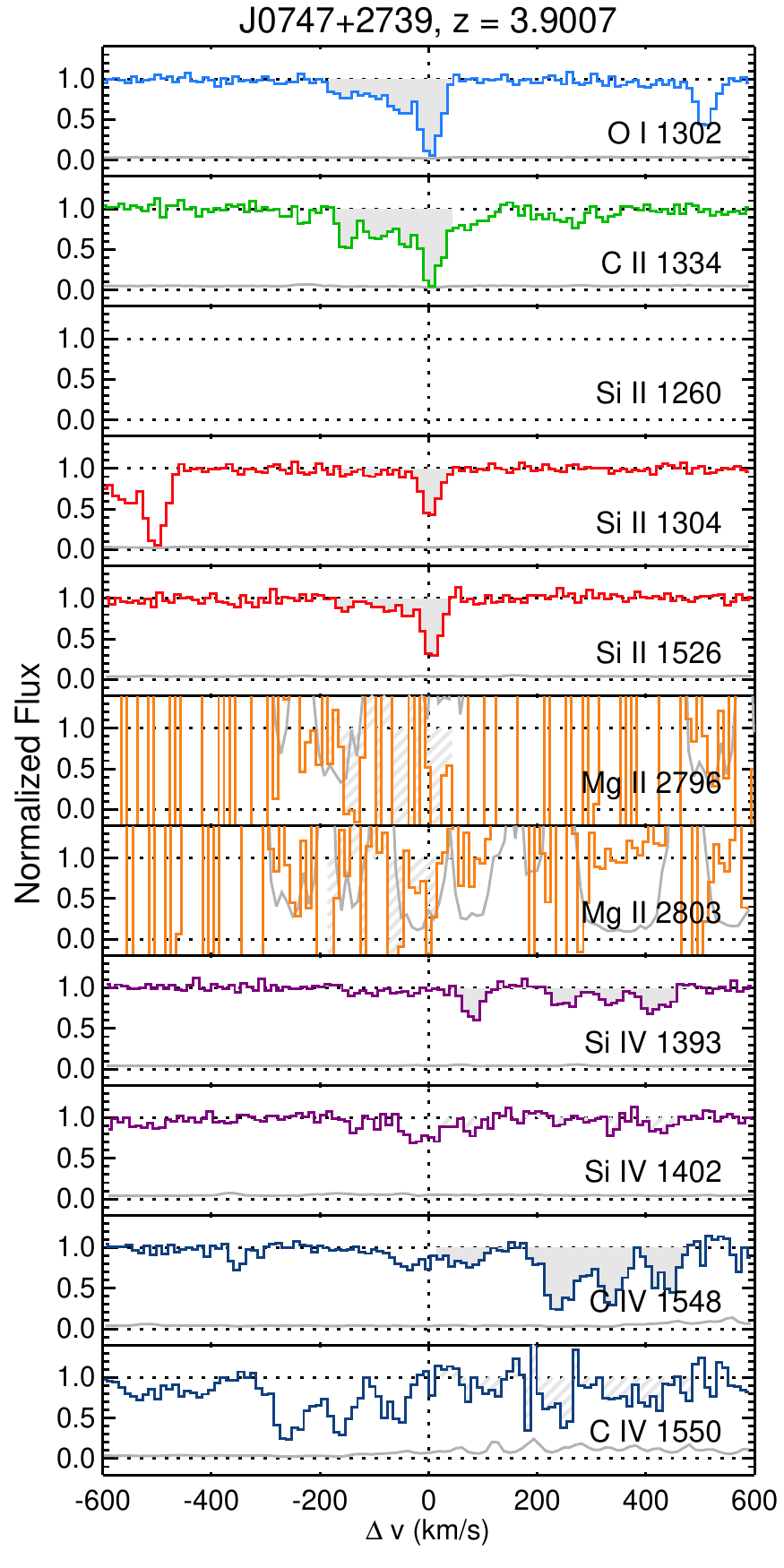}
   \caption{Stacked velocity plot for the $z=3.9007$ system towards J0747+2739.  Lines and shading are as described in Figure~\ref{fig:z3p3844720_J1018+0548}.\label{fig:z3p9006820_J0747+2739}}
\end{figure}
 
\begin{figure}[!b]
   \centering
   \includegraphics[height=0.40\textheight]{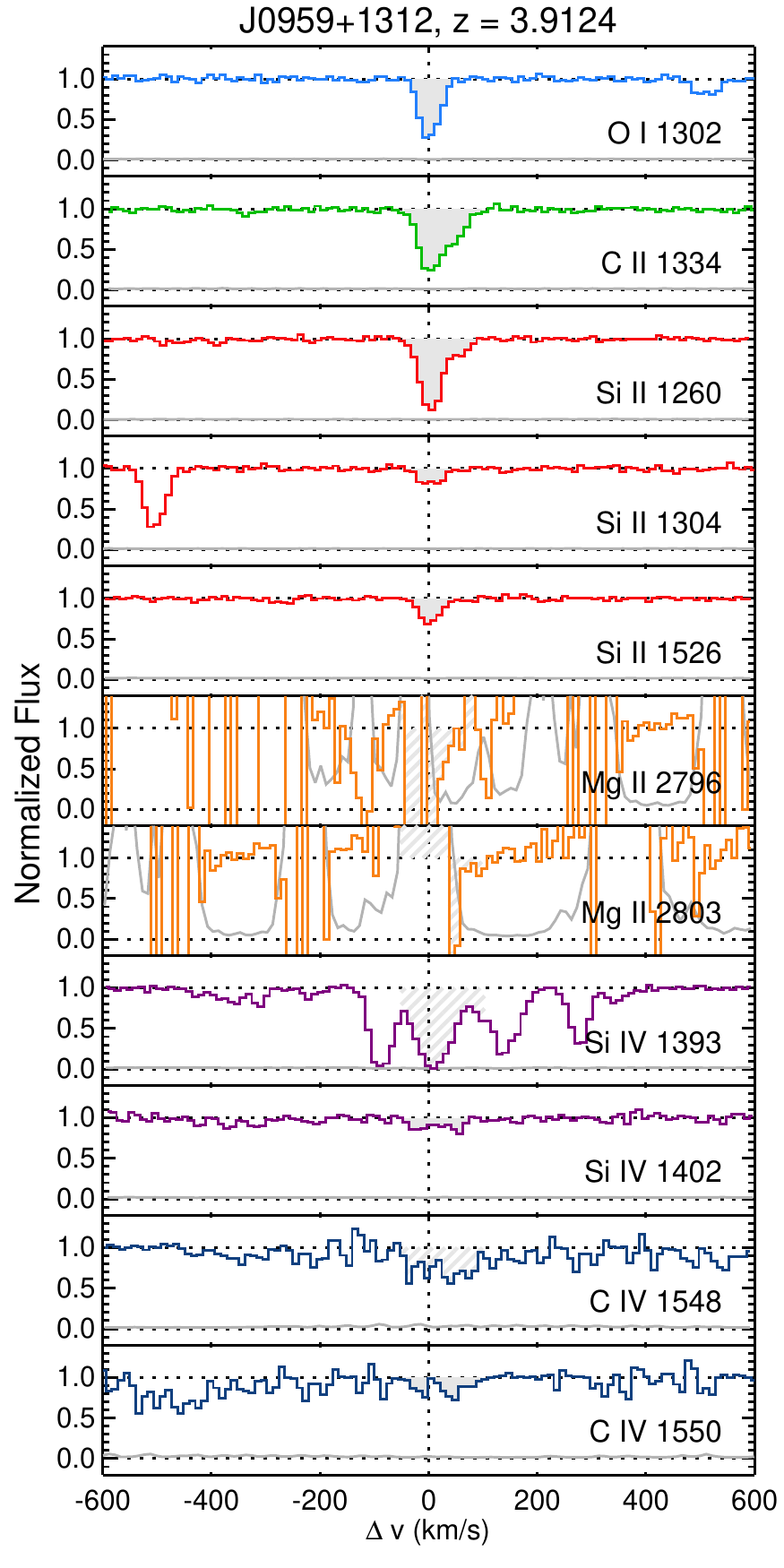}
   \caption{Stacked velocity plot for the $z=3.9124$ system towards J0959+1312.  Lines and shading are as described in Figure~\ref{fig:z3p3844720_J1018+0548}.\label{fig:z3p9124130_J0959+1312}}
\end{figure}
 
\clearpage
 
\begin{figure}[!t]
   \centering
   \includegraphics[height=0.40\textheight]{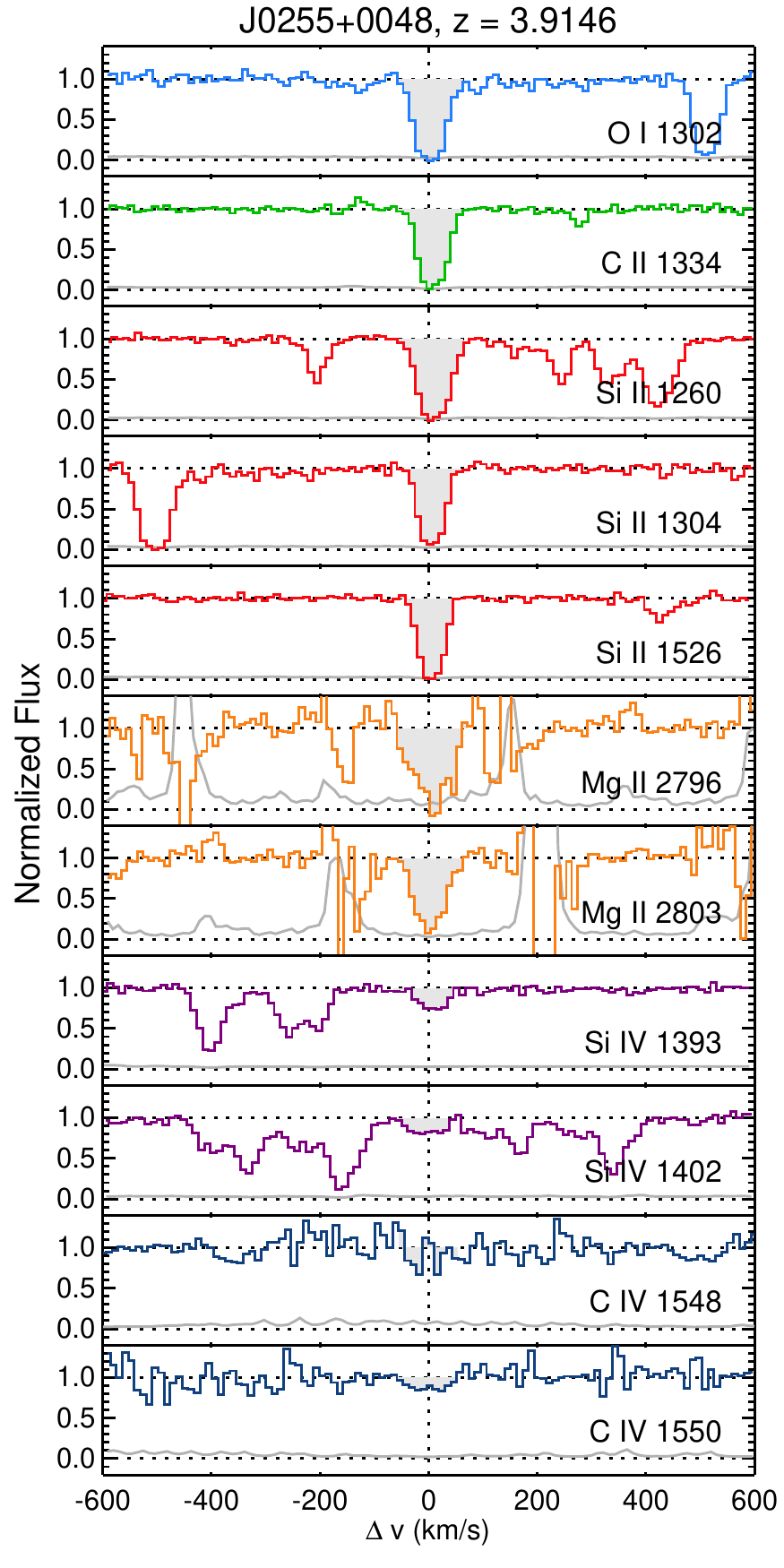}
   \caption{Stacked velocity plot for the $z=3.9146$ system towards J0255+0048.  Lines and shading are as described in Figure~\ref{fig:z3p3844720_J1018+0548}.\label{fig:z3p9146190_J0255+0048}}
\end{figure}
 
\begin{figure}[!b]
   \centering
   \includegraphics[height=0.40\textheight]{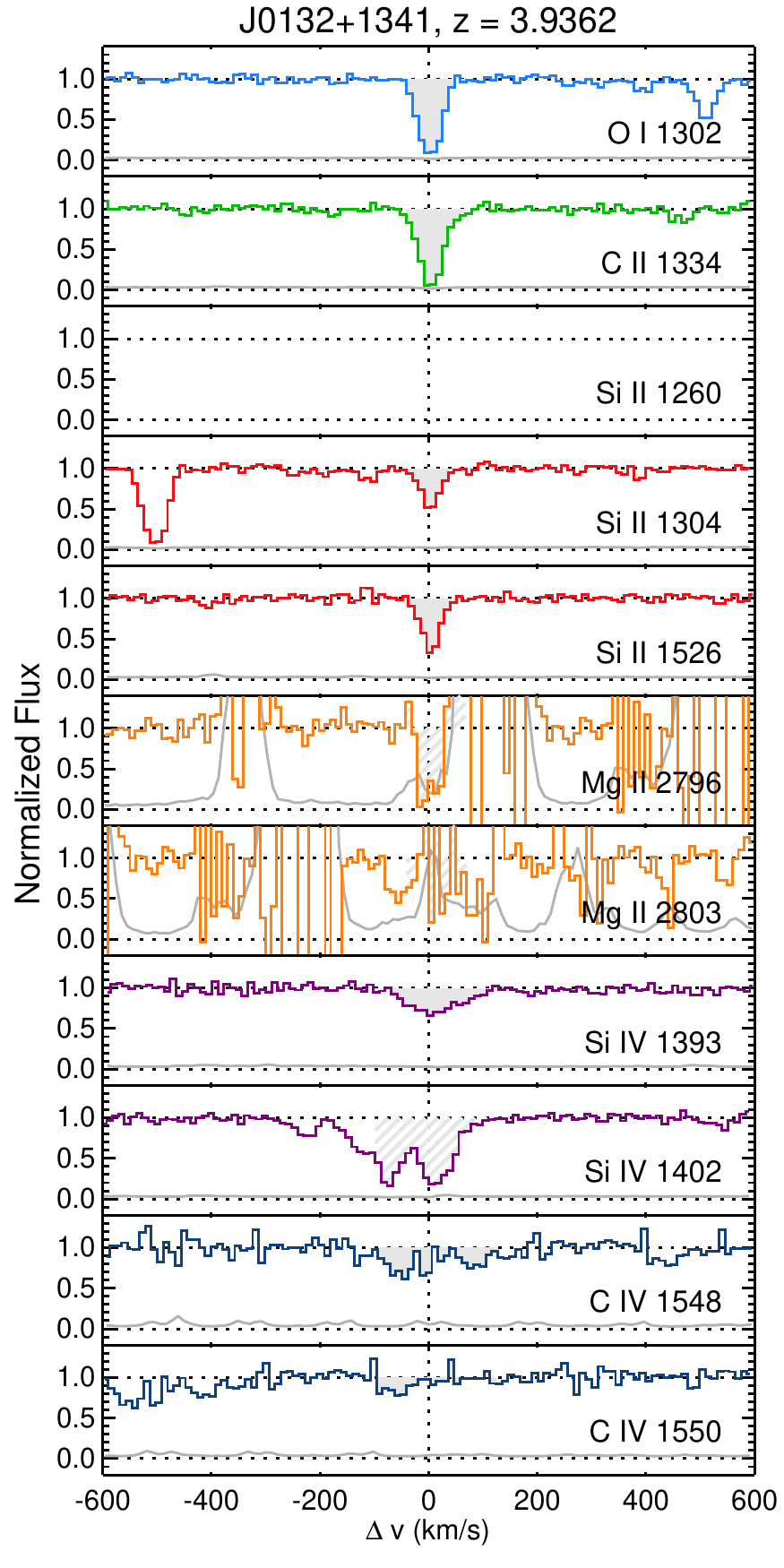}
   \caption{Stacked velocity plot for the $z=3.9362$ system towards J0132+1341.  Lines and shading are as described in Figure~\ref{fig:z3p3844720_J1018+0548}.\label{fig:z3p9362010_J0132+1341}}
\end{figure}
 
\begin{figure}[!t]
   \centering
   \includegraphics[height=0.40\textheight]{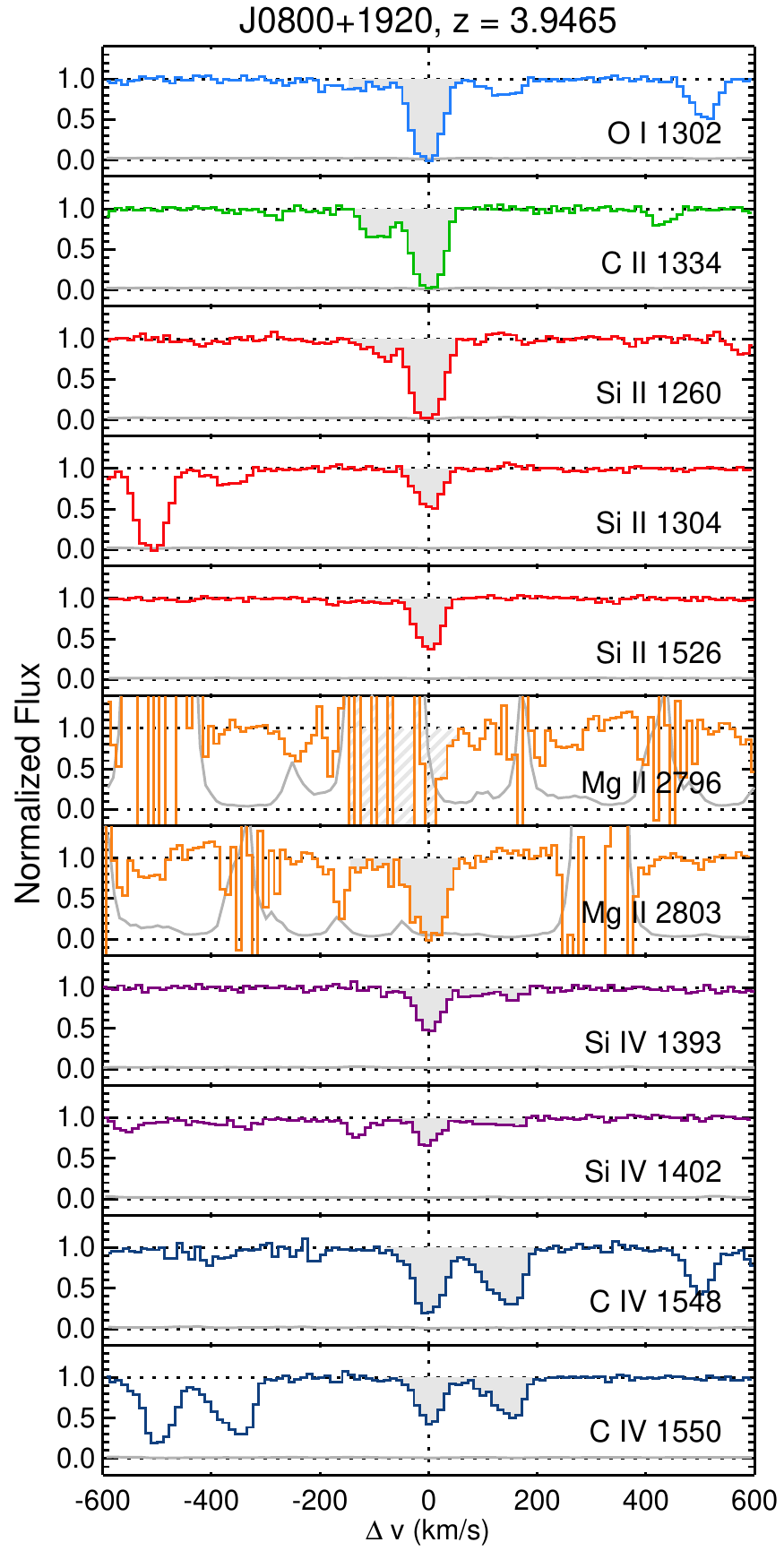}
   \caption{Stacked velocity plot for the $z=3.9465$ system towards J0800+1920.  Lines and shading are as described in Figure~\ref{fig:z3p3844720_J1018+0548}.\label{fig:z3p9464860_J0800+1920}}
\end{figure}
 
\begin{figure}[!b]
   \centering
   \includegraphics[height=0.40\textheight]{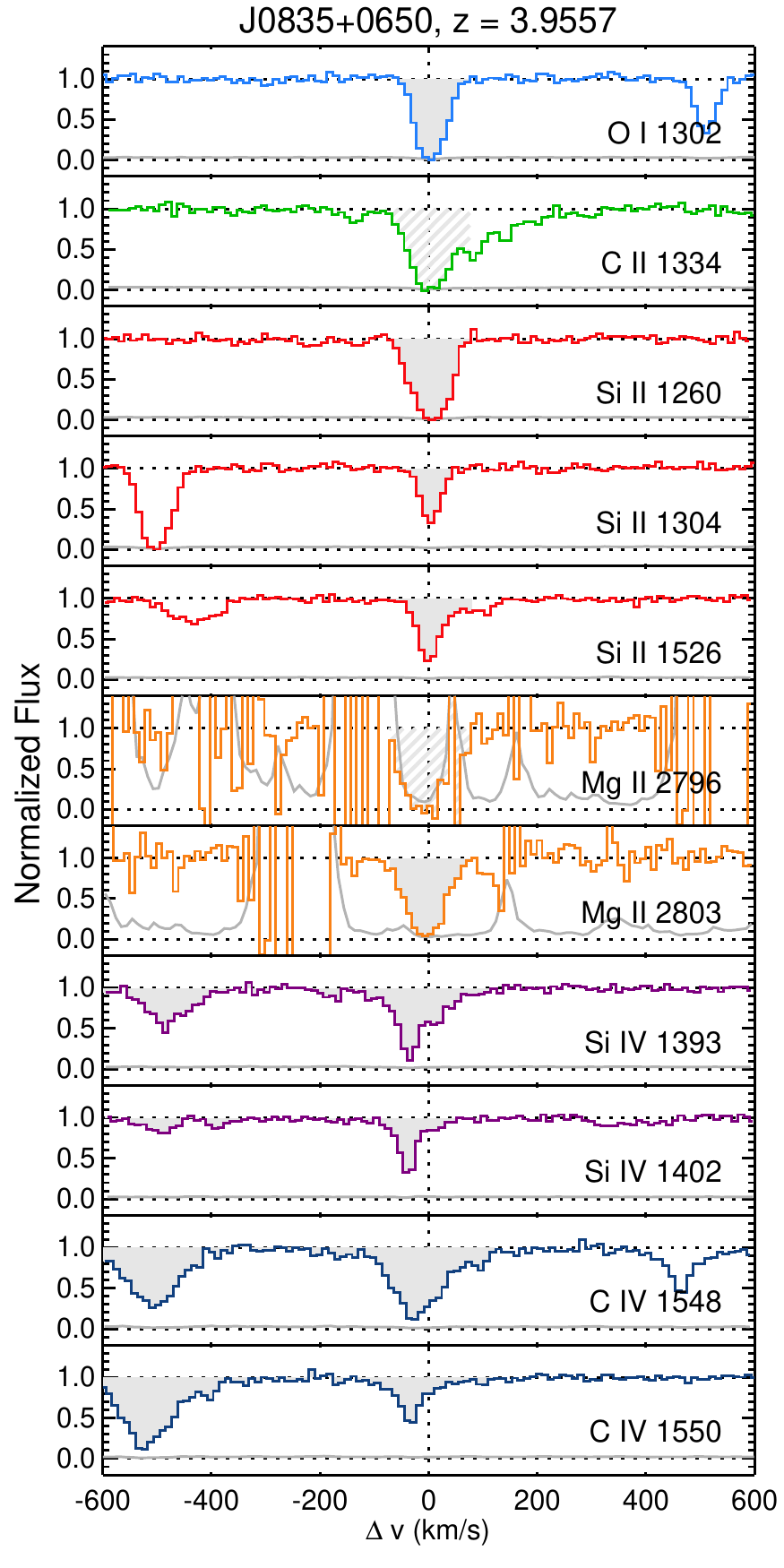}
   \caption{Stacked velocity plot for the $z=3.9557$ system towards J0835+0650.  Lines and shading are as described in Figure~\ref{fig:z3p3844720_J1018+0548}. See notes on this system in Appendix~\ref{app:details}.\label{fig:z3p9556800_J0835+0650}}
\end{figure}
 
\clearpage
 
\begin{figure}[!t]
   \centering
   \includegraphics[height=0.40\textheight]{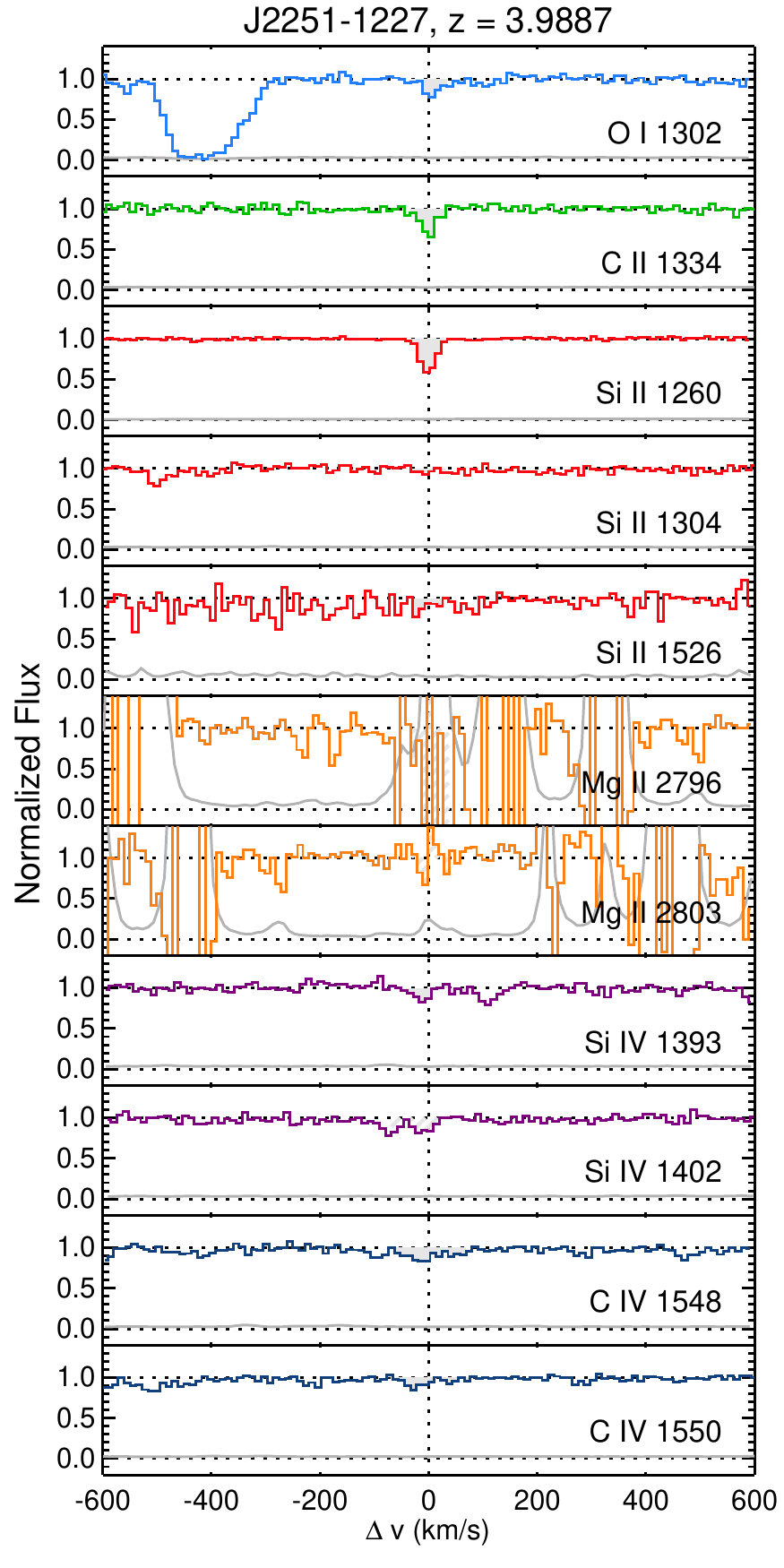}
   \caption{Stacked velocity plot for the $z=3.9887$ system towards J2251-1227.  Lines and shading are as described in Figure~\ref{fig:z3p3844720_J1018+0548}.\label{fig:z3p9886920_J2251-1227}}
\end{figure}
 
\begin{figure}[!b]
   \centering
   \includegraphics[height=0.40\textheight]{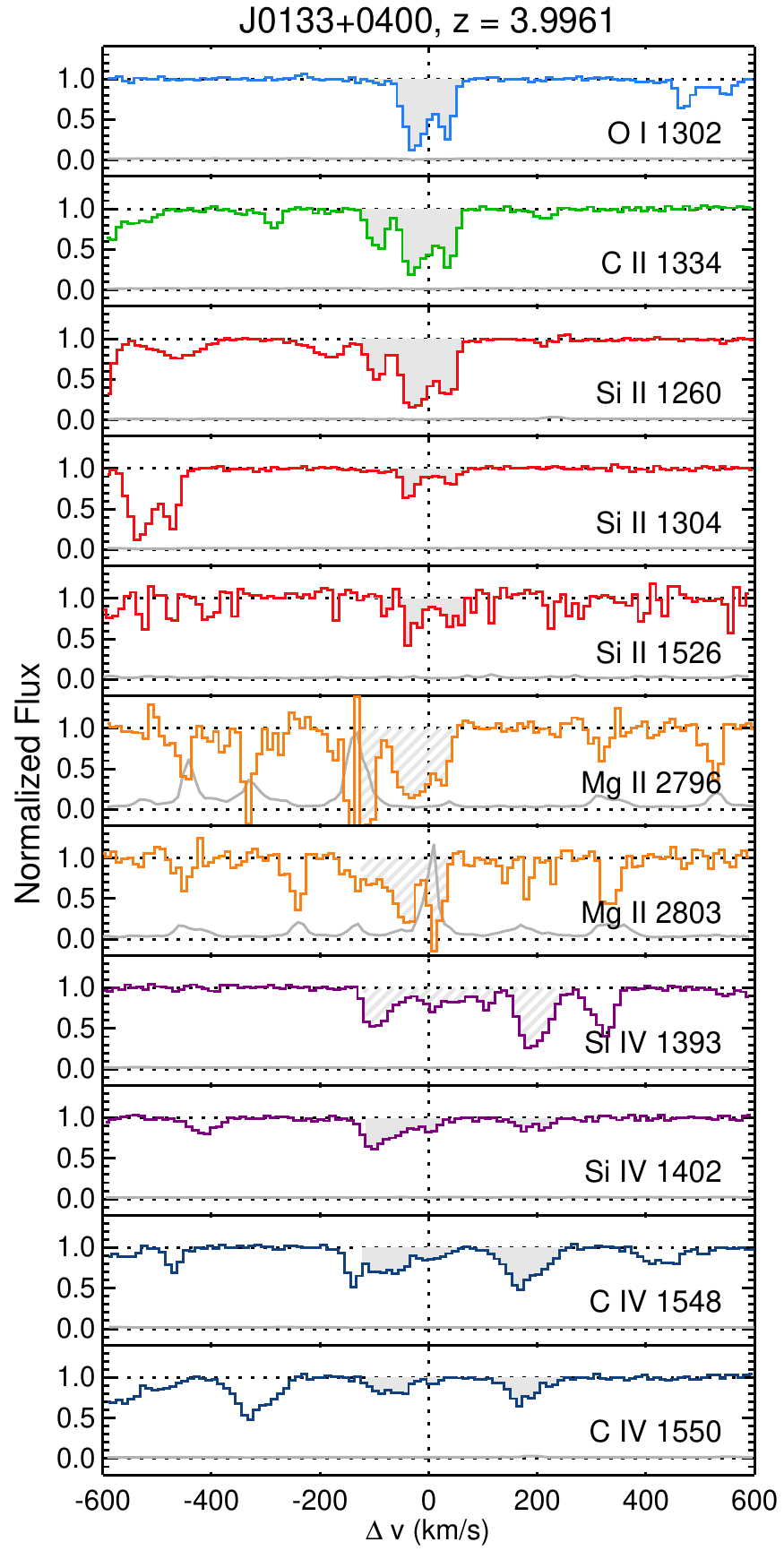}
   \caption{Stacked velocity plot for the $z=3.9961$ system towards J0133+0400.  Lines and shading are as described in Figure~\ref{fig:z3p3844720_J1018+0548}.\label{fig:z3p9960810_J0133+0400}}
\end{figure}
 
\begin{figure}[!t]
   \centering
   \includegraphics[height=0.40\textheight]{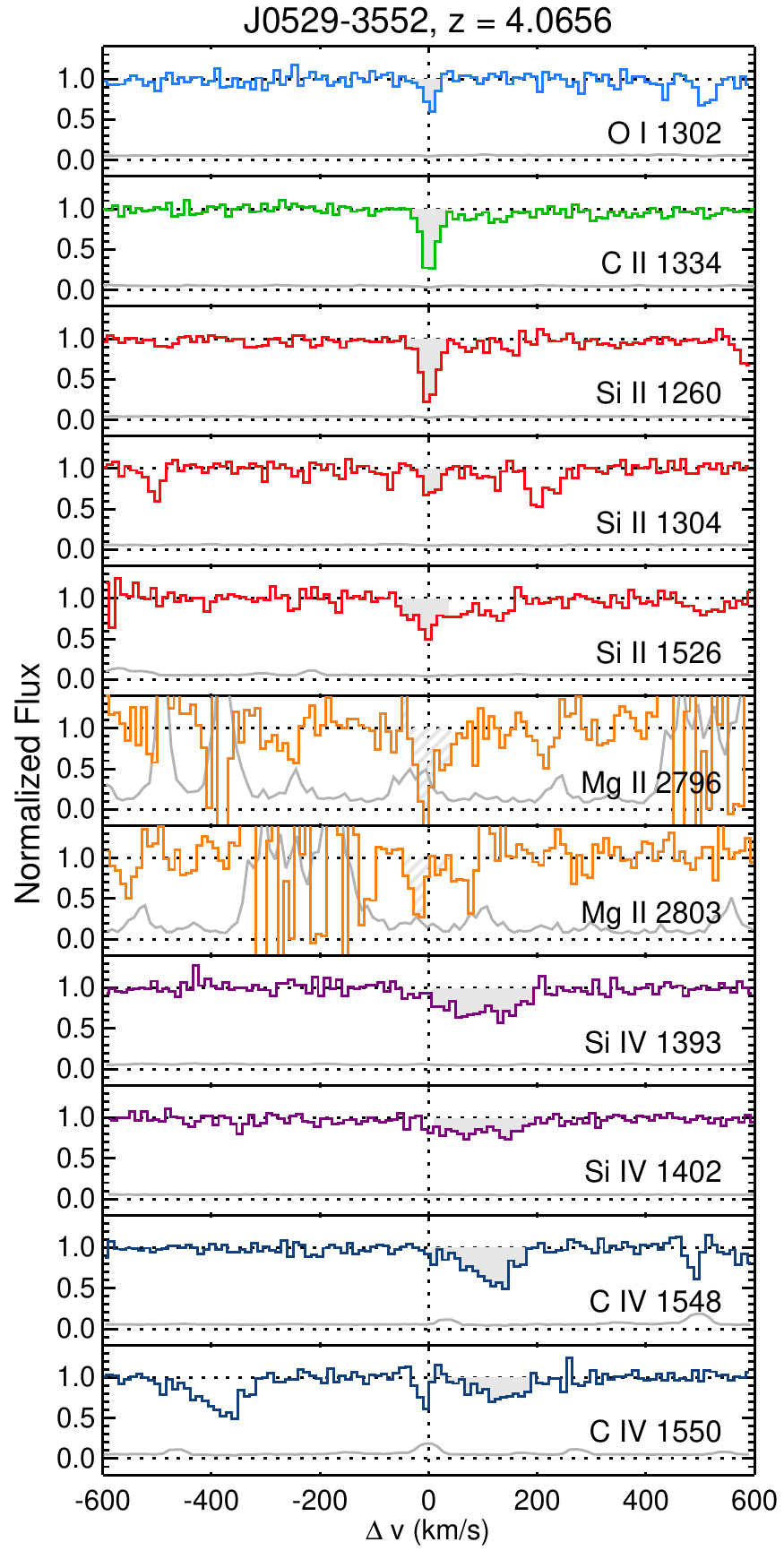}
   \caption{Stacked velocity plot for the $z=4.0656$ system towards J0529-3552.  Lines and shading are as described in Figure~\ref{fig:z3p3844720_J1018+0548}.\label{fig:z4p0655950_J0529-3552}}
\end{figure}
 
\begin{figure}[!b]
   \centering
   \includegraphics[height=0.40\textheight]{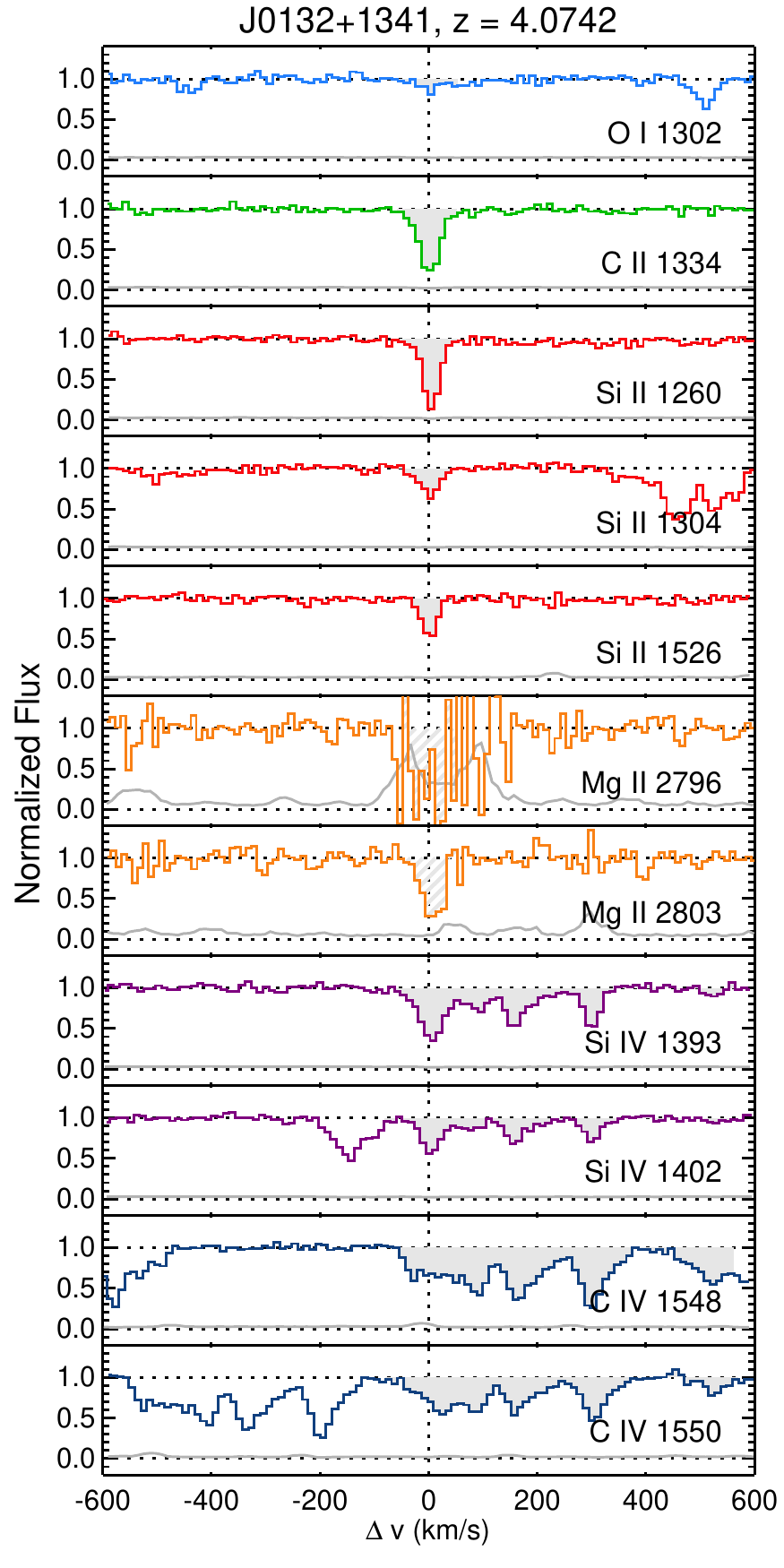}
   \caption{Stacked velocity plot for the $z=4.0742$ system towards J0132+1341.  Lines and shading are as described in Figure~\ref{fig:z3p3844720_J1018+0548}. See notes on this system in Appendix~\ref{app:details}.\label{fig:z4p0741980_J0132+1341}}
\end{figure}
 
\clearpage
 
\begin{figure}[!t]
   \centering
   \includegraphics[height=0.40\textheight]{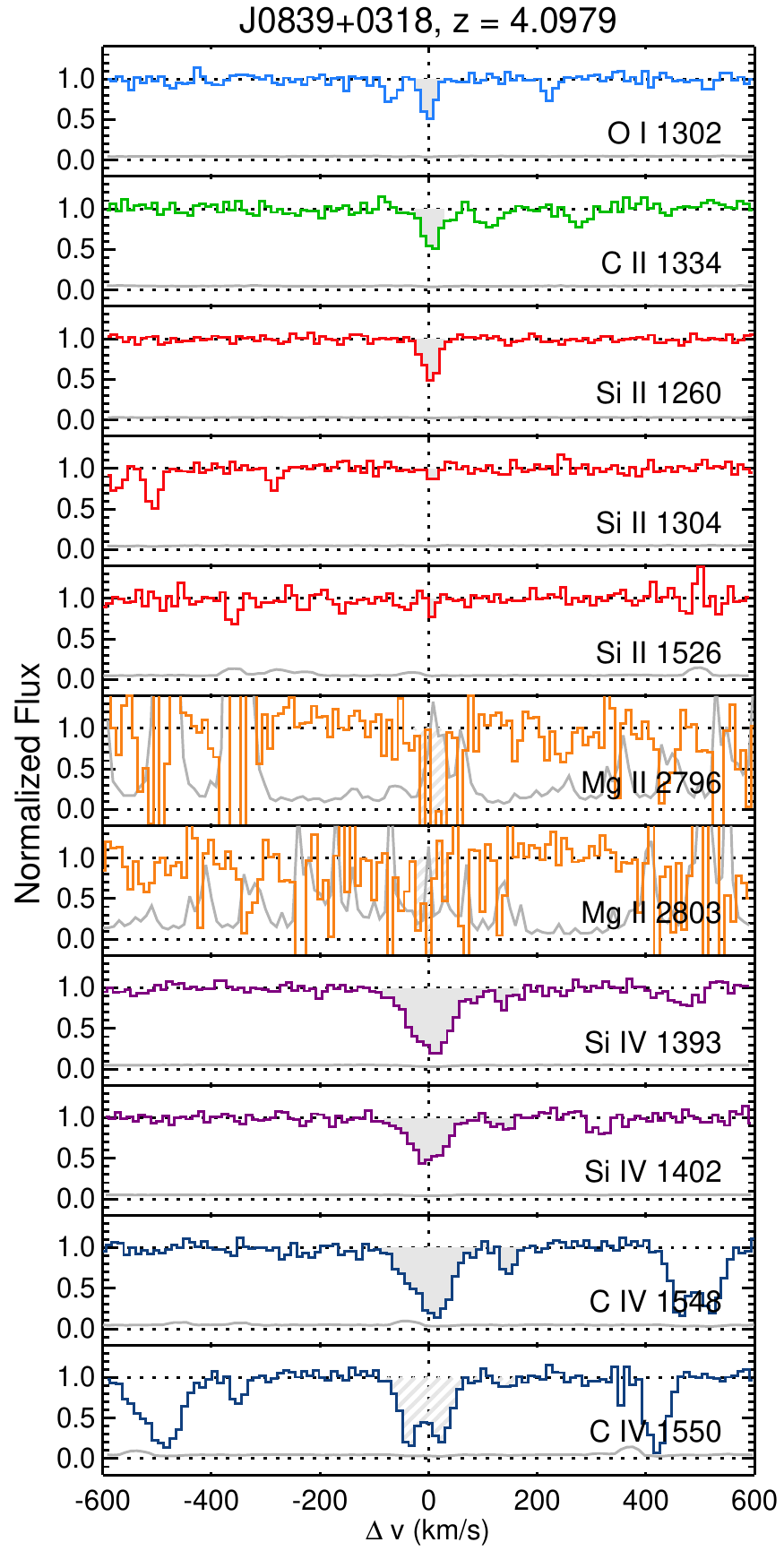}
   \caption{Stacked velocity plot for the $z=4.0979$ system towards J0839+0318.  Lines and shading are as described in Figure~\ref{fig:z3p3844720_J1018+0548}.\label{fig:z4p0979260_J0839+0318}}
\end{figure}
 
\begin{figure}[!b]
   \centering
   \includegraphics[height=0.40\textheight]{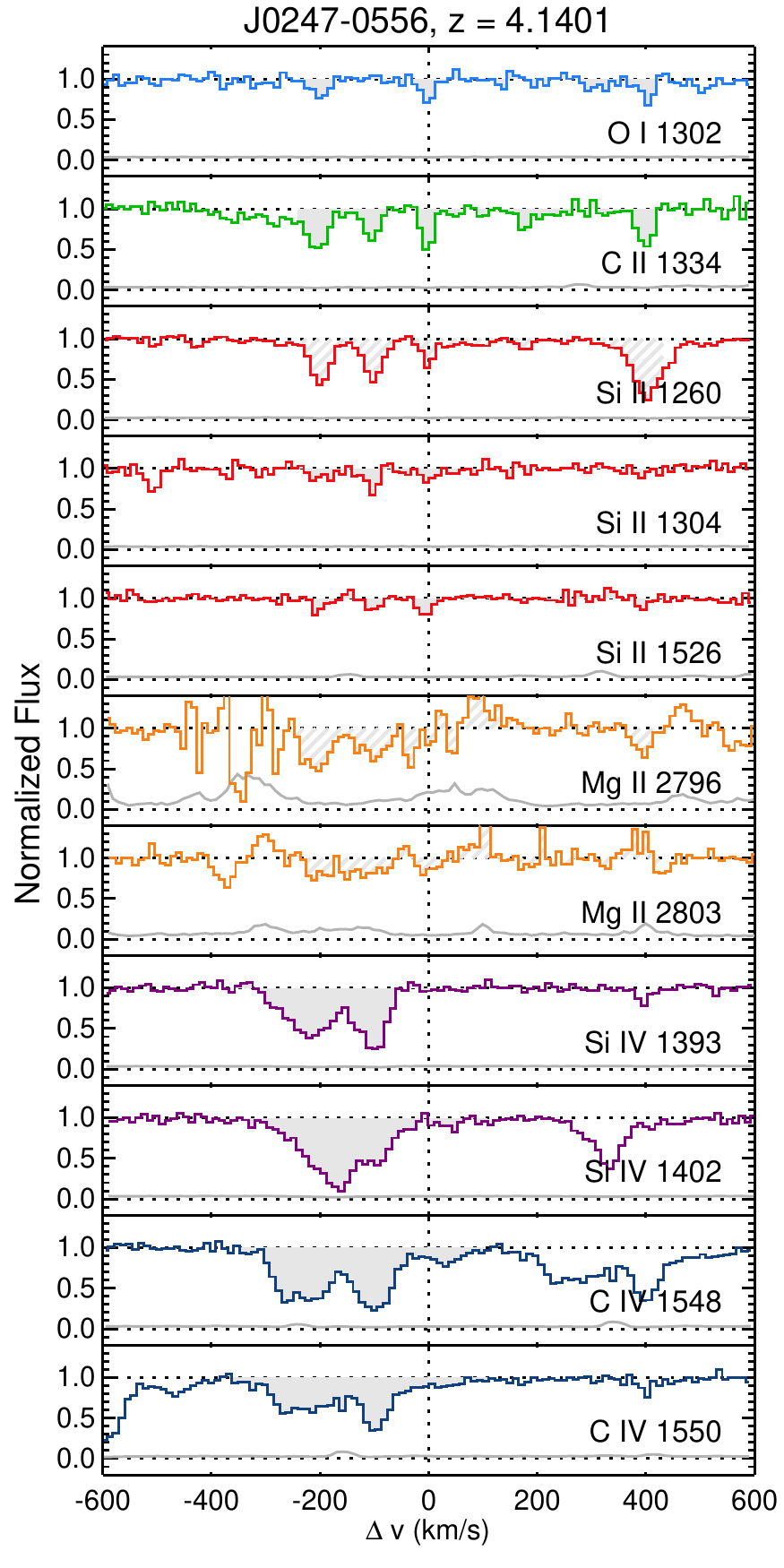}
   \caption{Stacked velocity plot for the $z=4.1401$ system towards J0247-0556.  Lines and shading are as described in Figure~\ref{fig:z3p3844720_J1018+0548}. See notes on this system in Appendix~\ref{app:details}.\label{fig:z4p1401050_J0247-0556}}
\end{figure}
 
\begin{figure}[!t]
   \centering
   \includegraphics[height=0.40\textheight]{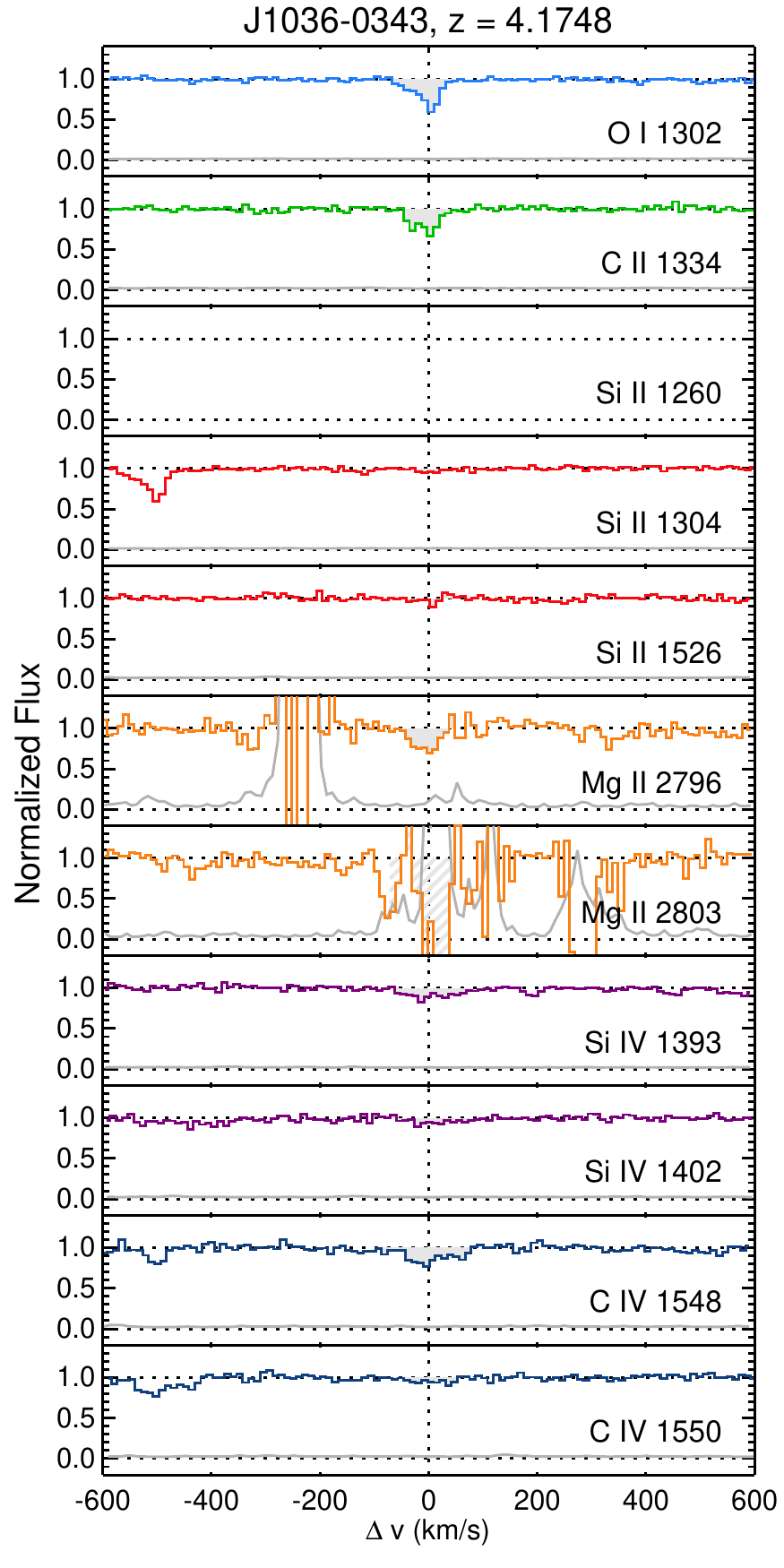}
   \caption{Stacked velocity plot for the $z=4.1748$ system towards J1036-0343.  Lines and shading are as described in Figure~\ref{fig:z3p3844720_J1018+0548}.\label{fig:z4p1747810_J1036-0343}}
\end{figure}
 
\begin{figure}[!b]
   \centering
   \includegraphics[height=0.40\textheight]{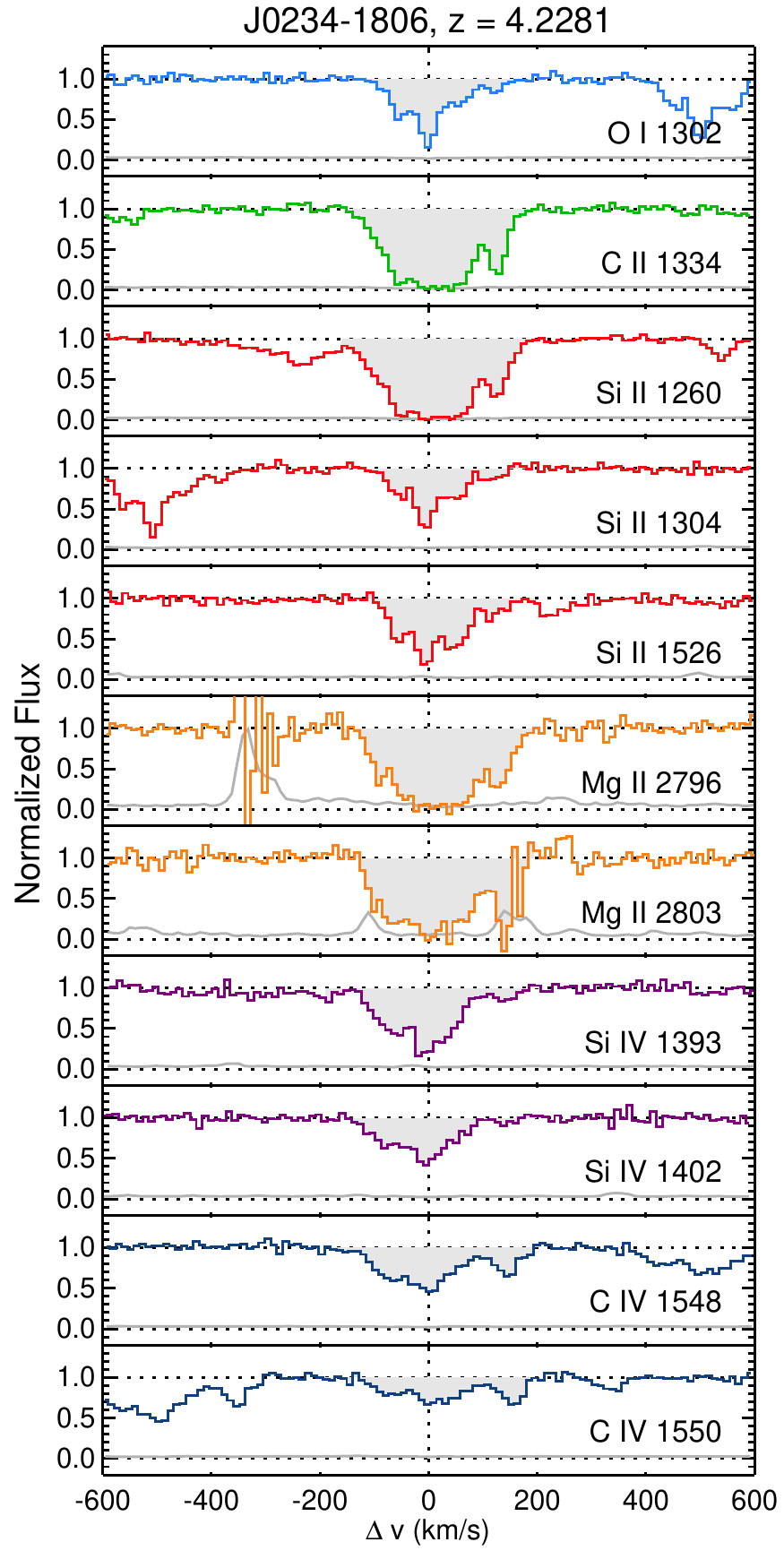}
   \caption{Stacked velocity plot for the $z=4.2281$ system towards J0234-1806.  Lines and shading are as described in Figure~\ref{fig:z3p3844720_J1018+0548}.\label{fig:z4p2281300_J0234-1806}}
\end{figure}
 
\clearpage
 
\begin{figure}[!t]
   \centering
   \includegraphics[height=0.40\textheight]{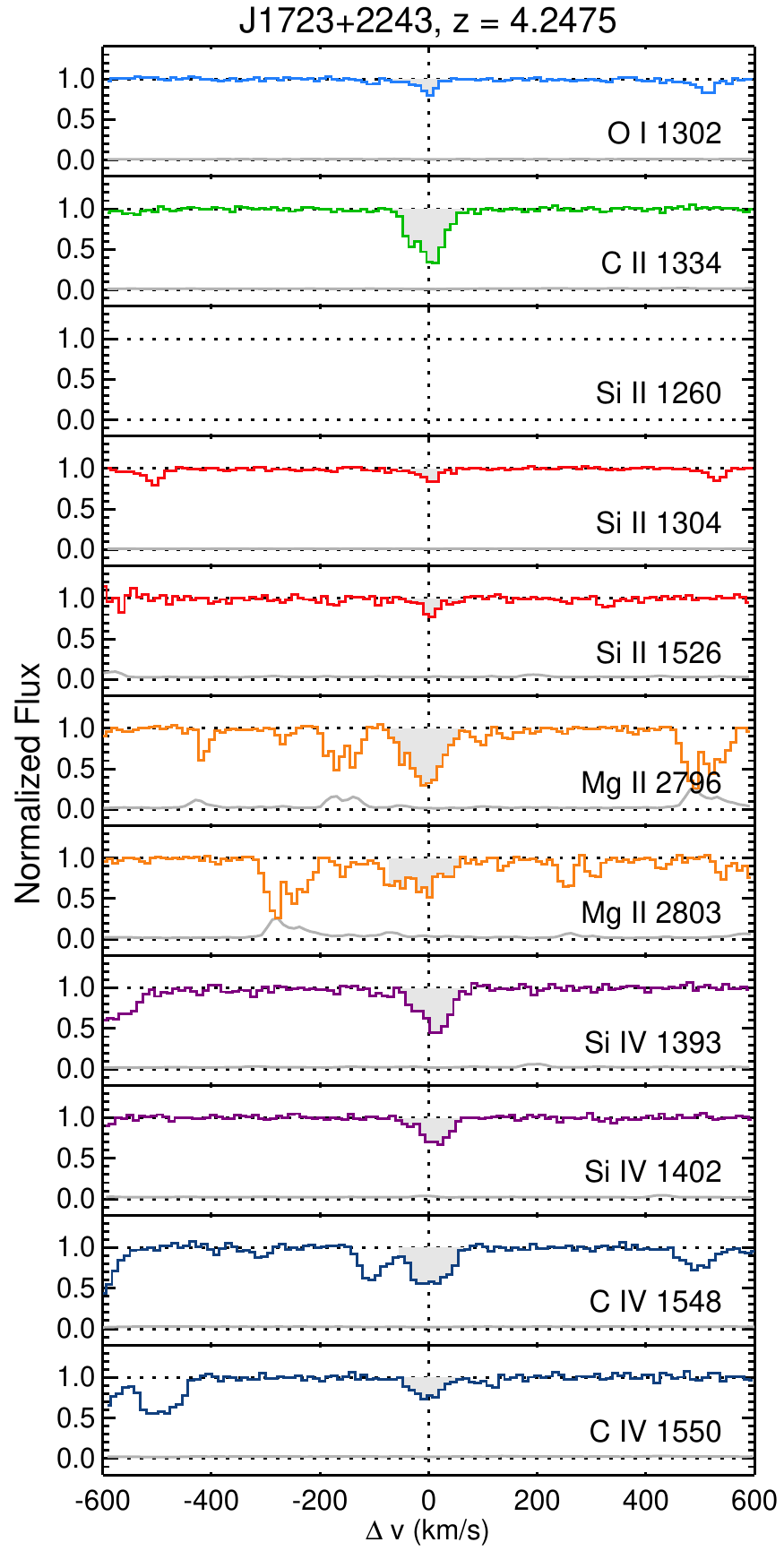}
   \caption{Stacked velocity plot for the $z=4.2475$ system towards J1723+2243.  Lines and shading are as described in Figure~\ref{fig:z3p3844720_J1018+0548}.\label{fig:z4p2474650_J1723+2243}}
\end{figure}
 
\begin{figure}[!b]
   \centering
   \includegraphics[height=0.40\textheight]{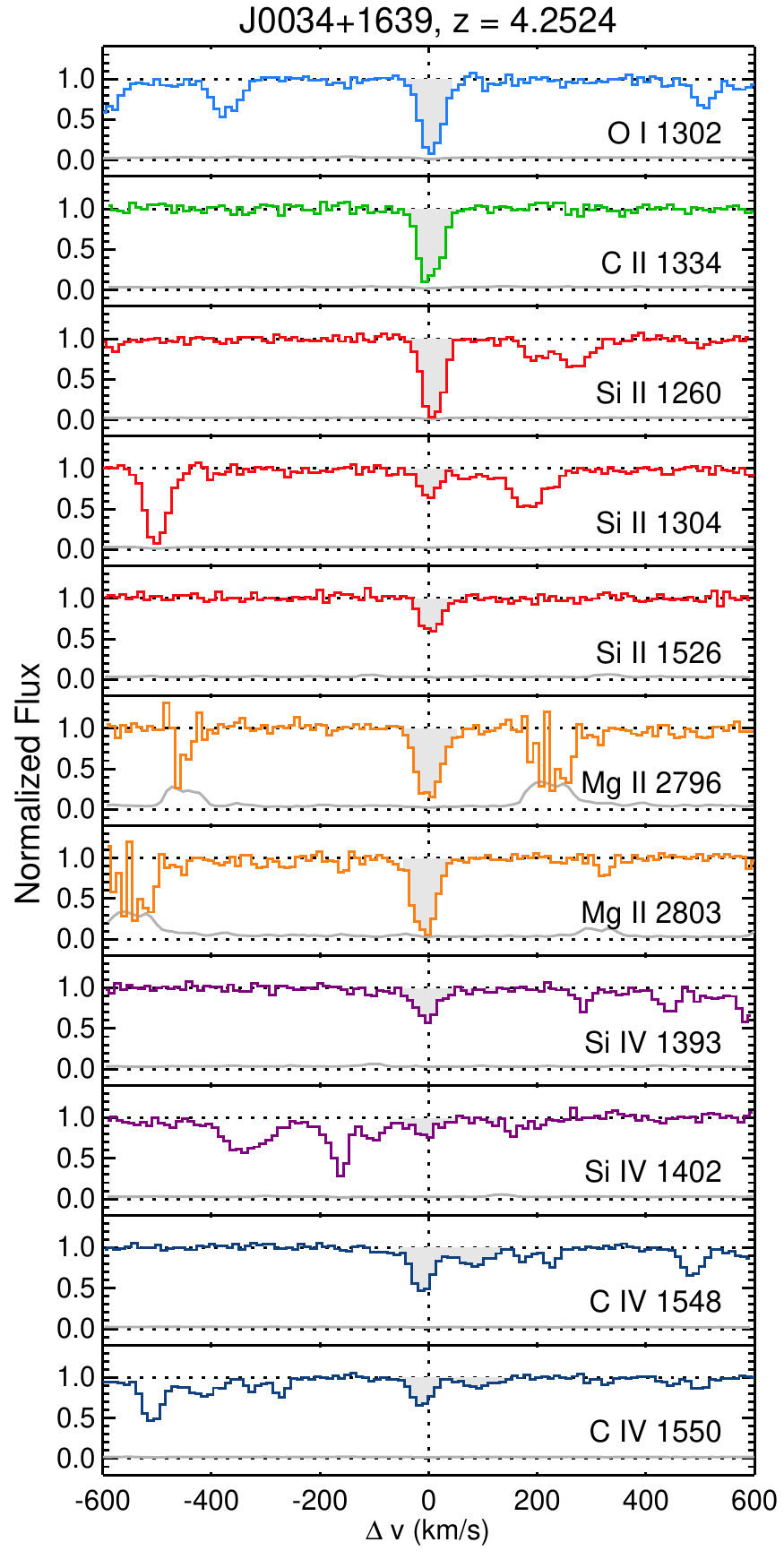}
   \caption{Stacked velocity plot for the $z=4.2524$ system towards J0034+1639.  Lines and shading are as described in Figure~\ref{fig:z3p3844720_J1018+0548}.\label{fig:z4p2524280_J0034+1639}}
\end{figure}
 
\begin{figure}[!t]
   \centering
   \includegraphics[height=0.40\textheight]{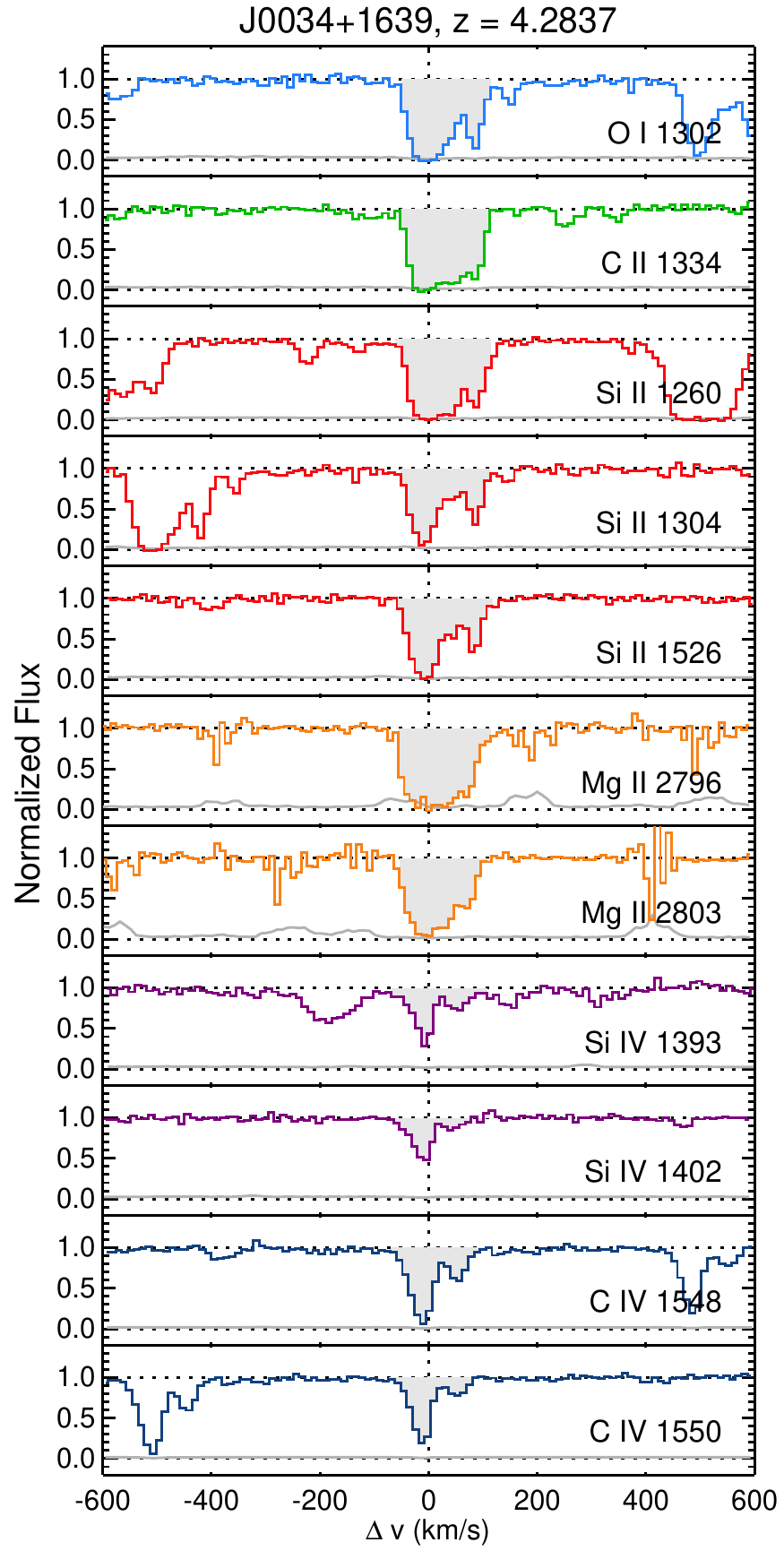}
   \caption{Stacked velocity plot for the $z=4.2837$ system towards J0034+1639.  Lines and shading are as described in Figure~\ref{fig:z3p3844720_J1018+0548}.\label{fig:z4p2836630_J0034+1639}}
\end{figure}
 
\begin{figure}[!b]
   \centering
   \includegraphics[height=0.40\textheight]{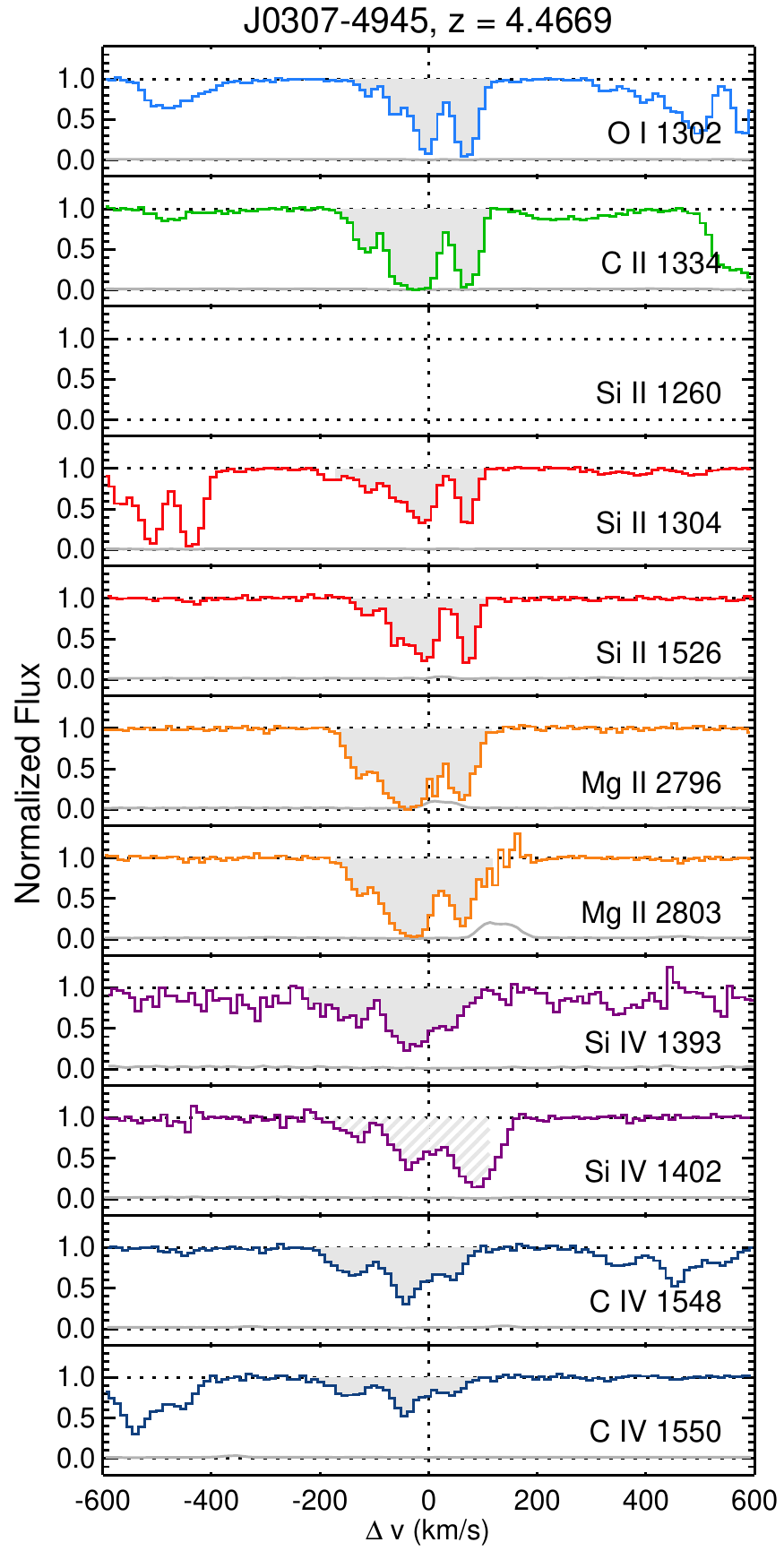}
   \caption{Stacked velocity plot for the $z=4.4669$ system towards J0307-4945.  Lines and shading are as described in Figure~\ref{fig:z3p3844720_J1018+0548}.\label{fig:z4p4668560_J0307-4945}}
\end{figure}
 
\clearpage
 
\begin{figure}[!t]
   \centering
   \includegraphics[height=0.40\textheight]{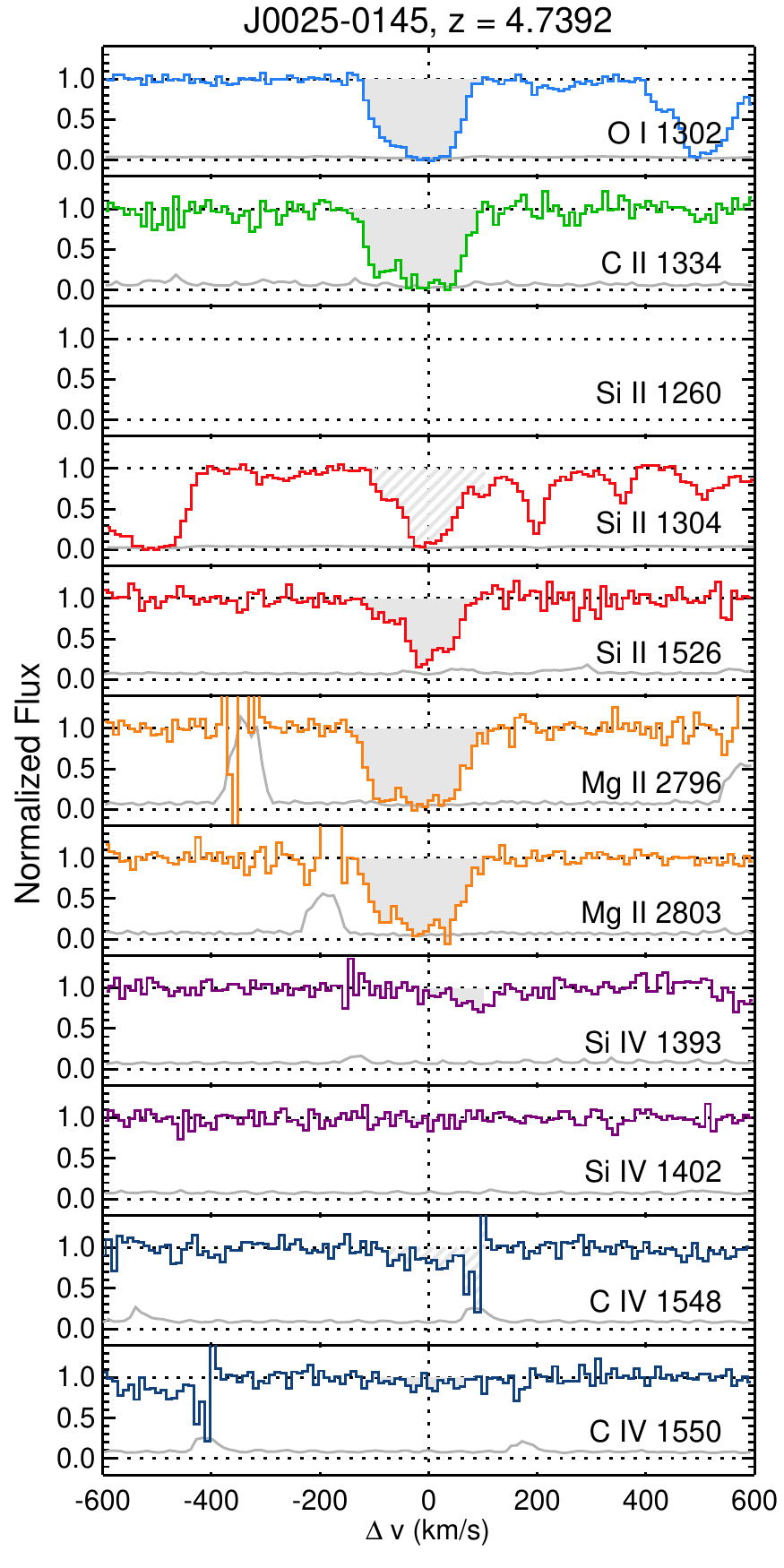}
   \caption{Stacked velocity plot for the $z=4.7392$ system towards J0025-0145.  Lines and shading are as described in Figure~\ref{fig:z3p3844720_J1018+0548}.\label{fig:z4p7391580_J0025-0145}}
\end{figure}
 
\begin{figure}[!b]
   \centering
   \includegraphics[height=0.40\textheight]{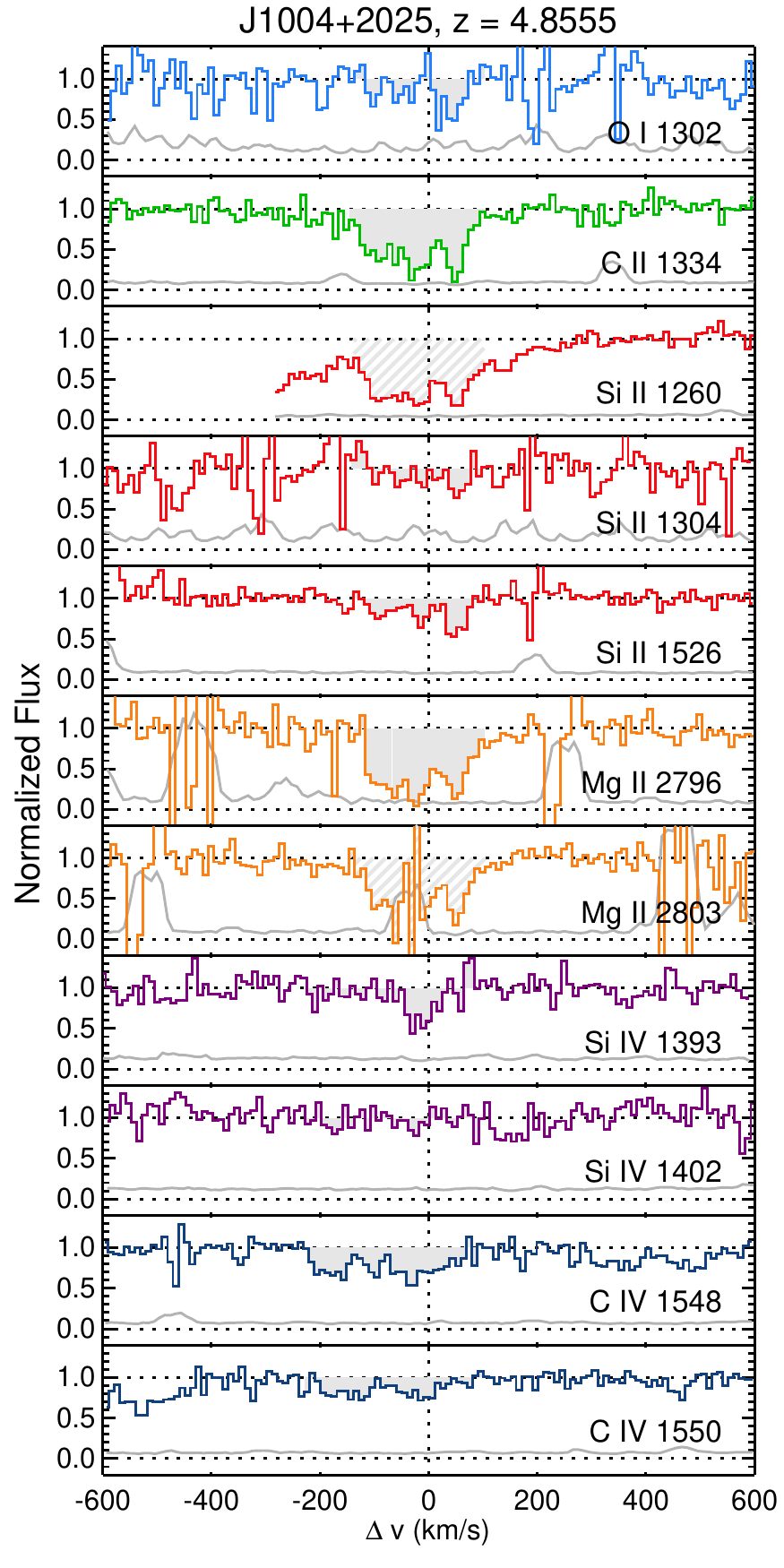}
   \caption{Stacked velocity plot for the $z=4.8555$ system towards J1004+2025.  Lines and shading are as described in Figure~\ref{fig:z3p3844720_J1018+0548}. See notes on this system in Appendix~\ref{app:details}.\label{fig:z4p8554890_J1004+2025}}
\end{figure}
 
\begin{figure}[!t]
   \centering
   \includegraphics[height=0.40\textheight]{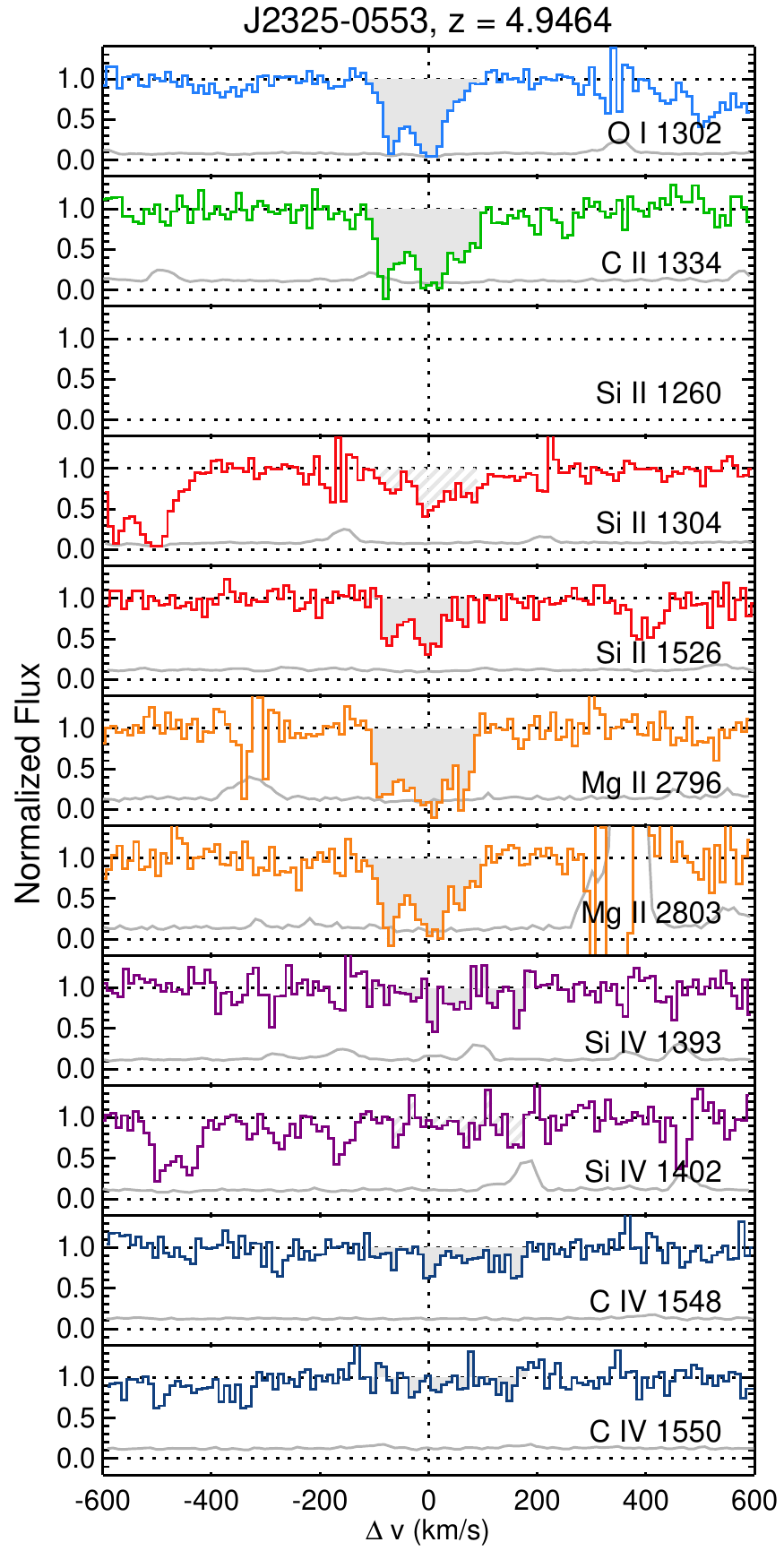}
   \caption{Stacked velocity plot for the $z=4.9464$ system towards J2325-0553.  Lines and shading are as described in Figure~\ref{fig:z3p3844720_J1018+0548}.\label{fig:z4p9463820_J2325-0553}}
\end{figure}
 
\begin{figure}[!b]
   \centering
   \includegraphics[height=0.40\textheight]{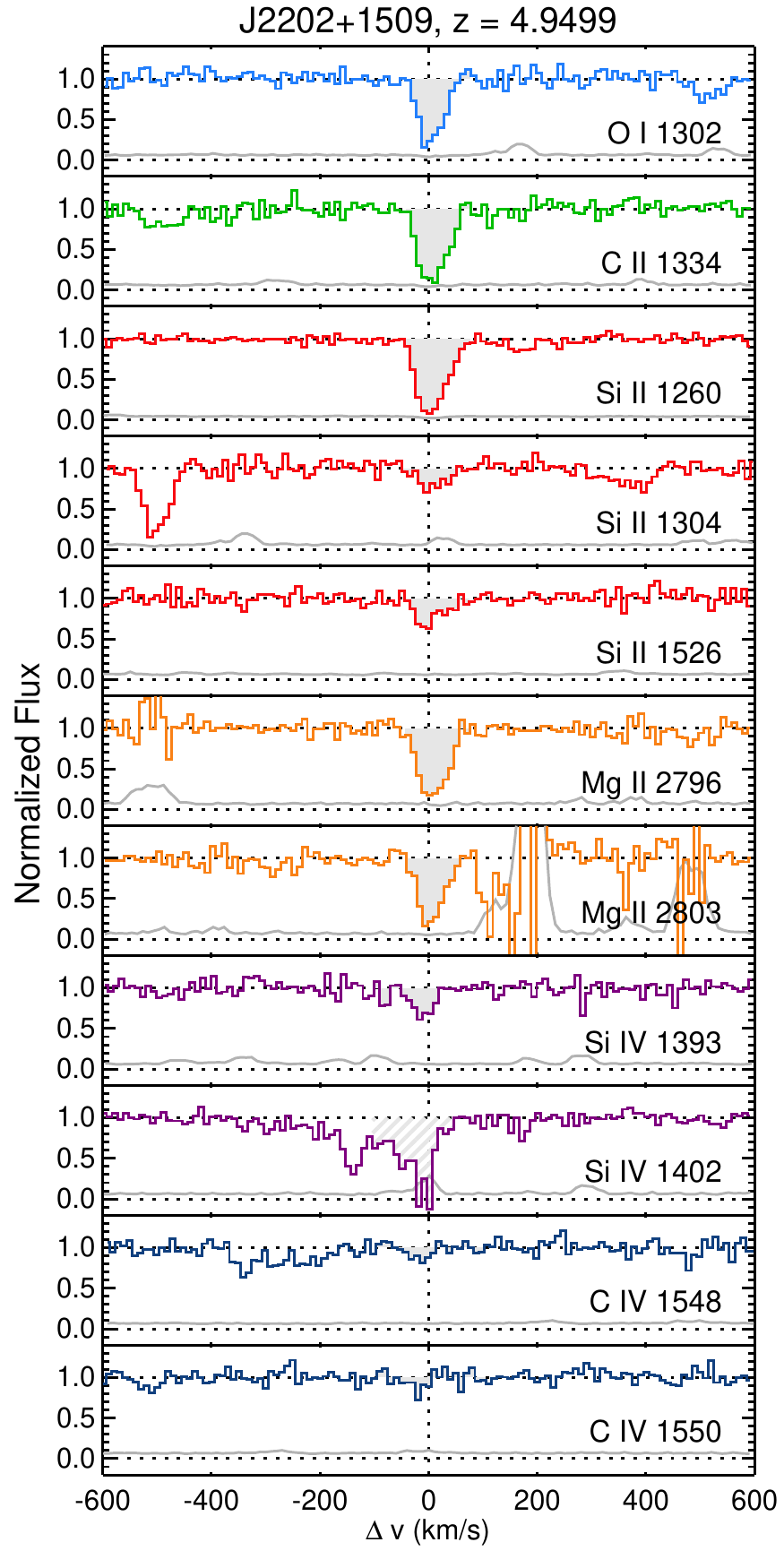}
   \caption{Stacked velocity plot for the $z=4.9499$ system towards J2202+1509.  Lines and shading are as described in Figure~\ref{fig:z3p3844720_J1018+0548}.\label{fig:z4p9499140_J2202+1509}}
\end{figure}
 
\clearpage
 
\begin{figure}[!t]
   \centering
   \includegraphics[height=0.40\textheight]{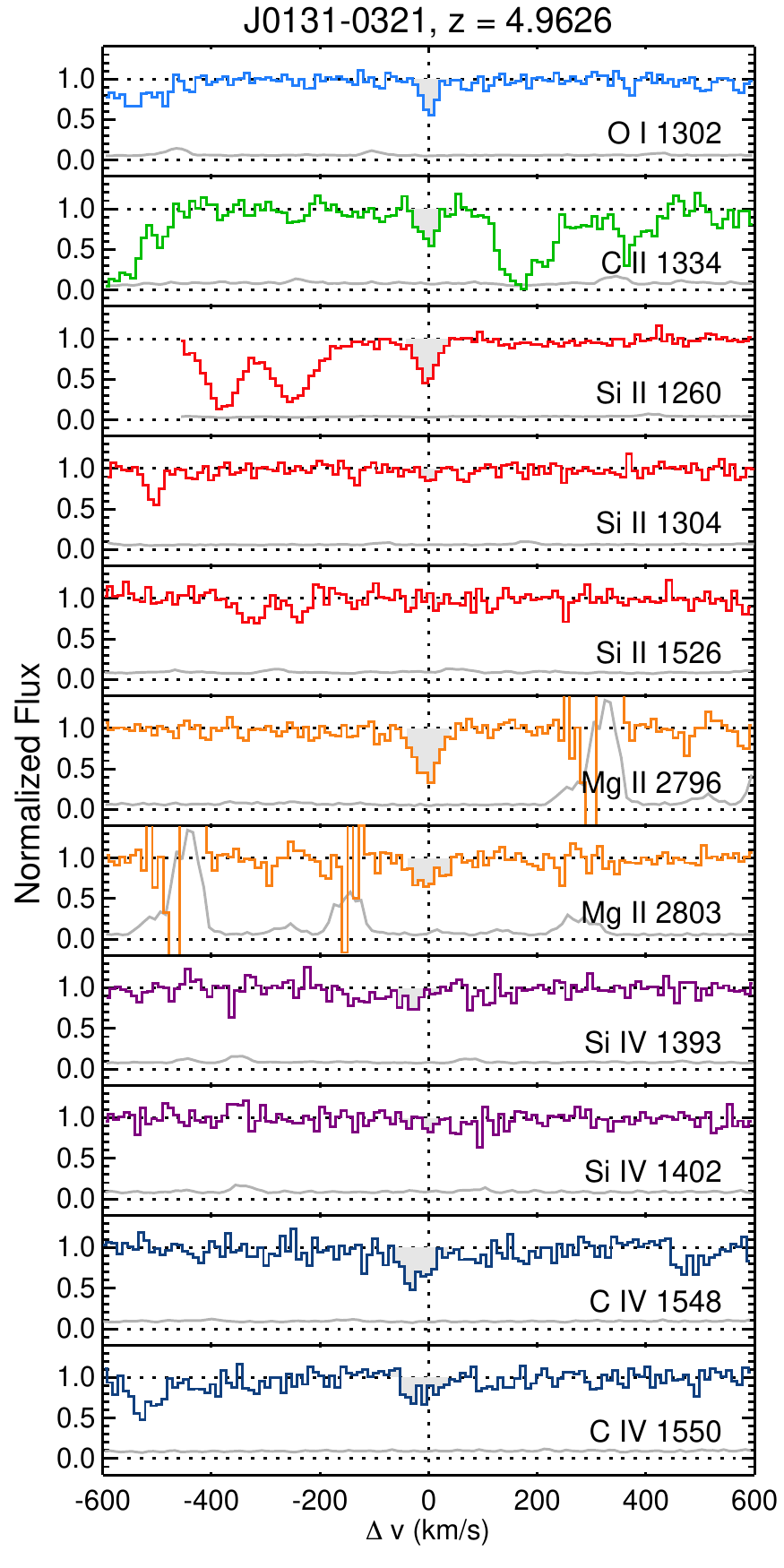}
   \caption{Stacked velocity plot for the $z=4.9626$ system towards J0131-0321.  Lines and shading are as described in Figure~\ref{fig:z3p3844720_J1018+0548}.\label{fig:z4p9625770_J0131-0321}}
\end{figure}
 
\begin{figure}[!b]
   \centering
   \includegraphics[height=0.40\textheight]{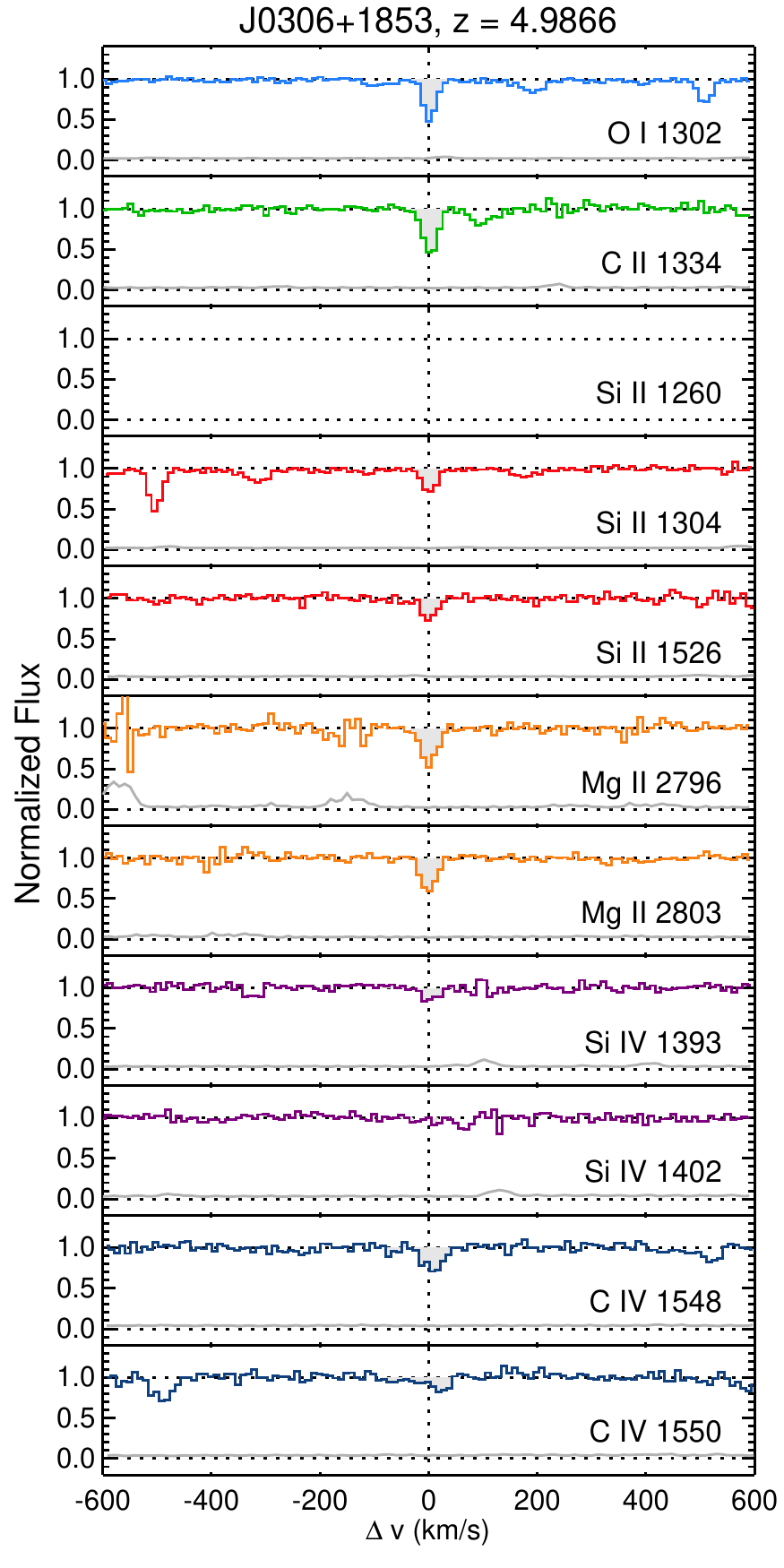}
   \caption{Stacked velocity plot for the $z=4.9866$ system towards J0306+1853.  Lines and shading are as described in Figure~\ref{fig:z3p3844720_J1018+0548}.\label{fig:z4p9865800_J0306+1853}}
\end{figure}
 
\begin{figure}[!t]
   \centering
   \includegraphics[height=0.40\textheight]{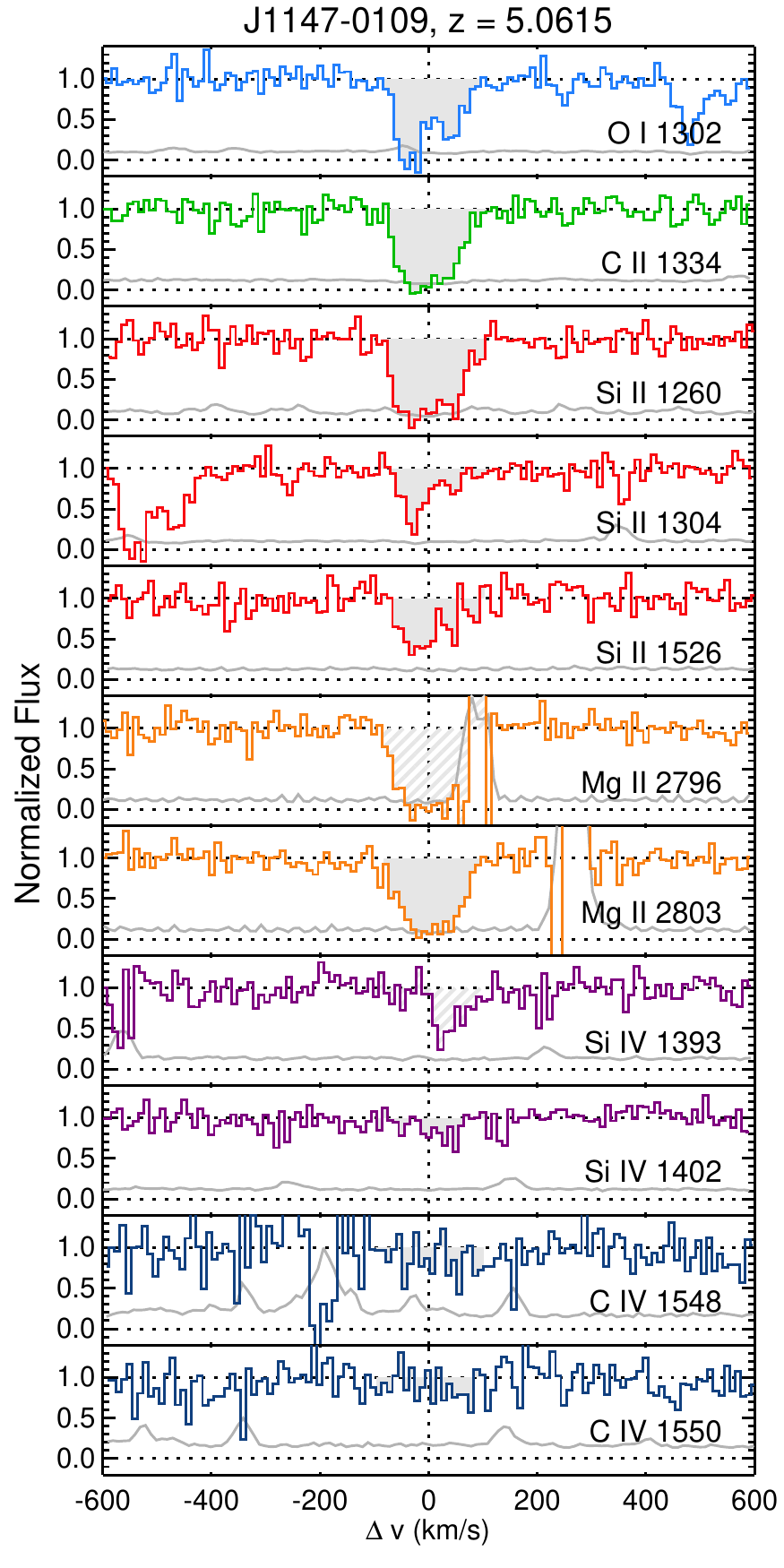}
   \caption{Stacked velocity plot for the $z=5.0615$ system towards J1147-0109.  Lines and shading are as described in Figure~\ref{fig:z3p3844720_J1018+0548}. See notes on this system in Appendix~\ref{app:details}.\label{fig:z5p0615260_J1147-0109}}
\end{figure}
 
\begin{figure}[!b]
   \centering
   \includegraphics[height=0.40\textheight]{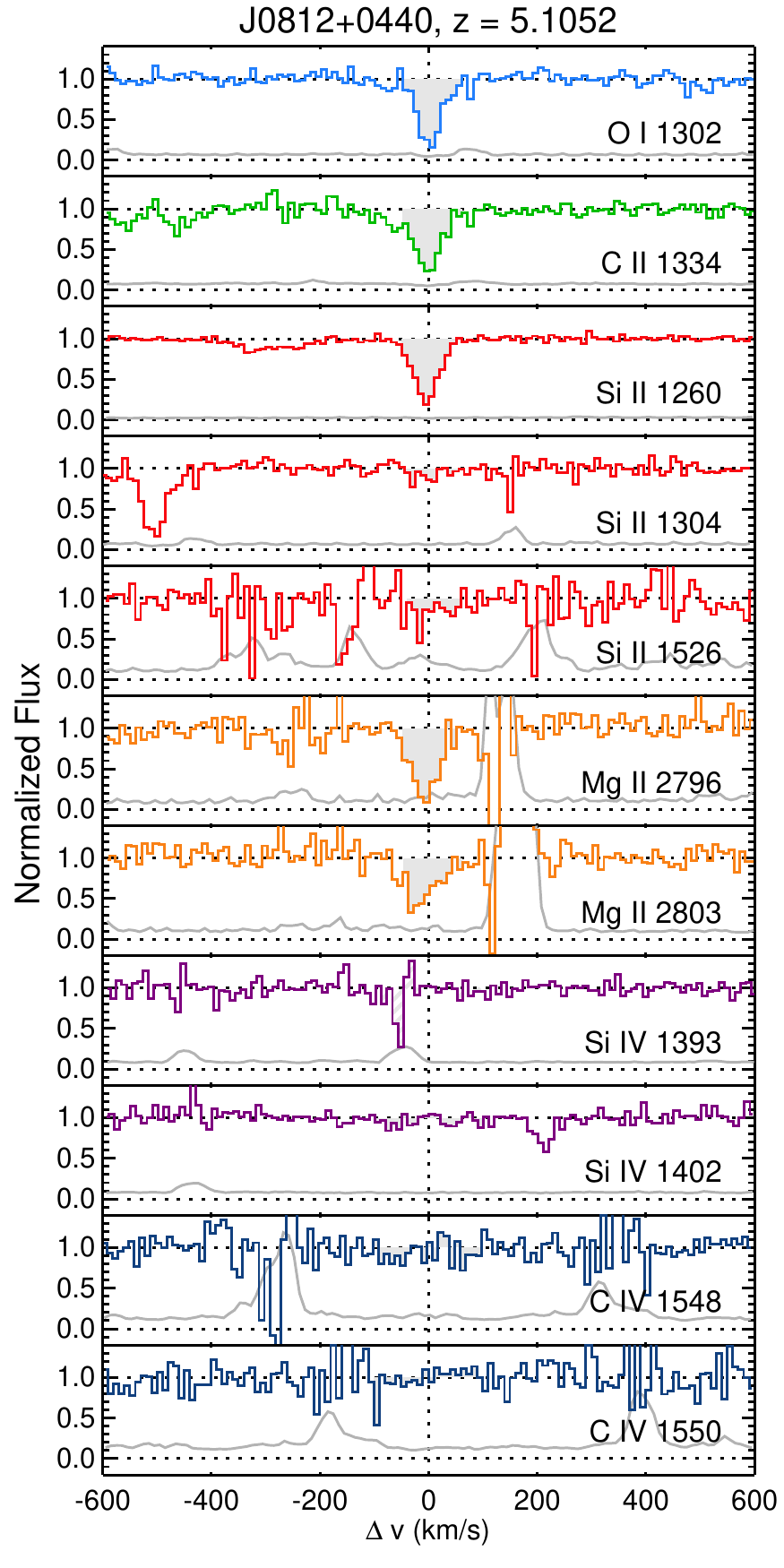}
   \caption{Stacked velocity plot for the $z=5.1052$ system towards J0812+0440.  Lines and shading are as described in Figure~\ref{fig:z3p3844720_J1018+0548}.\label{fig:z5p1052330_J0812+0440}}
\end{figure}
 
\clearpage
 
\begin{figure}[!t]
   \centering
   \includegraphics[height=0.40\textheight]{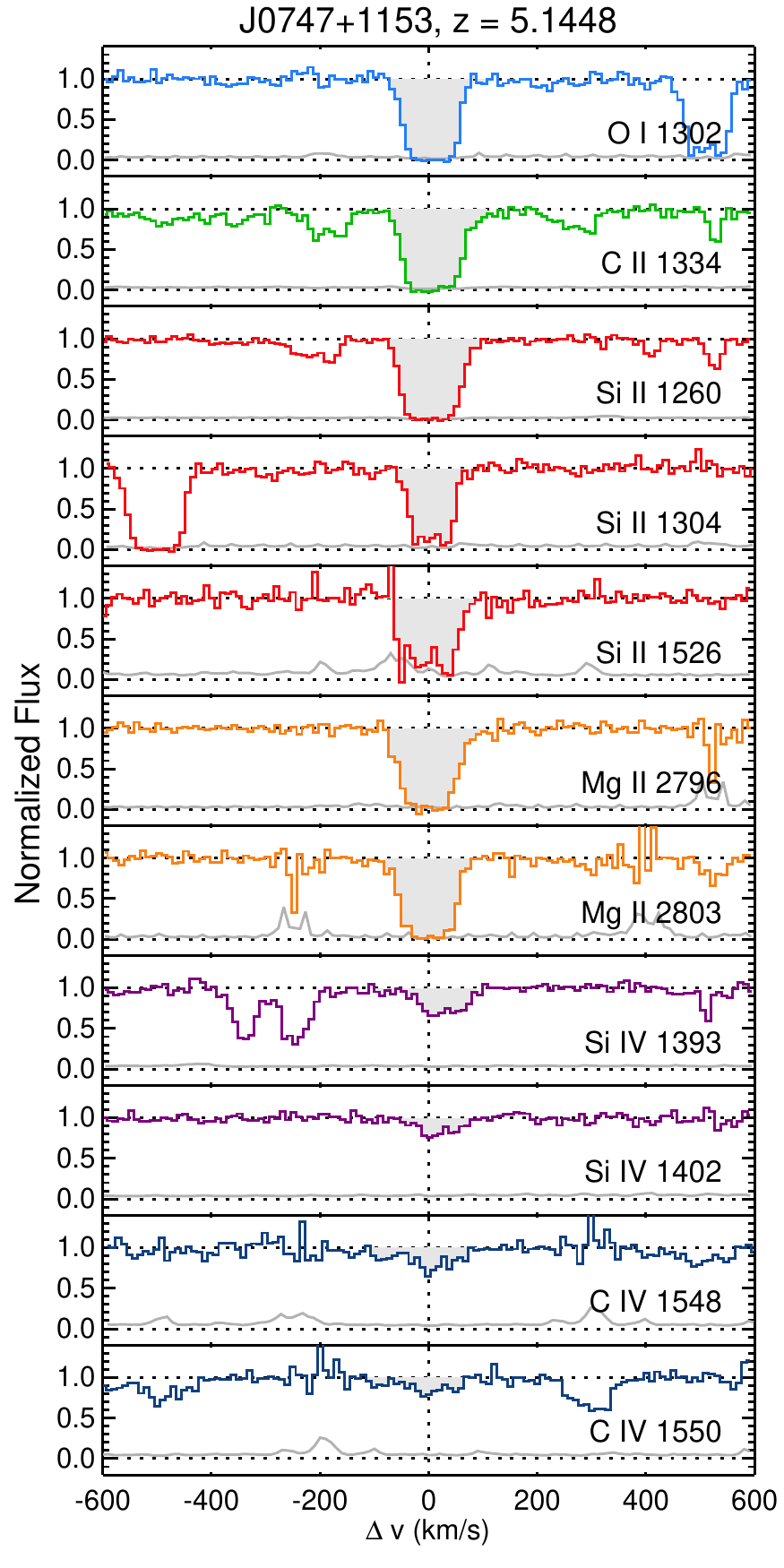}
   \caption{Stacked velocity plot for the $z=5.1448$ system towards J0747+1153.  Lines and shading are as described in Figure~\ref{fig:z3p3844720_J1018+0548}.\label{fig:z5p1447510_J0747+1153}}
\end{figure}
 
\begin{figure}[!b]
   \centering
   \includegraphics[height=0.40\textheight]{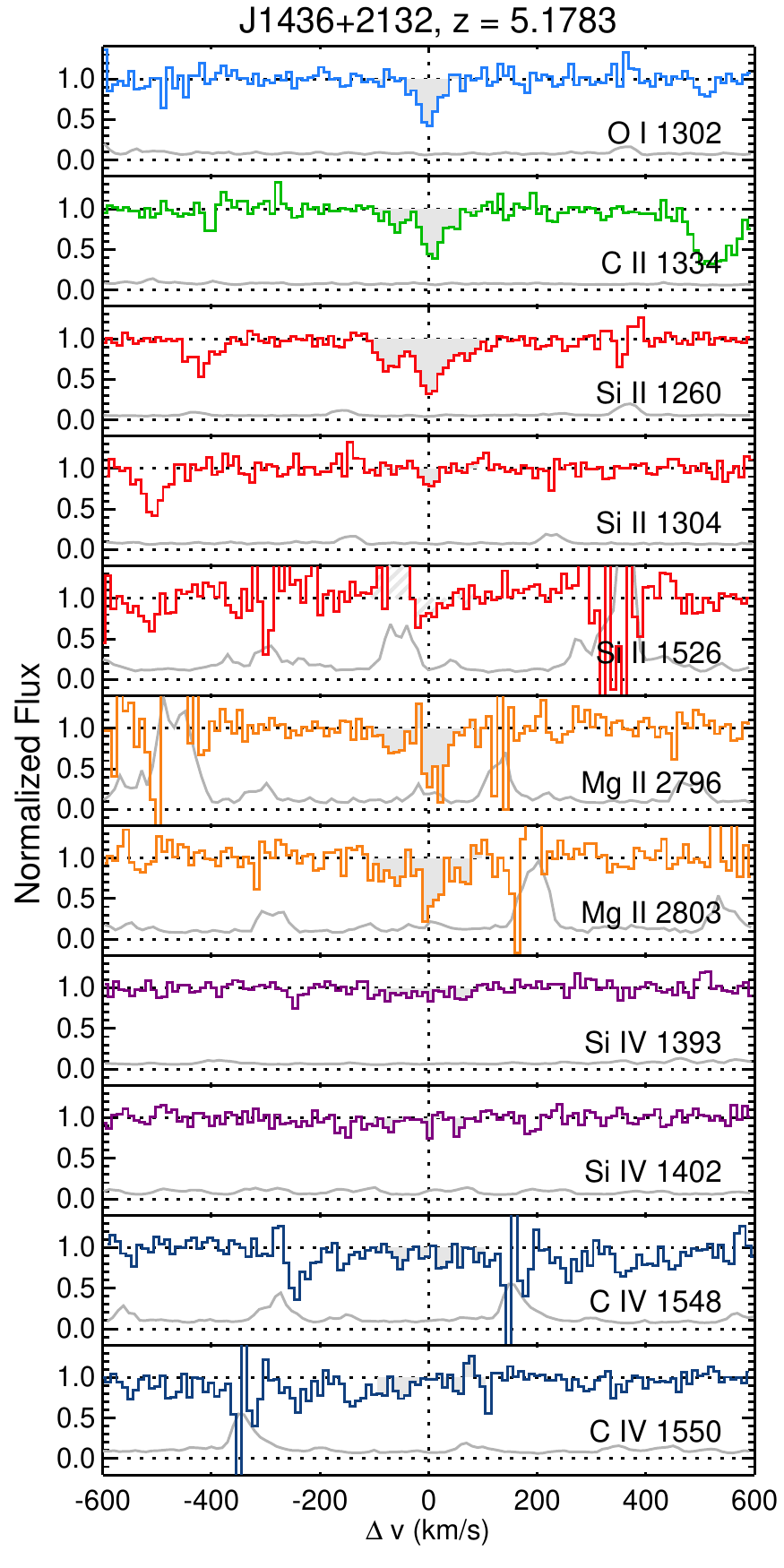}
   \caption{Stacked velocity plot for the $z=5.1783$ system towards J1436+2132.  Lines and shading are as described in Figure~\ref{fig:z3p3844720_J1018+0548}.\label{fig:z5p1782590_J1436+2132}}
\end{figure}
 
\begin{figure}[!t]
   \centering
   \includegraphics[height=0.40\textheight]{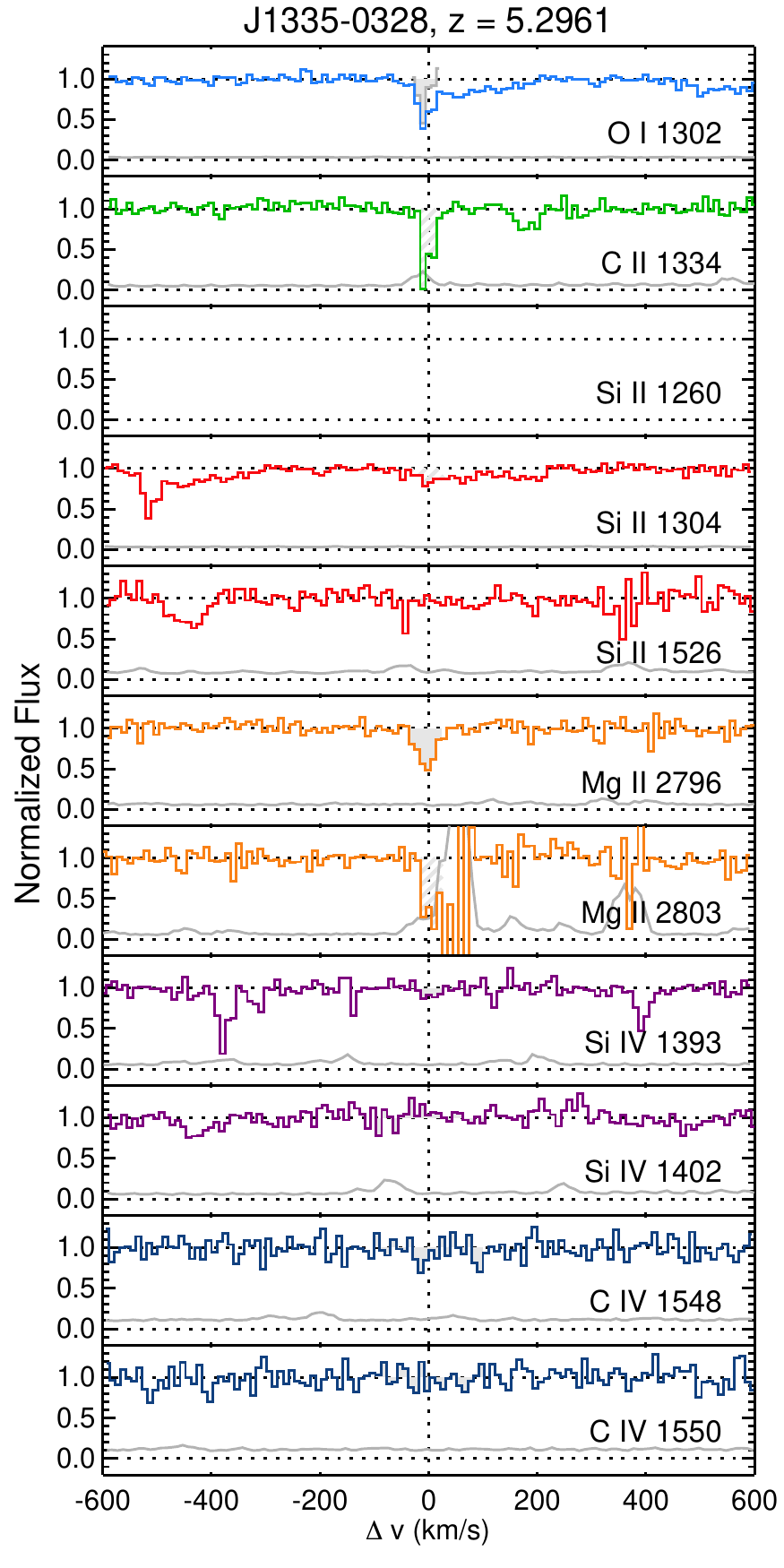}
   \caption{Stacked velocity plot for the $z=5.2961$ system towards J1335-0328.  Lines and shading are as described in Figure~\ref{fig:z3p3844720_J1018+0548}.  The grey histogram in the \oi~\lam1302 panel is the deblended flux. See notes on this system in Appendix~\ref{app:details}.\label{fig:z5p2960670_J1335-0328}}
\end{figure}
 
\begin{figure}[!b]
   \centering
   \includegraphics[height=0.40\textheight]{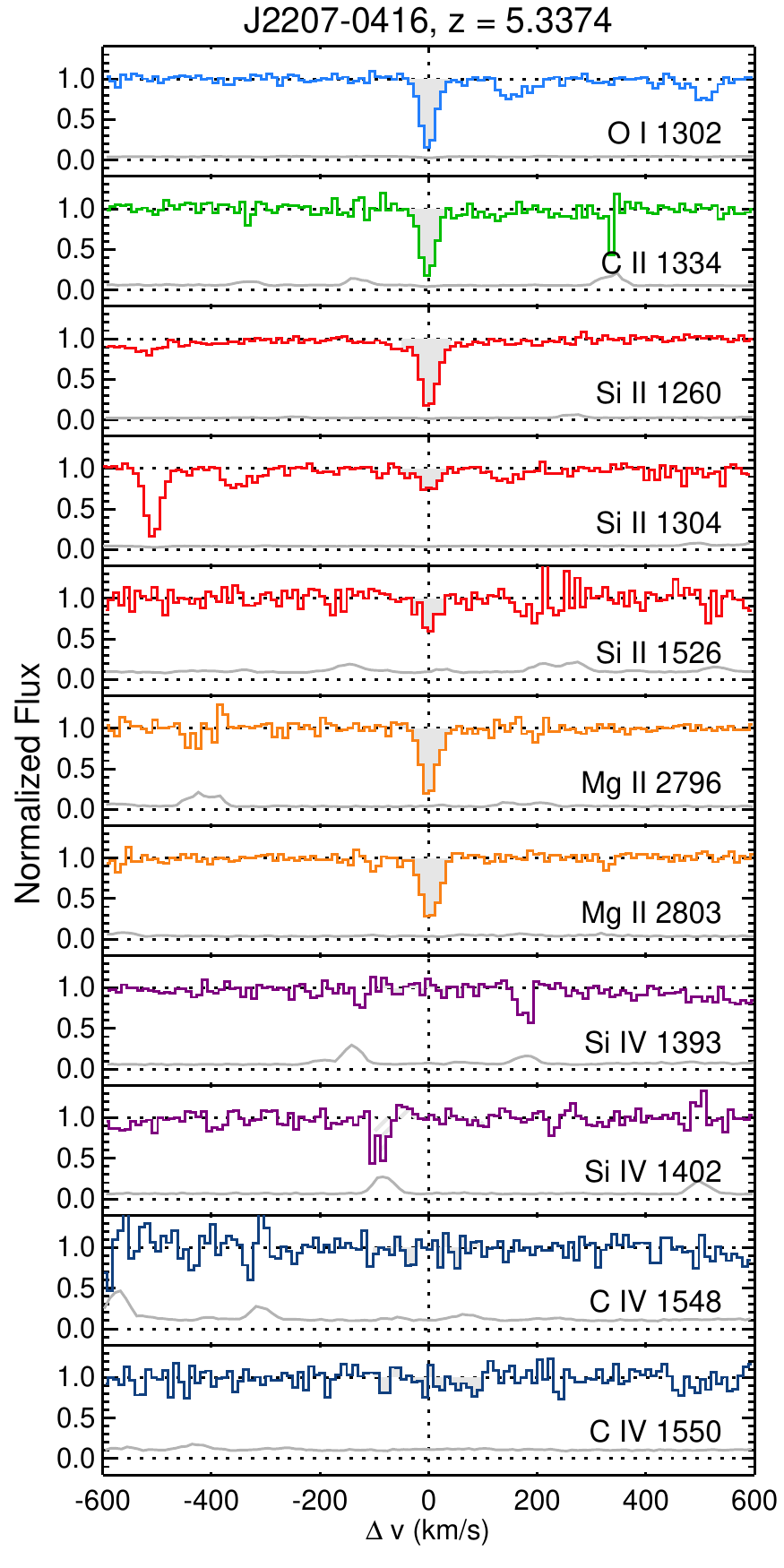}
   \caption{Stacked velocity plot for the $z=5.3374$ system towards J2207-0416.  Lines and shading are as described in Figure~\ref{fig:z3p3844720_J1018+0548}.\label{fig:z5p3374230_J2207-0416}}
\end{figure}
 
\clearpage
 
\begin{figure}[!t]
   \centering
   \includegraphics[height=0.40\textheight]{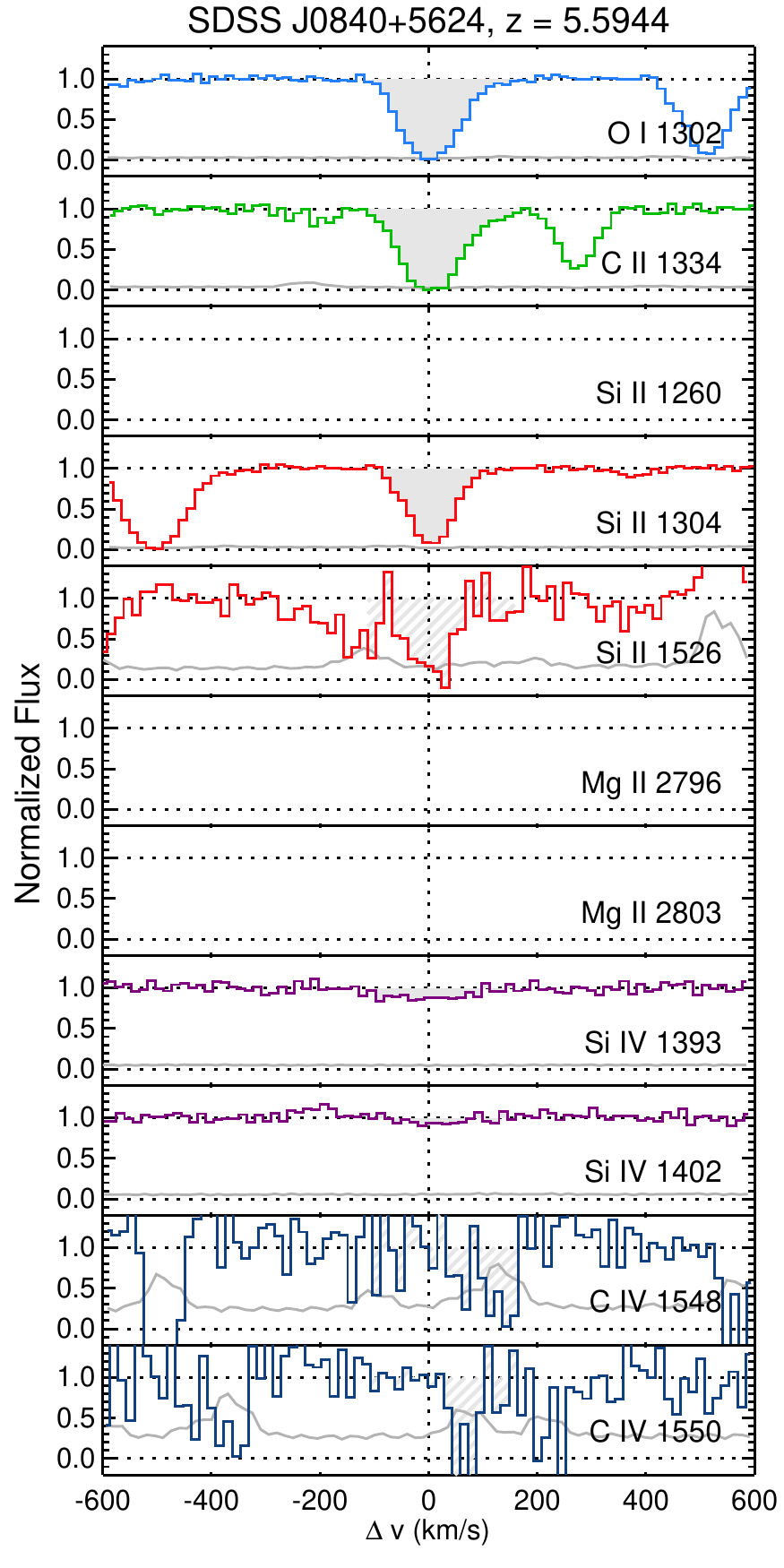}
   \caption{Stacked velocity plot for the $z=5.5944$ system towards SDSS J0840+5624.  Lines and shading are as described in Figure~\ref{fig:z3p3844720_J1018+0548}.\label{fig:z5p5944320_SDSSJ0840+5624}}
\end{figure}
 
\begin{figure}[!b]
   \centering
   \includegraphics[height=0.40\textheight]{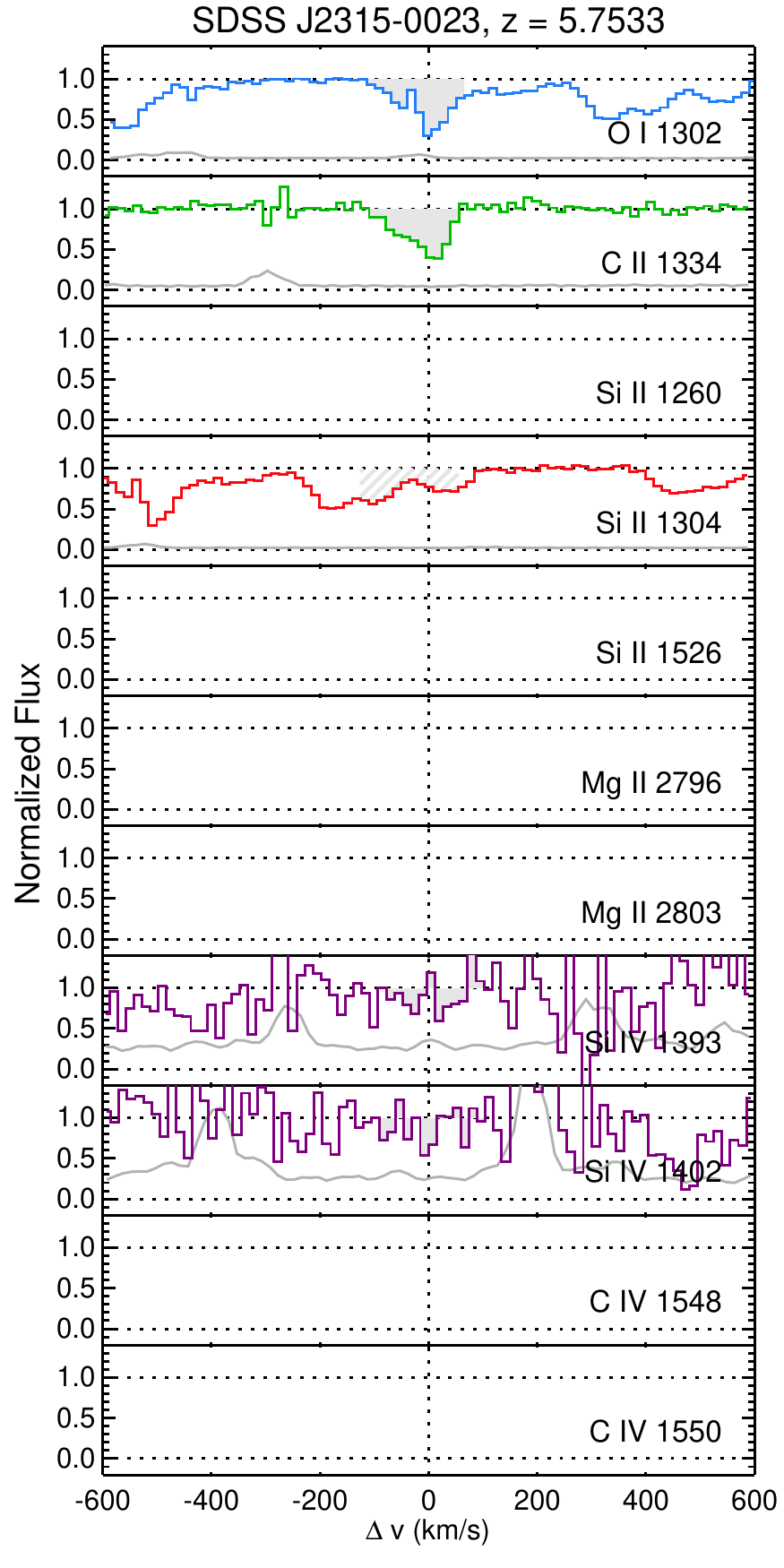}
   \caption{Stacked velocity plot for the $z=5.7533$ system towards SDSS J2315-0023.  Lines and shading are as described in Figure~\ref{fig:z3p3844720_J1018+0548}.\label{fig:z5p7532610_SDSSJ2315-0023}}
\end{figure}
 
\begin{figure}[!t]
   \centering
   \includegraphics[height=0.40\textheight]{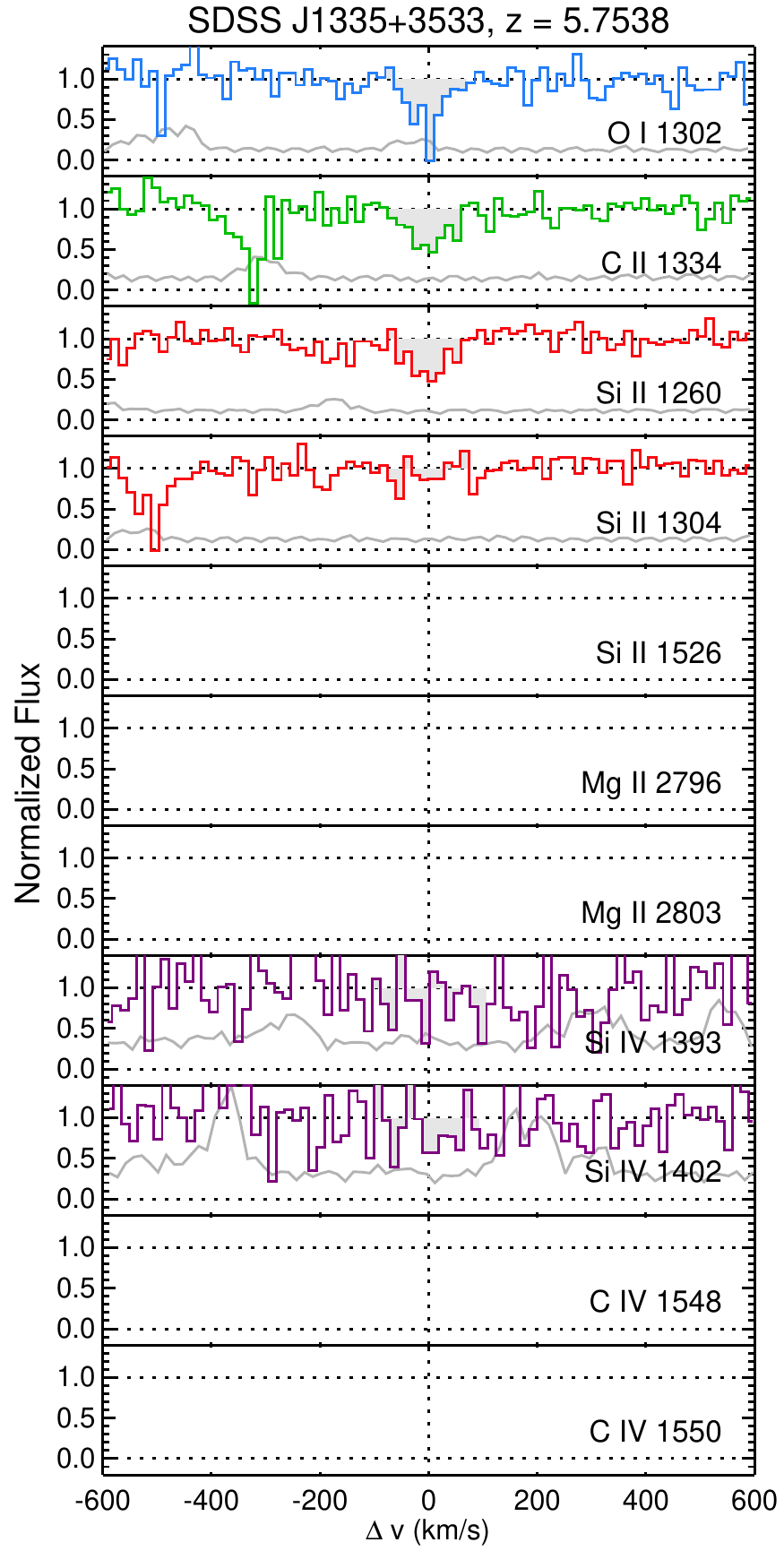}
   \caption{Stacked velocity plot for the $z=5.7538$ system towards SDSS J1335+3533.  Lines and shading are as described in Figure~\ref{fig:z3p3844720_J1018+0548}.\label{fig:z5p7538330_SDSSJ1335+3533}}
\end{figure}
 
\begin{figure}[!b]
   \centering
   \includegraphics[height=0.40\textheight]{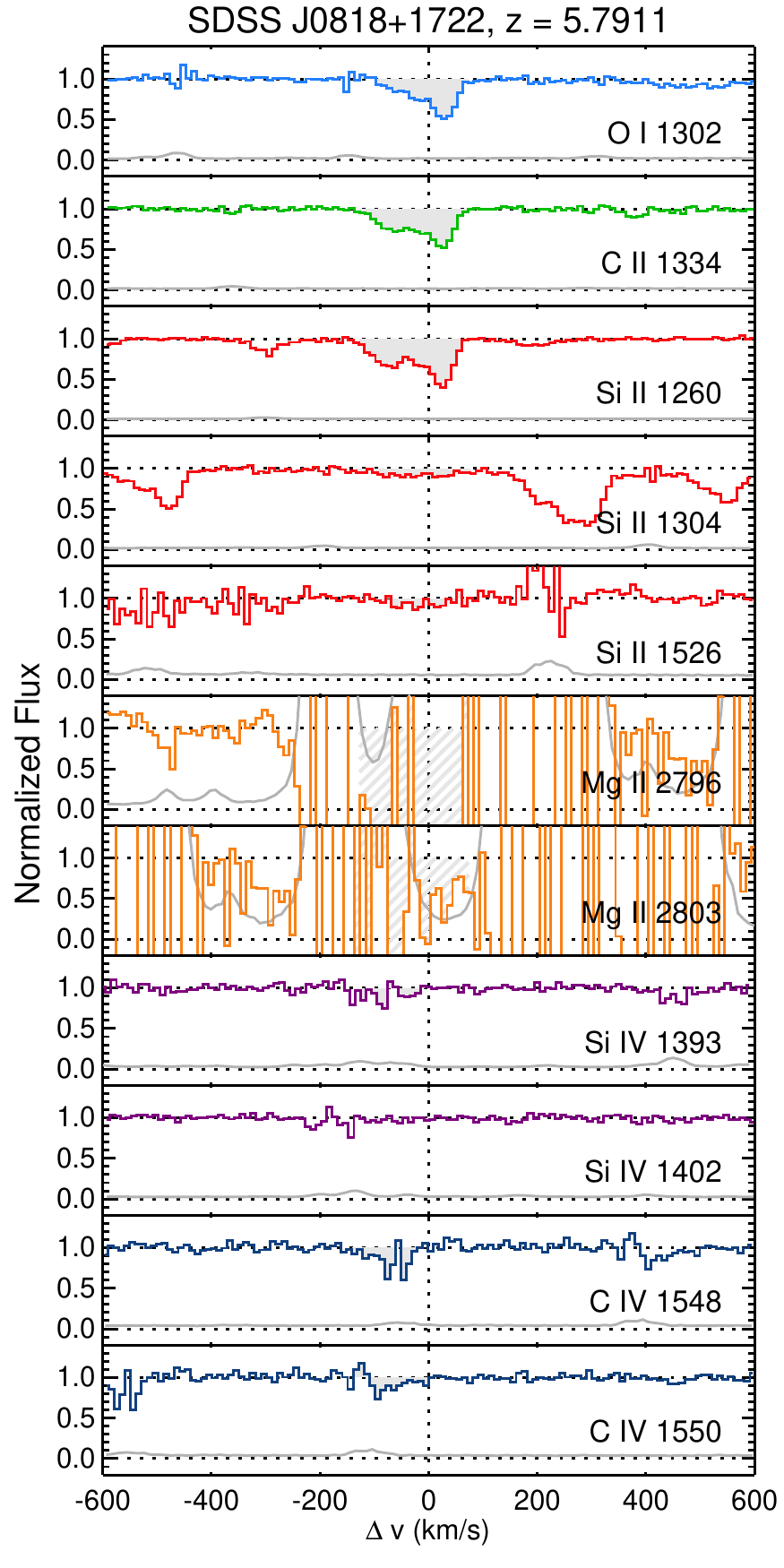}
   \caption{Stacked velocity plot for the $z=5.7911$ system towards SDSS J0818+1722.  Lines and shading are as described in Figure~\ref{fig:z3p3844720_J1018+0548}.\label{fig:z5p7911030_SDSSJ0818+1722}}
\end{figure}
 
\clearpage
 
\begin{figure}[!t]
   \centering
   \includegraphics[height=0.40\textheight]{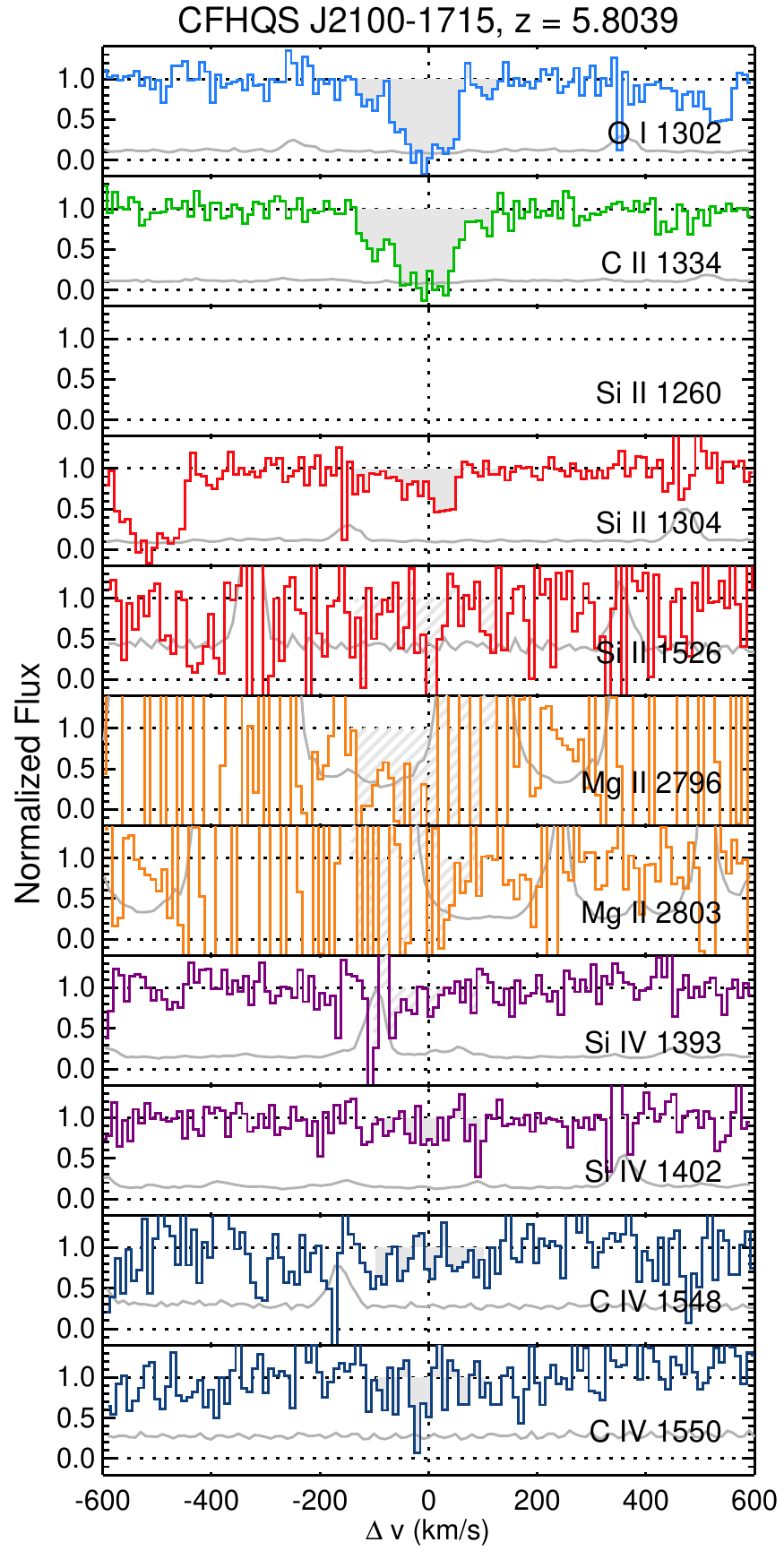}
   \caption{Stacked velocity plot for the $z=5.8039$ system towards CFHQS J2100-1715.  Lines and shading are as described in Figure~\ref{fig:z3p3844720_J1018+0548}.\label{fig:z5p8039450_CFHQSJ2100-1715}}
\end{figure}
 
\begin{figure}[!b]
   \centering
   \includegraphics[height=0.40\textheight]{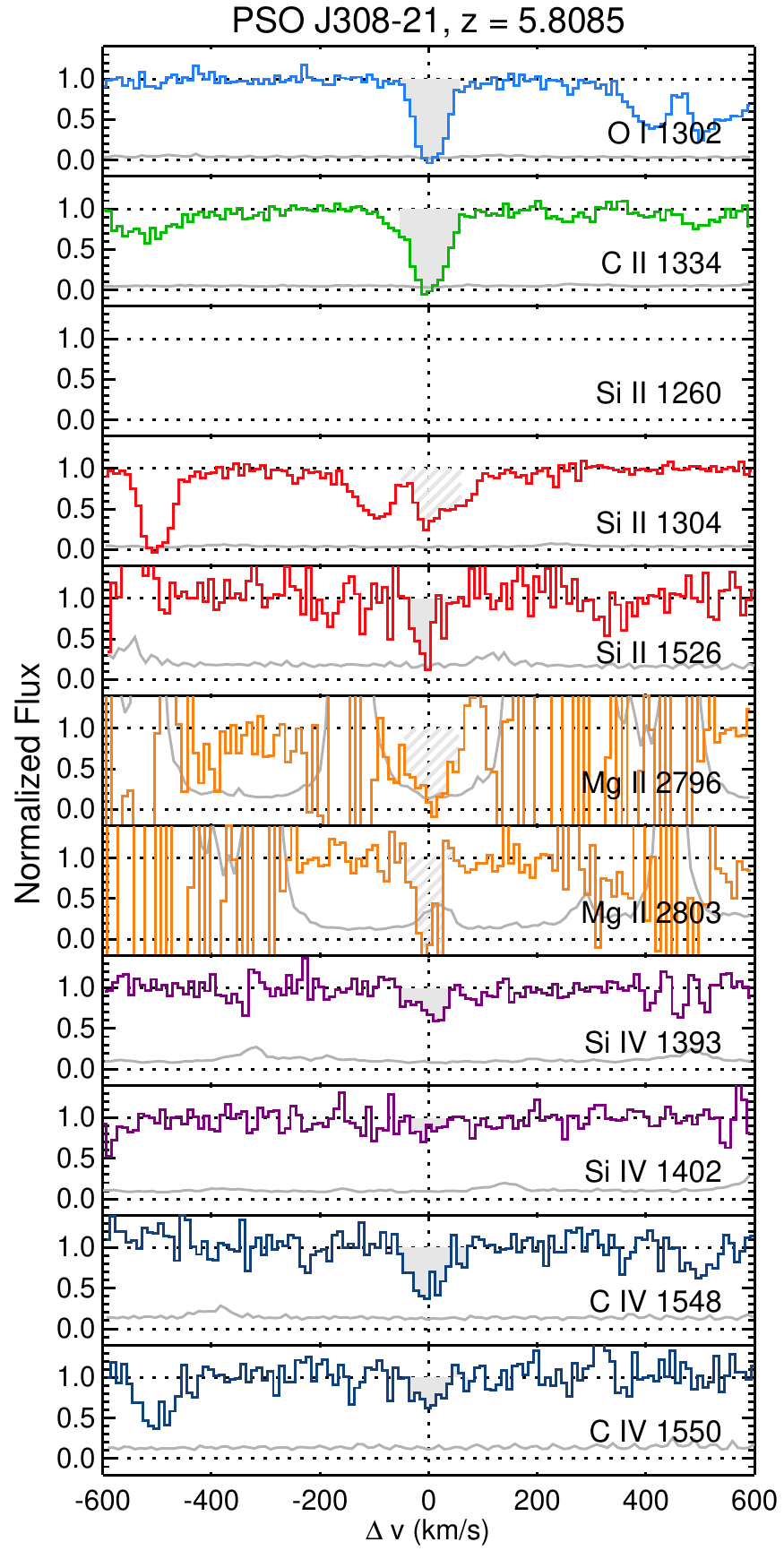}
   \caption{Stacked velocity plot for the $z=5.8085$ system towards PSO J308-21.  Lines and shading are as described in Figure~\ref{fig:z3p3844720_J1018+0548}.\label{fig:z5p8085010_PSOJ308-21}}
\end{figure}
 
\begin{figure}[!t]
   \centering
   \includegraphics[height=0.40\textheight]{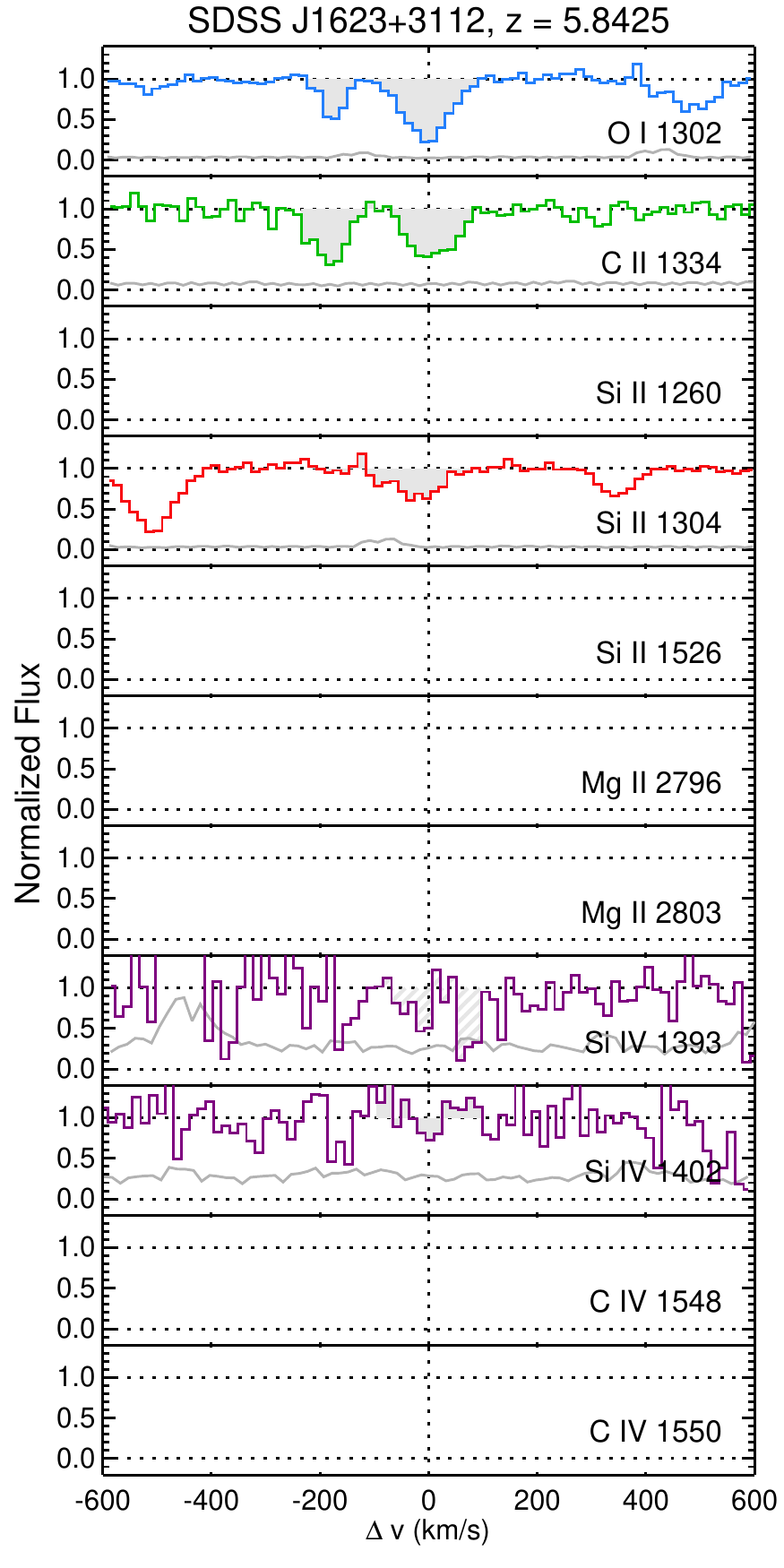}
   \caption{Stacked velocity plot for the $z=5.8425$ system towards SDSS J1623+3112.  Lines and shading are as described in Figure~\ref{fig:z3p3844720_J1018+0548}.\label{fig:z5p8424870_SDSSJ1623+3112}}
\end{figure}
 
\begin{figure}[!b]
   \centering
   \includegraphics[height=0.40\textheight]{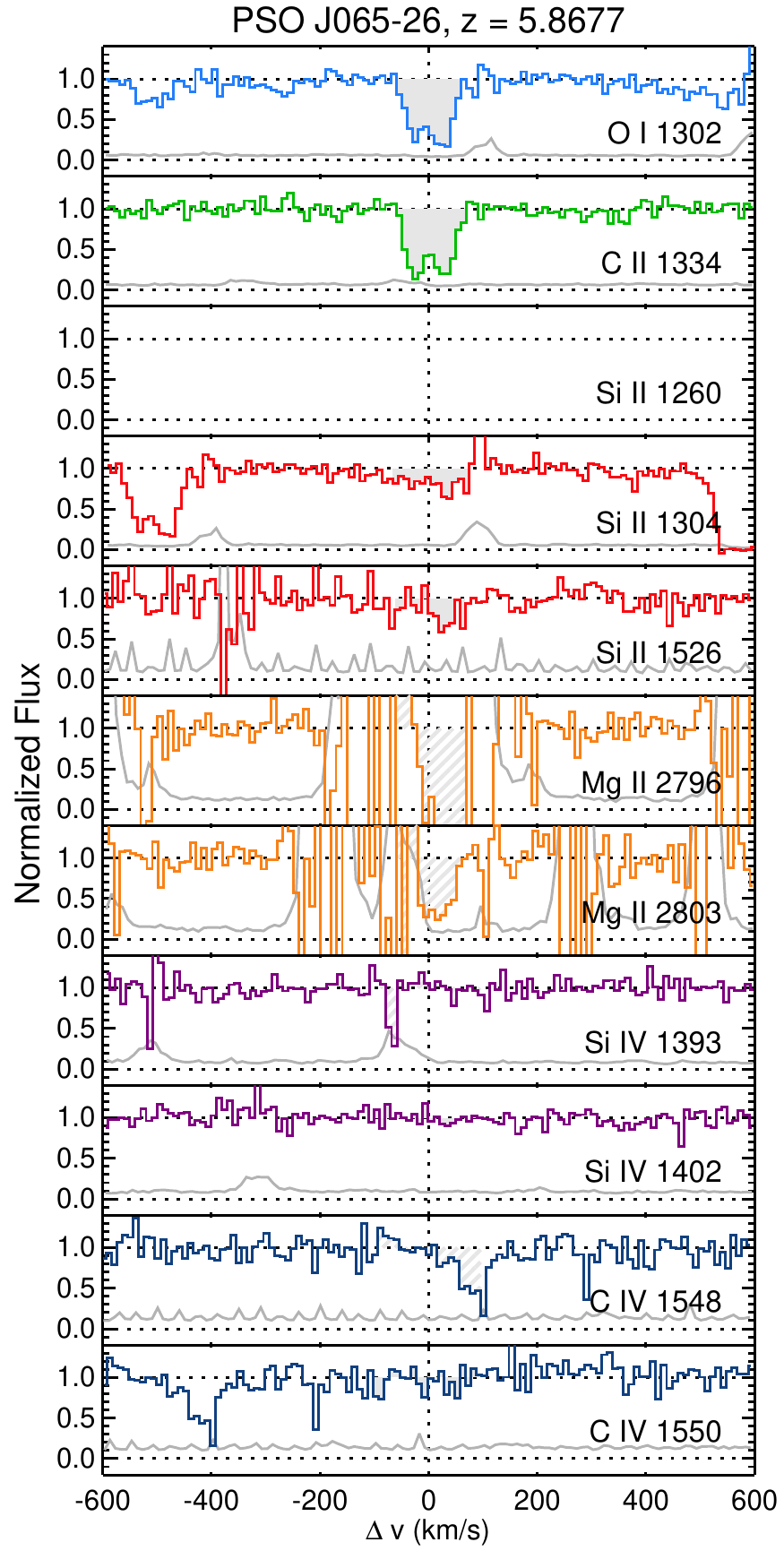}
   \caption{Stacked velocity plot for the $z=5.8677$ system towards PSO J065-26.  Lines and shading are as described in Figure~\ref{fig:z3p3844720_J1018+0548}.\label{fig:z5p8677130_PSOJ065-26}}
\end{figure}
 
\clearpage
 
\begin{figure}[!t]
   \centering
   \includegraphics[height=0.40\textheight]{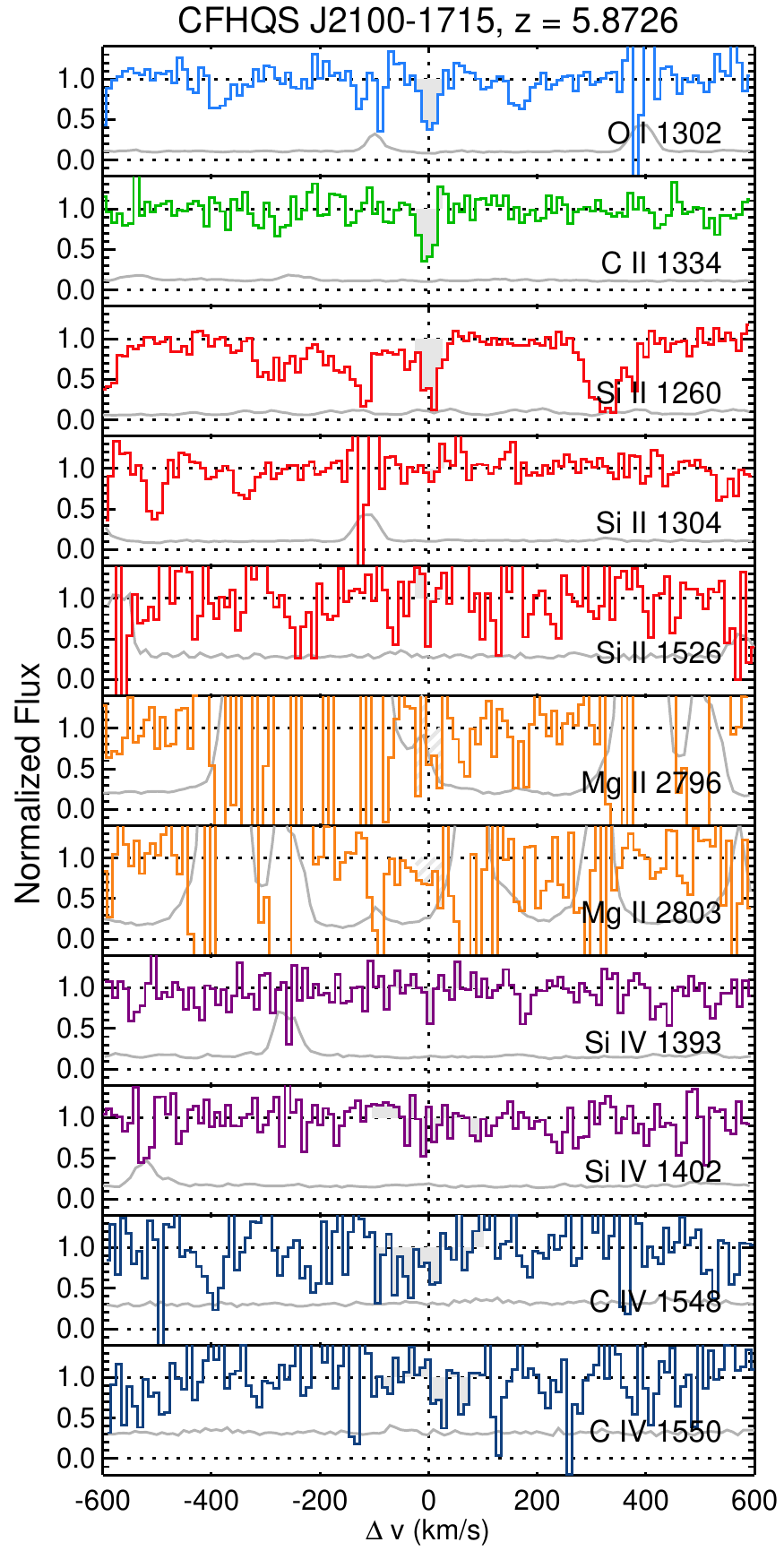}
   \caption{Stacked velocity plot for the $z=5.8726$ system towards CFHQS J2100-1715.  Lines and shading are as described in Figure~\ref{fig:z3p3844720_J1018+0548}.\label{fig:z5p8726170_CFHQSJ2100-1715}}
\end{figure}
 
\begin{figure}[!b]
   \centering
   \includegraphics[height=0.40\textheight]{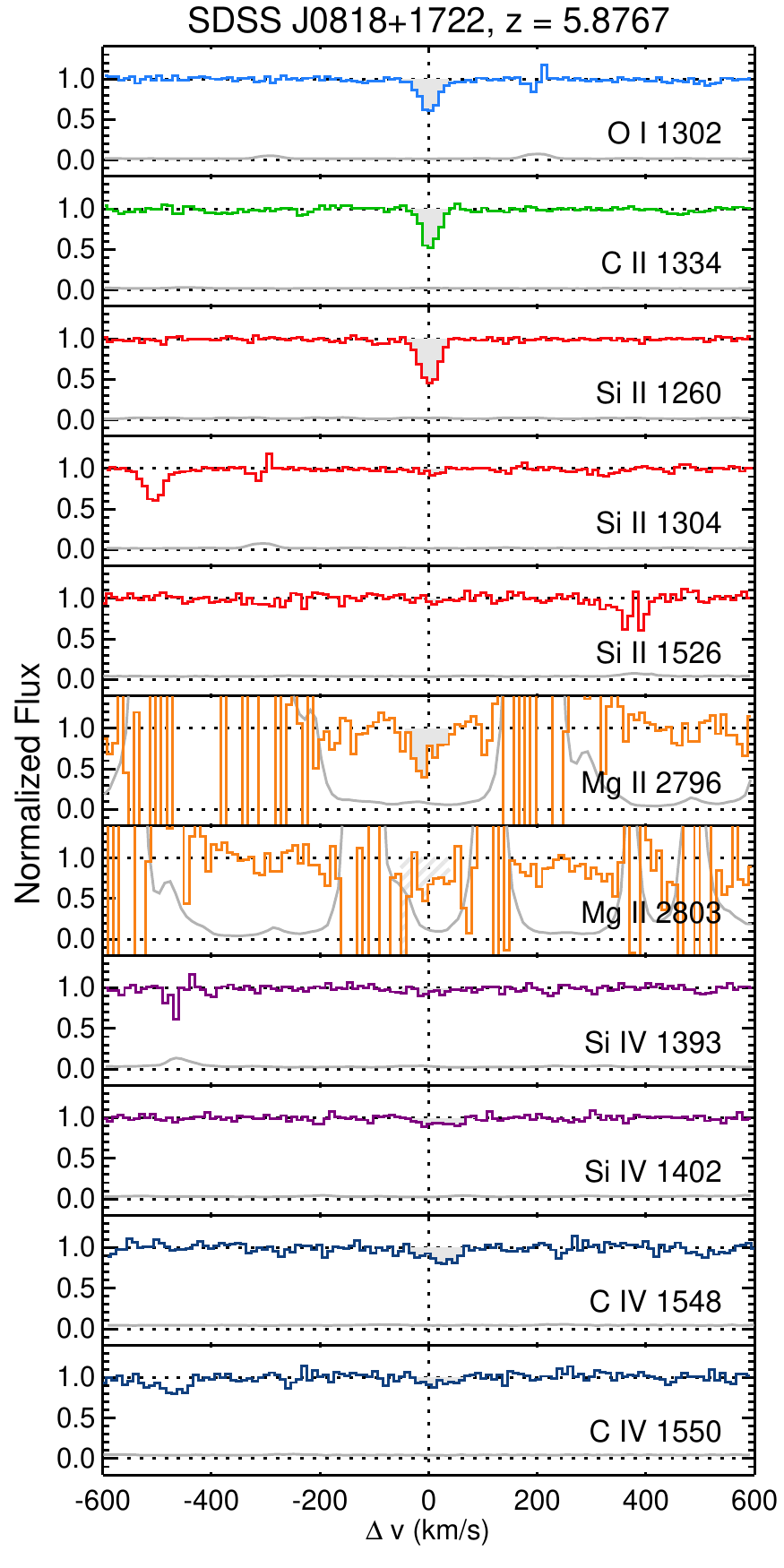}
   \caption{Stacked velocity plot for the $z=5.8767$ system towards SDSS J0818+1722.  Lines and shading are as described in Figure~\ref{fig:z3p3844720_J1018+0548}.\label{fig:z5p8767010_SDSSJ0818+1722}}
\end{figure}
 
\begin{figure}[!t]
   \centering
   \includegraphics[height=0.40\textheight]{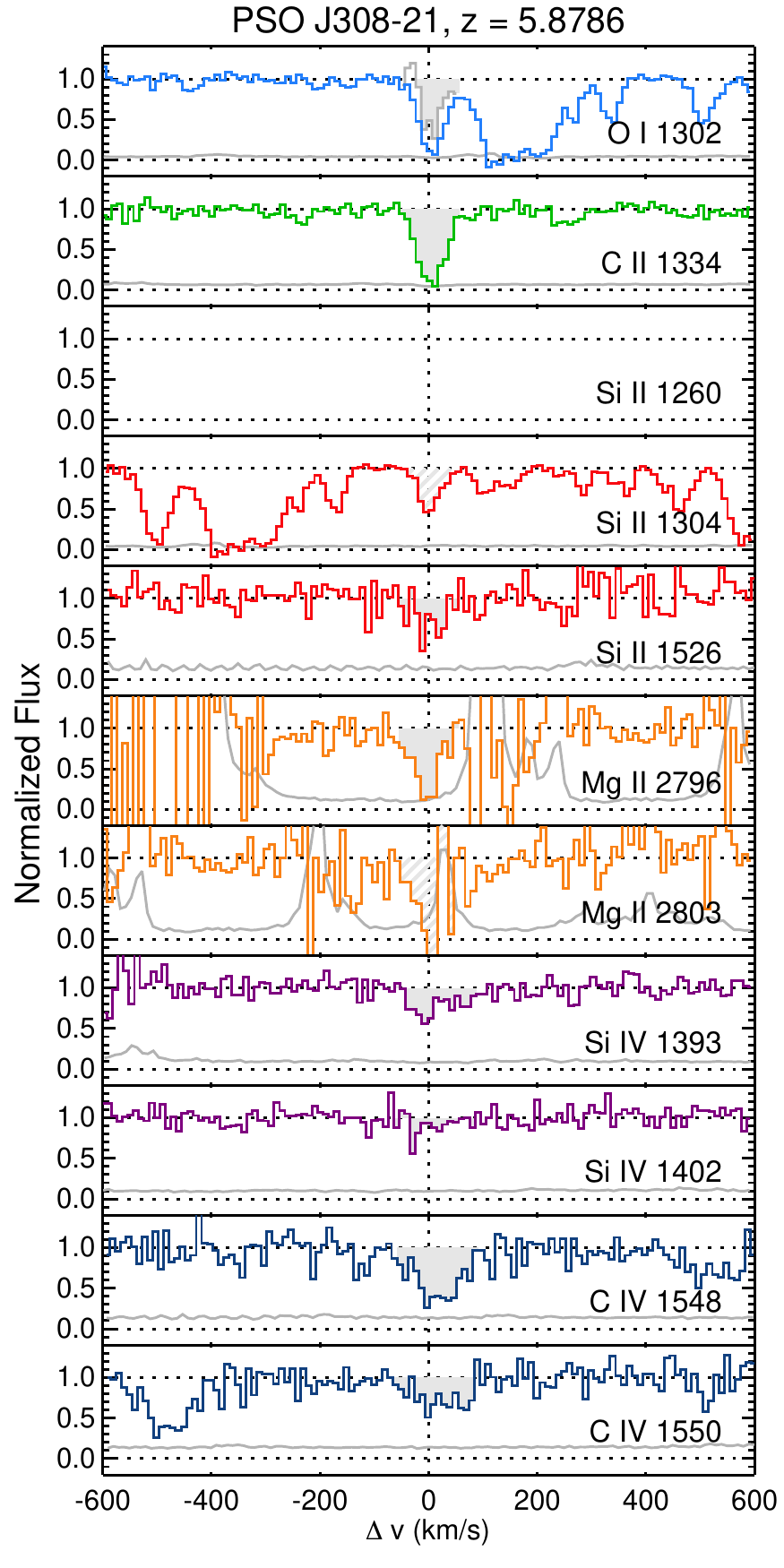}
   \caption{Stacked velocity plot for the $z=5.8786$ system towards PSO J308-21.  Lines and shading are as described in Figure~\ref{fig:z3p3844720_J1018+0548}.  The grey histogram in the \oi~\lam1302 panel is the deblended flux. See notes on this system in Appendix~\ref{app:details}.\label{fig:z5p8785790_PSOJ308-21}}
\end{figure}
 
\begin{figure}[!b]
   \centering
   \includegraphics[height=0.40\textheight]{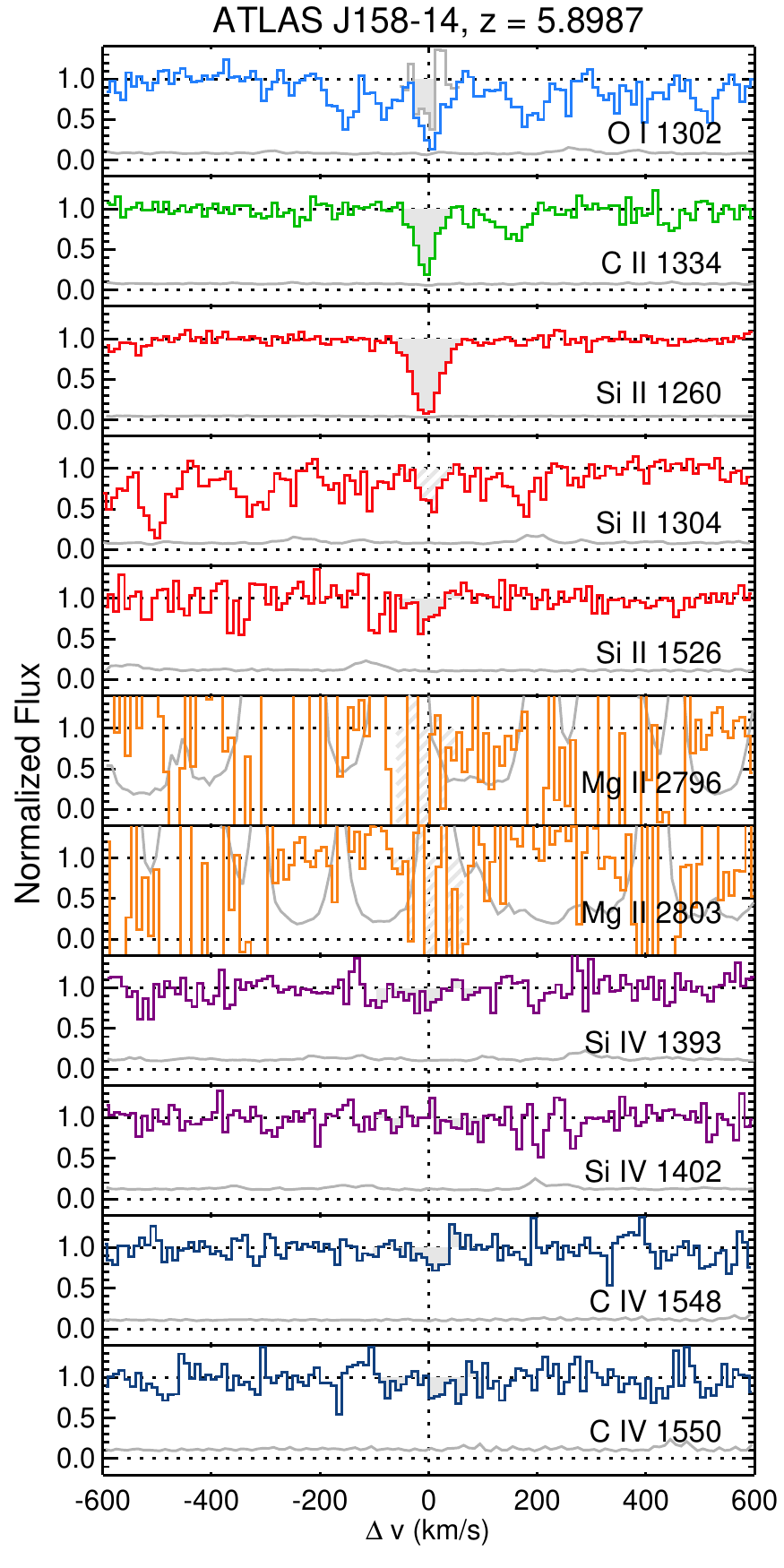}
   \caption{Stacked velocity plot for the $z=5.8987$ system towards ATLAS J158-14.  Lines and shading are as described in Figure~\ref{fig:z3p3844720_J1018+0548}.  The grey histogram in the \oi~\lam1302 panel is the deblended flux. See notes on this system in Appendix~\ref{app:details}.\label{fig:z5p8986670_ATLASJ158-14}}
\end{figure}
 
\clearpage
 
\begin{figure}[!t]
   \centering
   \includegraphics[height=0.40\textheight]{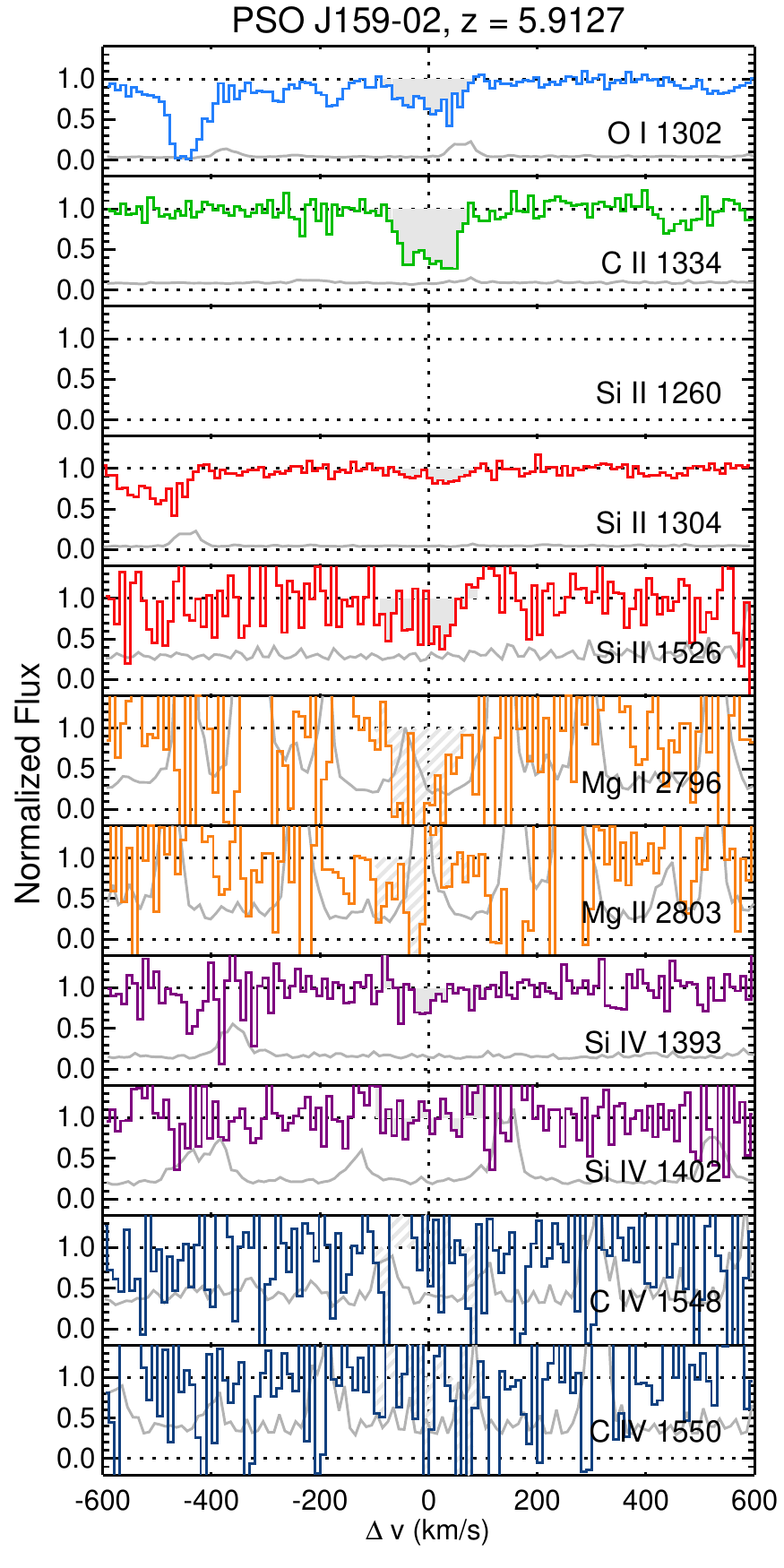}
   \caption{Stacked velocity plot for the $z=5.9127$ system towards PSO J159-02.  Lines and shading are as described in Figure~\ref{fig:z3p3844720_J1018+0548}.\label{fig:z5p9127050_PSOJ159-02}}
\end{figure}
 
\begin{figure}[!b]
   \centering
   \includegraphics[height=0.40\textheight]{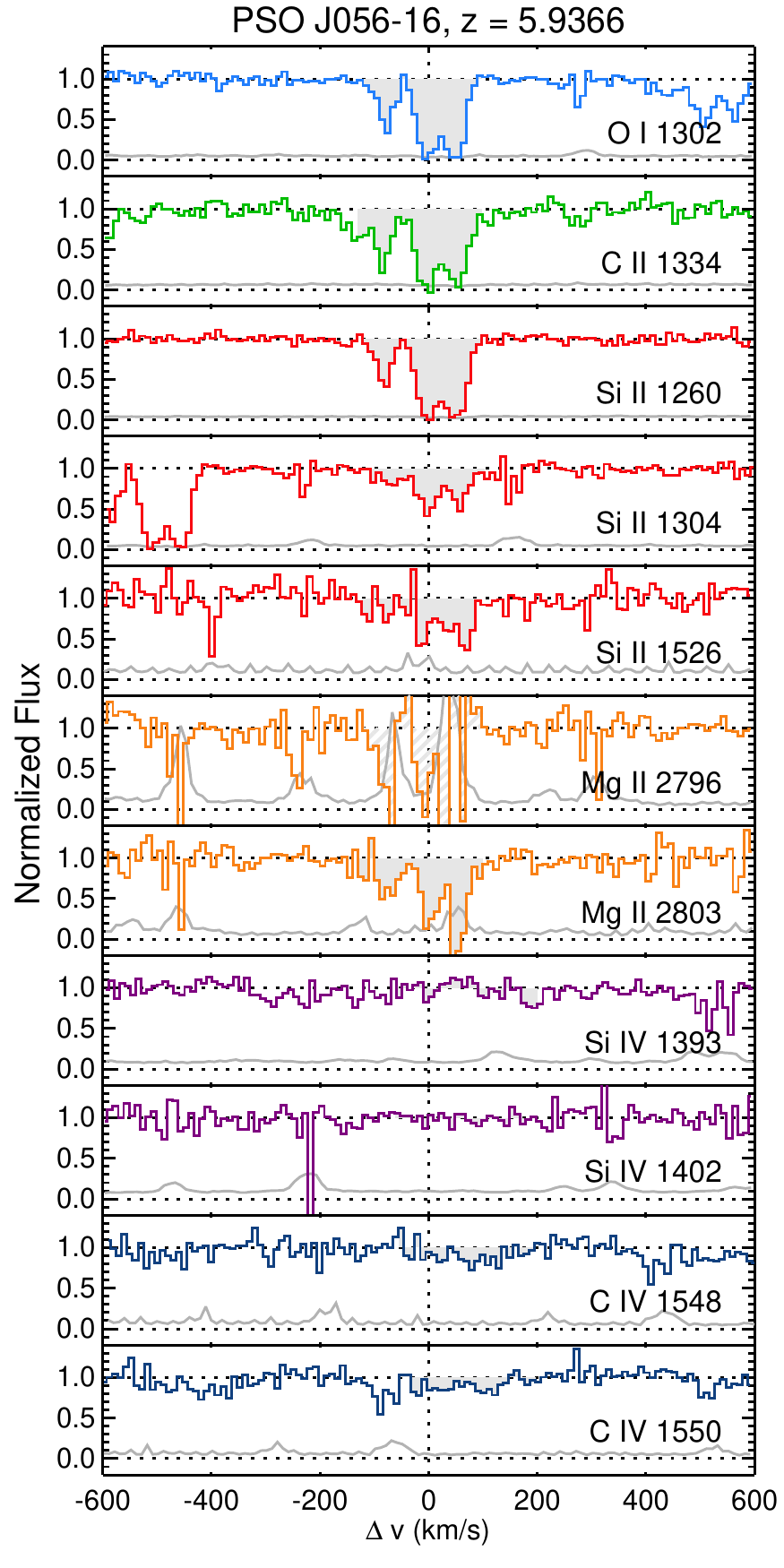}
   \caption{Stacked velocity plot for the $z=5.9366$ system towards PSO J056-16.  Lines and shading are as described in Figure~\ref{fig:z3p3844720_J1018+0548}.\label{fig:z5p9365800_PSOJ056-16}}
\end{figure}
 
\begin{figure}[!t]
   \centering
   \includegraphics[height=0.40\textheight]{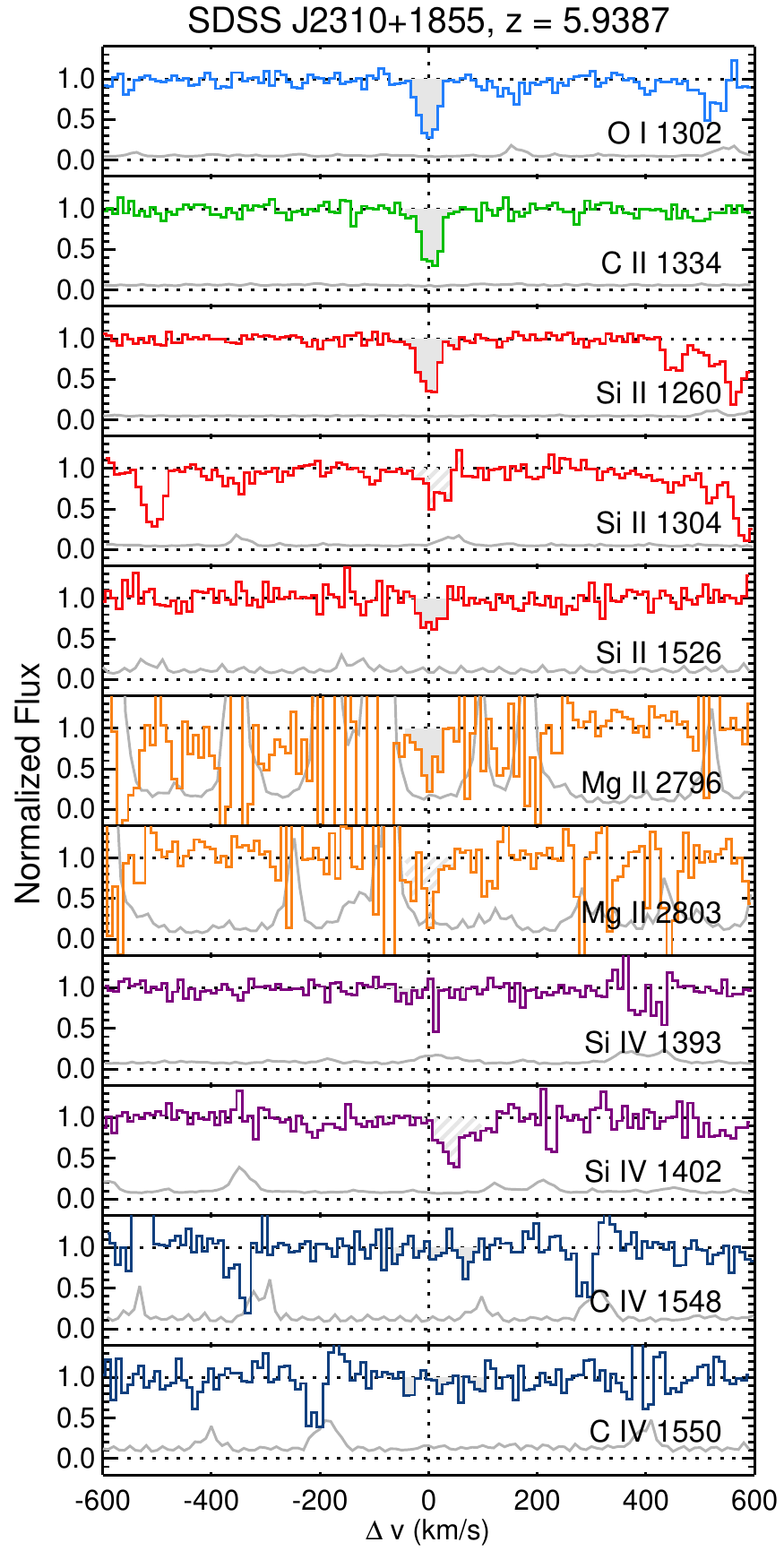}
   \caption{Stacked velocity plot for the $z=5.9387$ system towards SDSS J2310+1855.  Lines and shading are as described in Figure~\ref{fig:z3p3844720_J1018+0548}.\label{fig:z5p9387060_SDSSJ2310+1855}}
\end{figure}
 
\begin{figure}[!b]
   \centering
   \includegraphics[height=0.40\textheight]{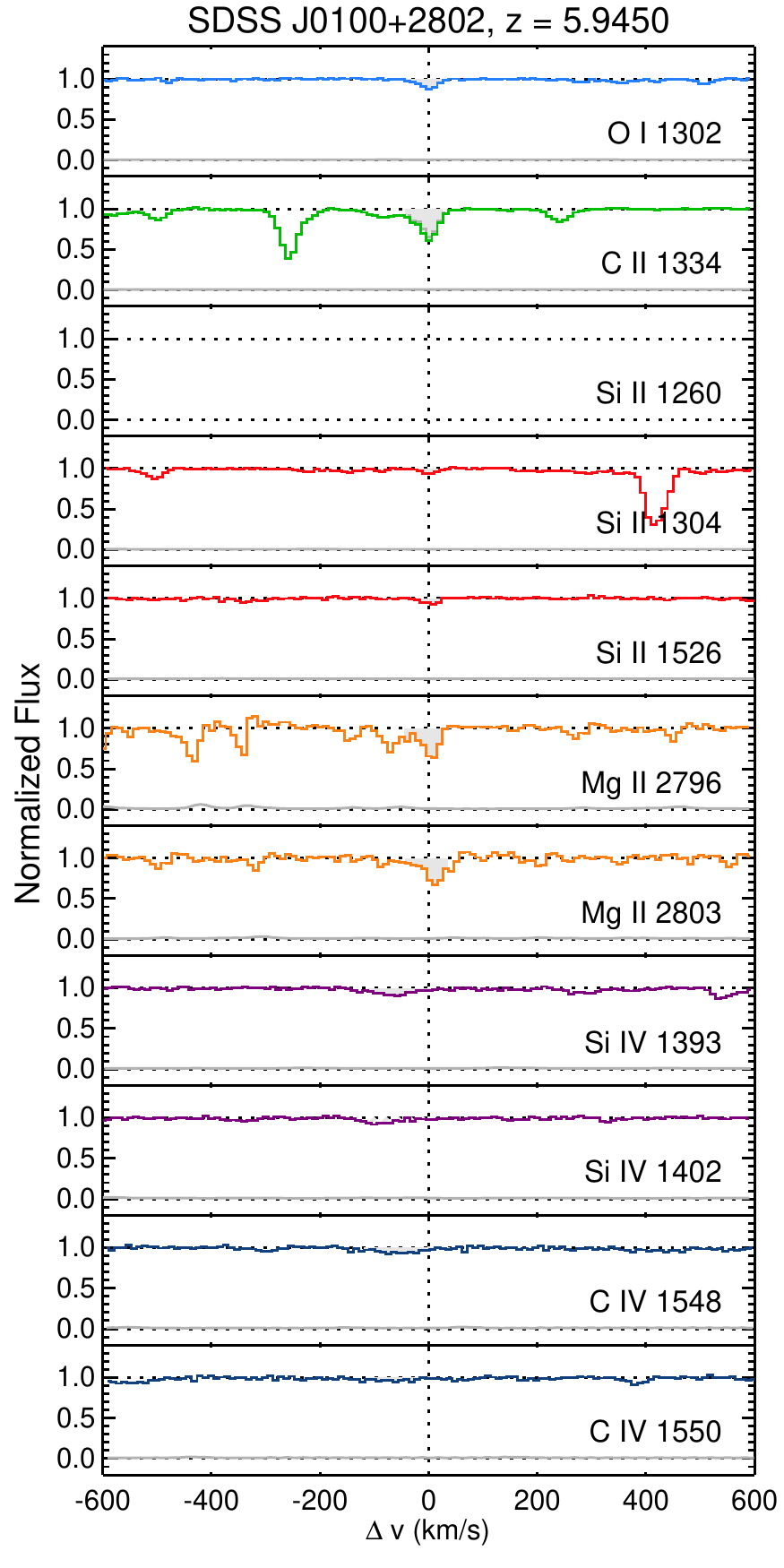}
   \caption{Stacked velocity plot for the $z=5.9450$ system towards SDSS J0100+2802.  Lines and shading are as described in Figure~\ref{fig:z3p3844720_J1018+0548}.  The grey histogram in the \cii~\lam1334 panel is the deblended flux. See notes on this system in Appendix~\ref{app:details}.\label{fig:z5p9449860_SDSSJ0100+2802}}
\end{figure}
 
\clearpage
 
\begin{figure}[!t]
   \centering
   \includegraphics[height=0.40\textheight]{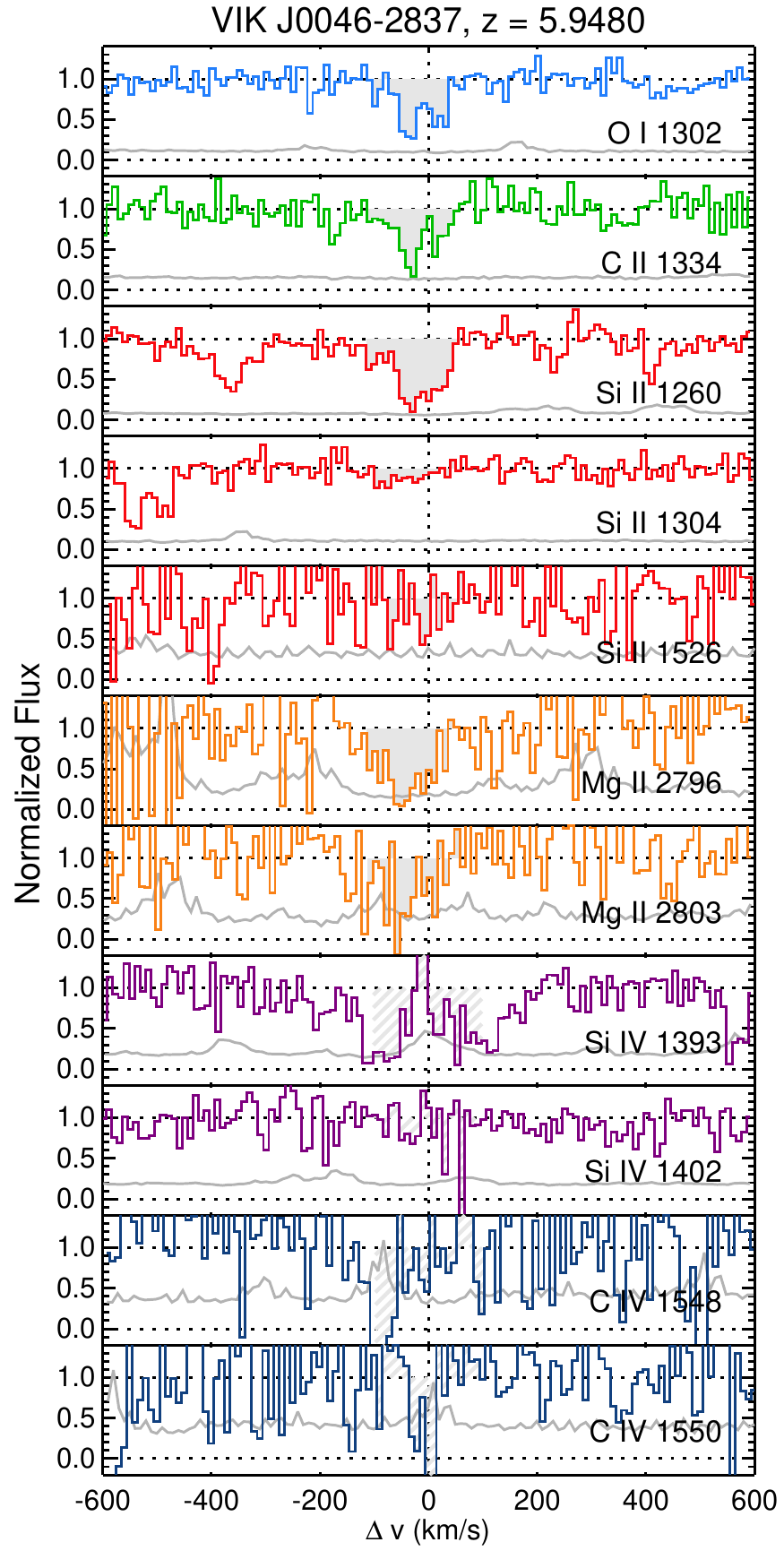}
   \caption{Stacked velocity plot for the $z=5.9480$ system towards VIK J0046-2837.  Lines and shading are as described in Figure~\ref{fig:z3p3844720_J1018+0548}.\label{fig:z5p9479820_VIKJ0046-2837}}
\end{figure}
 
\begin{figure}[!b]
   \centering
   \includegraphics[height=0.40\textheight]{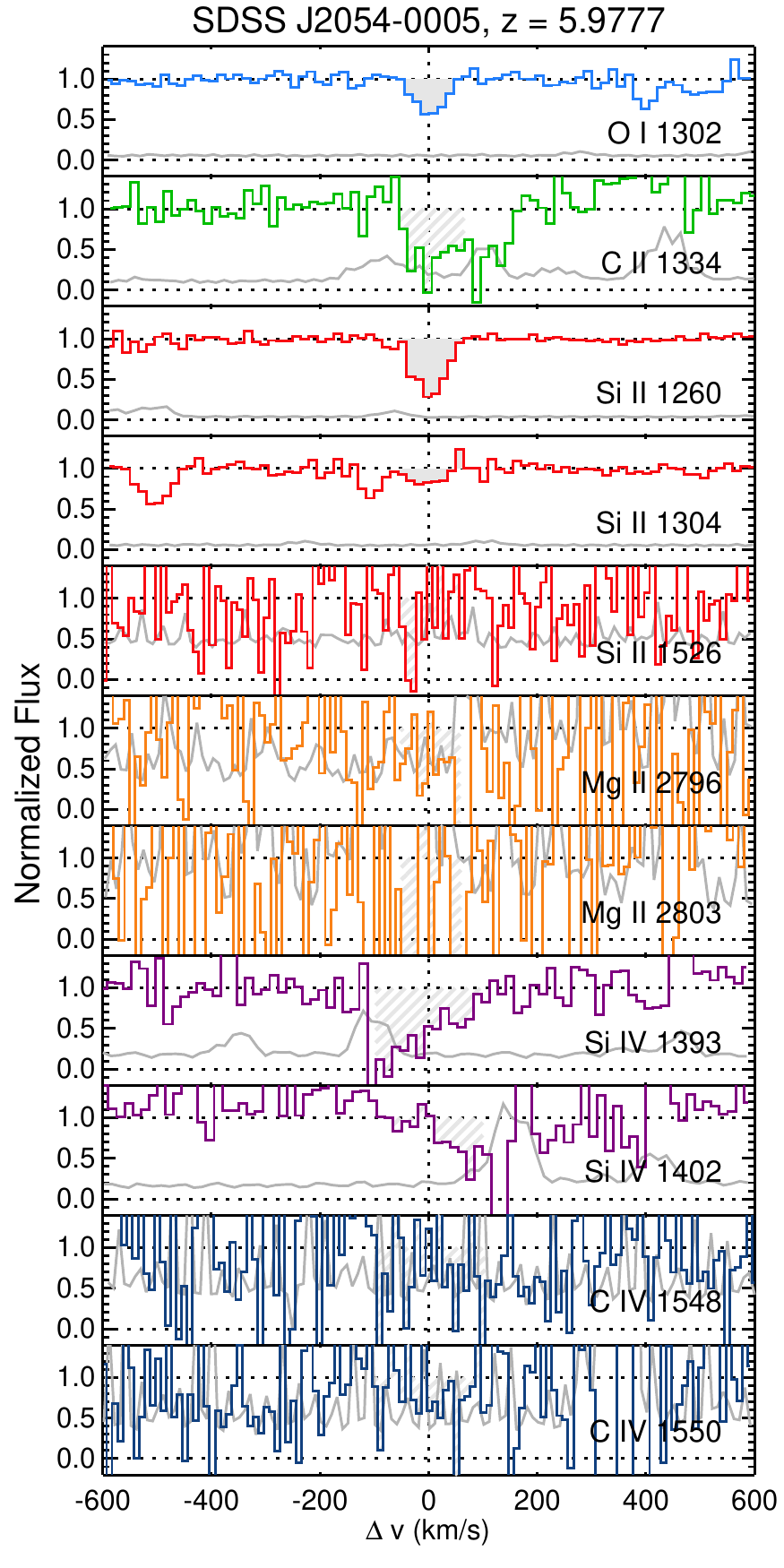}
   \caption{Stacked velocity plot for the $z=5.9777$ system towards SDSS J2054-0005.  Lines and shading are as described in Figure~\ref{fig:z3p3844720_J1018+0548}.\label{fig:z5p9776540_SDSSJ2054-0005}}
\end{figure}
 
\begin{figure}[!t]
   \centering
   \includegraphics[height=0.40\textheight]{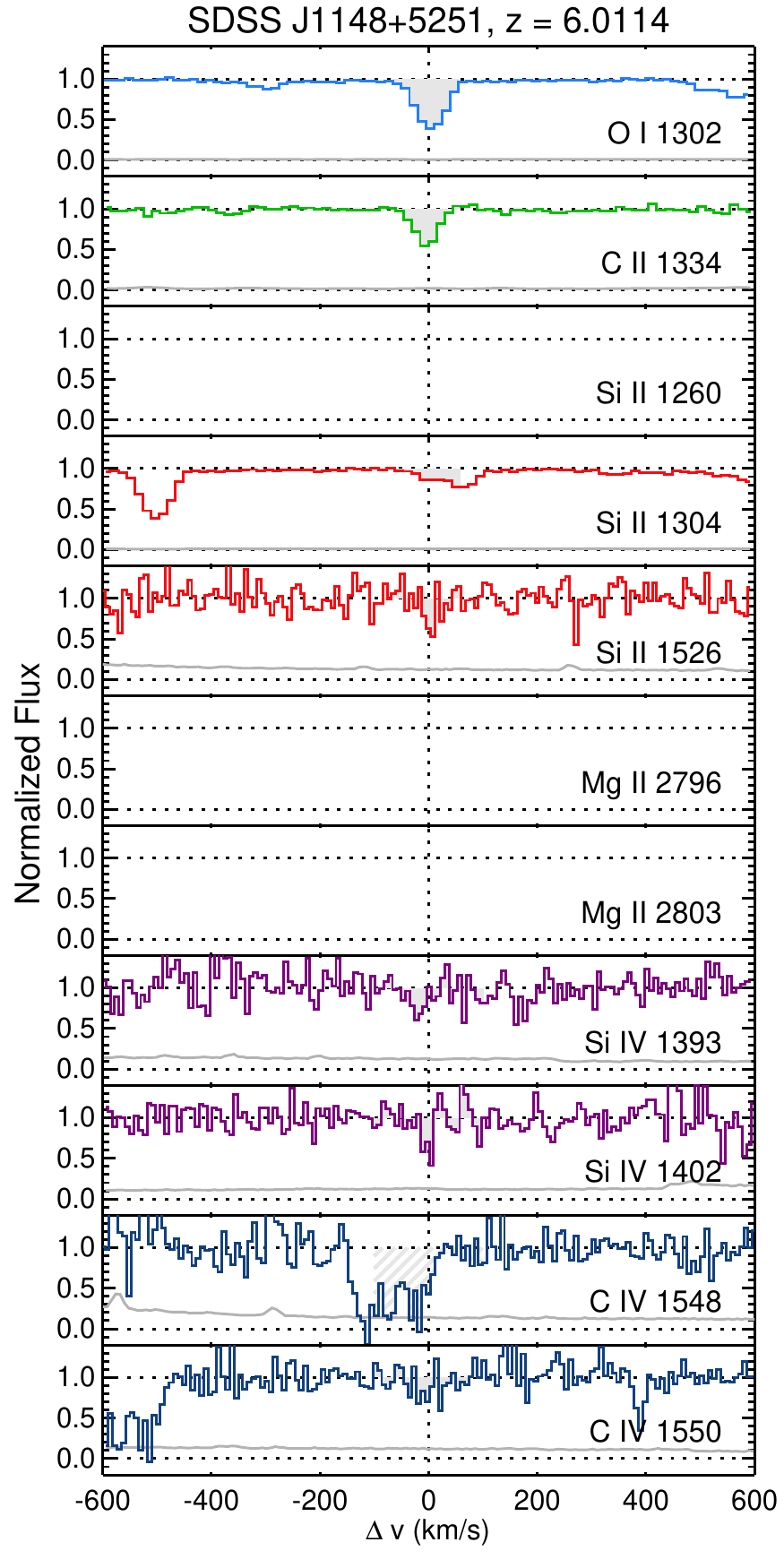}
   \caption{Stacked velocity plot for the $z=6.0114$ system towards SDSS J1148+5251.  Lines and shading are as described in Figure~\ref{fig:z3p3844720_J1018+0548}. See notes on this system in Appendix~\ref{app:details}.\label{fig:z6p0113980_SDSSJ1148+5251}}
\end{figure}
 
\begin{figure}[!b]
   \centering
   \includegraphics[height=0.40\textheight]{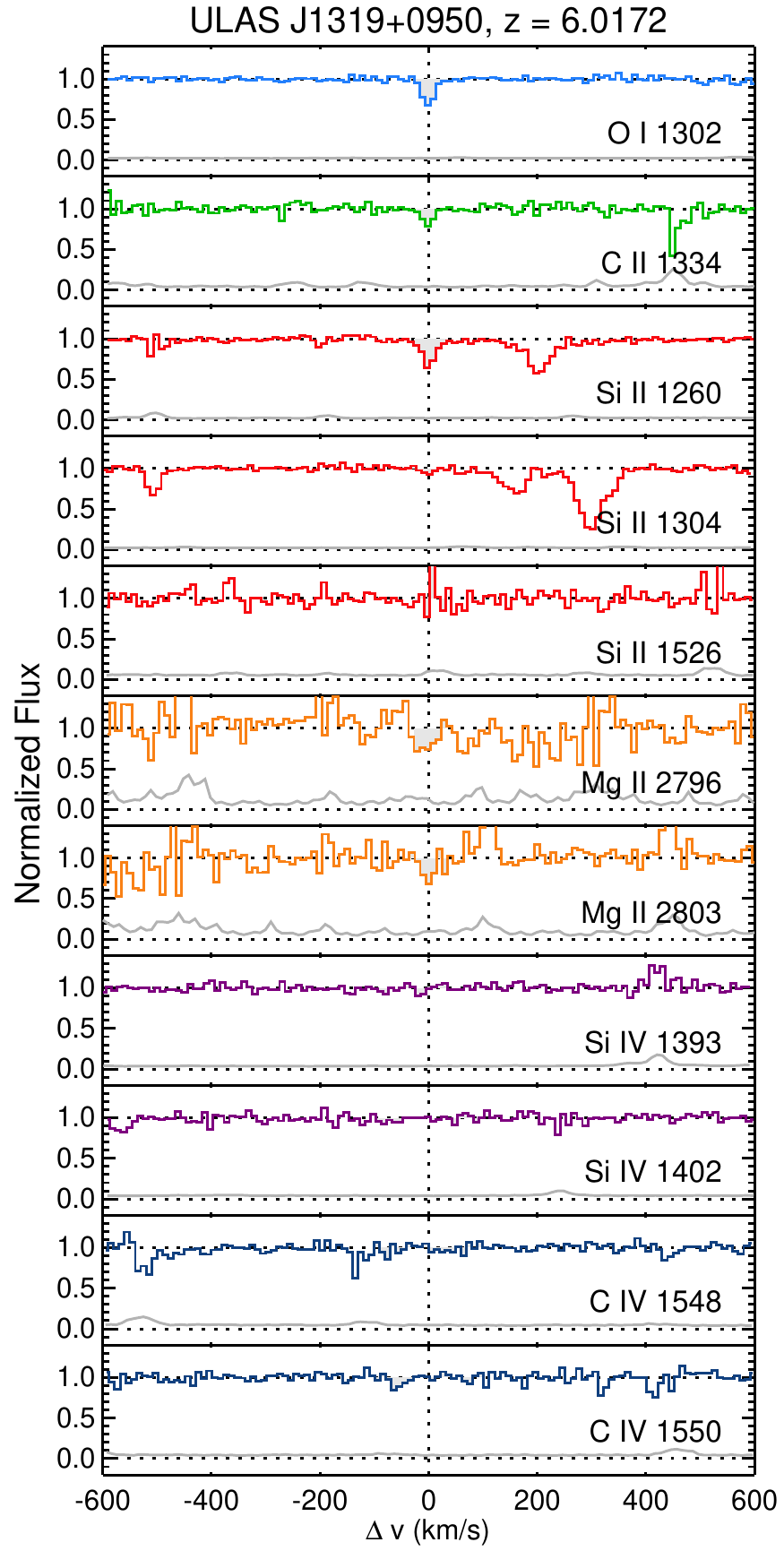}
   \caption{Stacked velocity plot for the $z=6.0172$ system towards ULAS J1319+0950.  Lines and shading are as described in Figure~\ref{fig:z3p3844720_J1018+0548}.\label{fig:z6p0172180_ULASJ1319+0950}}
\end{figure}
 
\clearpage
 
\begin{figure}[!t]
   \centering
   \includegraphics[height=0.40\textheight]{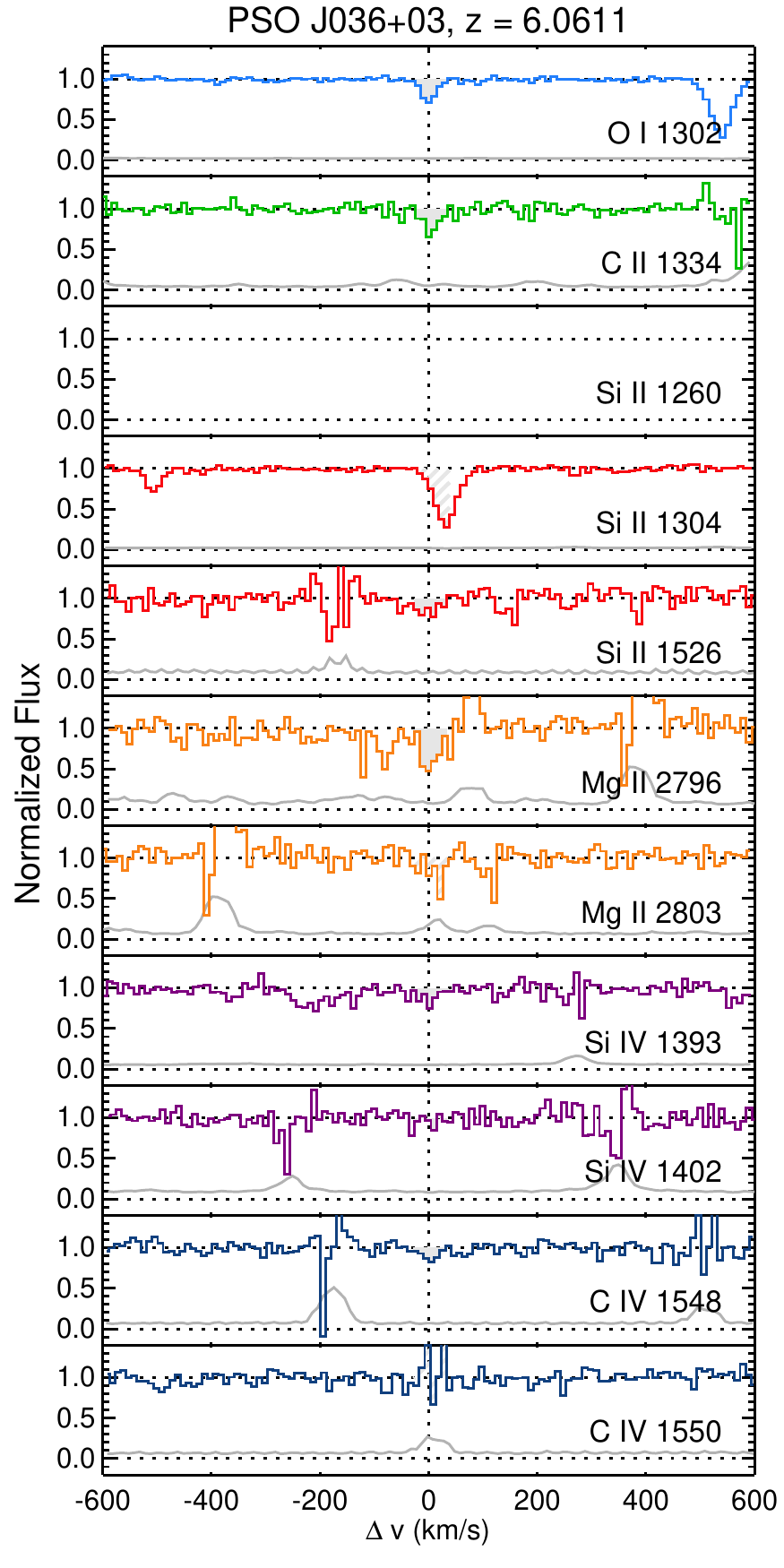}
   \caption{Stacked velocity plot for the $z=6.0611$ system towards PSO J036+03.  Lines and shading are as described in Figure~\ref{fig:z3p3844720_J1018+0548}.\label{fig:z6p0610940_PSOJ036+03}}
\end{figure}
 
\begin{figure}[!b]
   \centering
   \includegraphics[height=0.40\textheight]{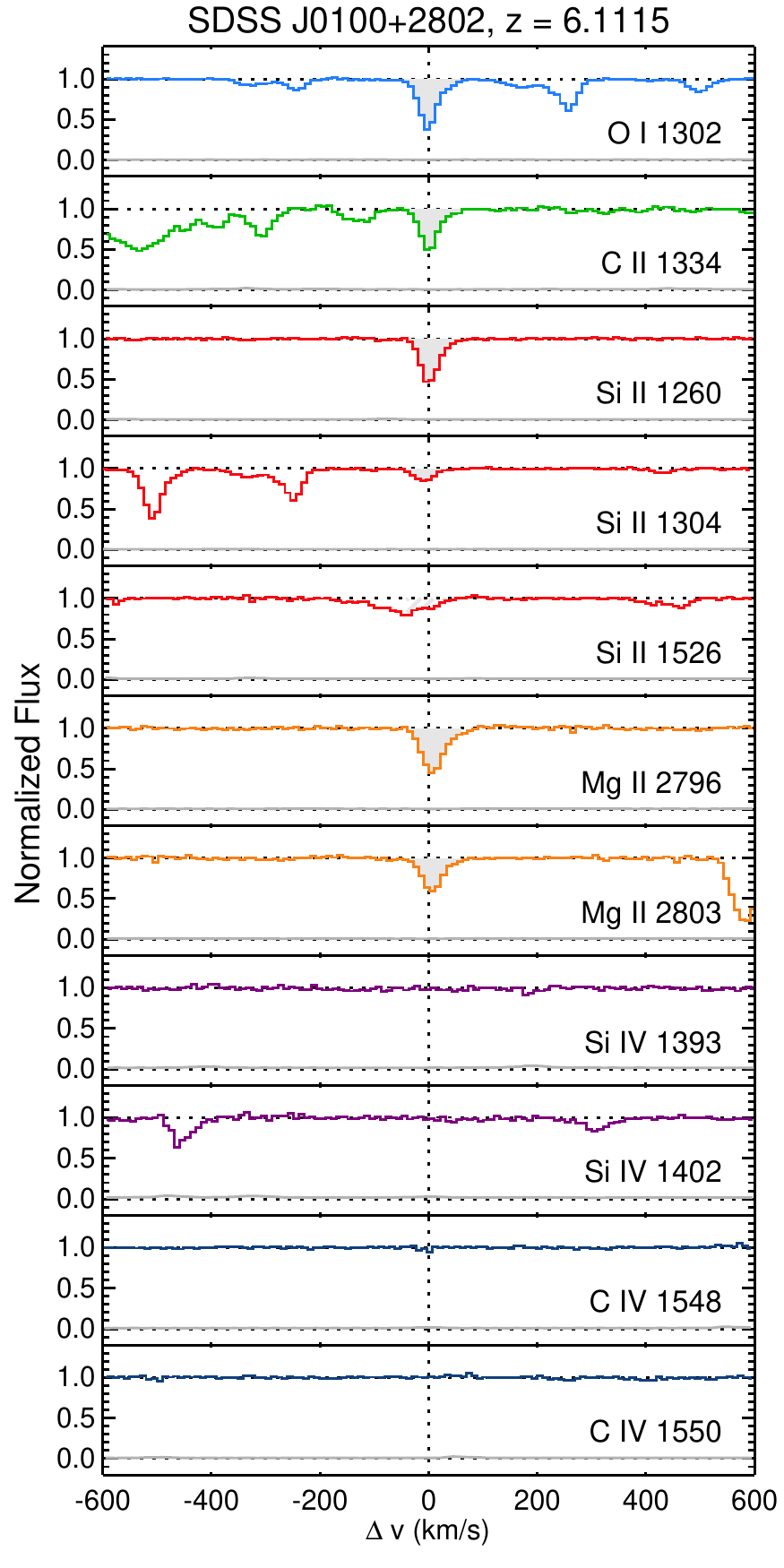}
   \caption{Stacked velocity plot for the $z=6.1115$ system towards SDSS J0100+2802.  Lines and shading are as described in Figure~\ref{fig:z3p3844720_J1018+0548}.\label{fig:z6p1115230_SDSSJ0100+2802}}
\end{figure}
 
\begin{figure}[!t]
   \centering
   \includegraphics[height=0.40\textheight]{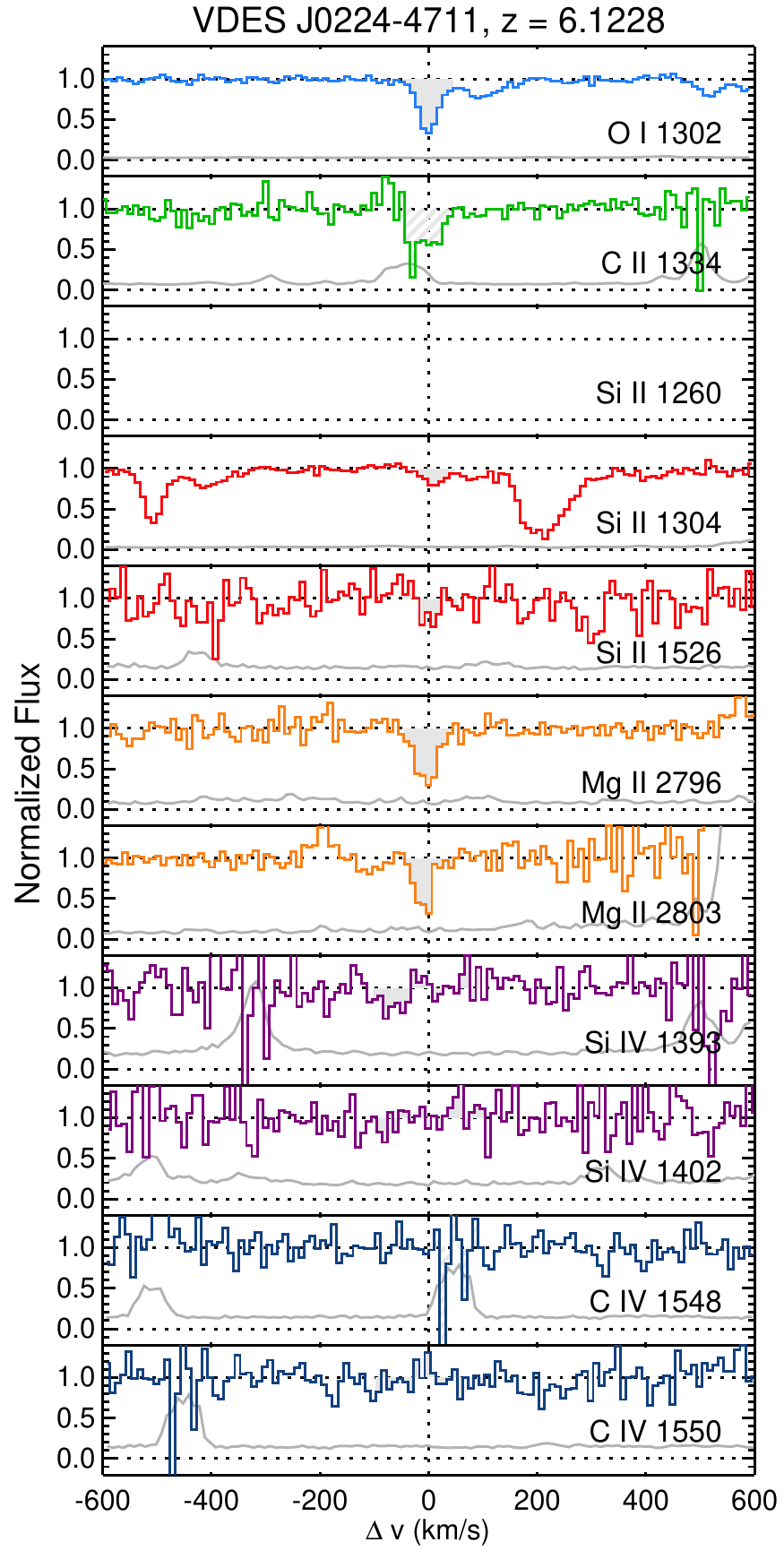}
   \caption{Stacked velocity plot for the $z=6.1228$ system towards VDES J0224-4711.  Lines and shading are as described in Figure~\ref{fig:z3p3844720_J1018+0548}.\label{fig:z6p1228350_VDESJ0224-4711}}
\end{figure}
 
\begin{figure}[!b]
   \centering
   \includegraphics[height=0.40\textheight]{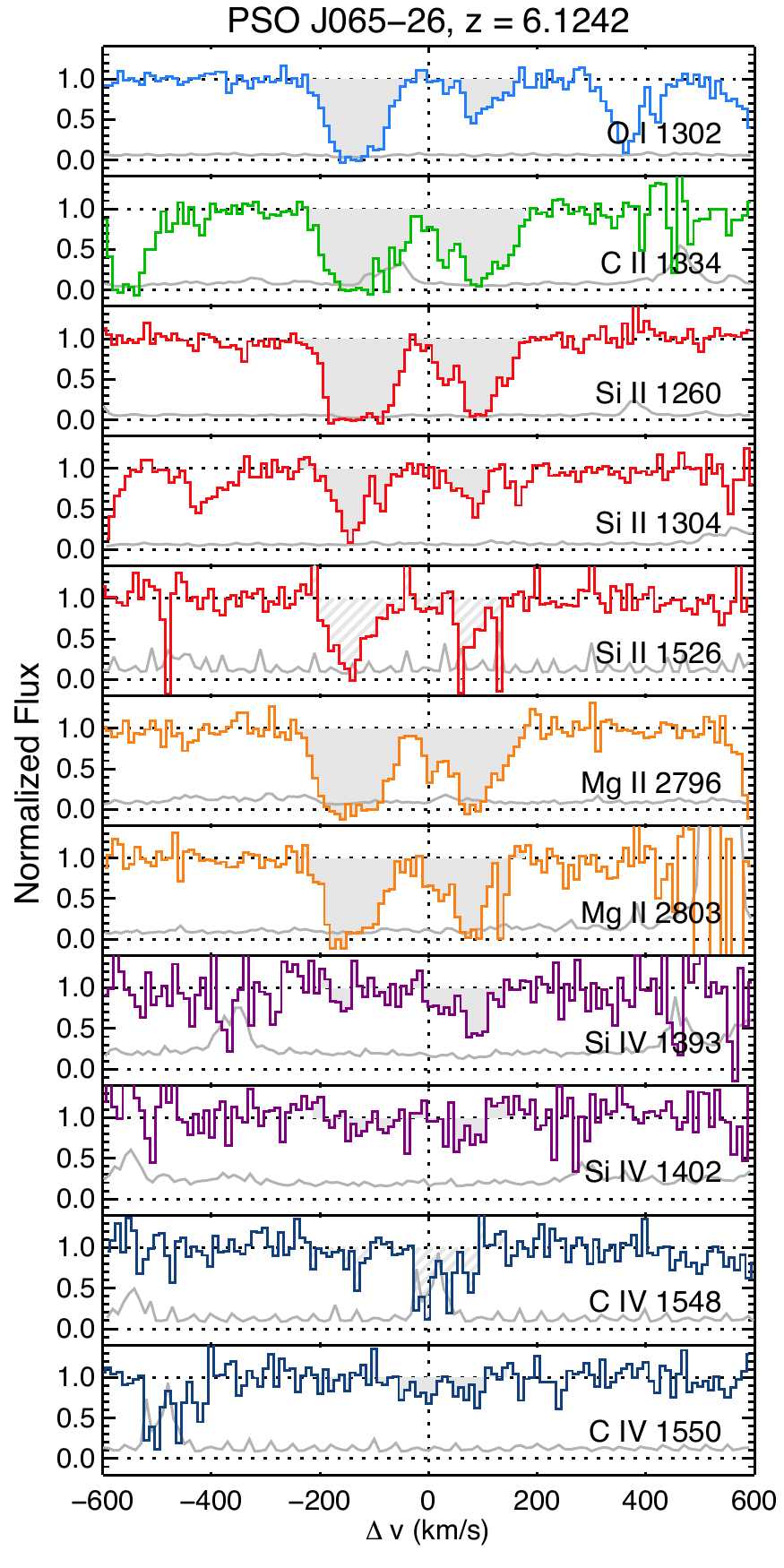}
   \caption{Stacked velocity plot for the $z=6.1242$ system towards PSO J065-26.  Lines and shading are as described in Figure~\ref{fig:z3p3844720_J1018+0548}. See notes on this system in Appendix~\ref{app:details}.\label{fig:z6p1242130_PSOJ065-26}}
\end{figure}
 
\clearpage
 
\begin{figure}[!t]
   \centering
   \includegraphics[height=0.40\textheight]{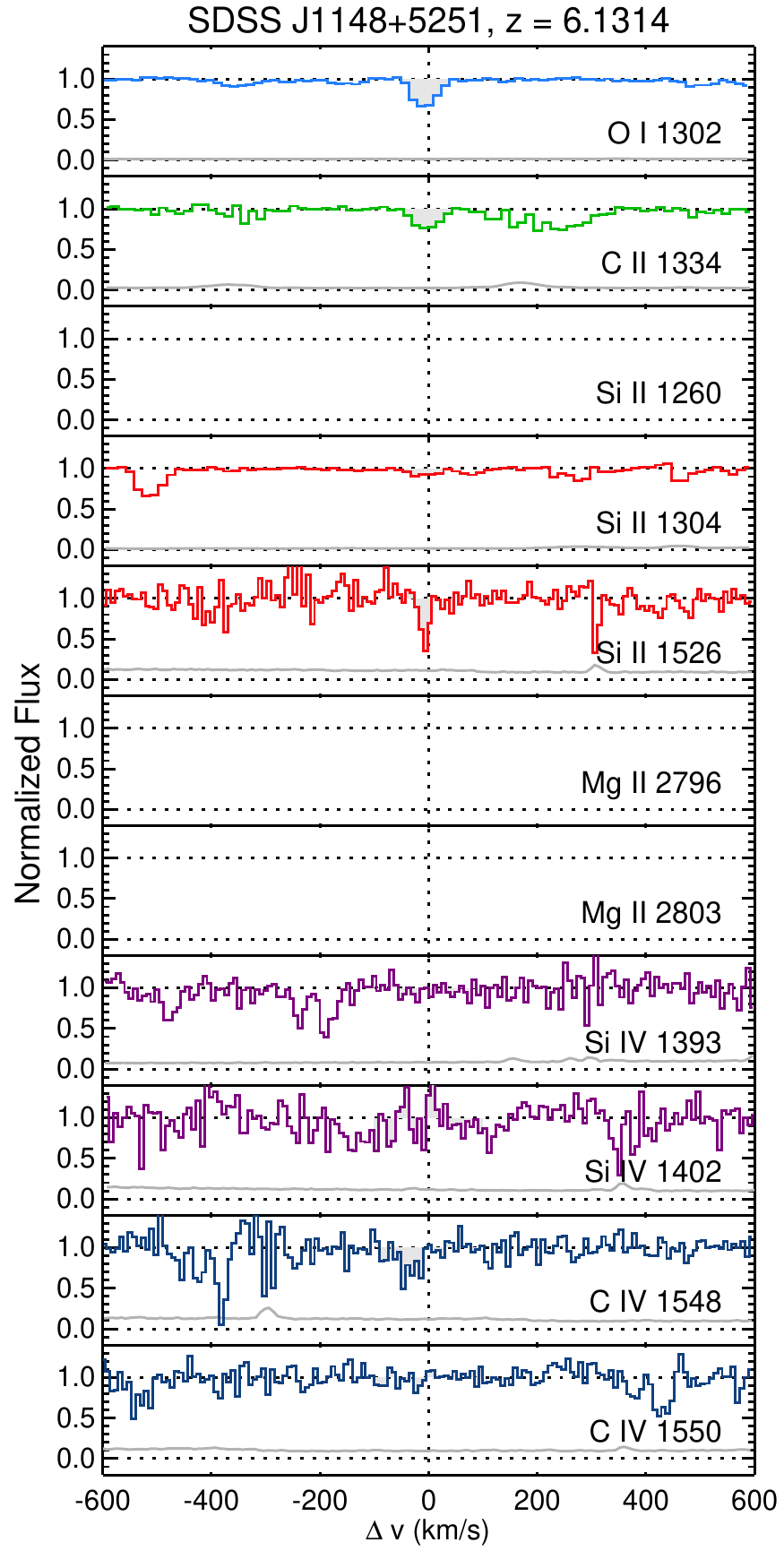}
   \caption{Stacked velocity plot for the $z=6.1314$ system towards SDSS J1148+5251.  Lines and shading are as described in Figure~\ref{fig:z3p3844720_J1018+0548}. See notes on this system in Appendix~\ref{app:details}.\label{fig:z6p1313540_SDSSJ1148+5251}}
\end{figure}
 
\begin{figure}[!b]
   \centering
   \includegraphics[height=0.40\textheight]{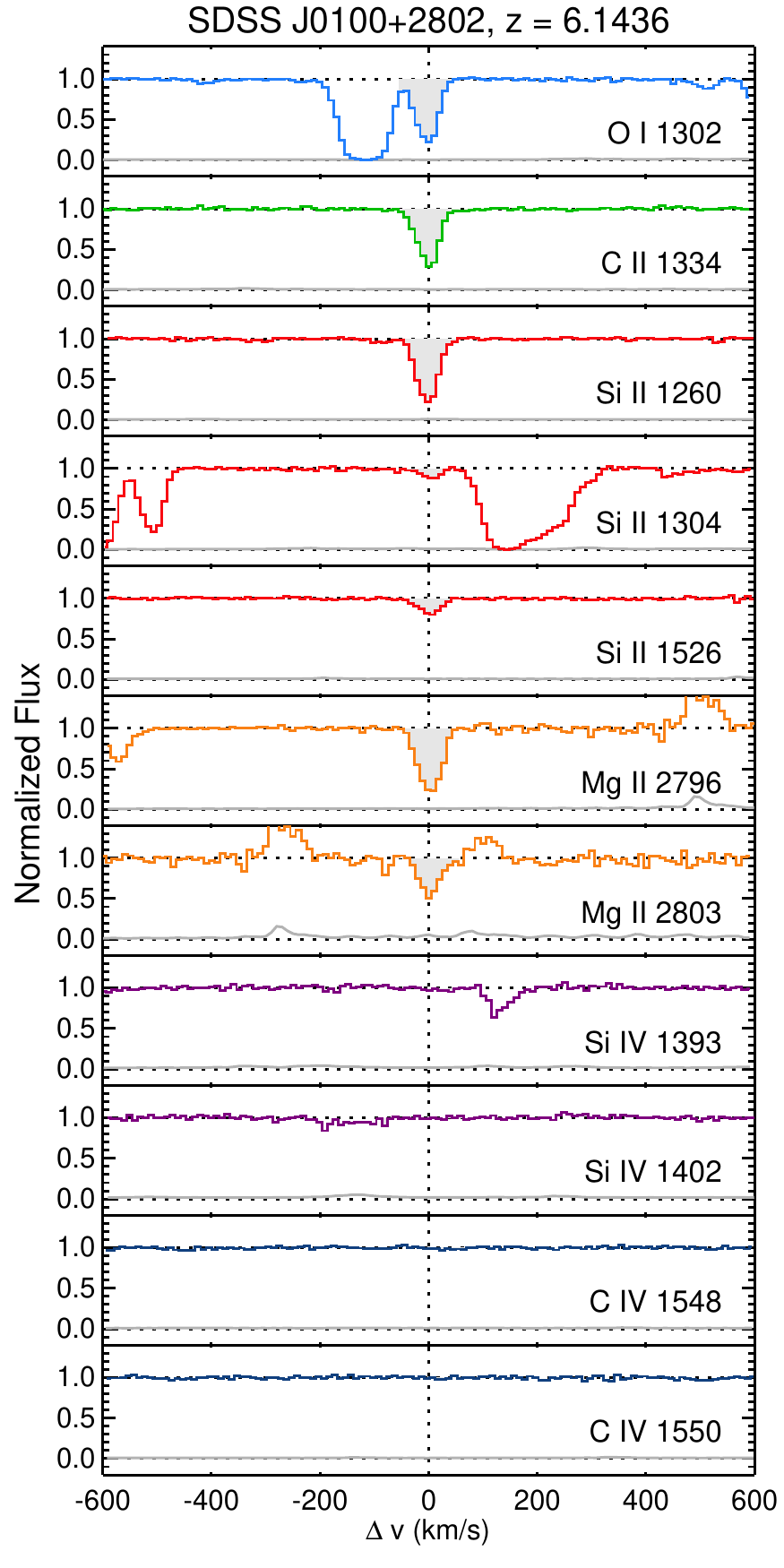}
   \caption{Stacked velocity plot for the $z=6.1436$ system towards SDSS J0100+2802.  Lines and shading are as described in Figure~\ref{fig:z3p3844720_J1018+0548}.\label{fig:z6p1435510_SDSSJ0100+2802}}
\end{figure}
 
\begin{figure}[!t]
   \centering
   \includegraphics[height=0.40\textheight]{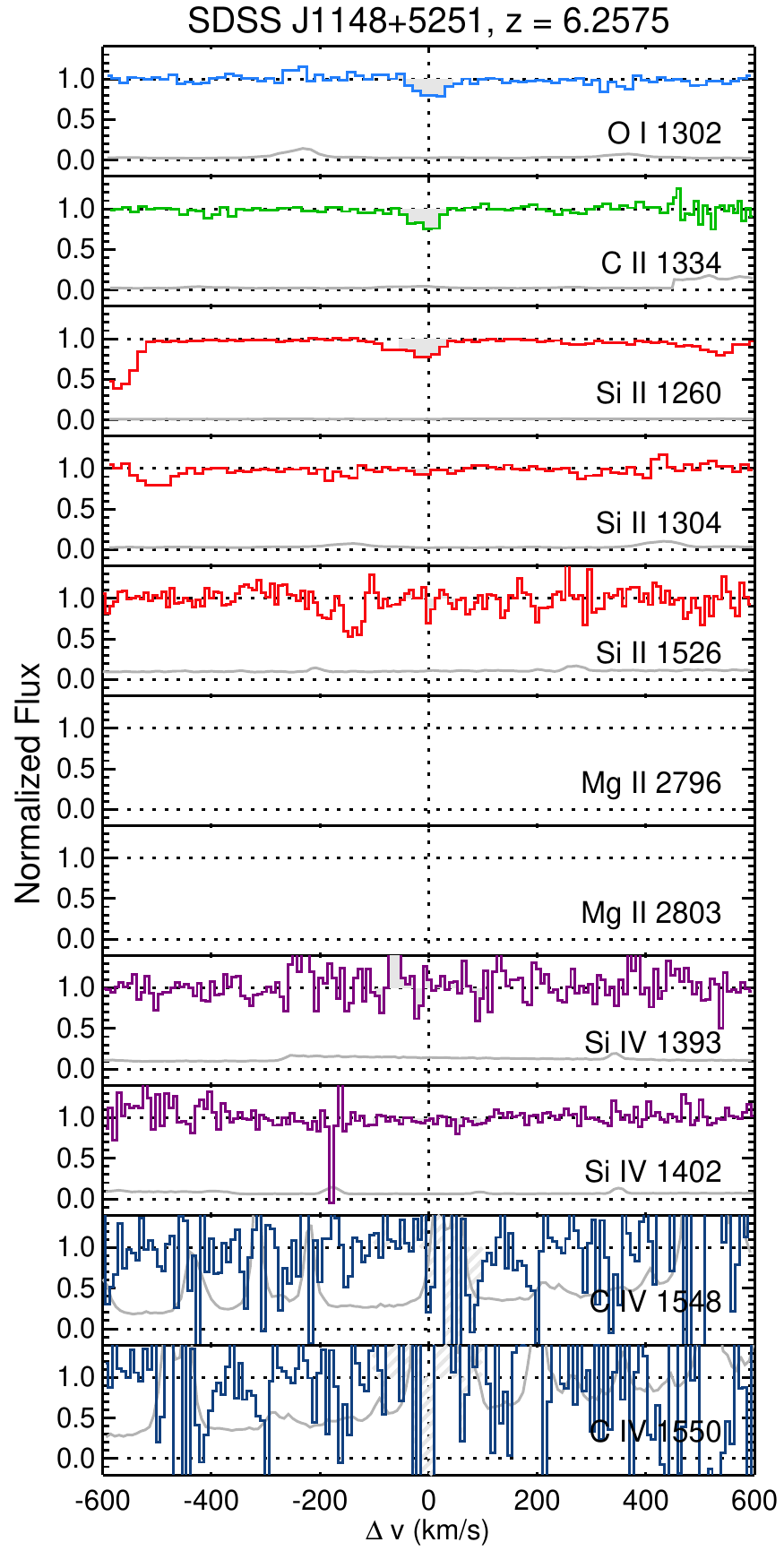}
   \caption{Stacked velocity plot for the $z=6.2575$ system towards SDSS J1148+5251.  Lines and shading are as described in Figure~\ref{fig:z3p3844720_J1018+0548}. See notes on this system in Appendix~\ref{app:details}.\label{fig:z6p2575350_SDSSJ1148+5251}}
\end{figure}
 
\begin{figure}[!b]
   \centering
   \includegraphics[height=0.40\textheight]{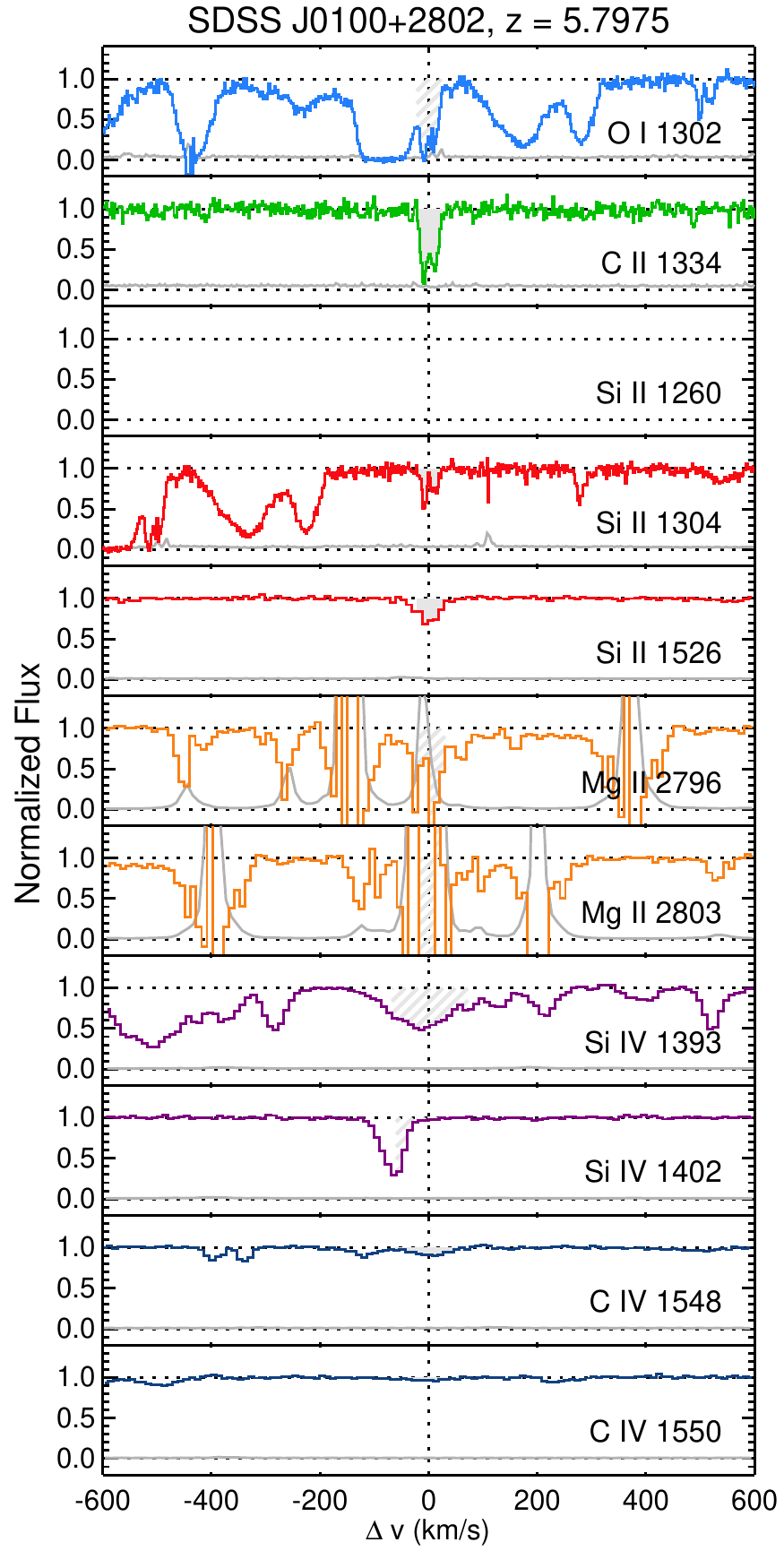}
   \caption{Stacked velocity plot for the $z=5.7975$ system towards SDSS J0100+2802.  Lines and shading are as described in Figure~\ref{fig:z3p3844720_J1018+0548}.  The data for \oi~\lam1302, \cii~\lam1334, and \siii~\lam1304 are from HIRES.  This system falls outside of the nominal survey range for this QSO and is not included in our statistical sample.\label{fig:z5p79750_SDSSJ0100+2802}}
\end{figure}
 
\clearpage